\documentclass[traditabstract, longauth]{aa}

\usepackage[stable]{footmisc}
\usepackage{graphicx}
\usepackage[varg]{txfonts}
\usepackage{hyperref}
\usepackage{natbib,twoopt}
\usepackage{lscape}
\usepackage{hyperref} 
\usepackage{subcaption}
\usepackage{nicefrac}
\bibpunct{(}{)}{;}{a}{}{,} 

\makeatother\bibpunct[, ]{(}{)}{;}{a}{}{,}

\newcommand{\farc}{\hbox{$.\!\!^{\prime\prime}$}} 
 
\newcommand{\erg}{$\rm{erg\,cm^{-2}\,s^{-1}}$} 
\newcommand{\kms}{$\rm{km\,s^{-1}}$}
\newcommand{\hb}{H$\beta$} 
\newcommand{\ha}{H$\alpha$}

\newcommand{\hii}{\mbox{H\,{\sc ii}}} 
\newcommand{\ebv}{$E_{B-V}\,$} 
\newcommand{\oh}{12+\log(\mathrm{O/H})}

\newcommand{\oii}{[\ion{O}{ii}]} 
\newcommand{\oiii}{[\ion{O}{iii}]}
\newcommand{\neiii}{[\ion{Ne}{iii}]}
\newcommand{\nii}{[\ion{N}{ii}]}

\newcommand{\Msun}{$M_\odot$}
\newcommand{\Msunyr}{$M_\odot\,\rm{yr}^{-1}$}
\newcommand\nodata{ ~$\cdots$~ }

\begin{document}
\title{GRB hosts through cosmic time\thanks{Based on observations at ESO, Program IDs: 084.A-0260, 084.A-0303, 
085.A-0009,  086.B-0954, 086.A-0533, 086.A-0874, 087.A-0055, 087.A-0451, 087.B-0737, 088.A-0051, 088.A-0644, 089.A-0067, 089.A-0120, 089.D-0256,  089.A-0868, 090.A-0088, 090.A-0760, 090.A-0825, 091.A-0342, 091.A-0703, 091.A-0877, 091.C-0934, 092.A-0076, 092.A-0124, 092.A-0231, 093.A-0069, 094.A-0593}}
\subtitle{VLT/X-Shooter emission-line spectroscopy\thanks{The reduced spectra are only available at the CDS via anonymous ftp to cdsarc.u-strasbg.fr (130.79.128.5) or via http://cdsarc.u-strasbg.fr/viz-bin/qcat?J/A+A/581/A125} of 96 $\gamma$-ray-burst-selected galaxies at $0.1 < z < 3.6$}
\titlerunning{VLT/X-Shooter spectroscopy of GRB hosts}

\author{T.~Kr\"{u}hler\inst{1,2}
\and D.~Malesani\inst{2}
\and J.~P.~U.~Fynbo\inst{2} 
\and O.~E.~Hartoog \inst{3}
\and J.~Hjorth \inst{2}
\and P.~Jakobsson \inst{4}
\and D.~A.~Perley \inst{5, 6}
\and A.~Rossi \inst{7, 8}
\and P.~Schady \inst{9} 
\and S.~Schulze \inst{10, 11}
\and N.~R.~Tanvir \inst{12}
\and S.~D.~Vergani \inst{13, 14}
\and K.~Wiersema \inst{12}
\and P.~M.~J.~Afonso \inst{15}
\and J.~Bolmer \inst{9}
\and Z.~Cano \inst{4}
\and S.~Covino \inst{14}
\and V.~D'Elia \inst{16, 17}
\and A.~de~Ugarte~Postigo \inst{18, 2}
\and R.~Filgas \inst{19}
\and M.~Friis \inst{4}
\and J.~F.~Graham \inst{9}
\and J.~Greiner \inst{9}
\and P.~Goldoni \inst{20}
\and A.~Gomboc \inst{21}
\and F.~Hammer \inst{13}
\and J.~Japelj \inst{21}
\and D.~A.~Kann \inst{8, 9}
\and L.~Kaper \inst{3}
\and S.~Klose \inst{8}
\and A.~J.~Levan \inst{22}
\and G.~Leloudas \inst{23, 2}
\and B.~Milvang-Jensen \inst{2}
\and A.~Nicuesa~Guelbenzu \inst{8}
\and E.~Palazzi \inst{7}
\and E.~Pian \inst{24, 25, 26}
\and S.~Piranomonte \inst{15}
\and R.~S\'{a}nchez-Ram\'{\i}rez \inst{27, 28, 18}
\and S.~Savaglio \inst{29, 25}
\and J.~Selsing \inst{2}
\and G.~Tagliaferri \inst{14}
\and P.~M.~Vreeswijk \inst{22, 30}
\and D.~J.~Watson \inst{2}
\and D.~Xu \inst{2}}

\institute{European Southern Observatory, Alonso de C\'{o}rdova 3107, Vitacura, Casilla 19001, Santiago 19, Chile 
\and Dark Cosmology Centre, Niels Bohr Institute, University of Copenhagen, Juliane Maries Vej 30, 2100 K\o benhavn \O, Denmark 
\and Astronomical Institute Anton Pannekoek, University of Amsterdam, Science Park 904, NL-1098 XH Amsterdam, the Netherlands
\and Centre for Astrophysics and Cosmology, Science Institute, University of Iceland, Dunhagi 5, IS-107 Reykjavik, Iceland
\and Department of Astronomy, California Institute of Technology, MC 249-17, 1200 East California Blvd, Pasadena CA 91125, USA
\and Hubble Fellow
\and INAF-IASF Bologna, Area della Ricerca CNR, via Gobetti 101, I-40129 Bologna, Italy 
\and Th\"uringer Landessternwarte Tautenburg, Sternwarte 5, 07778 Tautenburg, Germany
\and Max-Planck-Institut f\"{u}r extraterrestrische Physik, Giessenbachstra\ss e, 85748 Garching, Germany
\and Instituto de Astrof\'isica, Facultad de F\'isica, Pontificia Universidad Cat\'olica de Chile, Vicu\~{n}a Mackenna 4860, 7820436 Macul, Santiago, Chile
\and Millennium Institute of Astrophysics, Vicu\~{n}a Mackenna 4860, 7820436 Macul, Santiago, Chile
\and Department of Physics and Astronomy, University of Leicester, University Road, Leicester, LE1 7RH, UK
\and Laboratoire Galaxies Etoiles Physique et Instrumentation, Observatoire de Paris, 5 place Jules Janssen, 92195 Meudon, France
\and INAF, Osservatorio Astronomico di Brera, Via E. Bianchi 46, I-23807 Merate (LC), Italy
\and American River College, Physics and Astronomy Dpt., 4700 College Oak Drive, Sacramento, CA 95841, USA
\and INAF, Osservatorio Astronomico di Roma, Via Frascati 33, I-00040 Monteporzio Catone, Italy
\and ASI-Science Data Centre, Via Galileo Galilei, I-00044 Frascati, Italy
\and Instituto de Astrof\'{\i}sica de Andaluc\'{\i}a (IAA-CSIC), Glorieta de la Astronom\'{\i}a s/n, E-18008, Granada, Spain
\and Institute of Experimental and Applied Physics, Czech Technical University in Prague, Horska 3a/22, 128 00 Prague 2, Czech Republic
\and APC, Univ. Paris Diderot, CNRS/IN2P3, CEA/Irfu, Obs. de Paris, Sorbonne Paris Cite, France 
\and Faculty of Mathematics and Physics, University of Ljubljana, Jadranska ulica 19, 1000 Ljubljana, Slovenia
\and Department of Physics, University of Warwick, Coventry CV4 7AL, UK
\and Department of Particle Physics \& Astrophysics, Weizmann Institute of Science, Rehovot 76100, Israel.
\and INAF, Osservatorio Astronomico di Trieste, via G.B. Tiepolo 11, 34143 Trieste, Italy.
\and European Southern Observatory, Karl-Schwarzschild-Strasse 2, 85748 Garching bei M\"unchen, Germany.
\and Scuola Normale Superiore, Piazza dei Cavalieri 7, 56126 Pisa, Italy.
\and Unidad Asociada Grupo Ciencias Planetarias (UPV/EHU, IAA-CSIC), Departamento de F\'{\i}sica Aplicada I, E.T.S. Ingenier\'{\i}a, Universidad del Pa\'{\i}s Vasco (UPV/EHU), Alameda de Urquijo s/n, E-48013 Bilbao, Spain.
\and Ikerbasque, Basque Foundation for Science, Alameda de Urquijo 36-5, E-48008 Bilbao, Spain.
\and Physics Department, University of Calabria, via P. Bucci, I-87036 Arcavacata di Rende, Italy.
\and Benoziyo Fellow
}

\abstract
{We present data and initial results from VLT/X-Shooter emission-line spectroscopy of {96} galaxies selected by long $\gamma$-ray bursts (GRBs) at $0.1<z<3.6$, the largest sample of GRB host spectra available to date. Most of our GRBs were detected by \textit{Swift} and 76\% are at $0.5<z<2.5$ with a median $z_{\rm{med}}\sim1.6$. Based on Balmer and/or forbidden lines of oxygen, nitrogen, and neon, we measure systemic redshifts, star formation rates (SFR), visual attenuations ($A_V$), oxygen abundances ($\oh$), and emission-line widths ($\sigma$). We study GRB hosts up to $z\sim3.5$ and find a strong change in their typical physical properties with redshift. The median SFR of our GRB hosts increases from $SFR_{\rm{med}}\sim0.6$\,\Msunyr\,at $z\sim0.6$ up to $SFR_{\rm{med}}\sim15$\,\Msunyr\,at $z\sim2$. A higher ratio of \oiii/\oii\, at higher redshifts leads to an increasing distance of GRB-selected galaxies to the locus of local galaxies in the Baldwin-Phillips-Terlevich diagram. There is weak evidence for a redshift evolution in $A_V$ and $\sigma$, with the highest values seen at $z\sim1.5$ ($A_V$) or $z\sim2$ ($\sigma$). {Oxygen abundances of the galaxies are distributed between $\oh=7.9$ and $\oh=9.0$ with a median $\oh_{\rm{med}}\sim 8.5$. The fraction of GRB-selected galaxies with super-solar metallicities is $\sim 20\%$ at $z<1$ in the adopted metallicity scale.}  This is significantly less than the fraction of total star formation in similar galaxies, illustrating that GRBs are scarce in high metallicity environments. At $z\sim3$, sensitivity limits us to probing only the most luminous GRB hosts for which we derive metallicities of $Z\lesssim0.5\,Z_{\sun}$. Together with a high incidence of $Z\sim0.5\,Z_{\sun}$ galaxies at $z\sim1.5$, this indicates that a metallicity dependence at low redshift will not be dominant at $z\sim3$. Significant correlations exist between the hosts' physical properties. Oxygen abundance, for example, relates to $A_V$ ($\oh \propto 0.17\cdot A_V$), line width ($\oh \propto\sigma^{0.6}$), and SFR ($\oh\propto SFR^{0.2}$). In the last two cases, the normalization of the relations shift to lower metallicities at $z>2$ by $\sim0.4$~dex. 

These properties of GRB hosts and their evolution with redshift can be understood in a cosmological context of star-forming galaxies and a picture in which the hosts' properties at low redshift are influenced by the tendency of GRBs to {avoid the most metal-rich environments}.}

\maketitle

\section{Introduction}

The extreme luminosities of long $\gamma$-ray bursts (GRBs) and their afterglows \citep[e.g.,][for recent reviews]{2009ARA&A..47..567G, 2014arXiv1410.0679K} over the full electromagnetic spectrum make them powerful messengers from the early Universe, a potential that was identified very early on \citep[e.g.,][]{1998MNRAS.294L..13W, 2000ApJ...536....1L}. Coupled with the association of long GRBs with core-collapse supernovae of type Ic \citep[e.g.,][]{1998Natur.395..670G, 2003Natur.423..847H, 2006Natur.442.1011P}, and thus the formation of some massive stars, GRBs provide means of studying star-forming galaxies at the highest redshifts \citep[e.g.,][]{2007ApJ...671..272C, 2009ApJ...691..152C, 2012ApJ...754...46T}, faintest luminosities \citep[e.g.,][]{2001ApJ...546..672V, 2012ApJ...749L..38T}, as well as the global star formation rate (SFR) density up to the epoch of re-ionization \citep[e.g.,][]{2009ApJ...705L.104K, 2012ApJ...744...95R, 2012A&A...539A.113E}. In particular the information that is obtained through afterglow spectroscopy provides insights of unprecedented detail into the chemical composition of high-redshift galaxies \citep[e.g.,][]{2006A&A...451L..47F, 2009ApJ...691L..27P, 2013MNRAS.428.3590T, 2015ApJ...804...51C}.

Long $\gamma$-ray bursts, however, are rare objects \citep[e.g.,][]{2004ApJ...607L..17P, 2007ApJ...657L..73G}. The rate at which GRBs are accurately localized through satellites is around two per week. Number statistics of GRBs are thus orders of magnitude lower than those of other prominent probes of the early Universe, notably Lyman-break galaxies (LBGs, \citealp[e.g.,][]{1996ApJ...462L..17S}), and Damped-Lyman-$\alpha$ Absorbers (DLAs, \citealp[e.g.,][]{2005ARA&A..43..861W}) along the sightline to Quasi-Stellar Objects (QSOs). Quite similarly, the GRB explosion represents a very rare endpoint of stellar evolution, and could thus be subject to environmental factors enhancing or quenching the GRB rate with respect to star formation. For instance, metallicity often has been discussed as being likely to influence GRBs \citep[e.g.,][]{2006AcA....56..333S, 2008AJ....135.1136M}. Variations in the initial-mass function (IMF) or the fraction of massive stars in tight binaries could also affect the GRB rate \citep[][]{2014ApJ...789...23K}.

A physical understanding of GRBs and the conditions in which they form is thus necessary to put observations of high-redshift GRBs into a cosmological context. Arguably, one of the key pieces to the puzzle of understanding GRB progenitors and their role in probing distant star formation lies in the nature of their host galaxies \citep[e.g.,][for a review]{2014PASP..126....1L}. 

Seminal sample studies \citep[e.g.,][]{2003A&A...400..499L, 2004A&A...425..913C, 2004MNRAS.352.1073T, 2010AJ....139..694L} culminating in the compilation of GRB host properties of \citet{2009ApJ...691..182S} revealed a population of low-luminosity, low-mass, star-forming galaxies at $z\lesssim1$. The host's UV emission is highly concentrated around the GRB position and their morphologies appear more irregular than the hosts of core-collapse supernovae \citep{2006Natur.441..463F, 2010MNRAS.405...57S}. 

In line with this picture were Ly$\alpha$ properties \citep{2003A&A...406L..63F, 2012ApJ...756...25M} and early measurements of sub-solar GRB host metallicities \citep[e.g.,][]{2004ApJ...611..200P, 2005A&A...444..711G, 2005NewA...11..103S}. These data supported a strong dependence of GRB formation with metallicity \citep[e.g.,][]{2005A&A...443..581H, 2006A&A...460..199Y} as would be theoretically expected in the collapsar model \citep{1993ApJ...405..273W, 1999ApJ...524..262M}. 

More recently, however, this previously quite uniform picture of GRB hosts became somewhat more diverse: the offset of GRB-selected galaxies towards lower metallicities in the mass-metallicity relation \citep{2010AJ....140.1557L} could, for example, be partially explained with the dependence of the metallicity of star-forming galaxies on SFR \citep[e.g.,][]{2011MNRAS.414.1263M, 2011ApJ...735L...8K}. Additionally, several metal-rich GRB hosts were discovered \citep{2010ApJ...712L..26L, 2013A&A...556A..23E, 2015Patsubm}, and extensive observation in multi-band photometry revealed a population of red, high-mass, high-luminosity hosts, mostly associated with dust-extinguished afterglows \citep{2011A&A...534A.108K, 2012A&A...545A..77R, 2012ApJ...756..187H, 2013ApJ...778..128P}. Similar GRBs were underrepresented in previous studies, illustrating the need for dust-independent, X-ray-selected samples such as The Optically Unbiased GRB Host survey \citep[TOUGH,][]{2012ApJ...756..187H} or the \textit{Swift} GRB Host Galaxy Legacy Survey \citep[SHOALS,][]{2015arXiv150402482P}.

Because of these inherent uncertainties from sample selection and small number statistics, the question of how directly GRBs trace star formation, and how representatively they select star-forming galaxies remains a matter of debate \citep[e.g.,][]{2012ApJ...755...85M, 2014arXiv1409.7064V, 2015arXiv150304246S}: a metal-dependence on the GRB selection is shown by e.g., \citet{2013ApJ...774..119G} or \citet{2013ApJ...778..128P}, while \citet{2014A&A...565A.112H} advocate no strong evidence for GRBs to provide a biased census of star formation.
 
Elucidating the tight connection between star formation and GRBs also holds the promise of shedding more light on the nature of star-forming galaxies in general, especially at faint luminosities as well as their evolution at high redshift. Detailed spectroscopic observation of GRB hosts, however, remained challenging, in particular at $z>1$: prominent tracers of the physical conditions in the hot gas are redshifted into the NIR where spectroscopy traditionally is much less efficient. Spectroscopic data for $z > 1$ GRB hosts from emission lines is therefore available for only a handful of cases \citep[e.g.,][]{2012MNRAS.419.3039C, 2014arXiv1409.6315F, 2015Silviasubm}, and even at $z < 1$ the largest samples \citep{2009ApJ...691..182S, 2010AJ....139..694L, 2013ApJ...774..119G} contain only 10-15 events with detailed information on the host's gas properties.

Here, we present initial results from emission-line spectroscopy of 96 GRB-selected galaxies in the redshift range $0.1<z<3.6$, with the bulk of the targets (76\%) between $0.5<z<2.5$. In this work, we focus on strong recombination and nebular lines such as the Balmer series and the forbidden transitions of \oii\, or \oiii. The Ly$\alpha$ emission properties of the sample, a stacking analysis, and photometric follow-up observations will be discussed in detail elsewhere. 

Throughout the paper, we report line fluxes in units of $10^{-17}\,$\erg, magnitudes in the AB system, wavelengths and redshifts in vacuum and a heliocentric reference frame, and errors at 1$\sigma$ confidence levels. We assume concordance cosmology \citep[][$\Omega_{\rm{m}}=0.315$, $\Omega_\Lambda=0.685$, $H_0=67.3\,\rm{km}\,s^{-1}\, Mpc^{-1}$]{2014A&A...571A..16P}, and a solar oxygen abundance of $\oh=8.69$ \citep{2009ARA&A..47..481A}.

\section{Sample and observations}
\begin{longtab}
\begin{longtable}{cccccccccc}
\caption{X-shooter spectroscopic observations\label{tab:xsobs}}\\
\hline\hline
GRB host & Gal. $E_{B-V}$ & \multicolumn{3}{c}{Exposure time (s)} & \multicolumn{3}{c}{Slit width} & Obs. date & References \\ 
 & (mag) & UVB & VIS & NIR & UVB & VIS & NIR & & \\ 
\hline
\endfirsthead
\caption{X-shooter spectroscopic observations (continued)}\\
\hline\hline
GRB host & Gal. $E_{B-V}$ & \multicolumn{3}{c}{Exposure time (s)} & \multicolumn{3}{c}{Slit width} & Obs. date & References \\ 
 & (mag) & UVB & VIS & NIR & UVB & VIS & NIR & & \\ 
\hline
\endhead
\hline
\endfoot
GRB 050416A & 0.026 & $4\times 900$ & $4\times 900$ & $12\times 300$ & 1\farc{0} & 0\farc{9} & 0\farc{9} & 2011-Jan-19 & (1), (2) \\
GRB 050525A & 0.082 & $4\times 1050$ & $4\times 1050$ & $12\times 350$ & 1\farc{0} & 0\farc{9} & 0\farc{9}$JH$ & 2012-Sep-18 & (1) \\
 			& 		& $4\times 630$ & $4\times 664$ & $4\times 695$ & 1\farc{0} & 0\farc{9} & 0\farc{9}$JH$ & 2012-Sep-21 & \\
GRB 050714B & 0.048 & $12\times 940$ & $12\times 900$ & $36\times 325$ & 0\farc{8} & 0\farc{7} & 0\farc{6} & 2013-Mar-19 & (1) \\
GRB 050819  & 0.100 & $2\times 900$ & $2\times 900$ & $4\times 450$ & 1\farc{0} & 0\farc{9} & 0\farc{9} & 2010-Oct-29 & (1), (3) \\
GRB 050824  & 0.030 & $4\times 900$ & $4\times 900$ & $8\times 450$ & 1\farc{0} & 0\farc{9} & 0\farc{9}$JH$ & 2011-Sep-06 & (1) \\
GRB 050915A & 0.022 & $6\times 1200$ & $6\times 1200$ & $12\times 600$ & 1\farc{0} & 0\farc{9} & 0\farc{9}$JH$ & 2011-Sep-06 & (1), (3), (4), (5) \\
GRB 051001  & 0.013 & $2\times 1800$ & $2\times 1800$ & $4\times 900$ & 1\farc{0} & 0\farc{9} & 0\farc{9}$JH$ & 2010-Oct-30 & (1), (3) \\
GRB 051016B & 0.035 & $2\times 600$ & $2\times 600$ & $4\times 300$ & 1\farc{0} & 0\farc{9} & 0\farc{9} & 2009-Dec-23 & (1) \\
GRB 051022A & 0.052 & $2\times 1200$ & $2\times 1200$ & $4\times 600$ & 1\farc{6} & 1\farc{5} & 0\farc{9}$JH$ & 2012-Nov-13 & (6) \\
GRB 051117B & 0.048 & $1\times 720$ & $1\times 720$ & $2\times 360$ & 1\farc{0} & 0\farc{9} & 0\farc{9} & 2009-Dec-23 & (1) \\
GRB 060204B  & 0.015 & $4\times 900$ & $4\times 900$ & $9\times 300$ & 1\farc{0} & 0\farc{9} & 0\farc{9} & 2015-Mar-05 &  \\

GRB 060306  & 0.030 & $4\times 1800$ & $4\times 1800$ & $12\times 600$ & 1\farc{3} & 1\farc{2} & 1\farc{2} & 2011-Sep-01 & (1), (3), (5)\\
GRB 060604  & 0.037 & $4\times 1800$ & $4\times 1800$ & $8\times 900$ & 1\farc{0} & 0\farc{9} & 0\farc{9} & 2010-Oct-30 & (1), (3)\\
GRB 060707  & 0.019 & $4\times 860$ & $4\times 850$ & $12\times 300$ & 1\farc{3} & 1\farc{2} & 1\farc{2} & 2009-Oct-22 & (1), (7) \\
GRB 060719  & 0.007 & $4\times 1500$ & $4\times 1500$ & $12\times 500$ & 1\farc{0} & 0\farc{9} & 0\farc{9} & 2011-Oct-20 & (1), (3), (5)\\
GRB 060729  & 0.048 & $4\times 810$ & $4\times 844$ & $4\times 875$ & 1\farc{0} & 0\farc{9} & 0\farc{9}$JH$ & 2012-Sep-21 & (1)\\
 			& 		& $4\times 520$ & $4\times 554$ & $4\times 585$ & 1\farc{0} & 0\farc{9} & 0\farc{9}$JH$ & 2012-Sep-21 & \\
GRB 060805A & 0.021 & $10\times 940$ & $10\times 900$ & $30\times 320$ & 0\farc{8} & 0\farc{7} & 0\farc{6} & 2013-Mar-19 & (1), (3)\\
GRB 060814  & 0.034 & $2\times 1800$ & $2\times 1800$ & $6\times 600$ & 1\farc{0} & 0\farc{9} & 0\farc{9} & 2010-Feb-17 & (1), (3), (5) \\
GRB 060912A & 0.045 & $2\times 900$ & $2\times 934$ & $2\times 965$ & 1\farc{3} & 1\farc{2} & 1\farc{2} & 2012-Sep-21 & (1)\\
GRB 060923B & 0.129 & $2\times 1800$ & $2\times 1800$ & $6\times 600$ & 1\farc{0} & 0\farc{9} & 0\farc{9}$JH$ & 2013-Mar-20 & (1) \\
GRB 060926  & 0.137 & $4\times 1800$ & $4\times 1800$ & $12\times 600$ & 1\farc{0} & 0\farc{9} & 0\farc{9} & 2011-Apr-24 & (18) \\
GRB 061021  & 0.049 & $12\times 940$ & $12\times 900$ & $36\times 320$ & 1\farc{0} & 0\farc{9} & 0\farc{9}$JH$ & 2013-Mar-21 & (1)\\
GRB 061110A & 0.076 & $4\times 1800$ & $4\times 1800$ & $12\times 600$ & 1\farc{0} & 0\farc{9} & 0\farc{9}$JH$ & 2011-Oct-20 & (1)\\
GRB 061202  & 0.131 & $4\times 900$ & $4\times 900$ & $9\times 300$ & 1\farc{0} & 0\farc{9} & 0\farc{9} & 2014-Nov-29 &  \\
GRB 070103  & 0.058 & $2\times 900$ & $2\times 900$ & $4\times 450$ & 1\farc{0} & 0\farc{9} & 0\farc{9} & 2010-Oct-29 & (1), (3) \\
GRB 070110  & 0.013 & $4\times 1680$ & $4\times 1680$ & $8\times 840$ & 1\farc{0} & 0\farc{9} & 0\farc{9} & 2010-Oct-29 & (1), (7) \\
GRB 070129  & 0.120 & $4\times 900$ & $4\times 900$ & $12\times 300$ & 1\farc{0} & 0\farc{9} & 0\farc{9} & 2009-Dec-23 & (1)\\
GRB 070224  & 0.049 & $8\times 940$ & $8\times 900$ & $24\times 300$ & 1\farc{0} & 0\farc{9} & 0\farc{9} & 2013-Mar-20 & (1), (3) \\
 			& 		& $4\times 900$ & $4\times 920$ & $12\times 320$ & 1\farc{0} & 0\farc{9} & 0\farc{9} & 2013-Mar-20 & \\
GRB 070306  & 0.024 & $6\times 900$ & $6\times 900$ & $6\times 900$ & 1\farc{0} & 0\farc{9} & 0\farc{9} & 2010-Dec-09 & (1), (5), (8), (9) \\
GRB 070318  & 0.015 & $4\times 910$ & $4\times 885$ & $12\times 300$ & 1\farc{0} & 0\farc{9} & 0\farc{9} & 2010-Oct-06 & (1) \\
GRB 070328  & 0.031 & $7\times 844$ & $7\times 810$ & $8\times 875$ & 1\farc{0} & 0\farc{9} & 0\farc{9}$JH$ & 2012-Sep-21 & (1) \\
 			& 		& $1\times 589$ & $1\times 585$ & \nodata & 1\farc{0} & 0\farc{9} & 0\farc{9}$JH$ & 2012-Sep-21 & \\
GRB 070419B & 0.076 & $4\times 1800$ & $4\times 1800$ & $12\times 600$ & 1\farc{0} & 0\farc{9} & 0\farc{9} & 2011-Oct-20 & (1) \\
GRB 070521  & 0.023 & $4\times 900$ & $4\times 900$ & $12\times 300$ & 1\farc{0} & 0\farc{9} & 0\farc{9} & 2013-Apr-03 & (5), (10) \\
 			& 		& $4\times 900$ & $4\times 900$ & $12\times 300$ & 1\farc{0} & 0\farc{9} & 0\farc{9} & 2013-May-07 & \\
GRB 070802  & 0.023 & $4\times 1800$ & $4\times 1800$ & $8\times 900$ & 1\farc{0} & 0\farc{9} & 0\farc{9} & 2010-Oct-30 & (1), (5), (9), (11) \\
GRB 071021  & 0.057 & $4\times 1800$ & $4\times 1800$ & $12\times 600$ & 1\farc{0} & 0\farc{9} & 0\farc{9} & 2011-Sep-06 & (3), (5) \\
GRB 080207  & 0.020 & $4\times 1800$ & $4\times 1800$ & $12\times 600$ & 1\farc{0} & 0\farc{9} & 0\farc{9} & 2011-Apr-24 & (3), (5), (10) \\
GRB 080413B & 0.032 & $3\times 1800$ & $3\times 1800$ & $6\times 900$ & 1\farc{0} & 0\farc{9} & 0\farc{9} & 2010-Aug-15 & (23) \\
GRB 080602	& 0.025	& $4\times 900$	 & $4\times 900$ & $12\times 300$ &	1\farc{0} &	0\farc{9} &	0\farc{9} &	2013-Aug-18	& (10) \\%
		    &		& $4\times 900$	 & $4\times 900$ & $12\times 300$ &	1\farc{0} &	0\farc{9} &	0\farc{9} &	2013-Aug-19	& \\
GRB 080605  & 0.118 & $4\times885$ & $4\times910$ & $12\times300$ & 1\farc{0} & 0\farc{9} & 0\farc{9} & 2011-Apr-26 & (9), (12) \\
GRB 080804  & 0.014 & $4\times1350$ & $4\times1350$ & $12\times450$ & 1\farc{0} & 0\farc{9} & 0\farc{9} & 2012-Apr-19 & \\
GRB 080805  & 0.037 & $4\times885$ & $4\times910$ & $12\times300$ & 1\farc{0} & 0\farc{9} & 0\farc{9} & 2011-May-13 & (9) \\
GRB 081109  & 0.017 & $2\times1200$ & $2\times1200$ & $4\times600$ & 1\farc{6} & 1\farc{5} & 0\farc{9}$JH$ & 2012-Oct-13 & (5), (9)\\
 			& 		& $2\times1200$ & $2\times1200$ & $4\times600$ & 1\farc{6} & 1\farc{5} & 0\farc{9}$JH$ & 2012-Nov-14 & \\
GRB	081210	& 0.066	& $4\times900$	& $4\times900$	& $12\times300$	& 1\farc{0}	& 1\farc{0}	& 0\farc{9}	& 2014-Oct-19 &	\\
GRB 081221  & 0.019 & $4\times1800$ & $4\times1800$ & $12\times600$ & 1\farc{3} & 1\farc{2} & 1\farc{2} & 2011-Sep-01 & (5), (14) \\
GRB 090113  & 0.072 & $2\times1800$ & $2\times1800$ & $4\times900$ & 1\farc{0} & 0\farc{9} & 0\farc{9}$JH$ & 2011-Sep-06 & (3) \\
GRB 090201  & 0.059 & $2\times1200$ & $2\times1200$ & $8\times300$ & 1\farc{0} & 0\farc{9} & 0\farc{9} & 2012-Dec-25 & \\
 			& 		& $2\times1200$ & $2\times1200$ & $8\times300$ & 1\farc{0} & 0\farc{9} & 0\farc{9} & 2012-Dec-31 & \\
GRB 090323  & 0.021 & $3\times1560$ & $3\times1560$ & $9\times520$ & 1\farc{0} & 0\farc{9} & 0\farc{9} & 2011-Mar-11 & (15) \\
GRB 090407  & 0.059 & $4\times1800$ & $4\times1800$ & $8\times900$ & 0\farc{8} & 0\farc{7} & 0\farc{6} & 2010-Oct-30 & (3), (5) \\
GRB 090926B & 0.020 & $2\times1200$ & $2\times1200$ & $8\times300$ & 1\farc{0} & 0\farc{9} & 0\farc{9}$JH$ & 2012-Nov-14 & (9) \\
 			& 		& $2\times1200$ & $2\times1200$ & $8\times300$ & 1\farc{0} & 0\farc{9} & 0\farc{9}$JH$ & 2012-Dec-17 & \\
GRB 091018  & 0.025 & $4\times600$ & $4\times600$ & $4\times600$ & 1\farc{0} & 0\farc{9} & 0\farc{9} & 2009-Oct-19 & (16) \\
GRB 091127  & 0.033 & $4\times1500$ & $4\times1500$ & $8\times750$ & 1\farc{0} & 0\farc{9} & 0\farc{9} & 2009-Dec-02 & (17) \\
GRB 100316D & 0.101 & $4\times1200$ & $4\times1200$ & $12\times400$ & 0\farc{8} & 0\farc{7} & 0\farc{9} & 2011-Apr-03 & (18), (19) \\
GRB 100418A & 0.063 & $4\times1200$ & $4\times1200$ & $8\times600$ & 1\farc{0} & 0\farc{9} & 0\farc{9} & 2010-Apr-21 & (20), (21) \\
GRB 100424A & 0.029 & $8\times600$ & $8\times600$ & $8\times600$ & 1\farc{0} & 0\farc{9} & 0\farc{9} & 2013-Mar-11 & \\
GRB 100508A & 0.024 & $2\times1200$ & $2\times1200$ & $8\times300$ & 1\farc{0} & 0\farc{9} & 0\farc{9}$JH$ & 2012-Dec-25 & \\
GRB 100606A  & 0.023 & $4\times 900$ & $4\times 900$ & $9\times 300$ & 1\farc{0} & 0\farc{9} & 0\farc{9} & 2014-Nov-18 &  \\
		     &       & $4\times 900$ & $4\times 900$ & $9\times 300$ & 1\farc{0} & 0\farc{9} & 0\farc{9} & 2014-Nov-28 &  \\
GRB 100615A & 0.039 & $4\times1200$ & $4\times1200$ & $16\times300$ & 1\farc{0} & 0\farc{9} & 0\farc{9}$JH$ & 2013-Mar-05 & \\
GRB 100621A & 0.027 & $2\times1200$ & $2\times1200$ & $8\times300$ & 1\farc{0} & 0\farc{9} & 0\farc{9}$JH$ & 2012-Oct-16 & (9), (22) \\
GRB 100724A & 0.037 & $4\times1200$ & $4\times1200$ & $12\times400$ & 0\farc{8} & 0\farc{7} & 0\farc{9} & 2011-Apr-03 & \\
GRB 100728A & 0.151 & $2\times1200$ & $2\times1200$ & $8\times300$ & 1\farc{0} & 0\farc{9} & 0\farc{9} & 2012-Nov-15 & \\
GRB 100814A & 0.017 & $4\times1200$ & $4\times1200$ & $8\times600$ & 1\farc{0} & 0\farc{9} & 0\farc{9} & 2010-Aug-18 & \\
GRB 100816A & 0.074 & $4\times1200$ & $4\times1200$ & $8\times600$ & 1\farc{0} & 0\farc{9} & 0\farc{9} & 2010-Aug-17 & \\
GRB 110808A & 0.009 & $4\times600$ & $4\times600$ & $4\times600$ & 1\farc{0} & 0\farc{9} & 0\farc{9} & 2011-Aug-08 & \\
GRB 110818A & 0.030 & $4\times600$ & $4\times600$ & $4\times600$ & 1\farc{0} & 0\farc{9} & 0\farc{9} & 2013-May-08 & \\
 			& 		& $4\times600$ & $4\times600$ & $4\times600$ & 1\farc{0} & 0\farc{9} & 0\farc{9} & 2013-May-10 & \\
 			& 		& $8\times600$ & $8\times600$ & $8\times600$ & 1\farc{0} & 0\farc{9} & 0\farc{9} & 2013-Jul-04 & \\
GRB 110918A & 0.017 & $2\times1200$ & $2\times1200$ & $4\times600$ & 1\farc{6} & 1\farc{5} & 0\farc{9}$JH$ & 2012-Dec-17 & (23) \\
 			& 		& $2\times1200$ & $2\times1200$ & $4\times600$ & 1\farc{6} & 1\farc{5} & 0\farc{9}$JH$ & 2013-Jan-07 & \\
		 	& 		& $2\times1200$ & $2\times1200$ & $4\times600$ & 1\farc{6} & 1\farc{5} & 0\farc{9}$JH$ & 2013-Jan-16 & \\
GRB 111123A & 0.047 & $4\times600$ & $4\times600$ & $4\times600$ & 1\farc{0} & 0\farc{9} & 0\farc{9} & 2013-Apr-07 & \\
GRB 111129A & 0.036 & $4\times1200$ & $4\times1200$ & $16\times300$ & 1\farc{0} & 0\farc{9} & 0\farc{9} & 2011-Apr-29 & \\
GRB 111209A & 0.015 & $4\times2400$ & $4\times2400$ & $16\times600$ & 1\farc{0} & 0\farc{9} & 0\farc{9} & 2011-Dec-29 & (24) \\
GRB 111211A & 0.041 & $4\times600$ & $4\times600$ & $4\times600$ & 1\farc{0} & 0\farc{9} & 0\farc{9} & 2011-Dec-13 & \\
GRB 111228A & 0.028 & $4\times600$ & $4\times600$ & $4\times600$ & 1\farc{0} & 0\farc{9} & 0\farc{9} & 2011-Dec-29 & \\
GRB 120118B & 0.124 & $6\times600$ & $6\times600$ & $6\times600$ & 1\farc{0} & 1\farc{0} & 0\farc{9}$JH$ & 2013-Feb-13 & \\
GRB 120119A & 0.096 & $6\times600$ & $6\times600$ & $6\times600$ & 1\farc{0} & 1\farc{0} & 0\farc{6}$JH$ & 2013-Feb-26 &	\\
GRB 120211A & 0.026 & $4\times1200$ & $4\times1200$ & $16\times300$ & 1\farc{0} & 0\farc{9} & 0\farc{9} & 2013-Feb-17 & \\
GRB 120224A & 0.032 & $4\times600$ & $4\times600$ & $4\times600$ & 1\farc{0} & 0\farc{9} & 0\farc{9} & 2012-Feb-25 & \\
GRB 120422A & 0.030 & $4\times1200$ & $4\times1200$ & $16\times300$ & 1\farc{0} & 0\farc{9} & 0\farc{9} & 2012-Apr-22 & (25) \\
GRB 120624B & 0.048 & $1\times1200$ & $1\times1200$ & $4\times300$ & 1\farc{0} & 0\farc{9} & 0\farc{9} & 2012-Jul-12 & (26) \\
            &       & $3\times1200$ & $3\times1200$ & $12\times300$ & 1\farc{0} & 0\farc{9} & 0\farc{9} & 2012-Jul-13 & \\
GRB 120714B & 0.008 & $4\times1200$ & $4\times1200$ & $16\times300$ & 1\farc{0} & 0\farc{9} & 0\farc{9} & 2012-Jul-15 & \\
GRB 120722A & 0.042 & $2\times2400$ & $2\times2400$ & $8\times600$ & 1\farc{0} & 0\farc{9} & 0\farc{9} & 2012-Jul-23 & \\
GRB 120805A & 0.027 & $4\times900$ & $4\times900$ & $4\times900$ & 1\farc{0} & 1\farc{0} & 0\farc{9}$JH$ & 2012-Sep-15 & \\
GRB 120815A & 0.099 & $4\times600$ & $4\times600$ & $4\times600$ & 1\farc{0} & 0\farc{9} & 0\farc{9} & 2013-Apr-14 & (27) \\
 			&     	& $4\times600$ & $4\times600$ & $4\times600$ & 1\farc{0} & 0\farc{9} & 0\farc{9} & 2013-Apr-16 & \\
 			& 		& $4\times600$ & $4\times600$ & $4\times600$ & 1\farc{0} & 0\farc{9} & 0\farc{9} & 2013-Apr-21 & \\
GRB 121024A & 0.088 & $4\times600$ & $4\times600$ & $4\times600$ & 1\farc{0} & 0\farc{9} & 0\farc{9} & 2012-Oct-24 & (28) \\
GRB 121027A & 0.017 & $14\times600$ & $14\times600$ & $14\times600$ & 1\farc{0} & 0\farc{9} & 0\farc{9} & 2012-Nov-30 &  (23)\\
GRB 121201A  & 0.008 & $4\times 1200$ & $4\times 1200$ & $16\times 300$ & 1\farc{0} & 0\farc{9} & 0\farc{9}$JH$ & 2012-Dec-02 &  \\
GRB 121209A & 0.039 & $2\times600$ & $2\times600$ & $2\times600$ & 1\farc{0} & 0\farc{9} & 0\farc{9} & 2012-Nov-13 & \\
GRB 130131B & 0.029 & $8\times900$ & $8\times900$ & $24\times300$ & 1\farc{0} & 0\farc{9} & 0\farc{9}$JH$ & 2013-Mar-09 & \\
GRB 130427A & 0.017 & $2\times600$ & $2\times600$ & $2\times600$ & 1\farc{0} & 0\farc{9} & 0\farc{9}$JH$ & 2013-Apr-28 & (29) \\
GRB 130701A & 0.073 & $2\times600$ & $2\times600$ & $2\times600$ & 1\farc{0} & 0\farc{9} & 0\farc{9}$JH$ & 2013-Aug-01 & \\
GRB 130925A & 0.018 & $4\times1470$ & $4\times1500$ & $20\times300$ & 1\farc{0} & 0\farc{9} & 0\farc{9}$JH$ & 2013-Oct-25 & (30), (31) \\
GRB 131103A & 0.009 & $4\times600$ & $4\times600$ & $4\times600$ & 1\farc{0} & 0\farc{9} & 0\farc{9}$JH$ & 2013-Nov-05 & \\
GRB 131105A & 0.030 & $8\times600$ & $8\times600$ & $8\times600$ & 1\farc{0} & 0\farc{9} & 0\farc{9} & 2013-Nov-05 & \\
GRB 131231A & 0.022 & $4\times600$ & $4\times600$ & $4\times600$ & 1\farc{0} & 0\farc{9} & 0\farc{9}$JH$ & 2014-Feb-01 & \\
GRB 140114A & 0.014 & $6\times900$ & $6\times900$ & $18\times300$ & 1\farc{0} & 0\farc{9} & 0\farc{9}$JH$ & 2014-Mar-28 & \\
GRB 140213A & 0.131 & $2\times600$ & $2\times600$ & $2\times600$ & 1\farc{0} & 0\farc{9} & 0\farc{9}$JH$ & 2014-Feb-14 & \\
GRB 140301A & 0.026 & $12\times600$ & $12\times600$ & $12\times600$ & 1\farc{0} & 0\farc{9} & 0\farc{9}$JH$ & 2014-Mar-02 & \\
GRB 140430A & 0.117 & $2\times600$ & $2\times600$ & $2\times600$ & 1\farc{0} & 0\farc{9} & 0\farc{9} & 2014-Apr-30 & \\
GRB 140506A & 0.082 & $8\times600$ & $8\times600$ & $8\times600$ & 1\farc{0} & 0\farc{9} & 0\farc{9} & 2014-May-07 & (32) \\
 			& 		& $8\times600$ & $8\times600$ & $8\times600$ & 1\farc{0} & 0\farc{9} & 0\farc{9} & 2014-Jun-08 & \\
\end{longtable}
\tablebib{
(1)~\citet{2012ApJ...756..187H},
(2)~\citet{2007ApJ...661..982S},
(3)~\citet{2012ApJ...758...46K}, (4)~\citet{2007ApJ...662..294O}, (5)~\citet{2013ApJ...778..128P}, (6)~\citet{2010AJ....139..694L},
(7)~\citet{2012ApJ...756...25M}, (8)~\citet{2008ApJ...681..453J}, (9)~\citet{2011A&A...534A.108K}, (10)~\citet{2012A&A...545A..77R},
(11)~\citet{2009ApJ...697.1725E},
(12)~\citet{2011A&A...526A.113F}, (13)~\citet{2012A&A...546A...8K}, (14)~\citet{2012ApJ...749...68S}, (15)~\citet{2010A&A...516A..71M}, 
(16)~\citet{2012MNRAS.426....2W}, (17)~\citet{2011A&A...535A.127V},
(18)~\citet{2011MNRAS.411.2792S}, (19)~\citet{2011ApJ...739...23L}, (20)~\citet{2012PASJ...64..115N}, (21)~\citet{2011AN....332..297D},
(22)~\citet{2013A&A...560A..70G}, (23)~\citet{2013A&A...556A..23E}, (24)~\citet{2014ApJ...781...13L}, (25)~\citet{2014A&A...566A.102S},
(26)~\citet{2013A&A...557L..18D},
(27)~\citet{2013A&A...557A..18K}, (28)~\citet{2014arXiv1409.6315F}, (29)~\citet{2013ApJ...776...98X}, (30)~\citet{2014A&A...568A..75G},
(31)~\citet{2015Patsubm}, (32)~\citet{2014arXiv1409.4975F}
 }
\end{longtab}

\subsection{Targets}
\label{sec:tarob}

\begin{figure}
\includegraphics[angle=0, width=0.99\columnwidth]{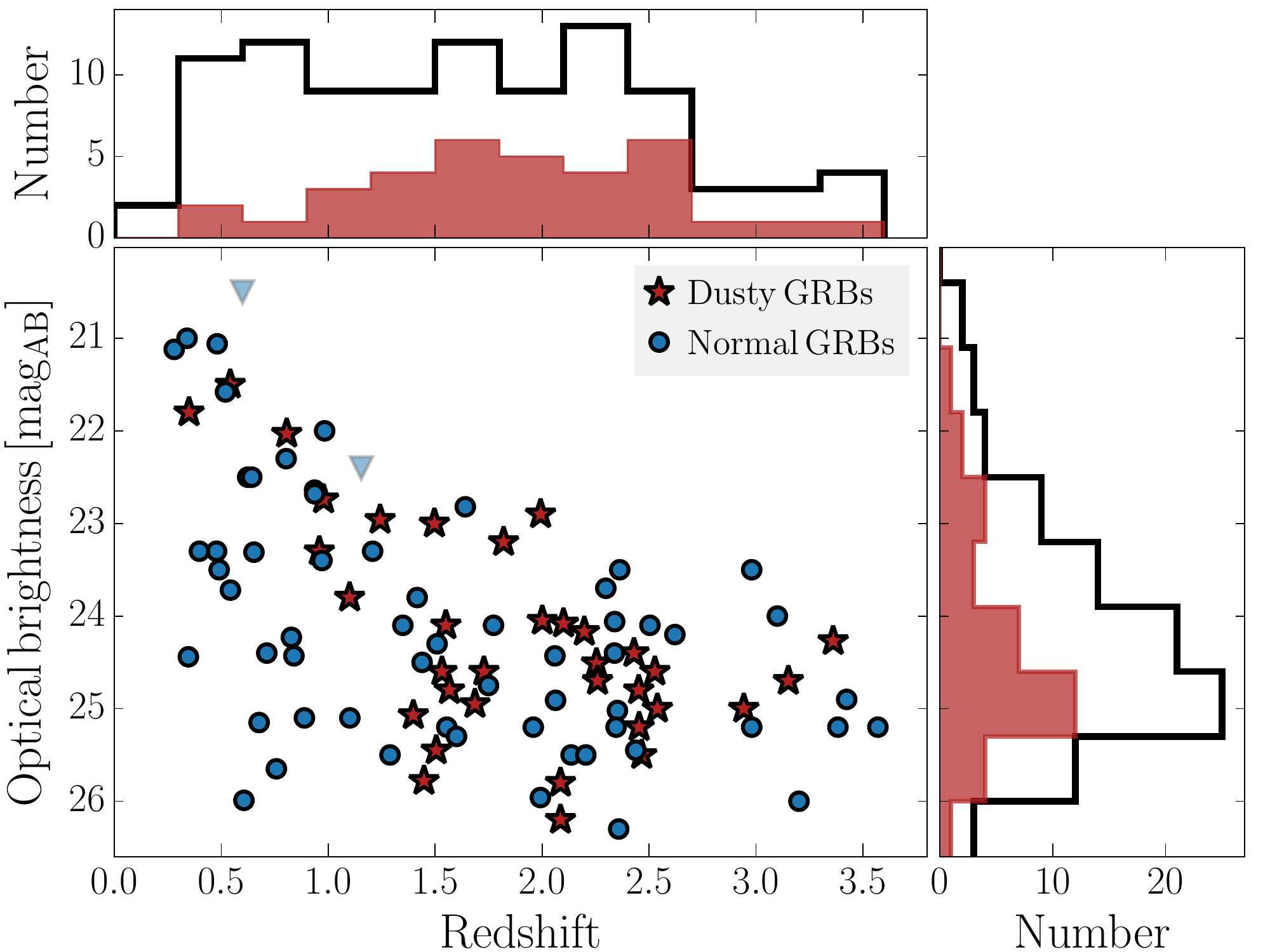}
\caption{Brightness and redshift distribution of the galaxies in our sample. Magnitudes are primarily measured in the $R/r$ filter (94\% of the sample). For the remaining hosts, $V$ or $I/i$-band magnitudes are shown. {Upper limits come from events for which the spectral continuum was dominated by a bright afterglow, they are denoted by downward triangles and are not included in the brightness histogram. Galaxies that hosted dusty GRBs (sightline $A^{\rm{GRB}}_V>1\,\rm{mag}$) are indicated by red stars and red-filled histograms.}}
\label{fig:brightness}
\end{figure}

 {To assemble our target list, we rely heavily on previous catalogs, in particular TOUGH, BAT6 \citep{2012ApJ...749...68S}, from GROND \citep{2011A&A...526A..30G, 2011A&A...534A.108K}, and SHOALS} for which we retrieve X-Shooter\footnote{X-Shooter is a medium-resolution, cross-dispersed echelle spectrograph mounted at ESO's Very Large Telescope and sensitive from the atmospheric UV cutoff ($\sim3000\,\AA$) up to the $K$-band.} \citep{2011A&A...536A.105V} optical/NIR spectroscopy from the ESO archive (PIs: de Ugarte Postigo, Flores, Fynbo, Kaper, Kr\"uhler, Malesani, Piranomonte, Rossi, Schady, Schulze). In addition, we also use GRB afterglow observations (PI: Fynbo) obtained with X-Shooter in target-of-opportunity mode. 

We immediately removed events from the analysis that belong to the class of short GRBs such as GRB~130603B \citep[e.g.,][]{2014A&A...563A..62D}. We include the host of GRB~100816A even though it cannot be uniquely attributed to either long or short category \citep{2010GCNR..300....1O}. To maintain a clean sample selected through genuinely long GRBs, this borderline case is not used in the discussion of galaxy properties in Sect.~\ref{sec:prop}. GRB~100316D at $z=0.0592$ \citep[e.g.,][]{2011MNRAS.411.2792S, 2012ApJ...753...67B} is the closest GRB in the sample. As the X-Shooter slit covers only a small fraction of the host light, we cannot infer galaxy-integrated properties. We therefore also omit the GRB~100316D host galaxy in Sect.~\ref{sec:prop}.

After an initial reduction and screening of the afterglow or host spectra, we excluded those events for which neither emission lines nor the stellar continuum of the host is detected: in the absence of a robust systemic redshift or in the presence of a very bright afterglow, no meaningful limits on emission-line fluxes can be established.  {We removed a total of nine host spectra from the sample in this step, primarily targeting very faint galaxies with $R\gtrsim25.5\,\rm{mag}$ (GRBs 050406, 060923A, 060923C, 060919, 090926A, 101219B, 110709B). Two excluded spectra are likely those of foreground objects (i.e., misidentifications, GRBs 070808, 081210). Out of these nine host spectra, two have accurate afterglow redshifts (GRBs~090926A at $z=2.11$, 101219B at $z=0.55$). GRB~060923A has a photo-$z$ of $z\sim2.6$ \citep{2008MNRAS.388.1743T, 2013ApJ...778..128P}.}

The total number of GRB host galaxy spectra presented here is 96, and nebular lines are detected for 91 of them. This sample advances in two crucial aspects from previous works: For the first time, it extends significantly beyond redshift of order unity owing to the sensitivity of X-Shooter up to $2.5\,\rm{\mu m}$. Our observations thus cover prominent tracers (\ha\,and \nii) of the ionized gas up to $z\sim2.5$, and \oii, \hb, and \oiii\, up to $z\sim3.5$. Secondly, number statistics are approximately a factor five larger than in previous spectroscopic samples. A log summarizing the X-Shooter observations is provided in Table~\ref{tab:xsobs}.

\subsection{The sample}

The presented X-Shooter observations were selected from the ESO archive, obtained by several groups and programs over the course of several years (2009-2015), and are hence diverse. They can however be divided into two main sub-categories: 

First, dedicated host spectroscopy, which aims on either measuring redshifts of GRBs for sample studies \citep[e.g.,][]{2012ApJ...758...46K, 2012ApJ...749...68S} or the physical properties of individual GRB hosts \citep[e.g.,][]{2011A&A...535A.127V, 2012A&A...546A...8K}. Second, we also discuss GRB afterglow observations in which host emission lines are detected above the afterglow continuum. 

Our GRB hosts are faint (21\,mag\,$\lesssim R\lesssim 26$\,mag with only a few exceptions) and their redshift distribution is broad extending up to $z\sim3.5$. It has a median of $z_{\rm{med}}=1.56$ and 35\% of the targets are at $0.5<z<1.5$, and another 41\% between $1.5<z<2.5$ (Fig.~\ref{fig:brightness}). {It is hence shifted significantly towards lower redshifts when compared against GRB (afterglow) redshift distributions \citep[][]{2011A&A...526A..30G, 2012ApJ...752...62J, 2012ApJ...749...68S, 2015arXiv150402482P}, an obvious result of a selection bias favoring $z<3.5$ galaxies for host spectroscopy. The brightness distribution of the galaxies peaks at around $R\sim25\,$mag with a median value of $R_{\rm med}\sim24.3\,\rm{mag}$. The basic properties of our galaxy sample, optical brightness and redshift are shown in Fig.~\ref{fig:brightness}.}

\subsection{Selection Effects}
\label{sec:seleffects}

\begin{figure}
\includegraphics[angle=0, width=0.99\columnwidth]{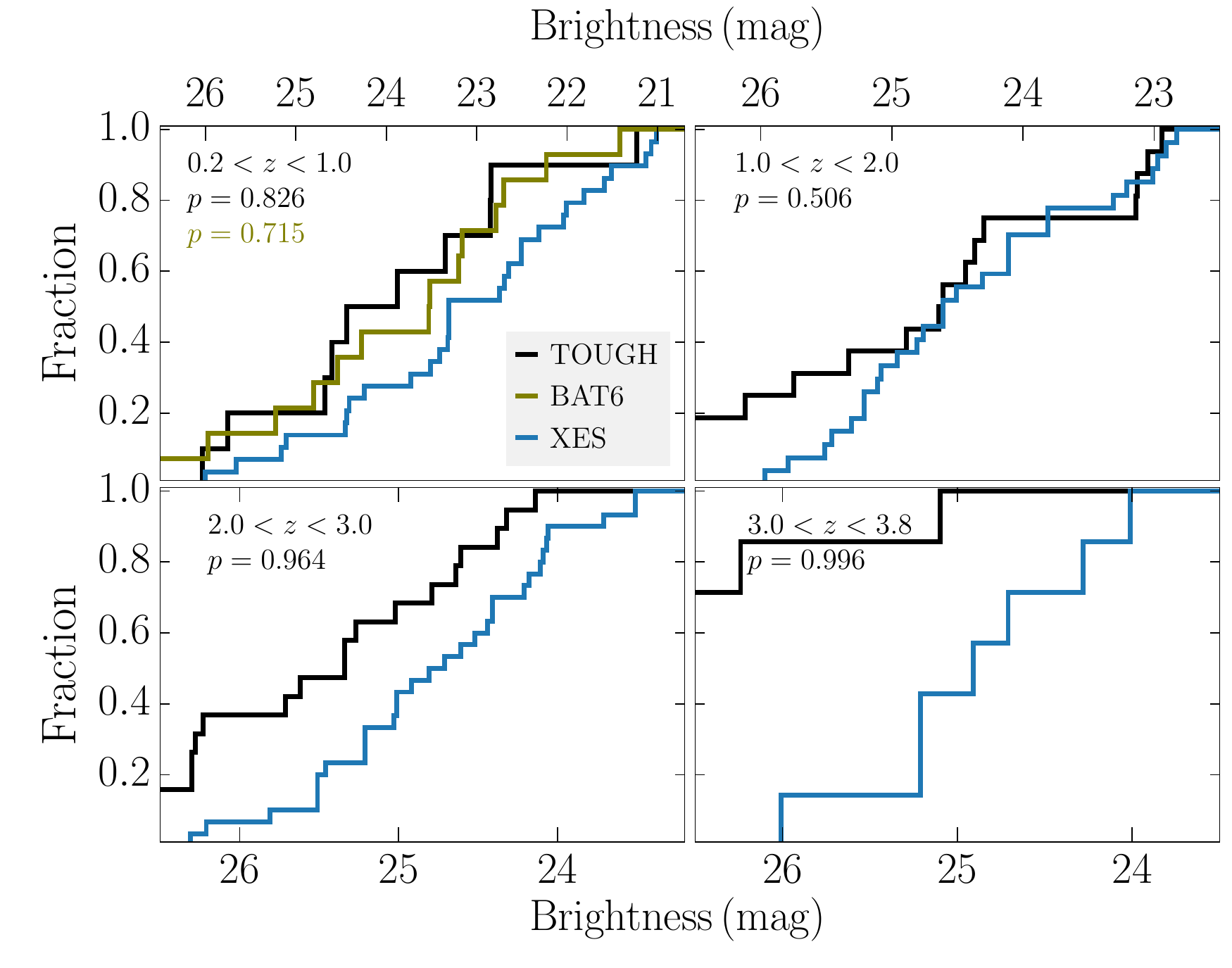}
\caption{Comparison of the cumulative brightness distributions of the host of our X-Shooter emission-line spectroscopy (XES) in blue against representative GRB host samples in black and olive lines \citep{2012ApJ...756..187H, 2014arXiv1409.7064V}. In the top left of each panel, we show the K.S.-test $p$-value to reject the null-hypothesis that both samples are drawn from a similar parent distribution.}
\label{fig:selection}
\end{figure}

 {Galaxy samples obtained from GRBs are different than those from flux-limited surveys. Since the galaxy is localized by an external high-energy trigger -- the GRB and its afterglow -- the selection is initially independent of the galaxies dust content, brightness or SFR. If we studied the hosts of all (or a representative subset of) GRBs, we could examine their properties irrespective of any flux limit.}

 {As we, however, heterogeneously draw our GRBs from a larger parent sample of \textit{Swift} GRBs, it is important to test how representative our GRB hosts really are. This implies understanding the selection effects on the GRBs itself as well as those introduced through the follow-up observations. Dedicated spectroscopic host observations favor, for example, optically brighter galaxies on average: a precise slit alignment requires a prior detection via broad-band imaging (Sect.~\ref{sec:bdist}).}

A further critical parameter is the amount of visual attenuation by dust along the GRB sight-line $A^{\rm{GRB}}_V$ \citep{2011A&A...534A.108K, 2012ApJ...756..187H, 2013ApJ...778..128P, 2015ApJ...801..102P} because it correlates with host properties -- galaxies hosting dusty\footnote{We prefer to work with \textit{dusty} GRBs instead of \textit{dark} GRBs. The definition of dark GRBs is not unique \citep{2004ApJ...617L..21J, 2009ApJ...699.1087V}, depends on the physics of the GRB shockwave and on the time and filter when the observations are performed \citep{2011A&A...526A..30G}. The rest-frame dust column density along the GRB sight line  $A^{\rm{GRB}}_V$ is thus a more physical quantity than the darkness of a afterglow.} GRBs have an order of magnitude higher stellar mass, luminosity, and SFR. 

The presence of GRBs with suppressed optical afterglows \citep[e.g.,][]{1998ApJ...493L..27G, 2006ApJ...647..471L} constituted a significant uncertainty in previous studies because they were underrepresented in redshift distributions (e.g., Fig.~9 in \citealp{2012ApJ...756..187H}). As X-Shooter is extremely efficient in providing galaxy redshifts \citep{2012ApJ...758...46K}, there is an above average fraction of dusty ($A^{\rm{GRB}}_V>1\,\rm{mag}$) GRBs in this work: it is $35\pm5\,\%$  here, but 20\,\% to 30\,\% in unbiased GRB samples \citep{2011A&A...526A..30G, 2012MNRAS.421.1265M, 2014arXiv1412.6530L} in a similar redshift range (Sect.~\ref{sec:dusty}).

\subsubsection{Absolute statistics}
\label{sec:absstat}

{First, we examine absolute number statistics at $z<1$, where the sample of GRBs with redshifts is highly complete\footnote{Because of the exquisite X-ray positions from \textit{Swift} and extensive follow-up efforts from the ground with e.g., TOUGH, GROND, or SHOALS, low-redshift GRBs are very unlikely to be missed if they have favorable observing conditions.}. In total, we have targeted $\sim$50\% (32) of all $z<1$ GRBs in the \textit{Swift}-era. The remaining 50\% were either too northern, too close to the Galactic plane, or not observable because of technical or telescope time limitations. From the initially targeted 32 galaxies at $z<1$, we detect emission lines for 30 (94 \%). The only two non-detections are GRB~110715A ($z=0.82$), which spectrum was dominated by a $R=18.5\,\rm{mag}$ afterglow \citep{2011GCN..12164...1P}, and GRB~101219B at $z=0.55$ \citep{2011ApJ...735L..24S}.}

\subsubsection{Brightness distributions}
\label{sec:bdist}

 {To test whether the pre-imaging introduces biases towards optically bright hosts, we compare the distribution of $R$-band magnitudes (Fig.~\ref{fig:brightness}) with those of two representative\footnote{\textit{Representative}, or \textit{unbiased}, samples fully exploit the advantages of the high-energy selection by GRBs. Since they are based on criteria that are not connected to the burst's physical environment, they do not depend on dust obscuration or galaxy luminosity. They allow us to study afterglows or hosts in a truly representative manner.} and redshift-complete samples \citep{2012ApJ...756..187H, 2014arXiv1409.7064V} which we use as a control group.}
 
 {Figure~\ref{fig:selection} shows the respective cumulative brightness distributions in four redshifts intervals. At low redshift ($z<2$), a K.S.-test returns no strong evidence that both samples draw from different parent distributions (K.S.-test $p$-values of 0.87 and 0.50 at $z<1$ and $1<z<2$, respectively) -- except possibly a $\sim$20\% fraction of very faint hosts in TOUGH at $1<z<2$ that is not present here. Above $z \sim 2$, however, a selection effect is clearly evident: our galaxies are substantially brighter than the reference sample (K.S.-test $p$-values of $p=0.96$ and $p=0.996$ at $2<z<3$ and $3<z<3.8$, respectively).}

\subsubsection{The Fraction of Dusty GRBs}
\label{sec:dusty}

\begin{figure}
\includegraphics[angle=0, width=0.99\columnwidth]{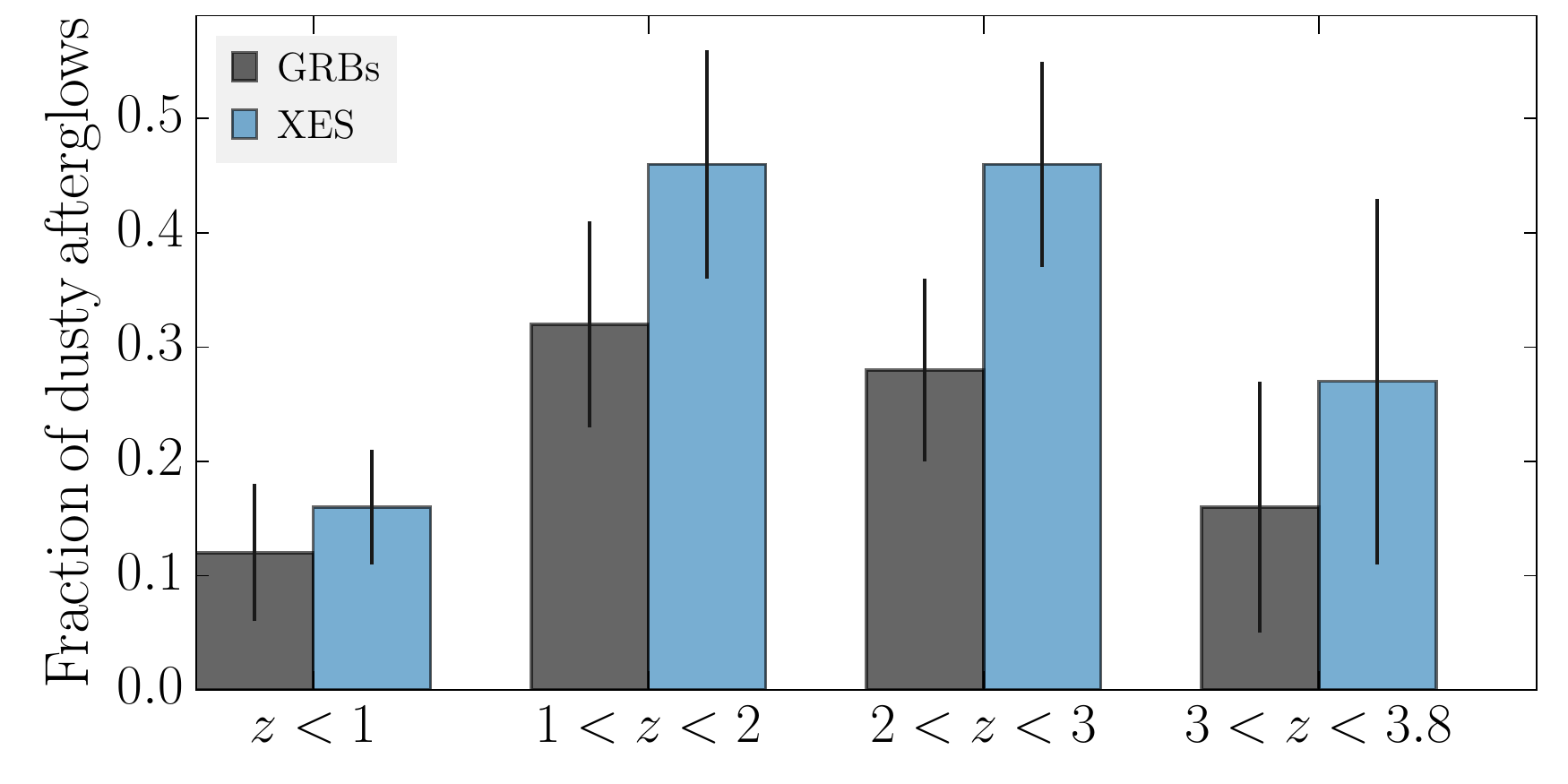}
\caption{Redshift dependent fractions with errors of dusty afterglows from our X-Shooter emission-line spectroscopy (XES) in blue against representative GRB samples in black \citep{2011A&A...526A..30G, 2013MNRAS.432.1231C}.}
\label{fig:dustyfrac}
\end{figure}

{Our $z<2$ galaxies therefore have a similar brightness distribution to unbiased GRB host samples. The over-proportionality of galaxies with dust-reddened afterglows in the sample, however, could nevertheless have an impact on their physical parameters.}

 {As representative comparison sample of the dust content towards GRBs ($A^{\rm{GRB}}_V$), we avail of the union of the afterglows from \citet{2011A&A...526A..30G} and \citet{2013MNRAS.432.1231C}. After excluding duplicates, and limiting the samples to the same redshifts as the host spectroscopy ($z<3.6$), 75 GRBs remain to constrain the fraction of dusty ($A^{\rm{GRB}}_V>1\,\rm{mag}$) GRBs.} {Figure~\ref{fig:dustyfrac} compares these values with those of our GRBs\footnote{Taken from the literature \citep{2009ApJ...693.1484C, 2011A&A...534A.108K, 2011A&A...532A.143Z, 2015arXiv150402482P} or derived from our own afterglow data. We assign GRBs to dusty afterglows in cases where we can constrain $A^{\rm{GRB}}_V>1\,\rm{mag}$ or when the optical/NIR afterglow is significantly underluminous with respect to the X-ray data. For example, GRB~120119A is dusty, although it had a bright optical counterpart \citep{2014MNRAS.440.1810M}, but GRB~100606A is normal (not-dusty), even though it did not \citep{2010GCN..10835...1N}.}} 

 {The ratio of dusty-to-normal afterglows is somewhat higher at $z\sim2$ than at $z<1$ (Figure~\ref{fig:dustyfrac}). In addition, our sample has a slightly higher fraction of dusty afterglows at all redshifts than uniform GRB control samples. The data from galaxies selected through dusty GRBs are highlighted by stars in the following plots to illustrate their influence on correlations and GRB host evolution.}

\subsection{A Stochastic Analysis}
\label{sec:statana}

 {To correct for the over-proportionality of dusty GRBs, we statistically create samples that have the right fraction of dusty-to-normal afterglows as follows: We use the initial 96 hosts and randomly draw $10^6$ sub-samples while using the previously established dusty GRB fraction as a strict prior. Specifically, this means applying a redshift-dependent probability to prevent a galaxy which hosted a dusty GRB to enter in sub-samples.}
 
 {The remaining analysis in this paper is performed with all $10^{6}$ sub-samples where each is a random representation of the input catalog but with a correct proportionality between dusty and normal GRBs. Nebular line fluxes are described by a Gaussian probability distribution with mean and standard deviation given by the raw measurement and error. Since the likelihood of keeping dusty GRBs has an error itself, we propagate in this way not only the statistical uncertainty due to sample size and the measurement error, but implicitly also the uncertainty in the prior, i.e., the dusty GRB fraction ($28\pm8\,\%$ at $2<z<3$, for example).}

 {The result of this procedure are $10^{6}$ different distributions of host parameters, correlations, and best-fit values. In the following, we then provide for all parameters or correlations the median of the a-posteriori probability and the 1$\sigma$-equivalent range as error to represent their distribution.}

 {Whenever we study GRB host parameters, we will plot two data sets: First, the more meaningful distribution restricted to $z<2$ (Sect.~\ref{sec:bdist}) and derived from the stochastic sample including error-bars always in blue. Second, we will also show the histogram of all measured values in black to illustrate the impact of the sample cuts. At $z>2$, we show the resulting parameters and distributions (without error regions) as reference. The selection effect towards optically bright galaxies at $z>2$, however, will often prevent us from deriving robust conclusions in this redshift range.}

\subsection{Misidentification of Hosts}

\begin{figure}
\includegraphics[angle=0, width=0.99\columnwidth]{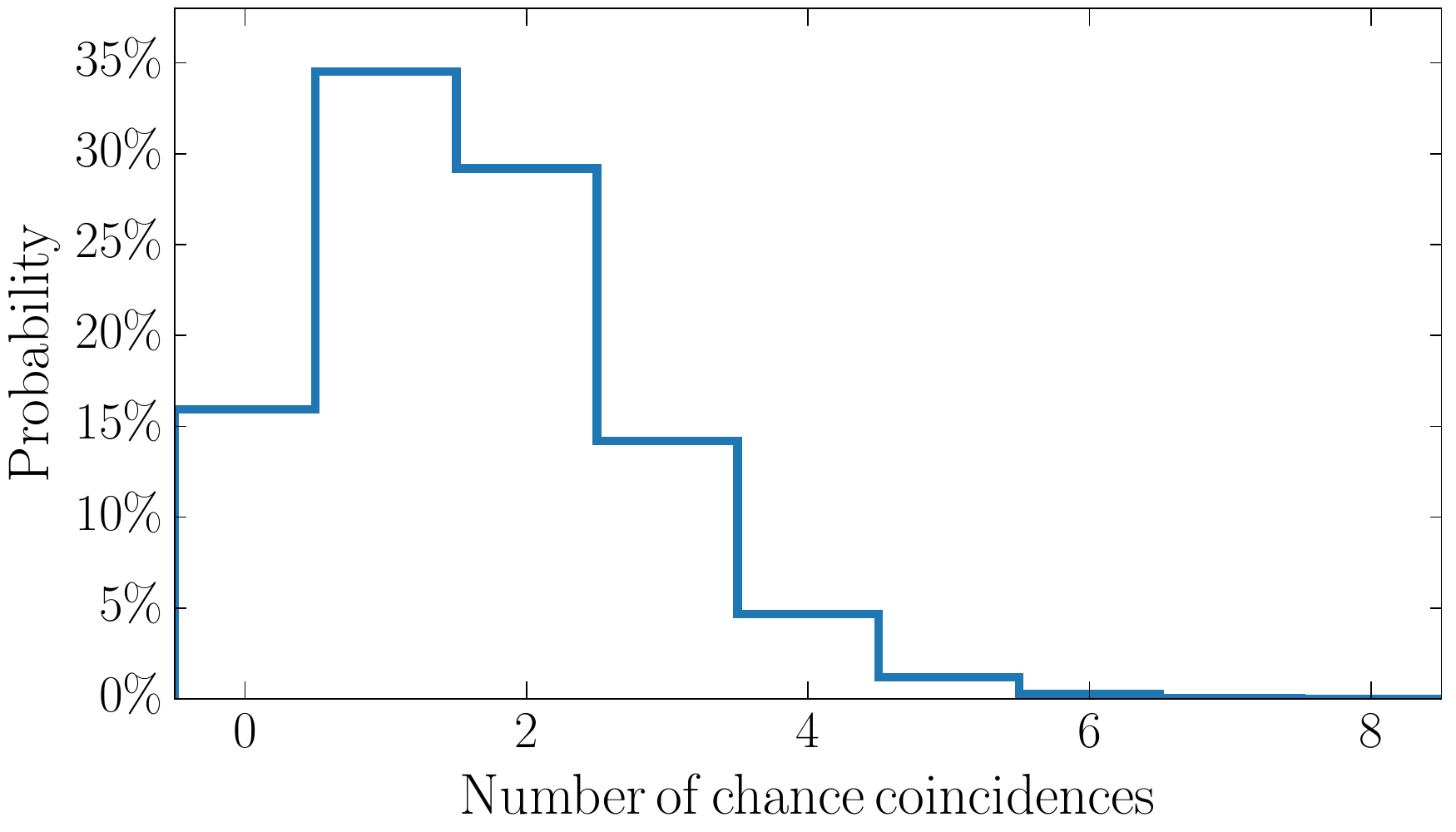}
\caption{Probability distribution of chance coincidence, or the likelihood of having a given number of galaxies in the total sample wrongly associated with the GRB. The probability that we have misidentified a total of 2 out of 96 galaxies is, for example, 30\%.}
\label{fig:misid}
\end{figure}

We use standard procedures \citep{2002AJ....123.1111B, 2009AJ....138.1690P} to assign galaxies to GRBs by using the afterglow position with the highest accuracy available. A crucial basis of this work are accurate GRB positions which are routinely available only since the launch of \textit{Swift} \citep{2004ApJ...611.1005G, 2005SSRv..120..165B}. 

In the best cases  {(16 of 96 GRBs)}, an afterglow redshift is set by fine-structure lines \citep[e.g.,][]{2007A&A...468...83V}, and a matching emission line redshift as well as a sub-arcsecond error-circle leave hardly any room for ambiguity.  {In total, 73 of the GRBs have localizations with an accuracy of around 0\farc{5} through radio, optical or \textit{Chandra} X-ray afterglow observations.}

In the remaining cases  {(23 of 96 GRBs)}, the host association relies on the X-ray position from \textit{Swift} which constrains the GRB position to a circle with $\sim$1\farc{5} radius at 90\% confidence. The probability of misidentification, i.e., the galaxy that we study is not related to the GRB of interest is then a function of positional accuracy, galaxy brightness, and density of galaxies for the given galaxy type, brightness, and coordinates \citep[e.g.,][]{2014A&A...572A..47R}. For average values of galaxy densities in empty fields \citep{2004AJ....127..180C, 2006ApJS..162....1G}, positional errors of 1\farc{5}, and galaxy brightnesses of $r\sim25$th magnitude, the chance of finding a random galaxy in the error circle is around 1-2\%. 

To quantify the total number of chance coincidences, or misidentifications in the full sample,  {we use the GRB host brightnesses from Fig.~\ref{fig:brightness} to assess how many similarly bright field galaxies \citep{2004AJ....127..180C, 2006ApJS..162....1G} are statistically expected within an area corresponding to the GRB's localization accuracy. After excluding those events for which the afterglow fine-structure/host redshift match, we calculate the probabilities for a given number of misidentifications through a Monte-Carlo method with $10^{6}$ trials. The resulting distribution is shown in Fig.~\ref{fig:misid}, and has a peak at around 1-2 events, with a tail out to a total of 5 misidentified galaxies. Very few field galaxies are thus interloping in our GRB host sample and their impact on the main conclusions will be limited.}X-Shooter

\section{X-Shooter Optical/NIR spectroscopy}
\label{xs}

X-Shooter operates in three dichroic-separated arms simultaneously: the ultra-violet, blue (UVB, 3000\,\AA\,-5600\,\AA), the visual (VIS, 5500\,\AA\,-10\,020\,\AA), and the near-infrared (NIR, 10\,000\AA\,\,-24\,800\,\AA) arm. Each of the three arms has its own slit, echelle grating, cross-disperser, and detector, and we refer to \citet{2011A&A...536A.105V} for a detailed description of the instrument. The resolving power of X-Shooter depends on the arm and slit width and is typically between $R=4000$ and $R=10000$ for the used setups\footnote{https://www.eso.org/sci/facilities/paranal/instruments/\\xshooter/inst.html}. Unless dictated by other observational constraints such as nearby bright objects, the observations were obtained in a nodding pattern in multiples of an ABBA sequence. The nod-length is constrained by the short slit-length of X-Shooter (11\farc{0}) and is usually 5\farc{0}. 

Whenever possible, slit loss due to atmospheric dispersion was minimized by observing at the parallactic angle and at minimum airmass. A large subset of the host observations were obtained before X-Shooter's ADC failure in August 2012 and most of the emission lines are located in the NIR part of the spectrum (e.g., the bluest of the lines of interest (\oii) is already above 10000\,\AA\,at $z>1.5$) where atmospheric dispersion is less severe. The lack of ADCs was not a primary concern in the analysis.

\subsection{Data reduction}

The basis of the data reduction was the X-Shooter pipeline supplied by ESO in its version \texttt{2.2.0} or higher \citep{2006SPIE.6269E..80G, 2010SPIE.7737E..56M}, which we used for flat-fielding, order tracing, rectification, and initial wavelength calibration and flexure compensation via arc-lamp frames (also Sect.~\ref{sec:wavecal}). These steps were applied to the individual frames of the individual arms of each set of spectra. During rectification, we chose a dispersion of 0.4\,\AA\,/px\,(UVB/VIS arm) and 0.6\,\AA\,/px\,(NIR arm). This minimizes correlated noise, while at the same time maintaining sufficient sampling for emission lines down to a velocity dispersion $\sigma$ of $\sim20\,$\kms. 

We use our own software and algorithms for bad-pixel and cosmic ray detection, as well as sky-subtraction and stacking of individual exposures. We further corrected the inter-order background for an apparent problem in the X-Shooter pipeline. In the UVB and VIS arm, we estimate the sky background locally in a small region around the spectral trace. For the NIR data with its abundant and strong sky-lines and high background, the frames for each set of observation are grouped in pairs of two such that they were taken in a different sky-position but as close in time as possible. Each two frames were then subtracted from each other for the background estimation.

A typical observation (Table~\ref{tab:xsobs}) consisting of four exposures in the UVB/VIS, and twelve in the NIR arm leads to four and six sky-subtracted, rectified, wavelength and flux-calibrated (Sect.~\ref{FluxCal}) frames in the UVB/VIS and NIR arm, respectively. These single frames are then averaged using a weight function from the variance or the signal-to-noise ratio (S/N). Previously detected bad-pixels and cosmic rays are masked in the stacking process.

From the combined two-dimensional frame we then optimally extract the one-dimensional spectrum. The weight function of the extraction is derived using the collapsed profile of either the emission lines themselves or the spectral continuum in case it is detected at a sufficient S/N level.

\subsection{Wavelength calibration}
\label{sec:wavecal}

X-Shooter is mounted on the VLT Cassegrain focus, so it suffers from instrument flexures and misalignments of the wavelength solution between day-time calibration and night-time science observations. We correct for these flexures first by using the ESO pipeline and frames supplied by X-Shooter's active flexure compensation. In a second step, we use a theoretical emission spectrum of the Paranal sky \citep{2012A&A...543A..92N}. In the VIS and NIR arm, we divide the spectrum into 10-15 slices, and cross-correlate the observed and theoretical sky-spectrum. This provides wavelength offsets averaged over the respective wavelength interval. A low-order polynomial was then fitted to the data, and the respective wavelength offset as derived from the fit was applied to the wavelength scale. Typical offsets are smaller than around 0.4\,\AA\,\ in the NIR arm and $<0.1$\AA\, in the UVB/VIS. Because of the lack of strong sky features in the UVB arm blueward of 5000\,\AA, we apply only a constant offset to the wavelength scale in the UVB arm.

\subsection{Flux calibration}
\label{FluxCal}

Correct flux scales of the emission line spectra are fundamental because they have a strong impact on the measurement of global galaxy characteristics. We first flux-calibrated each of the individual spectra with the nightly standard star taken with a slit-width of 5\farc{0}. Standard stars are different for each observation, and the full set is given at the ESO webpages\footnote{http://www.eso.org/sci/facilities/paranal/instruments\\/xshooter/tools/specphot\_list.html}. The resulting response curves were thoroughly validated against each other. To limit the effect of a time-dependent illumination by a time-varying flat-field lamp, particularly evident in the red part of the NIR arm, we used the same flat field for science and standard frames to correct for pixel-to-pixel variations.

Even with a well-defined instrumental response, the finite slit width inevitably leads to slit-loss. Slit-loss depends primarily on the ratio between slit-width and the convolution of the point-spread function delivered by the telescope with the spatial extent of the sources on the sky. In addition, the sky transparency varied significantly during the course of some of the observations due to clouds, leading to an additional decrease in flux compared to the standard star observation.

Accurate broadband photometry, however, is available to us providing means to estimate galaxy-integrated line-fluxes. The ratio between photometry and a synthetic brightness of the spectrum integrated over the band-pass of the respective filter yields a scale factor to estimate slit-loss (which we call $\tau$). In the ideal case, multiple independent photometric bands in each of X-Shooter's arms allow us to corroborate and check the intra- and inter-arm flux calibration. For many hosts, this is not the case and we assume that $\tau$ is constant in each arm. We then multiply the line flux by $\tau$ and propagate errors accordingly. $\tau$ is in the range between 1 and 3, and its distribution has a median and 1$\sigma$ bounds of $\tau_{\rm {med}} = 1.6_{-0.5}^{+0.7}$. 

Deriving $\tau$ for the individual arms is only possible for a bright continuum ($r<23$\,mag), i.e., luminous hosts or afterglows. For most of the fainter hosts ($r\sim25-26$\,mag), the continuum is detected in the bluer arms only leading to relatively large errors in $\tau$ and the line fluxes. In cases where no continuum is detected, or no comparison photometry is available in the given arm, we assume that $\tau$ is similar between all arms. In cases where we can measure $\tau$ for different arms or wavelengths, it is consistent within errors. This provides confidence that the scaling procedure is reliable within the quoted error bars.

Errors in $\tau$ come from the photon noise of the spectral continuum, the accuracy of the comparison photometry, and the scatter of different filters in each arm if available. Line ratios of adjacent lines (such as \ha\,and \nii\,or the \oiii\,doublet and \hb) and the physical properties that they are proxies for, have the same scaling factor in all cases. We associate the line-flux error from photon statistics with a statistical error, and the one derived from the scaling factor to a systematic one, and provide full details of both errors in the line-flux tables (Tables~\ref{tab:balmerlines} and \ref{tab:forbiddenlines}). All fluxes, spectra, and photometry are corrected for the Galactic foreground reddening according to \citet{2011ApJ...737..103S} assuming $R_V=3.08$ \citep{1992ApJ...395..130P}.

\subsection{Flux limits and sensitivities}
\label{sec:sens}

To better understand the fundamental properties and constraints of our observations, we look closer at the sensitivity limits. From a typical observational setup with an integration time of $4\times600$\,s, we derive the noise characteristics of the spectrum. Artificial emission lines of \ha\,and the \oii\,doublet at various redshifts with a typical intrinsic line width of $\sigma= 70$\,\kms\, (Sect.~\ref{sec:velo}) are then added to the noise spectrum. 

We record the emission-line flux that would have been detected with a combined significance of at least 6\,$\sigma$ over the noise level in the one-dimensional spectrum. The significance level is set conservatively because we have no a priori knowledge of the redshift in many cases\footnote{The look-elsewhere effect: It is for example very likely to find 3~$\sigma$ noise peaks in the medium-resolution spectrum of X-Shooter just because the parameter space to be searched is so large. If the redshift is known, the locations of the emission lines are fixed, and the probability of finding significant noise peaks by chance in the correct wavelength range drops substantially.}. Line-flux sensitivities for \ha\,and \oii\, are then converted into SFRs using the methods from Sect.~\ref{sec:newsc}.

\begin{figure}
\includegraphics[angle=0, width=0.99\columnwidth]{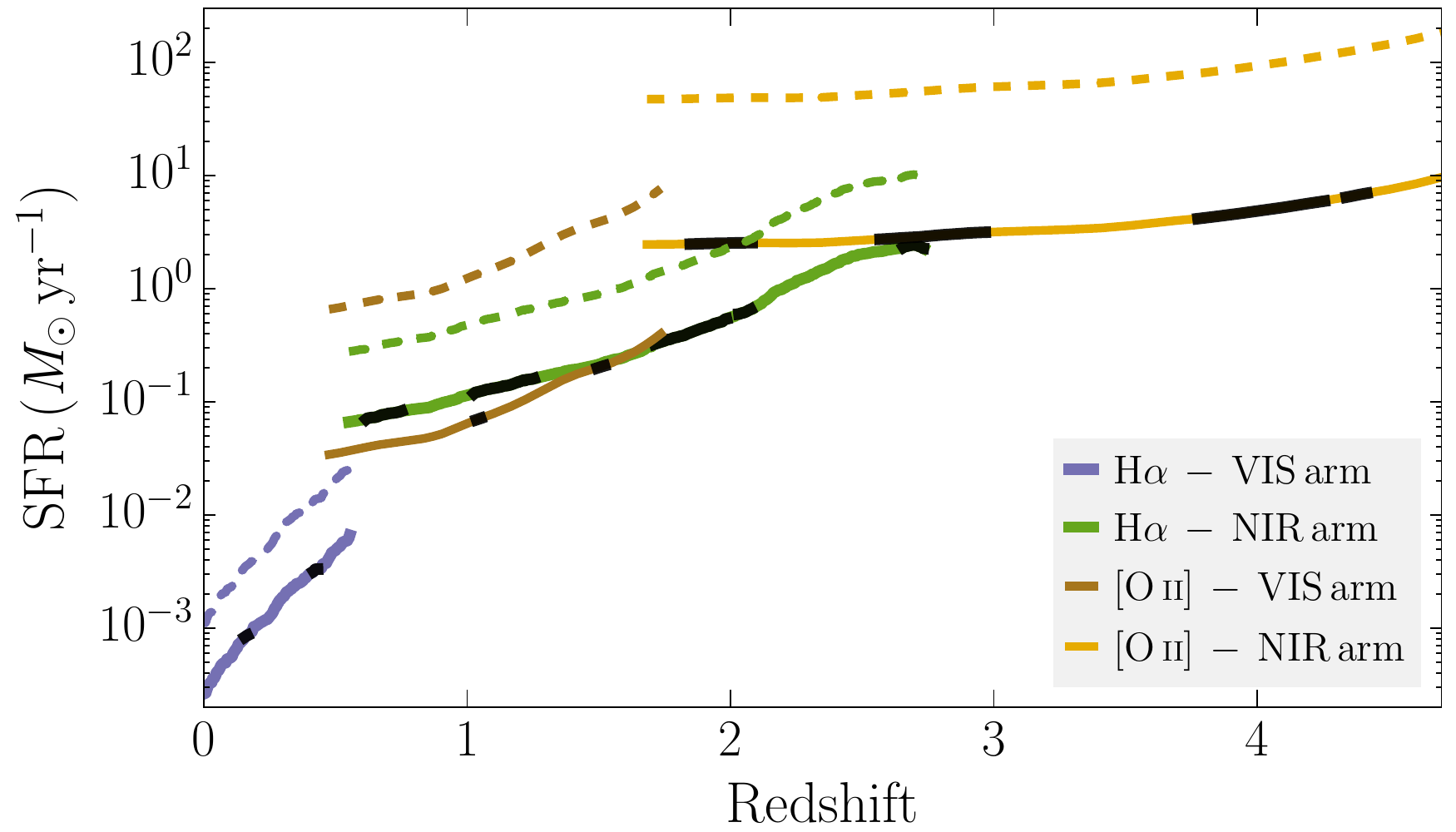}
\caption{Colored lines represent the sensitivity for \ha\,and \oii\,converted into a SFR. Solid/dashed lines denote the case of $A_V=0$\,mag and $A_V=2$\,mag, respectively. Black lines indicate redshift ranges where the line is located in windows of negligible atmospheric transmission. Skylines result in high SFR limits on small scales, but have been omitted to enhance clarity.}
\label{fig:sens}
\end{figure}

Figure~\ref{fig:sens} illustrates the redshift dependence of the detectable SFR. At $z<0.5$, \ha\, lies in the visual arm of X-Shooter with its sensitive CCD and low sky-background. The SFR limit is a few $10^{-4}$\,\Msunyr at $z\sim0.1$ and increases to $8\cdot10^{-3}$\,\Msunyr at $z\sim0.5$. In the non-extinguished case (solid lines in Fig.~\ref{fig:sens}), the SFR sensitivity stays below 1~\Msunyr\, up to $z\sim1.7$.

Above $z\sim2.5$, the sensitivity to unextinguished SFR is driven by \oii\,and stays at around a few \Msunyr up to $z\sim3.7$ above which it rises beyond 10\,\Msunyr. Adding dust increases these limits considerably, in particular for \oii\, (dashed lines in Fig.~\ref{fig:sens}). 

Noteworthy are redshift ranges where the SF-tracers are located in regions of prominent telluric absorption bands (black lines in the top panel of Fig.~\ref{fig:sens}). In a small window around $z\sim1$ and redshift ranges of $1.7<z<2.0$, $2.6<z<3.0$, and $z>3.7$, atmospheric absorption makes the detection of emission lines challenging from the ground. 

\subsection{Auxiliary photometry}
\label{sec:phot}

For most of the targets the photometry used to flux-calibrate the spectra is either provided by TOUGH \citep{2012ApJ...756..187H}, or the literature referenced in Table~\ref{tab:xsobs}. If not available elsewhere, dedicated observations were performed with GROND \citep{2008PASP..120..405G} at the 2.2~m MPG telescope on LaSilla, HAWK-I \citep{2008A&A...491..941K}, FORS2 \citep{1998Msngr..94....1A}, LRIS at Keck \citep{1995PASP..107..375O} or ALFOSC at the 2.5~m Nordic Optical Telescope (NOT).

The optical/NIR imaging data were reduced in a standard fashion using dedicated routines in pyraf/IRAF \citep{1993ASPC...52..173T} in the case of GROND, VLT, and NOT photometry or IDL for Keck/LRIS. Data reduction and photometry follow the procedures established in \citet{2008ApJ...685..376K} in the former case and is taken from the Keck GRB Host Survey \citep{2013EAS....61..391P} in the latter case. The photometric solution of each field was tied to the SDSS survey for those GRB fields which were covered by SDSS DR8 \citep{2011ApJS..193...29A}, zero-points derived from the observation of standard stars, or our own set of secondary calibrators in the field. Typically, secondary calibrators were again derived from observing SDSS fields close in time to the GRB observations. 

\section{Analysis and results}

\subsection{Line flux measurements}

\begin{figure*}
\begin{subfigure}{.33\textwidth}
  \includegraphics[width=0.999\linewidth]{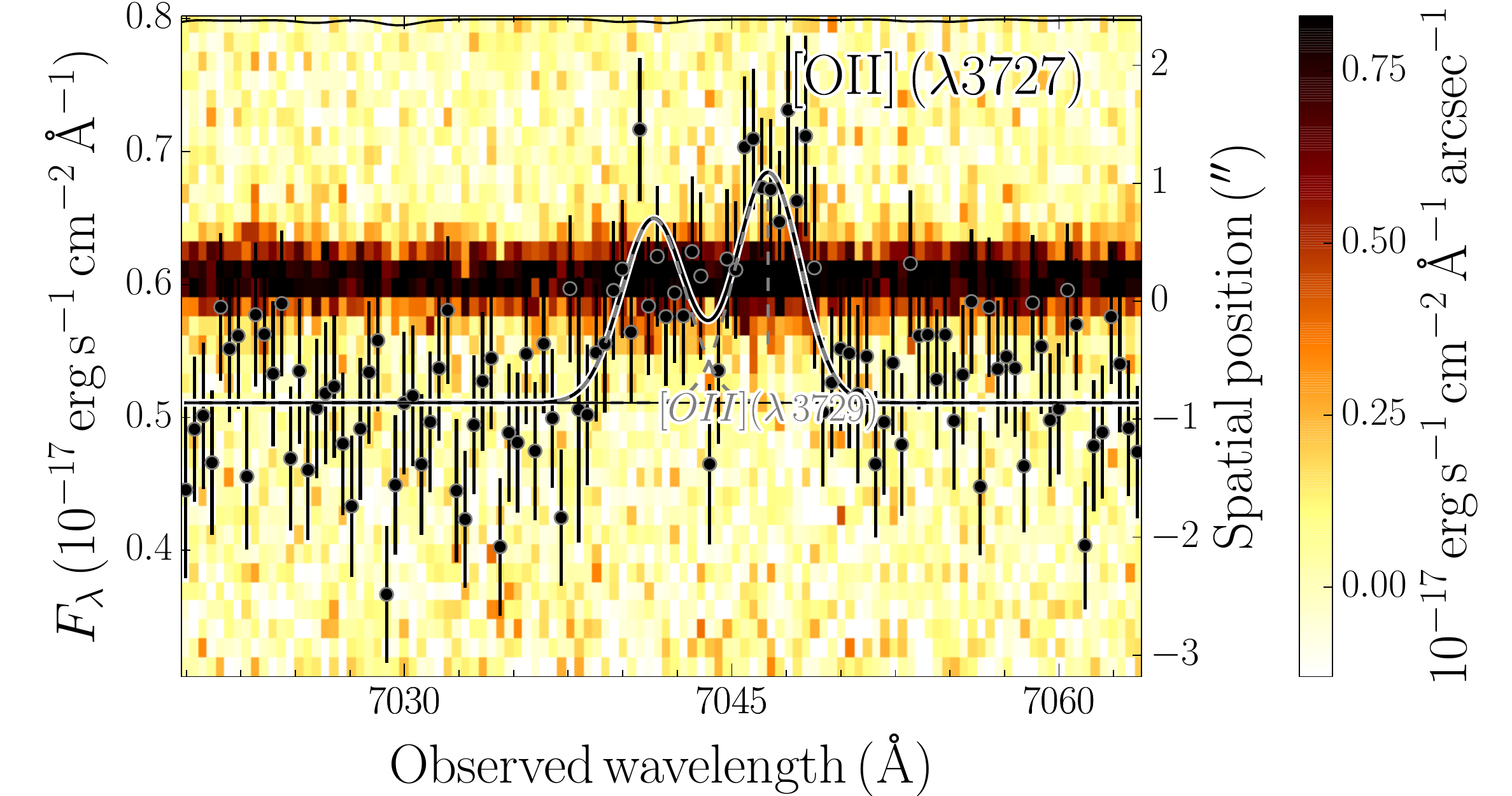}
\end{subfigure}
\begin{subfigure}{.33\textwidth}
  \includegraphics[width=0.999\linewidth]{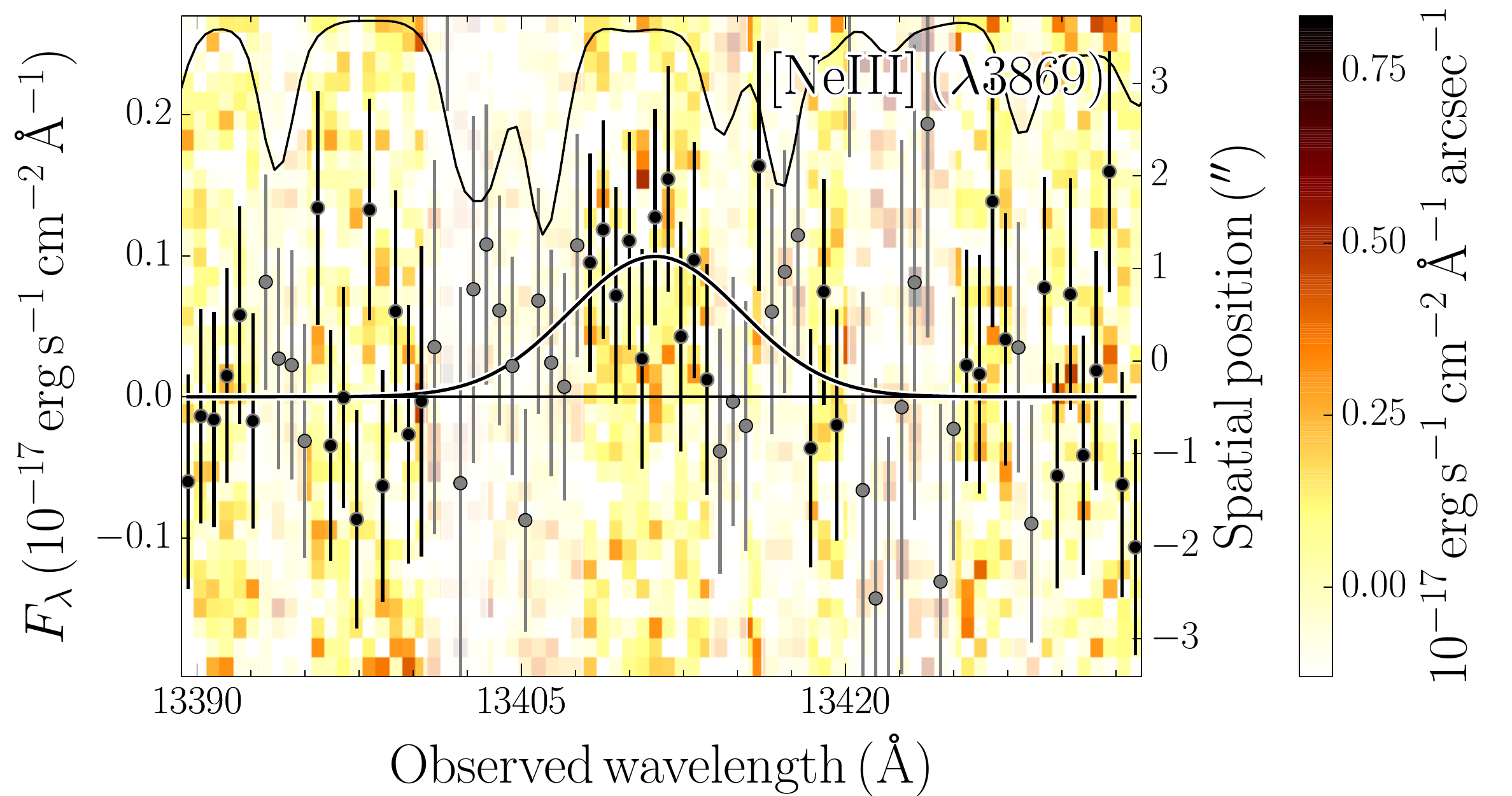}
\end{subfigure}
\begin{subfigure}{.33\textwidth}
  \includegraphics[width=0.999\linewidth]{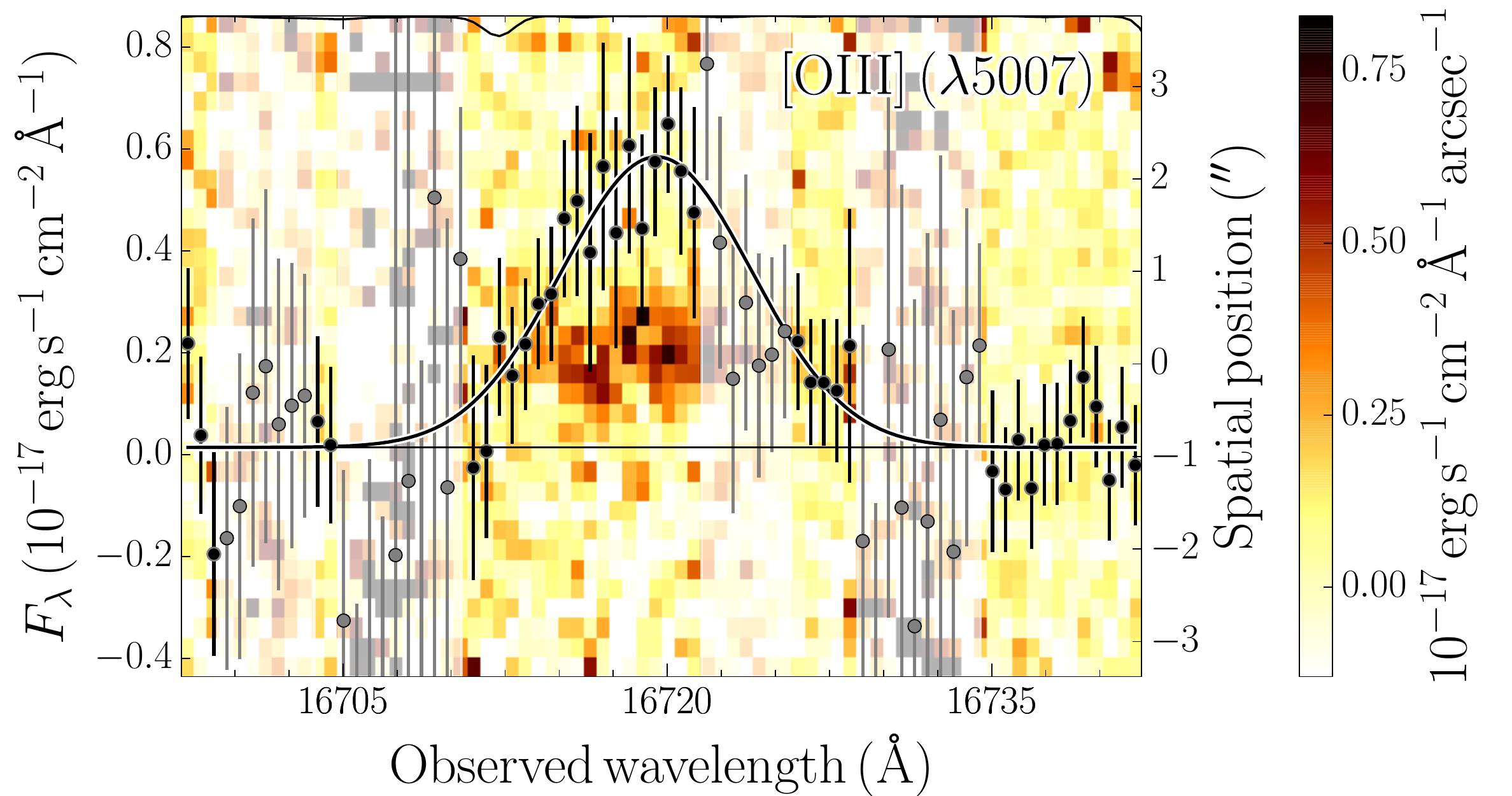}
\end{subfigure}
\begin{subfigure}{.33\textwidth}
  \includegraphics[width=0.999\linewidth]{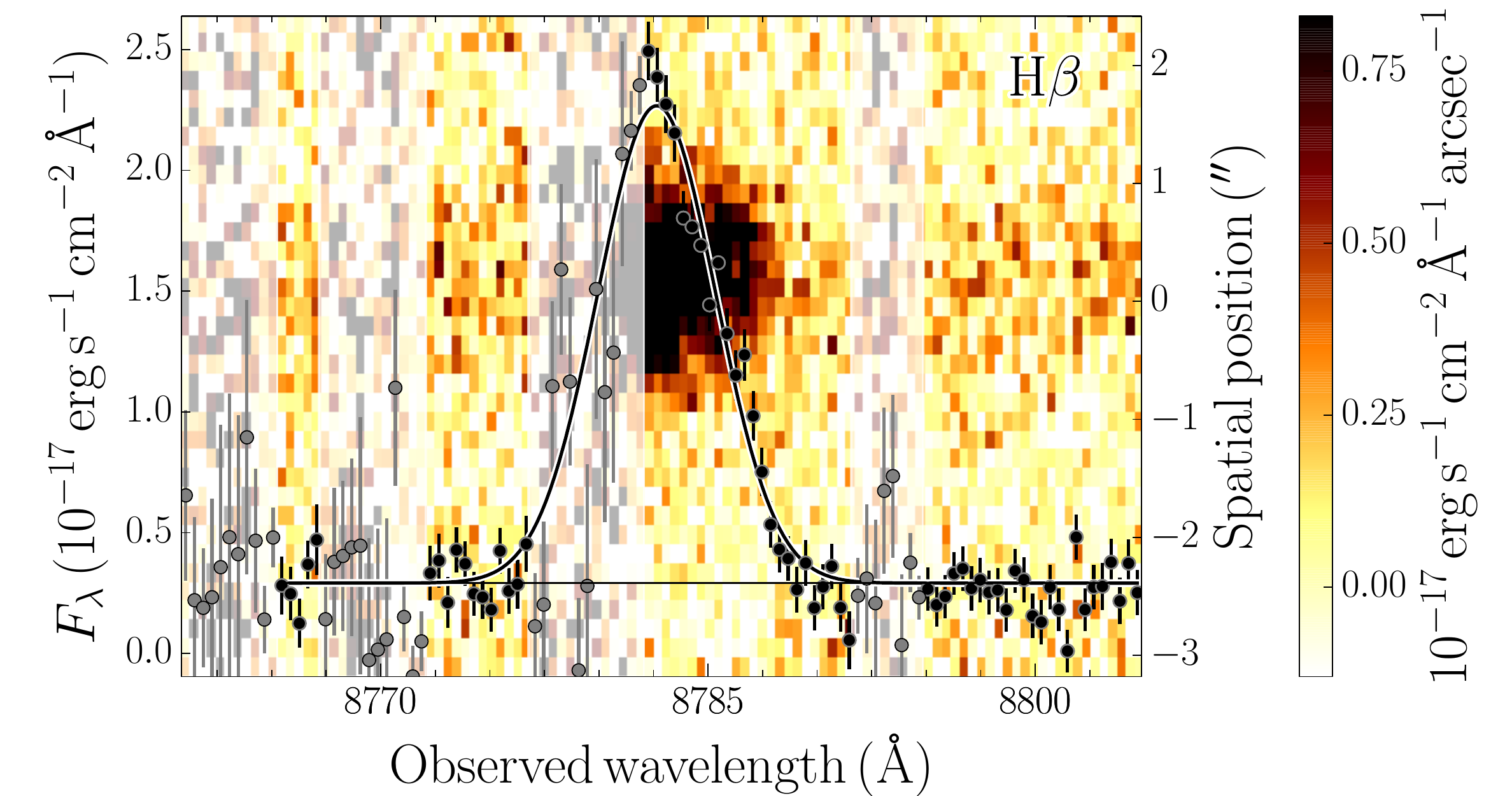}
\end{subfigure}
\begin{subfigure}{.33\textwidth}
  \includegraphics[width=0.999\linewidth]{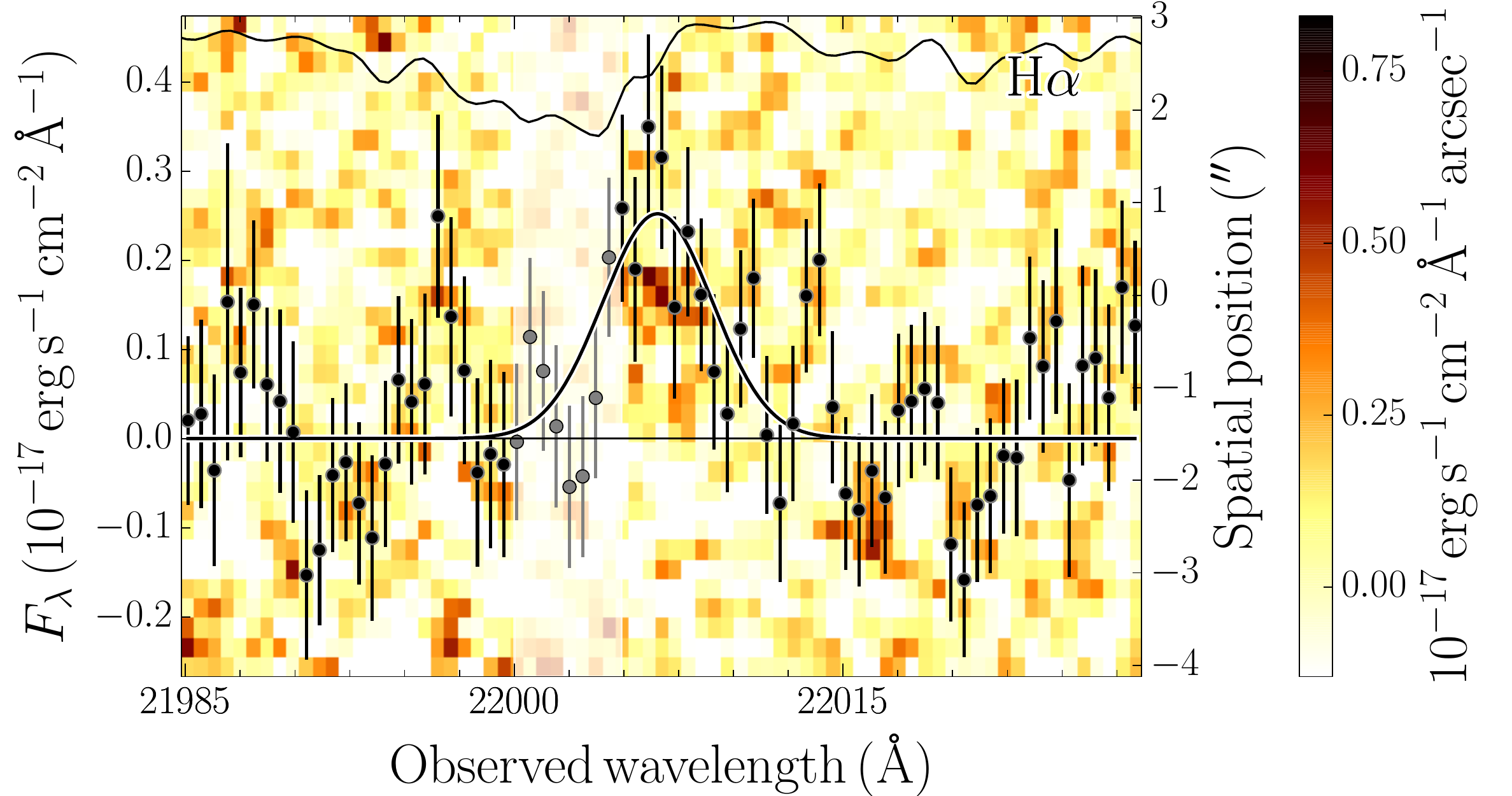}
\end{subfigure}
\begin{subfigure}{.33\textwidth}
  \includegraphics[width=0.999\linewidth]{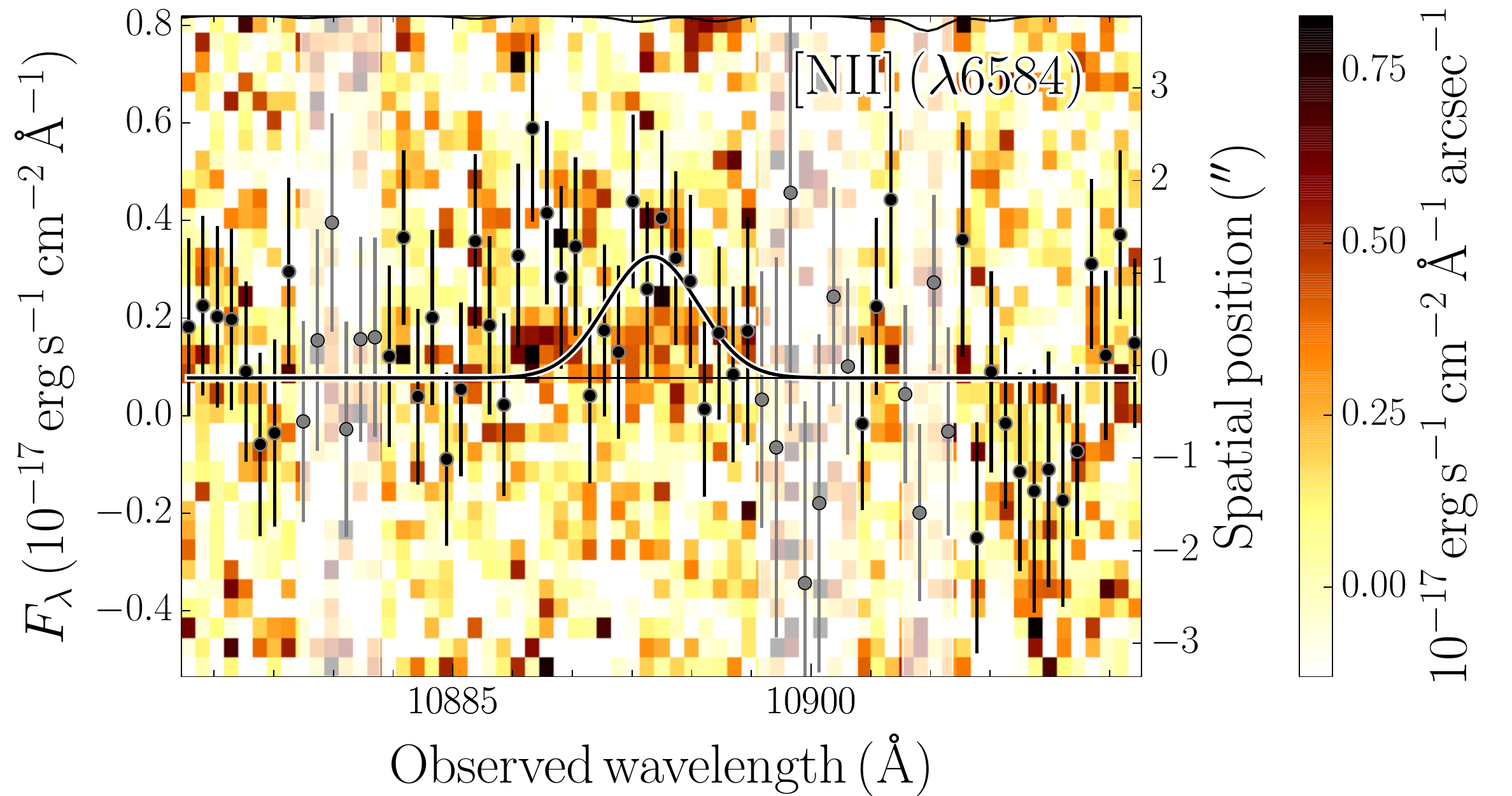}
\end{subfigure}
\caption{ {Examples of line-flux measurements for nebular lines at different signal-to-noise level. In each panel, the background is the two-dimensional spectrum, where the spatial direction is plotted against the right y-axis. The color coding is indicated by the color bar. The one-dimensional spectrum with a Gaussian fit are shown in black points and line. Gray data points are excluded from the fit because of telluric or skyline contamination. A telluric transmission spectrum is plotted in a thin black line at the very top of each panel. From top left to bottom right: \oii\, (GRB~140506A, with afterglow background), \neiii\, (GRB~100424A), \oiii\, (GRB~070129), \hb\, (GRB~051022), \ha\, (GRB~070110), \nii\,(GRB~050416A).}}
\label{fig:lineexamples}
\end{figure*}

Fluxes of emission lines were measured by fitting a superposition of Gaussian functions to the data, with the continuum set in small ($\pm40$\,\AA) regions around the wavelength of interest free of emission-, telluric-, and sky-lines. In most cases, a single Gaussian and constant continuum describes the line with sufficient accuracy (Fig.~\ref{fig:lineexamples}). We also measured the line-flux through numerical integration, and its result was used in cases where the lines show clear velocity structure.

For weak emission lines such as \neiii\,or \nii, we tied the Gaussian centroid (set by the galaxy redshift) and the line width to those of strong lines. 

Finally, a synthetic Paranal sky radiance and transmission spectrum \citep{2012A&A...543A..92N} was used to flag regions of telluric absorption and sky emission, and affected data were automatically discarded from the line fit (Fig.~\ref{fig:lineexamples}). The results of the line flux measurements are given in Tables~\ref{tab:balmerlines} and \ref{tab:forbiddenlines}. 

Many of the emission lines are not detected at high signal-to-noise ratios (S/N). If the spectral range is covered by X-Shooter, we decided to quote the actual measurement instead of providing an upper limit to maximize the information on e.g., line ratios under the given photon statistics.

\begin{longtab}
\begin{longtable}{c c c c c c}
\caption{Fluxes of Balmer lines for GRB hosts\label{tab:balmerlines}}\\
\hline\hline
{GRB host} & {Redshift} & $\rm{H}\delta$ & $\rm{H}\gamma$ & $\rm{H}\beta$ & $\rm{H}\alpha$ \\
\hline
\endfirsthead
\hline
\caption{Balmer lines fluxes (continued)}\\
\hline\hline
{GRB host} & {Redshift} & $\rm{H}\delta$ & $\rm{H}\gamma$ & $\rm{H}\beta$ & $\rm{H}\alpha$ \\
\hline
\endhead
GRB050416A & 0.6542 & $0.3 \pm 0.3 \pm 0.1$& $1.4 \pm 0.3 \pm 0.2$& $2.9 \pm 0.4 \pm 0.3$& $16.0 \pm 2.0 \pm 1.6$\\ 
GRB050525A & 0.6063 & $0.1 \pm 0.1 \pm 0.1$& $0.1 \pm 0.1 \pm 0.1$& $0.3 \pm 0.2 \pm 0.1$& $0.7 \pm 0.2 \pm 0.3$\\ 
GRB050714B & 2.4383 & \nodata & $0.6 \pm 0.3 \pm 0.2$& $0.5 \pm 0.2 \pm 0.2$& $3.4 \pm 0.6 \pm 0.8$\\ 
GRB050819A & 2.5042 & \nodata & $0.6 \pm 0.3 \pm 0.1$& $1.0 \pm 0.3 \pm 0.2$& \nodata \\ 
GRB050824 & 0.8277 & $0.7 \pm 0.1 \pm 0.1$& $1.1 \pm 0.1 \pm 0.2$& $2.6 \pm 0.4 \pm 0.5$& $6.8 \pm 0.6 \pm 1.3$\\ 
GRB050915A & 2.5275 & \nodata & $0.5 \pm 0.2 \pm 0.1$& $1.8 \pm 0.2 \pm 0.4$& \nodata \\ 
GRB051001 & 2.4295 & \nodata & $1.3 \pm 0.6 \pm 0.3$& $1.9 \pm 0.3 \pm 0.4$& $13 \pm 4 \pm 2$\\ 
GRB051016B & 0.9358 & $2.4 \pm 0.3 \pm 0.2$ & $6.3 \pm 0.4 \pm 0.5$ & $10.9 \pm 1.2 \pm 1.1$ & $40 \pm 6 \pm 4$\\ 
GRB051022A & 0.8061 & $2.8 \pm 0.5 \pm 0.3$& $8.1 \pm 0.8 \pm 0.7$& $22.1 \pm 1.3 \pm 1.7$& $110 \pm 4 \pm 15$\\ 
GRB051117B & 0.4805 & $-1.4 \pm 0.8 \pm 0.1$& $1.1 \pm 1.0 \pm 0.1$& $3.0 \pm 1.0 \pm 0.2$& $21.1 \pm 4.1 \pm 1.3$\\ 
GRB060204B & 2.3393 & \nodata & \nodata & $4.1 \pm 1.1 \pm 0.6$& $16.9 \pm 1.4 \pm 2.0$ \\ 
GRB060306 & 1.5597 & \nodata & \nodata & $1.4 \pm 2.0 \pm 0.4$& $8.9 \pm 3.4 \pm 1.4$\\ 
GRB060604 & 2.1355 & \nodata & \nodata & $0.4 \pm 0.1 \pm 0.1$& $1.8 \pm 0.3 \pm 0.5$\\ 
GRB060707 & 3.4246 & \nodata & \nodata & $1.1 \pm 1.8 \pm 0.8$& \nodata \\ 
GRB060719 & 1.5318 & \nodata & $0.1 \pm 0.5 \pm 0.1$& $0.8 \pm 0.5 \pm 0.2$& $3.8 \pm 0.4 \pm 0.7$\\ 
GRB060729 & 0.5429 & \nodata & $0.0 \pm 0.3 \pm 0.1$& $0.6 \pm 0.2 \pm 0.1$& $3.7 \pm 1.7 \pm 0.5$\\ 
GRB060805A & 2.3633 & \nodata & $0.5 \pm 0.4 \pm 0.2$& $1.4 \pm 0.3 \pm 0.4$& $3.7 \pm 0.5 \pm 0.9$\\ 
GRB060814 & 1.9223 & \nodata & \nodata & $8.3 \pm 3.0 \pm 1.5$& $28 \pm 4 \pm 4$\\ 
GRB060912A & 0.9362 & $0.5 \pm 0.3 \pm 0.1$& $2.5 \pm 0.3 \pm 0.2$& $5.9 \pm 0.5 \pm 0.5$& $15.8 \pm 3.8 \pm 1.6$\\ 
GRB060923B & 1.5094 & \nodata & \nodata & $-1.6 \pm 0.6 \pm 0.1$& $2.6 \pm 0.8 \pm 0.3$\\ 
GRB060926 & 3.2090 & $0.1 \pm 0.1 \pm 0.1$& \nodata & $0.3 \pm 0.2 \pm 0.2$& \nodata \\ 
GRB061021 & 0.3453 & $0.2 \pm 0.2 \pm 0.1$& $0.2 \pm 0.2 \pm 0.1$& $0.5 \pm 0.1 \pm 0.1$& $1.9 \pm 0.1 \pm 0.2$\\ 
GRB061110A & 0.7578 & $0.0 \pm 0.1 \pm 0.1$& \nodata & $0.3 \pm 0.1 \pm 0.1$& \nodata \\ 
GRB061202 & 2.2543 & \nodata & \nodata & $1.2 \pm 0.4 \pm 0.3$& $6.1 \pm 0.8 \pm 1.3$ \\ 
GRB070103 & 2.6208 & $1.1 \pm 1.9 \pm 0.6$& $3.9 \pm 1.6 \pm 1.2$& $4.3 \pm 0.8 \pm 1.1$& \nodata \\ 
GRB070110 & 2.3523 & \nodata & $0.6 \pm 0.4 \pm 0.2$& \nodata & $3.5 \pm 0.6 \pm 0.8$\\ 
GRB070129 & 2.3384 & \nodata & $0.7 \pm 0.6 \pm 0.2$& $1.5 \pm 0.9 \pm 0.4$& $6.5 \pm 1.1 \pm 1.1$\\ 
GRB070224 & 1.9922 & \nodata & $0.1 \pm 0.1 \pm 0.1$& $0.2 \pm 0.1 \pm 0.2$& \nodata \\ 
GRB070306 & 1.4965 & $1.7 \pm 1.3 \pm 0.1$& $7.7 \pm 3.6 \pm 0.8$& $11.6 \pm 1.2 \pm 0.7$& $53.5 \pm 1.4 \pm 3.7$\\ 
GRB070318 & 0.8401 & \nodata & $0.3 \pm 0.1 \pm 0.1$& $1.2 \pm 0.2 \pm 0.2$& $3.3 \pm 0.4 \pm 0.5$\\ 
GRB070328 & 2.0627 & $0.3 \pm 0.2 \pm 0.1$& $0.1 \pm 0.3 \pm 0.1$& $1.1 \pm 0.2 \pm 0.2$& \nodata \\ 
GRB070419B & 1.9586 & \nodata & \nodata & $0.9 \pm 0.3 \pm 0.3$& $4.4 \pm 0.6 \pm 1.1$\\ 
GRB070521 & 2.0865 & \nodata & \nodata & $-0.2 \pm 0.9 \pm 0.2$& $10 \pm 5 \pm 3$\\ 
GRB070802 & 2.4538 & \nodata & $0.4 \pm 0.2 \pm 0.1$& $1.4 \pm 0.2 \pm 0.3$& $5.1 \pm 0.7 \pm 1.0$\\ 
GRB071021 & 2.4515 & \nodata & $0.7 \pm 0.5 \pm 0.2$& $2.8 \pm 0.3 \pm 0.5$& $8.9 \pm 1.8 \pm 1.8$\\ 
GRB080207 & 2.0856 & \nodata & $1.0 \pm 0.6 \pm 0.3$& $1.3 \pm 0.5 \pm 0.3$& $10.9 \pm 1.9 \pm 2.2$\\ 
GRB080413B & 1.1012 & \nodata & $0.2 \pm 0.1 \pm 0.1$& \nodata & $2.6 \pm 1.0 \pm 0.8$\\ 
GRB080602 & 1.8204 & $1.6 \pm 0.9 \pm 0.3$& $2.1 \pm 0.7 \pm 0.3$& \nodata & $30 \pm 8 \pm 4$\\ 
GRB080605 & 1.6408 & \nodata & \nodata & $7.7 \pm 0.8 \pm 1.3$& $29.1 \pm 1.4 \pm 4.3$\\ 
GRB080804 & 2.2059 & \nodata & \nodata & $0.9 \pm 0.4 \pm 0.3$& $3.7 \pm 0.8 \pm 1.2$\\ 
GRB080805 & 1.5052 & \nodata & \nodata & $1.7 \pm 0.6 \pm 0.5$& $11.0 \pm 1.6 \pm 2.7$\\ 
GRB081109 & 0.9785 & \nodata & $1.8 \pm 0.4 \pm 0.3$ & $5.5 \pm 0.5 \pm 0.8$& $20.9 \pm 0.9 \pm 2.7$\\ 
GRB081210 & 2.0631 & $0.3 \pm 0.3 \pm 0.1$& $1.1 \pm 0.4 \pm 0.3$& $2.1 \pm 0.3 \pm 0.5$& \nodata \\ 
GRB081221 & 2.2590 & \nodata & \nodata & $2.4 \pm 1.3 \pm 1.7$& $9.5 \pm 1.9 \pm 5.3$\\ 
GRB090113 & 1.7494 & \nodata & $3.6 \pm 1.2 \pm 0.7$& $3.3 \pm 0.9 \pm 0.6$& $15 \pm 2 \pm 3$\\ 
GRB090201 & 2.1000 & \nodata & $3.8 \pm 1.8 \pm 1.1$& $6.0 \pm 1.4 \pm 1.4$& $23 \pm 2 \pm 4$\\ 
GRB090323 & 3.5832 & \nodata & $-0.3 \pm 0.4 \pm 0.1$& $0.0 \pm 0.5 \pm 0.1$& \nodata \\ 
GRB090407 & 1.4478 & $-0.2 \pm 0.3 \pm 0.1$& $-0.1 \pm 0.3 \pm 0.1$& $0.7 \pm 0.2 \pm 0.1$& $4.4 \pm 0.4 \pm 0.6$\\ 
GRB090926B & 1.2427 & $0.5 \pm 0.2 \pm 0.1$& $0.7 \pm 0.4 \pm 0.2$& $2.7 \pm 0.9 \pm 0.5$& $14.2 \pm 2.5 \pm 2.4$\\ 
GRB091018 & 0.9710 & \nodata & $0.0 \pm 0.2 \pm 0.1$& $1.5 \pm 0.8 \pm 0.1$& $4.2 \pm 0.6 \pm 0.3$\\ 
GRB091127 & 0.4904 & $0.7 \pm 0.2 \pm 0.1$& $0.7 \pm 0.2 \pm 0.1$& $1.6 \pm 0.2 \pm 0.1$& $5.5 \pm 0.3 \pm 0.5$\\ 
GRB100316D & 0.0592 & $14.0 \pm 0.7$& $27.8 \pm 1.4$& $62 \pm 3$& $212 \pm 11$\\ 
GRB100418A & 0.6235 & $1.6 \pm 1.3 \pm 0.3$& $4.5 \pm 0.5 \pm 0.5$& $10.0 \pm 0.7 \pm 1.0$& $35 \pm 3 \pm 4$\\ 
GRB100424A & 2.4656 & \nodata & $0.8 \pm 0.4 \pm 0.3$& $2.0 \pm 0.2 \pm 0.5$& $6.5 \pm 1.6 \pm 1.9$\\ 
GRB100508A & 0.5201 & \nodata & $1.0 \pm 0.9 \pm 0.1$& $7.2 \pm 0.6 \pm 0.5$& $24.9 \pm 1.8 \pm 1.6$\\ 
GRB100606A & 1.5545 & \nodata & \nodata & $1.7 \pm 1.0 \pm 0.7$& $4.7 \pm 0.5 \pm 1.4$ \\ 
GRB100615A & 1.3978 & \nodata & $0.2 \pm 0.4 \pm 0.1$& $1.0 \pm 0.4 \pm 0.2$& $4.8 \pm 0.5 \pm 0.7$\\ 
GRB100621A & 0.5426 & \nodata & $18.8 \pm 0.8 \pm 1.1$& $43.8 \pm 1.0 \pm 2.4$& $128 \pm 5 \pm 7$\\ 
GRB100724A & 1.2890 & \nodata & $0.2 \pm 0.3 \pm 0.1$& $1.1 \pm 0.7 \pm 0.5$& $3.8 \pm 0.5 \pm 1.1$\\ 
GRB100728A & 1.5670 & \nodata & $1.2 \pm 2.5 \pm 1.4$& $2.5 \pm 2.5 \pm 1.9$& $10.6 \pm 1.7 \pm 4.7$\\ 
GRB100814A & 1.4392 & \nodata & \nodata & $1.3 \pm 0.3 \pm 0.2$& $4.0 \pm 0.3 \pm 0.4$\\ 
GRB100816A & 0.8048 & \nodata & \nodata & $1.7 \pm 0.4 \pm 0.2$& $19 \pm 2 \pm 3$\\ 
GRB110808A & 1.3490 & \nodata & \nodata & $2.0 \pm 0.6 \pm 0.4$& $7.6 \pm 0.5 \pm 1.1$\\ 
GRB110818A & 3.3609 & \nodata & \nodata & $1.6 \pm 0.4 \pm 0.3$& \nodata \\ 
GRB110918A & 0.9843 & \nodata & $-0.1 \pm 1.6 \pm 0.1$& $10.3 \pm 2.4 \pm 0.6$& $42 \pm 11 \pm 4$\\ 
GRB111123A & 3.1513 & \nodata & \nodata & \nodata & \nodata \\ 
GRB111129A & 1.0796 & \nodata & $1.1 \pm 0.5 \pm 0.1$& $3.3 \pm 1.9 \pm 0.4$& \nodata \\ 
GRB111209A & 0.6770 & $0.2 \pm 0.1 \pm 0.1$& $0.3 \pm 0.1 \pm 0.1$& $0.7 \pm 0.1 \pm 0.1$& $2.4 \pm 0.6 \pm 0.5$\\ 
GRB111211A & 0.4786 & \nodata & $0.5 \pm 0.7 \pm 0.1$& $-0.5 \pm 0.4 \pm 0.1$& $2.3 \pm 0.4 \pm 0.2$\\ 
GRB111228A & 0.7164 & \nodata & $0.1 \pm 0.2 \pm 0.1$& $0.5 \pm 0.3 \pm 0.1$& \nodata \\ 
GRB120118B & 2.9428 & \nodata & $2.0 \pm 0.3 \pm 0.3$& $2.4 \pm 0.8 \pm 0.3$& \nodata \\ 
GRB120119A & 1.7291 & $0.2 \pm 1.6 \pm 0.3$& $1.5 \pm 0.7 \pm 0.4$& $4.9 \pm 1.0 \pm 1.1$& $19.2 \pm 1.1 \pm 3.7$\\ 
GRB120422A & 0.2826 & $3.2 \pm 0.3 \pm 0.1$& $6.4 \pm 0.5 \pm 0.3$& $14.8 \pm 0.7 \pm 0.6$& $57.5 \pm 1.1 \pm 2.1$\\ 
GRB120624B & 2.1974 & $-0.2 \pm 0.6 \pm 0.1$& \nodata & $2.9 \pm 1.4 \pm 1.0$ & $10.3 \pm 0.7 \pm 3.1$\\ 
GRB120714B & 0.3985 & $0.7 \pm 0.5 \pm 0.1$& $0.5 \pm 0.3 \pm 0.1$& $2.5 \pm 0.2 \pm 0.2$& $7.5 \pm 0.3 \pm 0.5$\\ 
GRB120722A & 0.9590 & $1.4 \pm 0.2 \pm 0.2$& $2.6 \pm 0.2 \pm 0.3$& $8.7 \pm 1.2 \pm 1.0$& $32.0 \pm 1.5 \pm 3.4$\\ 
GRB120815A & 2.3587 & \nodata & \nodata & $0.3 \pm 0.1 \pm 0.1$& $0.8 \pm 0.2 \pm 0.2$\\ 
GRB121024A & 2.3012 & \nodata & \nodata & $7.3 \pm 0.6 \pm 0.6$& $17.7 \pm 1.2 \pm 1.5$\\ 
GRB121027A & 1.7732 & \nodata & \nodata & \nodata & \nodata \\ 
GRB121201A & 3.3830 & $0.6 \pm 0.4 \pm 0.1$ & \nodata & \nodata & \nodata \\ 
GRB130131B & 2.5393 & \nodata & \nodata & $0.5 \pm 0.4 \pm 0.2$& \nodata \\ 
GRB130427A & 0.3401 & $-0.7 \pm 1.0 \pm 0.1$& $0.3 \pm 1.1 \pm 0.1$& $4.8 \pm 0.9 \pm 0.4$& $14.6 \pm 1.3 \pm 1.1$\\ 
GRB130701A & 1.1548 & \nodata & \nodata & $1.3 \pm 2.4 \pm 0.2$& \nodata \\ 
GRB130925A & 0.3483 & $1.2 \pm 1.0 \pm 0.1$& $5.9 \pm 1.0 \pm 0.3$& $11.7 \pm 0.7 \pm 0.6$& $54 \pm 2 \pm 3$\\ 
GRB131103A & 0.5960 & $3.1 \pm 1.1 \pm 0.6$& $6.0 \pm 1.3 \pm 1.0$& $20.2 \pm 1.1 \pm 2.8$& $51 \pm 5 \pm 8$\\ 
GRB131105A & 1.6854 & $1.0 \pm 0.6 \pm 0.3$& $0.7 \pm 0.6 \pm 0.3$& $1.7 \pm 0.5 \pm 0.5$& $10.1 \pm 0.7 \pm 2.3$\\ 
GRB131231A & 0.6427 & $1.1 \pm 0.3 \pm 0.1$& $2.4 \pm 0.3 \pm 0.2$& $3.9 \pm 0.5 \pm 0.4$& $14.1 \pm 1.1 \pm 1.4$\\ 
GRB140213A & 1.2079 & \nodata & \nodata & $0.6 \pm 0.7 \pm 0.1$& $1.2 \pm 0.5 \pm 0.1$\\ 
GRB140301A & 1.4155 & $2.1 \pm 0.6 \pm 0.2$& $1.4 \pm 0.6 \pm 0.2$& $5.0 \pm 0.6 \pm 0.5$& $31.7 \pm 1.5 \pm 2.7$\\ 
GRB140430A & 1.6019 & \nodata & \nodata & \nodata & $5.8 \pm 0.8 \pm 0.6$\\ 
GRB140506A & 0.8893 & \nodata & \nodata & \nodata & $1.2 \pm 0.4 \pm 0.2$\\ 

\hline\noalign{\smallskip}
\end{longtable}
\centering
\begin{minipage}{5.8in}
\tablefoot{
Measurements are in units of $10^{-17}$\,\erg, and are corrected for the Galactic foreground reddening. A correction for slit-loss based on broad-band photometry has been applied to the measurements as described in the text. These values thus represent host-integrated measurements except for GRB~100316D (see below). Line fluxes are not corrected for host-intrinsic extinction. The first error represents the statistical error due to photon statistics and line-flux measurement. The second error is the systematic error in the absolute flux calibration due to slit-loss and scaling to photometry. Redshifts are given in a heliocentric reference frame. No data means that either the wavelength range of the respective line is not covered, all data in that wavelength range have been excluded from the automated fitting procedure, or no meaningful constraints could be obtained for the given line.\\
$^{a}$ GRB~100316D has the lowest redshift in the sample ($z=0.0592$). The fraction of the host that is covered by the slit is too small to derive representative host-integrated measurements. These values are thus as derived from the observed spectrum and not scaled by photometry. No systematic error on the line fluxes is given in this case.}
\end{minipage}

\end{longtab}

\begin{longtab}
\begin{landscape}
\begin{longtable}{c c c c c c c c}
\caption{Fluxes of forbidden lines for GRB hosts\label{tab:forbiddenlines}}\\
\hline\hline
{GRB host} & {Redshift} & \multicolumn{2}{c}{$\left[\ion{O}{ii}\right]$} & $\left[\ion{Ne}{iii}\right]$ & \multicolumn{2}{c}{$\left[\ion{O}{iii}\right]$} & $\left[\ion{N}{ii}\right]$  \\
           &            & $\lambda$3726 & $\lambda$3729 & $\lambda$3869 & $\lambda$4959 & $\lambda$5007 & $\lambda$6584 \\
\hline
\endfirsthead
\hline
\caption{Fluxes of forbidden lines (continued)}\\
\hline\hline
{GRB host} & {Redshift} & \multicolumn{2}{c}{$\left[\ion{O}{ii}\right]$} & $\left[\ion{Ne}{iii}\right]$ & \multicolumn{2}{c}{$\left[\ion{O}{iii}\right]$} & $\left[\ion{N}{ii}\right]$  \\
           &            & $\lambda$3726 & $\lambda$3729 & $\lambda$3869 & $\lambda$4959 & $\lambda$5007 & $\lambda$6584 \\
\hline
\endhead
GRB050416A & 0.6542 & $4.9 \pm 0.4 \pm 0.5$& $4.5 \pm 0.5 \pm 0.4$& $0.7 \pm 0.4 \pm 0.1$& $2.5 \pm 0.3 \pm 0.3$& $7.8 \pm 0.6 \pm 0.8$& $1.9 \pm 0.6 \pm 0.2$\\ 
GRB050525A$^{\rm{b}}$ & 0.6063 & \multicolumn{2}{c}{$0.2 \pm 0.2 \pm 0.2$} & $0.1 \pm 0.1 \pm 0.1$& $0.4 \pm 0.1 \pm 0.2$& $0.8 \pm 0.1 \pm 0.3$& $-0.2 \pm 0.3 \pm 0.1$\\ 
GRB050714B$^{\rm{b}}$ & 2.4383 & \multicolumn{2}{c}{$2.2 \pm 1.0 \pm 0.5$} & $0.1 \pm 0.3 \pm 0.1$& $0.0 \pm 0.3 \pm 0.1$& $1.0 \pm 0.4 \pm 0.3$& $-1.4 \pm 0.6 \pm 0.1$\\ 
GRB050819A & 2.5042 & $1.0 \pm 0.5 \pm 0.2$& $1.4 \pm 0.4 \pm 0.3$& \nodata & \nodata & $5.2 \pm 0.7 \pm 0.9$& $-1.9 \pm 0.8 \pm 0.2$\\ 
GRB050824 & 0.8277 & $2.0 \pm 0.2 \pm 0.4$& $2.6 \pm 0.2 \pm 0.5$& $1.1 \pm 0.1 \pm 0.2$& $6.5 \pm 0.3 \pm 1.3$& $18.5 \pm 0.8 \pm 3.6$& $0.3 \pm 1.0 \pm 0.1$\\ 
GRB050915A$^{\rm{b}}$ & 2.5275 & \multicolumn{2}{c}{$3.0 \pm 0.5 \pm 0.5$} & \nodata & $1.0 \pm 0.1 \pm 0.2$& $2.9 \pm 0.5 \pm 0.6$& \nodata \\ 
GRB051001$^{\rm{b}}$  & 2.4295 & \multicolumn{2}{c}{$4.6 \pm 2.0 \pm 0.8$} & $1.0 \pm 0.5 \pm 0.2$ & $-0.3 \pm 0.6 \pm 0.1$& $5.3 \pm 0.8 \pm 1.0$& $-3.0 \pm 1.5 \pm 0.2$\\ 
GRB051016B & 0.9358 & $13.0 \pm 0.6 \pm 1.2$& $15.6 \pm 0.7 \pm 1.4$& $3.7 \pm 0.4 \pm 0.3$ & $15.9 \pm 1.1 \pm 1.5$& $53 \pm 2 \pm 5$& \nodata \\ 
GRB051022A & 0.8061 & $23.2 \pm 1.2 \pm 1.8$& $30.9 \pm 1.2 \pm 2.4$& $4.1 \pm 0.8 \pm 0.4$& $21.3 \pm 1.3 \pm 1.7$& $63 \pm 2 \pm 5$& $17 \pm 3 \pm 3$\\ 
GRB051117B & 0.4805 & $1.4 \pm 0.7 \pm 0.1$& $1.7 \pm 1.0 \pm 0.1$& $-1.0 \pm 1.2 \pm 0.1$& $1.1 \pm 1.3 \pm 0.1$& $1.2 \pm 0.9 \pm 0.1$& $7.1 \pm 1.6 \pm 0.4$\\ 
GRB060204B & 2.3393 & $5.6 \pm 0.5 \pm 0.7$& $3.8 \pm 0.5 \pm 0.5$& $1.2 \pm 0.9 \pm 0.3$ & $5.7 \pm 0.6 \pm 0.8$& $16.9 \pm 1.4 \pm 2.2$& $0.7 \pm 0.8 \pm 0.2$\\ 
GRB060306 & 1.5597 & $0.7 \pm 0.3 \pm 0.1$& $1.0 \pm 0.2 \pm 0.1$& $1.1 \pm 1.4 \pm 0.3$& \nodata & $3.3 \pm 4.5 \pm 0.9$& $2.4 \pm 1.4 \pm 0.4$\\ 
GRB060604$^{\rm{b}}$ & 2.1355 & \multicolumn{2}{c}{$0.5 \pm 0.5 \pm 0.2$} & \nodata & $1.0 \pm 0.2 \pm 0.3$& $4.0 \pm 0.3 \pm 0.9$& \nodata \\ 
GRB060707$^{\rm{b}}$ & 3.4246 & \multicolumn{2}{c}{$1.8 \pm 1.0 \pm 0.6$} & \nodata & $-0.4 \pm 2.0 \pm 0.4$& $4 \pm 4 \pm 2$& \nodata \\ 
GRB060719$^{\rm{b}}$ & 1.5318 & \multicolumn{2}{c}{$1.4 \pm 0.4 \pm 0.2$} & $0.0 \pm 0.3 \pm 0.1$& \nodata & $1.7 \pm 0.7 \pm 0.4$& $0.5 \pm 0.2 \pm 0.1$\\ 
GRB060729$^{\rm{b}}$ & 0.5429 & \multicolumn{2}{c}{$1.2 \pm 1.0 \pm 0.2$} & $-0.2 \pm 0.5 \pm 0.1$& \nodata & $1.1 \pm 0.4 \pm 0.1$& \nodata \\ 
GRB060805A & 2.3633 & $0.7 \pm 0.4 \pm 0.2$& $1.4 \pm 0.6 \pm 0.4$& $0.2 \pm 0.4 \pm 0.1$& $1.1 \pm 0.2 \pm 0.3$& $5.0 \pm 0.6 \pm 1.1$& $-0.1 \pm 0.5 \pm 0.1$\\ 
GRB060814$^{\rm{b}}$ & 1.9223 & \multicolumn{2}{c}{$26.3 \pm 2.0 \pm 3.7$} & \nodata & $8.4 \pm 1.3 \pm 1.3$& $31 \pm 6 \pm 5$& \nodata \\ 
GRB060912A & 0.9362 & $7.4 \pm 0.5 \pm 0.7$& $10.7 \pm 0.5 \pm 0.9$& $1.1 \pm 0.3 \pm 0.1$& $3.5 \pm 0.6 \pm 0.3$& $12.2 \pm 1.2 \pm 1.1$& \nodata \\ 
GRB060923B & 1.5094 & \multicolumn{2}{c}{\nodata} & $0.1 \pm 0.1 \pm 0.1$& $-0.7 \pm 0.4 \pm 0.1$& $1.0 \pm 0.4 \pm 0.1$& $-0.2 \pm 0.2 \pm 0.1$\\ 
GRB060926 & 3.2090 & $0.4 \pm 0.2 \pm 0.2$& $0.4 \pm 0.3 \pm 0.2$& $0.1 \pm 0.1 \pm 0.1$& $2.2 \pm 0.4 \pm 1.0$& $4.0 \pm 0.6 \pm 1.6$& \nodata \\ 
GRB061021 & 0.3453 & $0.7 \pm 0.1 \pm 0.1$& $0.9 \pm 0.1 \pm 0.1$& $0.0 \pm 0.1 \pm 0.1$& $0.6 \pm 0.1 \pm 0.1$& $1.5 \pm 0.2 \pm 0.2$& $0.1 \pm 0.1 \pm 0.1$\\ 
GRB061110A$^{\rm{b}}$ & 0.7578 & \multicolumn{2}{c}{$0.5 \pm 0.4 \pm 0.2$} & $0.1 \pm 0.1 \pm 0.1$& $0.4 \pm 0.1 \pm 0.2$& $1.3 \pm 0.1 \pm 0.5$& $0.1 \pm 0.3 \pm 0.1$\\ 
GRB061202 & 2.2543 & $1.2 \pm 0.6 \pm 0.3$& $1.6 \pm 0.6 \pm 0.4$& \nodata & $1.3 \pm 0.2 \pm 0.3$& $3.4 \pm 0.3 \pm 0.7$& $-0.6 \pm 0.4 \pm 0.1$\\ 
GRB070103 & 2.6208 & \nodata & \nodata & \nodata & $5.3 \pm 1.1 \pm 1.3$& $10 \pm 3 \pm 3$& \nodata \\ 
GRB070110$^{\rm{b}}$ & 2.3523 & \multicolumn{2}{c}{$1.7 \pm 0.6 \pm 0.4$} & $0.1 \pm 0.5 \pm 0.1$& $1.9 \pm 0.2 \pm 0.4$& $5.9 \pm 0.3 \pm 1.2$& $1.0 \pm 0.5 \pm 0.3$\\ 
GRB070129 & 2.3384 & $1.8 \pm 0.6 \pm 0.4$& $2.1 \pm 0.7 \pm 0.4$& \nodata & $2.3 \pm 0.4 \pm 0.4$& $6.3 \pm 1.1 \pm 1.1$& $0.7 \pm 1.4 \pm 0.3$\\ 
GRB070224$^{\rm{b}}$ & 1.9922 & \multicolumn{2}{c}{$0.9 \pm 0.4 \pm 0.5$} & $0.3 \pm 0.5 \pm 0.3$& $0.5 \pm 0.2 \pm 0.3$& $1.0 \pm 0.2 \pm 0.6$& \nodata \\ 
GRB070306 & 1.4965 & $9.1 \pm 0.6 \pm 0.7$& $7.7 \pm 0.8 \pm 0.6$& $1.9 \pm 0.3 \pm 0.1$& $15.5 \pm 0.6 \pm 1.1$& $46.0 \pm 1.6 \pm 3.2$& $6.4 \pm 0.3 \pm 0.4$\\ 
GRB070318 & 0.8401 & $1.0 \pm 0.2 \pm 0.2$& $1.8 \pm 0.2 \pm 0.3$& $0.3 \pm 0.1 \pm 0.1$& $1.5 \pm 0.2 \pm 0.3$& $4.8 \pm 0.3 \pm 0.7$& $-0.3 \pm 0.5 \pm 0.1$\\ 
GRB070328 & 2.0627 & $0.7 \pm 0.4 \pm 0.2$& $0.5 \pm 0.4 \pm 0.2$& $0.5 \pm 0.3 \pm 0.1$& $0.5 \pm 0.2 \pm 0.1$& $2.2 \pm 0.4 \pm 0.4$& \nodata \\ 
GRB070419B$^{\rm{b}}$ & 1.9586 & \multicolumn{2}{c}{$1.7 \pm 0.4 \pm 0.4$} & \nodata & $1.0 \pm 0.3 \pm 0.3$& $2.2 \pm 0.6 \pm 0.6$& \nodata \\ 
GRB070521$^{\rm{b}}$ & 2.0865 & \multicolumn{2}{c}{$-2.0 \pm 2.0 \pm 0.2$} & \nodata & \nodata & $0.2 \pm 0.5 \pm 0.2$& $-0.2 \pm 0.9 \pm 0.2$\\ 
GRB070802 & 2.4538 & $1.3 \pm 0.3 \pm 0.3$& $2.1 \pm 0.3 \pm 0.4$& $0.3 \pm 0.2 \pm 0.1$& $1.9 \pm 0.1 \pm 0.4$& $4.7 \pm 0.3 \pm 0.9$& $1.2 \pm 0.9 \pm 0.4$\\ 
GRB071021 & 2.4515 & $1.3 \pm 0.8 \pm 0.3$& $2.1 \pm 0.7 \pm 0.4$& $0.3 \pm 0.8 \pm 0.2$& $2.2 \pm 0.5 \pm 0.4$& $5.0 \pm 0.6 \pm 0.9$& $0.2 \pm 1.0 \pm 0.2$\\ 
GRB080207$^{\rm{b}}$ & 2.0856 & \multicolumn{2}{c}{$1.0 \pm 1.0 \pm 0.4$} & $0.4 \pm 0.6 \pm 0.2$& $1.7 \pm 0.3 \pm 0.4$& $5.6 \pm 0.3 \pm 1.0$& $2.5 \pm 0.5 \pm 0.5$\\ 
GRB080413B & 1.1012 & $0.6 \pm 0.1 \pm 0.1$& $0.8 \pm 0.1 \pm 0.2$& $0.2 \pm 0.1 \pm 0.1$& \nodata & $2.8 \pm 0.5 \pm 0.7$& \nodata \\ 
GRB080602 & 1.8204 & $7.9 \pm 2.3 \pm 1.2$& $7.7 \pm 1.1 \pm 1.0$& $0.0 \pm 0.6 \pm 0.1$& \nodata & $17.0 \pm 2.0 \pm 2.2$& \nodata \\ 
GRB080605 & 1.6408 & $7.9 \pm 0.7 \pm 0.8$& $9.2 \pm 1.1 \pm 1.0$& \nodata & $10.3 \pm 0.8 \pm 1.6$& $29.6 \pm 1.5 \pm 4.4$& $4.0 \pm 0.4 \pm 0.6$\\ 
GRB080804$^{\rm{b}}$ & 2.2059 & \multicolumn{2}{c}{$0.9 \pm 0.9 \pm 0.4$} & $0.7 \pm 0.7 \pm 0.4$& \nodata & $4.9 \pm 0.9 \pm 1.6$& \nodata\\ 
GRB080805$^{\rm{b}}$ & 1.5052 & \multicolumn{2}{c}{$2.2 \pm 0.6 \pm 0.4$} & \nodata & \nodata & $3.5 \pm 0.8 \pm 0.9$& $0.8 \pm 0.2 \pm 0.2$\\ 
GRB081109 & 0.9785 & $6.4 \pm 0.5 \pm 0.8$& $10.8 \pm 0.5 \pm 1.4$& \nodata & \nodata & $5.7 \pm 0.6 \pm 0.8$& $5.7 \pm 0.5 \pm 0.8$\\ 
GRB081210 & 2.0631 & $0.9 \pm 0.5 \pm 0.3$& $1.2 \pm 0.5 \pm 0.3$& $0.8 \pm 0.5 \pm 0.2$& $1.7 \pm 0.3 \pm 0.4$& $4.3 \pm 0.5 \pm 0.9$& \nodata \\ 
GRB081221 & 2.2590 & $1.2 \pm 1.0 \pm 1.1$& $0.9 \pm 1.4 \pm 1.0$& \nodata & $1.1 \pm 0.4 \pm 0.7$& $3.6 \pm 0.9 \pm 2.1$& $1.3 \pm 1.0 \pm 1.1$\\ 
GRB090113 & 1.7494 & $7.0 \pm 2.9 \pm 1.5$& $7.9 \pm 2.9 \pm 1.6$& $0.7 \pm 1.5 \pm 0.3$& \nodata & \nodata & \nodata \\ 
GRB090201 & 2.1000 & $4.2 \pm 2.3 \pm 1.2$& $5.9 \pm 2.7 \pm 1.6$& $1.0 \pm 1.6 \pm 0.5$& $3.0 \pm 0.9 \pm 0.7$& $9.7 \pm 0.7 \pm 2.0$& $-0.4 \pm 1.2 \pm 0.2$\\ 
GRB090323 & 3.5832 & $0.8 \pm 0.3 \pm 0.2$& $1.0 \pm 0.2 \pm 0.2$& $0.1 \pm 0.3 \pm 0.1$& \nodata & \nodata & \nodata \\ 
GRB090407$^{\rm{b}}$ & 1.4478 & \multicolumn{2}{c}{$1.3 \pm 0.2 \pm 0.2$} & $0.2 \pm 0.1 \pm 0.1$& $0.0 \pm 0.3 \pm 0.1$& $1.0 \pm 0.4 \pm 0.2$& $1.1 \pm 0.2 \pm 0.2$\\ 
GRB090926B & 1.2427 & $4.8 \pm 0.3 \pm 0.7$& $6.2 \pm 0.3 \pm 0.9$& $1.1 \pm 0.2 \pm 0.2$& $4.4 \pm 1.3 \pm 0.8$& $13.8 \pm 1.5 \pm 2.1$& $1.3 \pm 1.2 \pm 0.3$\\ 
GRB091018$^{\rm{b}}$ & 0.9710 & \multicolumn{2}{c}{$4.2 \pm 0.8 \pm 0.4$} & \nodata & $1.6 \pm 0.5 \pm 0.1$& $4.1 \pm 0.9 \pm 0.3$& $1.3 \pm 0.5 \pm 0.1$\\ 
GRB091127 & 0.4904 & $3.0 \pm 0.5 \pm 0.3$& $4.5 \pm 0.5 \pm 0.4$& $0.3 \pm 0.4 \pm 0.1$& $2.6 \pm 0.2 \pm 0.2$& $8.0 \pm 0.4 \pm 0.7$& $0.4 \pm 0.3 \pm 0.1$\\ 
GRB100316D & 0.0592 & $63 \pm 3$& $92 \pm 5$& $18.1 \pm 0.9$& $83 \pm 4$& $253 \pm 13$& $12.2 \pm 0.6$\\ 
GRB100418A$^{\rm{c}}$ & 0.6235 & \multicolumn{2}{c}{$33.6 \pm 1.8 \pm 3.4$} & $1.4 \pm 0.6 \pm 0.2$& $8.9 \pm 0.6 \pm 0.9$& $26.3 \pm 1.5 \pm 2.7$& $4.1 \pm 1.8 \pm 0.6$\\ 
GRB100424A & 2.4656 & $1.1 \pm 0.6 \pm 0.4$& $2.4 \pm 0.4 \pm 0.7$& $1.4 \pm 0.4 \pm 0.4$& $4.5 \pm 0.3 \pm 1.1$& $12.5 \pm 0.7 \pm 3.1$& \nodata \\ 
GRB100508A & 0.5201 & $9.2 \pm 2.2 \pm 0.7$& $11.7 \pm 1.8 \pm 0.8$& $3.0 \pm 1.3 \pm 0.3$& $2.7 \pm 0.7 \pm 0.2$& \nodata & $4.7 \pm 1.5 \pm 0.4$\\ 
GRB100606A & 1.5545 & \nodata & \nodata & \nodata & \nodata & \nodata & $0.9 \pm 0.3 \pm 0.3$\\ 
GRB100615A & 1.3978 & $1.4 \pm 0.1 \pm 0.2$& $1.8 \pm 0.1 \pm 0.3$& $0.4 \pm 0.1 \pm 0.1$& $3.0 \pm 0.4 \pm 0.5$& $6.5 \pm 0.5 \pm 1.0$& $0.6 \pm 0.3 \pm 0.1$\\ 
GRB100621A & 0.5426 & $41 \pm 2 \pm 2$& $50 \pm 2 \pm 3$& $12.4 \pm 1.1 \pm 0.7$& $51 \pm 4 \pm 3$& $150 \pm 3 \pm 8$& $7.6 \pm 11.2 \pm 1.0$\\ 
GRB100724A & 1.2890 & $0.8 \pm 0.1 \pm 0.2$& $1.4 \pm 0.1 \pm 0.4$& $0.1 \pm 0.1 \pm 0.1$& \nodata & $2.3 \pm 0.9 \pm 0.8$& $0.3 \pm 0.4 \pm 0.2$\\ 
GRB100728A & 1.5670 & $1.6 \pm 0.4 \pm 0.8$& $3.3 \pm 0.5 \pm 1.5$& $0.3 \pm 0.7 \pm 0.4$& \nodata & $3.4 \pm 1.6 \pm 1.9$& \nodata \\ 
GRB100814A & 1.4392 & $0.9 \pm 0.1 \pm 0.1$& $1.4 \pm 0.1 \pm 0.2$& $0.7 \pm 0.1 \pm 0.1$& $2.0 \pm 0.4 \pm 0.3$& $5.5 \pm 0.7 \pm 0.6$& $0.3 \pm 0.2 \pm 0.1$\\ 
GRB100816A$^{\rm{b}}$ & 0.8048 & \multicolumn{2}{c}{$5.2 \pm 1.0 \pm 0.5$} & $0.5 \pm 1.3 \pm 0.1$& $0.3 \pm 0.9 \pm 0.1$& $1.5 \pm 0.6 \pm 0.2$& $6.6 \pm 2.2 \pm 1.4$\\ 
GRB110808A & 1.3490 & $1.7 \pm 0.3 \pm 0.3$& $3.2 \pm 0.7 \pm 0.5$& $1.6 \pm 0.2 \pm 0.3$& \nodata & $14.3 \pm 1.0 \pm 2.1$& $-0.1 \pm 0.3 \pm 0.1$\\ 
GRB110818A & 3.3609 & $2.1 \pm 0.4 \pm 0.3$& $2.3 \pm 0.3 \pm 0.3$& $0.6 \pm 0.2 \pm 0.1$& $3.7 \pm 0.5 \pm 0.5$& $11.0 \pm 0.9 \pm 1.5$& \nodata \\ 
GRB110918A$^{\rm{c}}$ & 0.9843 & \multicolumn{2}{c}{$17.6 \pm 2.7 \pm 1.0$} & $2.0 \pm 2.7 \pm 0.2$& $2.6 \pm 2.2 \pm 0.2$& $7.1 \pm 1.9 \pm 0.4$& $14.1 \pm 4.0 \pm 1.2$\\ 
GRB111123A & 3.1513 & $4.0 \pm 0.5 \pm 0.6$& $3.2 \pm 0.4 \pm 0.5$& $1.8 \pm 0.4 \pm 0.3$& $1.3 \pm 1.0 \pm 0.3$& $7.9 \pm 1.6 \pm 1.3$& \nodata \\ 
GRB111129A$^{\rm{b}}$ & 1.0796 & \multicolumn{2}{c}{$7.7 \pm 1.8 \pm 0.8$} & $-0.4 \pm 0.5 \pm 0.1$& \nodata & $7.7 \pm 3.5 \pm 0.9$& \nodata \\ 
GRB111209A & 0.6770 & $1.1 \pm 0.2 \pm 0.2$& $1.4 \pm 0.2 \pm 0.2$& $0.3 \pm 0.1 \pm 0.1$& $1.1 \pm 0.2 \pm 0.1$& $2.9 \pm 0.2 \pm 0.3$& $-0.2 \pm 0.3 \pm 0.1$\\ 
GRB111211A & 0.4786 & $1.6 \pm 0.9 \pm 0.2$& $0.7 \pm 0.8 \pm 0.1$& $-0.4 \pm 0.8 \pm 0.1$& $0.7 \pm 0.4 \pm 0.1$& $2.7 \pm 0.7 \pm 0.2$& \nodata \\ 
GRB111228A & 0.7164 & $0.7 \pm 0.3 \pm 0.1$& $1.2 \pm 0.3 \pm 0.1$& $-0.1 \pm 0.2 \pm 0.1$& $0.8 \pm 0.2 \pm 0.1$& $1.2 \pm 0.3 \pm 0.1$& \nodata \\ 
GRB120118B & 2.9428 & $2.5 \pm 0.6 \pm 0.4$& $4.2 \pm 1.5 \pm 0.8$& $2.4 \pm 0.3 \pm 0.4$& \nodata & $32 \pm 3 \pm 5$& \nodata \\ 
GRB120119A$^{\rm{b}}$ & 1.7291 & \multicolumn{2}{c}{$1.6 \pm 2.6 \pm 0.5$} & $0.8 \pm 0.8 \pm 0.3$& \nodata & \nodata & $2.6 \pm 0.3 \pm 0.5$\\ 
GRB120422A & 0.2826 & $28.7 \pm 0.8 \pm 1.1$& $42.8 \pm 0.9 \pm 1.6$& $2.7 \pm 0.3 \pm 0.1$& $8.6 \pm 0.5 \pm 0.3$& $30.6 \pm 1.2 \pm 1.1$& $8.4 \pm 0.3 \pm 0.3$\\ 
GRB120624B & 2.1974 & $4.7 \pm 0.9 \pm 1.6$& $3.8 \pm 0.8 \pm 1.3$& $1.1 \pm 0.8 \pm 0.5$& $6.0 \pm 1.5 \pm 2.1$& $16.7 \pm 1.3 \pm 5.1$& $2.7 \pm 1.4 \pm 1.2$\\ 
GRB120714B & 0.3985 & $4.1 \pm 0.4 \pm 0.3$& $5.7 \pm 0.3 \pm 0.4$& $1.1 \pm 0.2 \pm 0.1$& $2.9 \pm 0.3 \pm 0.2$& $7.7 \pm 0.5 \pm 0.5$& $0.6 \pm 0.2 \pm 0.1$\\ 
GRB120722A & 0.9590 & $10.3 \pm 0.4 \pm 1.1$& $11.7 \pm 0.4 \pm 1.2$& $1.2 \pm 0.2 \pm 0.2$& $7.1 \pm 0.8 \pm 0.8$& $27.2 \pm 1.3 \pm 2.9$& $4.1 \pm 0.8 \pm 0.5$\\ 
GRB120815A$^{\rm{b}}$ & 2.3587 & \multicolumn{2}{c}{$0.5 \pm 0.4 \pm 0.2$} & $0.3 \pm 0.2 \pm 0.1$& $0.6 \pm 0.1 \pm 0.2$& $1.5 \pm 0.1 \pm 0.4$& $0.1 \pm 0.2 \pm 0.1$\\ 
GRB121024A & 2.3012 & $6.7 \pm 0.7 \pm 0.6$& $7.2 \pm 0.7 \pm 0.6$& \nodata & $12.2 \pm 0.6 \pm 1.0$& $31.9 \pm 1.3 \pm 2.6$& $1.8 \pm 0.9 \pm 0.2$\\ 
GRB121027A & 1.7732 & $4.0 \pm 3.2 \pm 0.6$& $4.2 \pm 2.1 \pm 0.5$& \nodata & $3.5 \pm 1.0 \pm 0.4$& $14.2 \pm 3.1 \pm 1.5$& \nodata \\ 
GRB121201A & 3.3830 & $1.0 \pm 0.4 \pm 0.1$ & $1.7 \pm 0.7 \pm 0.2$ & $1.6 \pm 0.7 \pm 0.2$ & \nodata & \nodata & \nodata\\ 
GRB130131B & 2.5393 & $0.7 \pm 0.3 \pm 0.2$& $0.6 \pm 0.4 \pm 0.2$& \nodata & $0.7 \pm 0.3 \pm 0.2$& $1.5 \pm 0.2 \pm 0.3$& \nodata \\ 
GRB130427A & 0.3401 & $9.1 \pm 1.7 \pm 0.7$& $11.1 \pm 1.5 \pm 0.9$& $0.0 \pm 1.1 \pm 0.1$& $0.9 \pm 0.8 \pm 0.1$& $6.5 \pm 1.0 \pm 0.5$& $1.5 \pm 0.6 \pm 0.2$\\ 
GRB130701A$^{\rm{b}}$ & 1.1548 & \multicolumn{2}{c}{$1.3 \pm 1.8 \pm 0.2$}  & $0.4 \pm 0.9 \pm 0.1$& $0.8 \pm 4.7 \pm 0.3$& $11.0 \pm 6.6 \pm 1.1$& \nodata \\ 
GRB130925A$^{\rm{c}}$ & 0.3483 & \multicolumn{2}{c}{$25.8 \pm 0.8 \pm 1.6$} & $1.2 \pm 0.7 \pm 0.1$& $5.6 \pm 0.8 \pm 0.3$& $18.7 \pm 0.7 \pm 0.9$& $11.7 \pm 0.5 \pm 0.6$\\ 
GRB131103A$^{\rm{c}}$ & 0.5960 & \multicolumn{2}{c}{$40.4 \pm 4.0 \pm 5.9$} & $5.5 \pm 3.6 \pm 1.2$& $22.8 \pm 1.6 \pm 3.3$& $66 \pm 4 \pm 11$& $2.6 \pm 1.2 \pm 0.5$\\ 
GRB131105A$^{\rm{b}}$ & 1.6854 & \multicolumn{2}{c}{$7.5 \pm 1.6 \pm 1.0$} & $1.1 \pm 0.8 \pm 0.4$& $0.8 \pm 0.8 \pm 0.3$& $3.2 \pm 1.6 \pm 1.0$& $1.4 \pm 0.5 \pm 0.4$\\ 
GRB131231A & 0.6427 & $2.9 \pm 0.6 \pm 0.3$& $4.4 \pm 0.6 \pm 0.4$& $1.0 \pm 0.4 \pm 0.1$& $6.6 \pm 0.5 \pm 0.6$& $17.0 \pm 0.9 \pm 1.4$& $1.5 \pm 0.5 \pm 0.2$\\ 
GRB140213A & 1.2079 & \nodata & \nodata & \nodata & $1.0 \pm 0.8 \pm 0.2$& $0.8 \pm 0.5 \pm 0.1$& \nodata \\ 
GRB140301A & 1.4155 & $3.3 \pm 0.5 \pm 0.3$& $3.9 \pm 0.3 \pm 0.3$& \nodata & $1.2 \pm 0.8 \pm 0.2$& $4.1 \pm 0.7 \pm 0.4$& $10.9 \pm 0.5 \pm 0.9$\\ 
GRB140430A$^{\rm{b}}$ & 1.6019 & \multicolumn{2}{c}{$1.8 \pm 1.2 \pm 0.2$} & \nodata & \nodata & $10.7 \pm 1.8 \pm 1.0$& $1.3 \pm 0.6 \pm 0.2$\\ 
GRB140506A & 0.8893 & $0.6 \pm 0.1 \pm 0.1$& $0.7 \pm 0.1 \pm 0.1$& $0.1 \pm 0.1 \pm 0.1$& \nodata & \nodata & $-0.0 \pm 0.2 \pm 0.1$\\ 

\hline\noalign{\smallskip}
\end{longtable}
\centering
\begin{minipage}{8.0in}
\tablefoot{Measurements are in units of $10^{-17}$\,\erg, and are corrected for the Galactic foreground reddening. A correction for slit-loss based on broad-band photometry has been applied to the measurements as described in the text. These values thus represent host-integrated measurements except for GRB~100316D (see below). Line fluxes are not corrected for host-intrinsic extinction. The first error represents the statistical error due to photon statistics and line-flux measurement. The second error is the systematic error in the absolute flux calibration due to slit-loss and scaling to photometry. Redshifts are given in a heliocentric reference frame. No data means that either the wavelength range of the respective line is not covered, all data in that wavelength range have been excluded from the automated fitting procedure, or no meaningful constraints could be obtained for the given line.\\
$^{a}$ GRB~100316D has the lowest redshift in the sample ($z=0.0592$). The fraction of the host that is covered by the slit is too small to derive representative host-integrated measurements. These values are thus as derived from the observed spectrum and not scaled by photometry. No systematic error on the line fluxes is given in this case.\\
$^{b}$ In these cases, we can not constrain the individual components of the  \oii\,doublet well, either because of telluric absorption, skylines or low signal-to-noise level. We constrain the ratio between the two components to be between \nicefrac{2}{3} and \nicefrac{3}{2}, and provide the measurement for the combined fit of \oii($\lambda$3726) and \oii($\lambda$3729).\\
$^{c}$ These galaxies have non-trivial velocity structure. We are not able to uniquely disentangle the individual components of the \oii\,doublet, so we provide the total \oii\,flux instead.}
\end{minipage}
\end{landscape}
\end{longtab}

We have not corrected the Balmer line fluxes for the underlying Balmer stellar absorption even though we note that it can have a measurable influence on the line fluxes and thus the derived physical properties of the emitting gas \citep{2011MNRAS.414.2793W}. The S/N of the emission lines and stellar continuum is not high enough to resolve and fit the absorption component separately. Generally, the Balmer absorption correction is small for galaxies with masses below $10^{10}$\,\Msun. 

At $z\sim0.8$, for example, it has an equivalent width (EW) of 1\,\AA\,for galaxies of $10^{10}$\,\Msun\, at \hb\,\citep{2011ApJ...730..137Z}. The EW of the \ha\, line is high enough $(\rm{EW}_{\rm{rest}} > 20\,\AA)$ to make the correction negligible, but it can have an effect on the other Balmer line-fluxes. We estimate $\lesssim$5\% for \hb\, on average, well below the average statistical error of the line flux measurements (Table~\ref{tab:balmerlines}).
 
In cases where \oiii$\lambda$5007 is not well constrained by the data, but \oiii$\lambda$4959 is, or vice versa, we assume that $f_{[\ion{O}{iii}(\lambda 4959)]} = \nicefrac{1}{3} \cdot f_{[\ion{O}{iii}(\lambda 5007)]}$ \citep{2000MNRAS.312..813S}.

\subsection{Emission-line widths}

We measure the intrinsic line-width ($\sigma$) of the hot gas using nebular lines of highest S/N, usually \ha, or \oiii. If the line-shape is adequately represented with a Gaussian function, we use the instrumental resolution $\Delta$V and subtract it quadratically from the fitted FWHM, giving $\sigma$ as $\sigma = \sqrt{\rm{FWHM}^2 - \Delta V^2} / \left(2\sqrt{2\ln 2}\right)$. If the line-shape is clearly asymmetric, we directly measure FWHM in the data. Errors on $\sigma$ are derived via the error-spectrum and propagated accordingly.

The physical line-width is well-resolved through the medium resolution of X-Shooter (typically $\Delta V \sim 35\,$\kms\, in the visual arm, or $\Delta V \sim 50\,$\kms\, in the NIR arm). The line-width measurements are provided in Table~\ref{tab:physprop}.

\begin{longtab}
\begin{longtable}{c c c c c c c}
\caption{Physical properties of GRB hosts\label{tab:physprop}}\\
\hline\hline
{GRB host$^{a}$} & {Redshift} & $E_{B-V}$ & SF-tracer & SFR$^{b}$  & $\sigma$ & $Z$ \\[1.5pt]
\hline
{}         &    {}      & (mag)     &   & (\Msunyr) & (\kms) & $12+\log(\rm{O/H})$ \\[1.5pt]
\hline
\endfirsthead
\hline 
\caption{Physical properties of GRB hosts (continued)}\\
\hline\hline
{GRB host} & {Redshift} & $E_{B-V}$ & SF-tracer & SFR$^{a}$  & $\sigma$ & $Z$ \\[1.5pt]
\hline
{}         &    {}      & (mag)     &    & (\Msunyr) & (\kms) & $12+\log(\rm{O/H})$ \\[1.5pt]
\hline
\endhead
GRB050416A & 0.6542 & $0.49_{-0.11}^{+0.11}$ & \ha & $4.5_{-1.2}^{+1.6}$ & $47\pm{4}$ & $8.46_{-0.11}^{+0.11}$ \\ [1.5pt] 
GRB050525A & 0.6063 & $0.10_{-0.10}^{+0.62}$ & \ha & $0.07_{-0.05}^{+0.21}$ & $26\pm{5}$ & \nodata \\ [1.5pt] 
GRB050714B & 2.4383 & $0.21_{-0.21}^{+0.28}$ & \ha & $12.9_{-5.3}^{+14.0}$ & $35\pm{16}$ & \nodata \\ [1.5pt] 
GRB050819A & 2.5042 & $0.34_{-0.34}^{+0.84}$ & \hb & $22_{-15}^{+426}$ & $50\pm{6}$ & \nodata \\ [1.5pt] 
GRB050824 & 0.8277 & $0.00_{-0.00}^{+0.07}$ & \ha & $1.20_{-0.26}^{+0.30}$ & $48\pm{5}$ & $8.11_{-0.20}^{+0.18}$ \\ [1.5pt] 
GRB050915A & 2.5275 & $0.84_{-0.60}^{+0.61}$ & \hb & $196_{-174}^{+1563}$ & $87\pm{10}$ & \nodata \\ [1.5pt] 
GRB051001 & 2.4295 & $0.58_{-0.28}^{+0.28}$ & \ha & $110_{-59}^{+124}$ & $67\pm{8}$ & \nodata \\ [1.5pt] 
GRB051016B & 0.9358 & $0.05_{-0.05}^{+0.07}$ & \ha & $10.2_{-2.0}^{+2.6}$ & $58\pm{4}$ & $8.27_{-0.20}^{+0.15}$ \\ [1.5pt] 
GRB051022A & 0.8061 & $0.56_{-0.04}^{+0.05}$ & \ha & $60_{-10}^{+12}$ & $88\pm{5}$ & $8.49_{-0.09}^{+0.09}$ \\ [1.5pt] 
GRB051117B & 0.4805 & $0.72_{-0.24}^{+0.27}$ & \ha & $4.7_{-2.2}^{+4.9}$ & $85\pm{11}$ & $9.00_{-0.16}^{+0.16}$ \\ [1.5pt] 
GRB060204B & 2.3393 & $0.34_{-0.23}^{+0.29}$ & \ha & $78_{-34}^{+85}$ & $85\pm{8}$ & \nodata \\ [1.5pt] 
GRB060306 & 1.5597 & $0.44_{-0.40}^{+0.69}$ & \ha & $17.6_{-11.0}^{+83.6}$ & $60\pm{14}$ & \nodata \\ [1.5pt] 
GRB060604 & 2.1355 & $0.38_{-0.27}^{+0.32}$ & \ha & $7.2_{-3.6}^{+9.4}$ & $68\pm{13}$ & $8.10_{-0.35}^{+0.28}$ \\ [1.5pt] 
GRB060707 & 3.4246 & \nodata & \oii & $19.9_{-14.3}^{+48.0}$ & \nodata & \nodata \\ [1.5pt] 
GRB060719 & 1.5318 & $0.40_{-0.36}^{+0.52}$ & \ha & $7.1_{-3.9}^{+18.9}$ & $42\pm{5}$ & $8.61_{-0.24}^{+0.20}$ \\ [1.5pt] 
GRB060729 & 0.5429 & $0.71_{-0.47}^{+0.43}$ & \ha & $0.96_{-0.69}^{+2.21}$ & $66\pm{16}$ & \nodata \\ [1.5pt] 
GRB060805A & 2.3633 & $0.00_{-0.00}^{+0.16}$ & \ha & $9.0_{-2.5}^{+3.9}$ & $71\pm{7}$ & \nodata \\ [1.5pt] 
GRB060814 & 1.9223 & $0.17_{-0.17}^{+0.39}$ & \ha & $54_{-19}^{+89}$ & $132\pm{11}$ & \nodata \\ [1.5pt] 
GRB060912A & 0.9362 & $0.16_{-0.09}^{+0.10}$ & \ha & $5.1_{-1.6}^{+2.1}$ & $62\pm{5}$ & $8.61_{-0.12}^{+0.11}$ \\ [1.5pt] 
GRB060923B & 1.5094 & \nodata & \ha & $3.0_{-1.5}^{+2.9}$ & $46\pm{15}$ & \nodata \\ [1.5pt] 
GRB060926 & 3.2090 & \nodata & \oiii & $26_{-17}^{+47}$ & $60\pm{8}$ & \nodata \\ [1.5pt] 
GRB061021 & 0.3453 & $0.11_{-0.11}^{+0.20}$ & \ha & $0.05_{-0.01}^{+0.03}$ & $16.1\pm{4.8}$ & \nodata \\ [1.5pt] 
GRB061110A & 0.7578 & \nodata & \hb & $0.23_{-0.15}^{+0.38}$ & $31\pm{4}$ & \nodata \\ [1.5pt] 
GRB061202 & 2.2543 & $0.58_{-0.27}^{+0.34}$ & \ha & $43_{-22}^{+60}$ & $64\pm{7}$ & \nodata \\ [1.5pt] 
GRB070103 & 2.6208 & $0.00_{-0.00}^{+0.50}$ & \hb & $43_{-17}^{+162}$ & $124\pm{30}$ & \nodata \\ [1.5pt] 
GRB070110 & 2.3523 & $0.00_{-0.00}^{+0.38}$ & \ha & $8.9_{-2.8}^{+10.9}$ & $24\pm{4}$ & \nodata \\ [1.5pt] 
GRB070129 & 2.3384 & $0.17_{-0.17}^{+0.35}$ & \ha & $20_{-7}^{+28}$ & $76\pm{11}$ & \nodata \\ [1.5pt] 
GRB070224 & 1.9922 & \nodata & \oiii & $3.2_{-2.3}^{+6.5}$ & $37\pm{9}$ & \nodata \\ [1.5pt] 
GRB070306 & 1.4965 & $0.43_{-0.07}^{+0.08}$ & \ha & $101_{-18}^{+24}$ & $121\pm{55}$ & $8.54_{-0.09}^{+0.09}$ \\ [1.5pt] 
GRB070318 & 0.8401 & $0.15_{-0.14}^{+0.16}$ & \ha & $0.79_{-0.24}^{+0.44}$ & $53\pm{5}$ & \nodata \\ [1.5pt] 
GRB070328 & 2.0627 & $0.16_{-0.16}^{+0.79}$ & \hb & $8.4_{-4.2}^{+130.7}$ & $93\pm{14}$ & \nodata \\ [1.5pt] 
GRB070419B & 1.9586 & $0.56_{-0.30}^{+0.39}$ & \ha & $21_{-11}^{+35}$ & $86\pm{10}$ & \nodata \\ [1.5pt] 
GRB070521 & 2.0865 & \nodata & \ha & $26_{-17}^{+34}$ & $249\pm{108}$ & \nodata \\ [1.5pt] 
GRB070802 & 2.4538 & $0.31_{-0.12}^{+0.12}$ & \ha & $24_{-8}^{+11}$ & $57\pm{5}$ & \nodata \\ [1.5pt] 
GRB071021 & 2.4515 & $0.19_{-0.17}^{+0.16}$ & \ha & $32_{-12}^{+20}$ & $100\pm{17}$ & \nodata \\ [1.5pt] 
GRB080207 & 2.0856 & $0.66_{-0.25}^{+0.28}$ & \ha & $77_{-38}^{+86}$ & $136\pm{18}$ & $8.74_{-0.15}^{+0.15}$ \\ [1.5pt] 
GRB080413B & 1.1012 & $0.43_{-0.32}^{+0.30}$ & \ha & $2.1_{-1.2}^{+3.1}$ & $39\pm{5}$ & $8.29_{-0.30}^{+0.32}$ \\ [1.5pt] 
GRB080602 & 1.8204 & $0.58_{-0.26}^{+0.29}$ & \ha & $125_{-65}^{+145}$ & $91\pm{13}$ & \nodata \\ [1.5pt] 
GRB080605 & 1.6408 & $0.26_{-0.10}^{+0.11}$ & \ha & $47_{-12}^{+17}$ & $80\pm{6}$ & $8.54_{-0.09}^{+0.09}$ \\ [1.5pt] 
GRB080804 & 2.2059 & $0.38_{-0.35}^{+0.51}$ & \ha & $15.2_{-8.7}^{+41.2}$ & $50\pm{9}$ & \nodata \\ [1.5pt] 
GRB080805 & 1.5052 & $0.78_{-0.31}^{+0.39}$ & \ha & $45_{-26}^{+79}$ & $54\pm{12}$ & $8.49_{-0.14}^{+0.13}$ \\ [1.5pt] 
GRB081109 & 0.9785 & $0.36_{-0.10}^{+0.11}$ & \ha & $11.8_{-2.9}^{+4.1}$ & $108\pm{6}$ & $8.75_{-0.09}^{+0.09}$ \\ [1.5pt] 
GRB081210 & 2.0631 & $0.13_{-0.13}^{+0.61}$ & \hb & $15.3_{-7.0}^{+111.7}$ & $118\pm{12}$ & \nodata \\ [1.5pt] 
GRB081221 & 2.2590 & $0.31_{-0.31}^{+0.55}$ & \ha & $35_{-22}^{+106}$ & $93\pm{12}$ & \nodata \\ [1.5pt] 
GRB090113 & 1.7494 & $0.01_{-0.01}^{+0.22}$ & \ha & $17.9_{-4.8}^{+10.1}$ & $70\pm{9}$ & \nodata \\ [1.5pt] 
GRB090201 & 2.1000 & $0.11_{-0.11}^{+0.19}$ & \ha & $48_{-14}^{+30}$ & $171\pm{12}$ & \nodata \\ [1.5pt] 
GRB090323 & 3.5832 & \nodata & \oii & $24_{-17}^{+53}$ & $60\pm{13}$ & \nodata \\ [1.5pt] 
GRB090407 & 1.4478 & $0.69_{-0.26}^{+0.34}$ & \ha & $13.8_{-6.7}^{+18.8}$ & $109\pm{8}$ & $8.85_{-0.13}^{+0.13}$ \\ [1.5pt] 
GRB090926B & 1.2427 & $0.63_{-0.18}^{+0.20}$ & \ha & $26_{-11}^{+19}$ & $65\pm{4}$ & $8.34_{-0.17}^{+0.15}$ \\ [1.5pt] 
GRB091018 & 0.9710 & $0.06_{-0.06}^{+0.56}$ & \ha & $1.29_{-0.32}^{+3.46}$ & $57\pm{10}$ & $8.78_{-0.19}^{+0.18}$ \\ [1.5pt] 
GRB091127 & 0.4904 & $0.16_{-0.08}^{+0.09}$ & \ha & $0.37_{-0.07}^{+0.10}$ & $30\pm{5}$ & $8.07_{-0.20}^{+0.18}$ \\ [1.5pt] 
GRB100418A & 0.6235 & $0.17_{-0.07}^{+0.06}$ & \ha & $4.2_{-0.8}^{+1.0}$ & $56\pm{4}$ & $8.52_{-0.10}^{+0.10}$ \\ [1.5pt] 
GRB100424A & 2.4656 & $0.13_{-0.13}^{+0.18}$ & \hb & $21_{-8}^{+20}$ & $87\pm{5}$ & $7.93_{-0.18}^{+0.25}$ \\ [1.5pt] 
GRB100508A & 0.5201 & $0.29_{-0.09}^{+0.09}$ & \ha & $2.6_{-0.5}^{+0.7}$ & $80\pm{11}$ & $8.68_{-0.10}^{+0.10}$ \\ [1.5pt] 
GRB100606A & 1.5545 & $0.05_{-0.05}^{+0.60}$ & \ha & $4.9_{-1.8}^{+12.9}$ & $107\pm{36}$ & $8.71_{-0.21}^{+0.19}$ \\ [1.5pt] 
GRB100615A & 1.3978 & $0.48_{-0.28}^{+0.38}$ & \ha & $8.6_{-4.4}^{+13.9}$ & $45\pm{5}$ & $8.40_{-0.13}^{+0.12}$ \\ [1.5pt] 
GRB100621A & 0.5426 & $0.05_{-0.03}^{+0.03}$ & \ha & $8.7_{-0.8}^{+0.8}$ & $82\pm{4}$ & $8.52_{-0.10}^{+0.10}$ \\ [1.5pt] 
GRB100724A & 1.2890 & $0.24_{-0.24}^{+0.37}$ & \ha & $3.2_{-1.4}^{+5.1}$ & $58\pm{7}$ & \nodata \\ [1.5pt] 
GRB100728A & 1.5670 & $0.23_{-0.23}^{+0.69}$ & \ha & $14.5_{-8.0}^{+60.6}$ & $57\pm{8}$ & \nodata \\ [1.5pt] 
GRB100814A & 1.4392 & $0.08_{-0.08}^{+0.26}$ & \ha & $3.2_{-0.7}^{+2.9}$ & $31\pm{5}$ & \nodata \\ [1.5pt] 
GRB100816A & 0.8048 & $1.32_{-0.22}^{+0.24}$ & \ha & $58_{-26}^{+51}$ & $111\pm{30}$ & $8.75_{-0.18}^{+0.16}$ \\ [1.5pt] 
GRB110808A & 1.3490 & $0.30_{-0.25}^{+0.34}$ & \ha & $8.3_{-3.6}^{+11.3}$ & $40\pm{4}$ & $7.93_{-0.23}^{+0.31}$ \\ [1.5pt] 
GRB110818A & 3.3609 & \nodata & \hb & $44_{-26}^{+62}$ & $89\pm{8}$ & $8.25_{-0.25}^{+0.17}$ \\ [1.5pt] 
GRB110918A & 0.9843 & $0.35_{-0.31}^{+0.31}$ & \ha & $23_{-11}^{+28}$ & $126\pm{18}$ & $8.93_{-0.11}^{+0.11}$ \\ [1.5pt] 
GRB111123A & 3.1513 & \nodata & \oii & $77_{-52}^{+163}$ & $135\pm{21}$ & $8.01_{-0.28}^{+0.28}$ \\ [1.5pt] 
GRB111129A & 1.0796 & \nodata & \oii & $5.1_{-3.4}^{+10.8}$ & $117\pm{35}$ & \nodata \\ [1.5pt] 
GRB111209A & 0.6770 & $0.16_{-0.16}^{+0.20}$ & \ha & $0.35_{-0.13}^{+0.26}$ & $35\pm{5}$ & $7.95_{-0.17}^{+0.30}$ \\ [1.5pt] 
GRB111211A & 0.4786 & $0.00_{-0.00}^{+0.57}$ & \ha & $0.12_{-0.03}^{+0.29}$ & $38\pm{8}$ & \nodata \\ [1.5pt] 
GRB111228A & 0.7164 & \nodata & \hb & $0.32_{-0.22}^{+0.56}$ & $19.7\pm{5.5}$ & \nodata \\ [1.5pt] 
GRB120118B & 2.9428 & $0.00_{-0.00}^{+0.16}$ & \hb & $28_{-11}^{+21}$ & $193\pm{8}$ & $7.89_{-0.17}^{+0.23}$ \\ [1.5pt] 
GRB120119A & 1.7291 & $0.35_{-0.14}^{+0.16}$ & \ha & $43_{-14}^{+24}$ & $104\pm{17}$ & $8.60_{-0.14}^{+0.14}$ \\ [1.5pt] 
GRB120422A & 0.2826 & $0.27_{-0.03}^{+0.03}$ & \ha & $1.38_{-0.12}^{+0.13}$ & $25\pm{4}$ & $8.39_{-0.09}^{+0.09}$ \\ [1.5pt] 
GRB120624B & 2.1974 & $0.21_{-0.21}^{+0.50}$ & \ha & $30_{-13}^{+73}$ & $77\pm{6}$ & $8.43_{-0.27}^{+0.20}$ \\ [1.5pt] 
GRB120714B & 0.3985 & $0.10_{-0.08}^{+0.08}$ & \ha & $0.27_{-0.05}^{+0.07}$ & $34\pm{4}$ & $8.39_{-0.11}^{+0.11}$ \\ [1.5pt] 
GRB120722A & 0.9590 & $0.46_{-0.05}^{+0.05}$ & \ha & $22_{-4}^{+4}$ & $56\pm{4}$ & $8.48_{-0.10}^{+0.10}$ \\ [1.5pt] 
GRB120815A & 2.3587 & $0.06_{-0.06}^{+0.34}$ & \ha & $2.3_{-1.0}^{+2.7}$ & $28\pm{5}$ & \nodata \\ [1.5pt] 
GRB121024A & 2.3012 & $0.00_{-0.00}^{+0.12}$ & \ha & $37_{-4}^{+4}$ & $88\pm{4}$ & $8.41_{-0.12}^{+0.11}$ \\ [1.5pt] 
GRB121027A & 1.7732 & \nodata & \oiii & $24_{-15}^{+41}$ & $119\pm{75}$ & \nodata \\ [1.5pt] 
GRB121201A & 3.3830 & \nodata & \oii & $30_{-21}^{+68}$ & $86\pm{17}$ & \nodata \\ [1.5pt] 
GRB130131B & 2.5393 & \nodata & \oiii & $8.0_{-5.0}^{+13.4}$ & $73\pm{29}$ & \nodata \\ [1.5pt] 
GRB130427A & 0.3401 & $0.06_{-0.06}^{+0.19}$ & \ha & $0.34_{-0.06}^{+0.20}$ & $40\pm{5}$ & $8.57_{-0.13}^{+0.12}$ \\ [1.5pt] 
GRB130701A & 1.1548 & \nodata & \oii & $0.78_{-0.60}^{+2.03}$ & $82\pm{42}$ & \nodata \\ [1.5pt] 
GRB130925A & 0.3483 & $0.41_{-0.06}^{+0.06}$ & \ha & $2.9_{-0.4}^{+0.5}$ & $49\pm{5}$ & $8.73_{-0.08}^{+0.08}$ \\ [1.5pt] 
GRB131103A & 0.5960 & $0.06_{-0.06}^{+0.07}$ & \ha & $4.4_{-0.9}^{+1.2}$ & $87\pm{7}$ & $8.48_{-0.12}^{+0.10}$ \\ [1.5pt] 
GRB131105A & 1.6854 & $0.53_{-0.18}^{+0.21}$ & \ha & $31_{-13}^{+25}$ & $52\pm{11}$ & $8.61_{-0.20}^{+0.17}$ \\ [1.5pt] 
GRB131231A & 0.6427 & $0.02_{-0.02}^{+0.08}$ & \ha & $1.38_{-0.20}^{+0.28}$ & $33\pm{4}$ & $8.45_{-0.12}^{+0.11}$ \\ [1.5pt] 
GRB140213A & 1.2079 & $0.06_{-0.06}^{+0.72}$ & \ha & $0.72_{-0.34}^{+2.65}$ & $34\pm{14}$ & \nodata \\ [1.5pt] 
GRB140301A & 1.4155 & $0.75_{-0.10}^{+0.11}$ & \ha & $106_{-25}^{+36}$ & $117\pm{6}$ & $8.89_{-0.09}^{+0.09}$ \\ [1.5pt] 
GRB140430A & 1.6019 & \nodata & \ha & $8.5_{-3.8}^{+7.1}$ & $40\pm{7}$ & $8.67_{-0.19}^{+0.18}$ \\ [1.5pt] 
GRB140506A & 0.8893 & \nodata & \ha & $0.35_{-0.19}^{+0.35}$ & $61\pm{9}$ & \nodata \\ [1.5pt] 

\hline\noalign{\smallskip}
\end{longtable}
\centering
\begin{minipage}{5.3in}
\tablefoot{
$^{a}$ The physical parameters presented here are integrated and thus averaged over the entire galaxy. We do not perform a resolved analysis (neither spatially, nor in velocity space). In some cases \citep[e.g., GRB~120422A,][]{2014A&A...566A.102S}, a spatially resolved analysis leads to somewhat different results and interpretation of the galaxy properties.
$^{b}$ The quoted error on SFR is logarithmic because it contains the error in the dust correction. The derived SFR also has a lower limit because of the physical condition that $A_V > 0$\,mag. This lower limit is given by SFR$_{\rm min} = 4.8 \times \rm{F_{{H}\alpha, 42}}$.\\}
\end{minipage}

\end{longtab}

\subsection{New GRB redshifts}
\label{sec:newzs}

To derive GRB redshifts from host galaxy data (Table~\ref{tab:newz}), we use both the emission lines as detailed in \citet{2012ApJ...758...46K}, spectral breaks in the stellar continuum \citep{2012ApJ...752...62J} as well as standard photo-$z$ techniques in \textit{LePhare} \citep{1999MNRAS.310..540A, 2006A&A...457..841I} from supplementary multi-band photometry. For galaxies with more than one detected emission line, the redshift solution is accurate, unique, and robust. For the other, somewhat more ambiguous cases we provide more information in Appendix~\ref{app:newz}. 

In particular, we provide redshifts for six GRBs (GRB~050714B at $z=2.438$, GRB~060306\footnote{The redshift of GRB~060306 has been discussed in \citet{2012ApJ...752...62J} and \citet{2013ApJ...778..128P}, already suggesting $z=1.55$ as the likely redshift based on a VLT/FORS spectrum in the former, and extensive photometry in the latter case. The detection of multiple emission lines unambiguously confirms $z=1.56$ as the correct redshift.} at $z=1.56$\, GRB~060805A\footnote{The association of GRB~060805A is ambiguous because two galaxies are detected in the \textit{Swift} XRT error-circle. Here, we provide the redshift for the fainter object. The brighter object has a redshift of $z=0.603$ \citep{2012ApJ...752...62J}.} at $z=2.363$, GRB~060923B at $z=1.509$, GRB~070224 at $z=1.992$, and GRB~070328 at $z=2.063$) in the TOUGH sample \citep{2012ApJ...756..187H}. This brings the redshift completeness of TOUGH to 87\% (60 out of 69 events), with a median/average redshift of $z_{\rm{med}} = 2.10\pm0.18$ and $z_{\rm{avg}}= 2.20\pm0.17$, respectively. GRB~070328, GRB~070521 ($z=2.087$), GRB~080602 ($z=1.820$), and GRB~090201 $z=2.100$ are part of BAT6, so its completeness is now 97\% (56 out of 58 events) with $z_{\rm{med}} = 1.67\pm0.15$ and $z_{\rm{avg}}= 1.75\pm0.13$.

In addition, we measure or constrain the redshifts of seven GRBs based on emission lines or spectral breaks in the stellar continuum  (Table~\ref{tab:newz}). 

\begin{table}[!ht]
\caption{New GRB redshifts\label{tab:newz}}
\centering
\begin{tabular}{ccc}
\hline
\hline\noalign{\smallskip}
{GRB host} & {Redshift} & Spectral features \\
\hline\noalign{\smallskip}
GRB~050714B & 2.4383  & \oiii(5007), \ha  \\
GRB~060306\tablefootmark{a}  & 1.5585 / 1.5597 & \oii, \ha \\
GRB~060805A\tablefootmark{b}  & 2.3633 & \hb, \oiii(5007), \ha \\
GRB~060923B & 1.5094 & \oiii(5007), \ha \\
GRB~070224  & 1.9922 & \oiii(4959,5007) \\
GRB~070328  & 2.0627 & \oii, \hb, \oiii \\
\noalign{\smallskip}\hline\noalign{\smallskip}
GRB~070521  & 2.0865 & \ha \\
GRB~080602  & 1.8204 & \oii, \oiii, \ha \\
GRB~090201  & 2.1000 & \oii, \hb, \oiii, \ha \\
\noalign{\smallskip}\hline\noalign{\smallskip}
GRB~100508A  & 0.5201 & \oii, \hb, \oiii, \ha \\
GRB~111129A & 1.0796 & \oii \\
GRB~120211A & $2.4\pm0.1$ & Ly-$\alpha$ break \\
GRB~120224A & $1.1\pm0.2$ & Balmer break \\
GRB~120805A & $3.1\pm0.1$ & Ly-$\alpha$ break \\
GRB~121209A & $2.1\pm0.3$ & Balmer break \\
GRB~140114A & $3.0\pm0.1$ & Ly-$\alpha$ break \\
\hline\noalign{\smallskip}

\end{tabular}
\tablefoot{The horizontal lines denote the separation between TOUGH GRBs (upper part), BAT6 (middle part) and others. 
\tablefoottext{a}{Emission lines of the host of GRB~060306 have a double-peaked profile, and we thus provide two redshifts.}
\tablefoottext{b}{Ambiguous association because two galaxies are present in the XRT error-circle. This is the redshift for the fainter of the two objects, for details see \citet{2012ApJ...752...62J}.}} 
\end{table}

\subsection{Scaling relations for star formation rates}
\label{sec:newsc}

\begin{figure}
\includegraphics[angle=0, width=0.99\columnwidth]{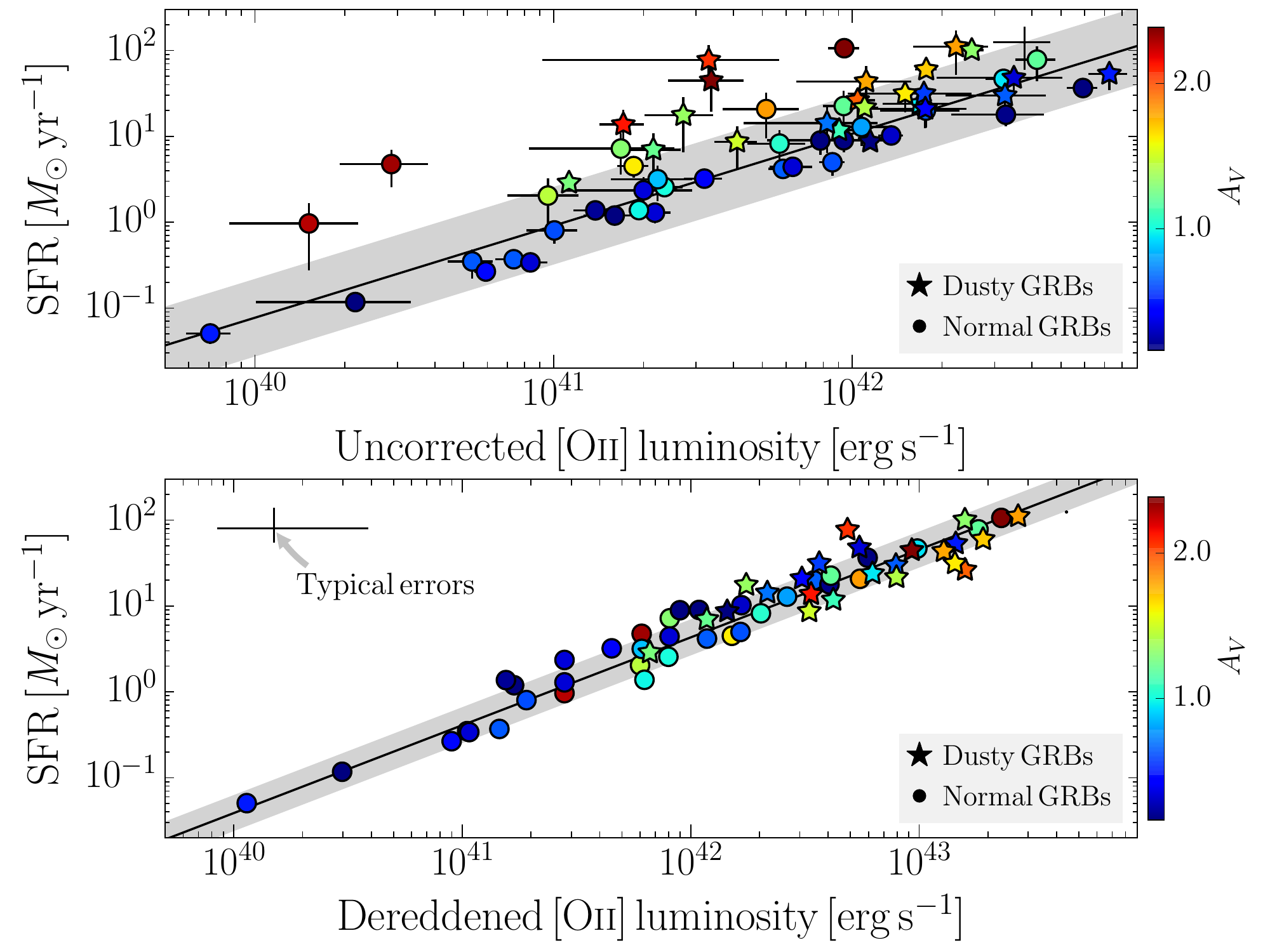}
\caption{Scaling relations between emission-line luminosity and SFR. Uncorrected/de-reddened \oii\,luminosity against SFR in the upper and lower panel, respectively. Error-bars in the lower panel are omitted to enhance clarity (typical errors in the top left corner). The color-coding of the individual measurements corresponds to $A_V$ as indicated by the color bar. The solid black lines show the best-fit scaling relations from Equation~\ref{eq:oiiscale}. The gray-shaded area denotes the $1\sigma$ scatter in the relation - 0.45 dex in the upper and 0.19 dex in the lower panel.}
\label{fig:haoii}
\end{figure}

Since \ha, the most reliable tracer of star formation in the rest-frame optical wavelength range, is not visible in all of the spectra, we use our sample to derive scaling relations between the luminosity of other strong nebular lines and the dust-corrected SFR$_{\rm{H}\alpha}$. To establish the comparison sample, we avail of those galaxies for which both \ha\,and \ebv from the Balmer decrement (Sect.~\ref{sec:av}) can be constrained. We convert the dust-corrected \ha\,luminosity into a SFR based on the widely used \citet{1998ARA&A..36..189K} relation adopting a \citet{2003PASP..115..763C} IMF. This provides us with a reference sample with reddening-corrected SFRs based on \ha.

We then plot other strong emission lines against SFR (e.g., Fig.~\ref{fig:haoii}) and minimize the scatter of the data against scaling relations in the form of $\log(SFR_{\rm{H}\alpha}) = a + b\cdot\log\left(L_{42}\right)$. Here, $10^{a}$ is the SFR in \Msunyr\,for a benchmark line luminosity of $L_{42}=10^{42}$\,erg\,{s}$^{-1}$, while $b$ takes into account a possible evolution between SFR and line flux with galaxy properties. These scaling relations then constrain the SFR for hosts for which other nebular lines than \ha\,are available.

A slope of $b=1$ provides a direct proportionality between e.g., \oii\,luminosity and SFR \citep{1998ARA&A..36..189K, 2009ApJ...691..182S}. $b$ significantly different from unity indicates that the relation between SFR and line flux changes with galaxy properties, for instance metallicities, dust attenuation or ionization \citep[e.g.,][]{2004AJ....127.2002K}. One would thus ideally add a further dependence in the relation based on e.g., metallicity, stellar mass or luminosity for accurate SFRs \citep{2004AJ....127.2002K, 2010MNRAS.405.2594G}. In the absence of further information on the galaxy apart from line flux, however, all the physical changes in the SFR to line-flux relation are implicit in the empirical factor $b$.

\oii\, for example, is strongly correlated with SFR (Figure~\ref{fig:haoii}) and based on  {60 hosts at $0.3 < z < 2.5$} we derive
\begin{equation}
\label{eq:oiiscale}
\begin{split}
\log({SFR_{\rm{H}\alpha}}) = 1.02 + 1.07\cdot\log\left(L{\oii_{\rm{,uncor,42}}}\right)\\
\log({SFR_{\rm{H}\alpha}}) = 0.63 + 1.03\cdot\log\left(L{\oii_{\rm{,dered,42}}}\right),
\end{split}
\end{equation}
where $L{\oii_{\rm{,uncor,42}}}$, $L{\oii_{\rm{,dered,42}}}$ are the measured and dereddened \oii\,luminosity in units of $10^{42}$\,erg\,{s}$^{-1}$, respectively.  The 1$\sigma$ dispersions in these relations are 0.44~dex for the dust-uncorrected and 0.19~dex for the dust-corrected luminosity. Even though the scatter is much lower in the latter case, the former is useful in most practical cases. For a reliable dust-correction, the measurement of at least two Balmer lines is required, which in itself provides a much more reliable SFR than through Equation~\ref{eq:oiiscale}. 

 {Figure \ref{fig:haoii} also illustrates the limitation of this procedure. Since \oii($\lambda$3727\,\AA) is strongly attenuated by dust, there is considerable scatter in the relation using the dust-uncorrected luminosity. Even though the calibration sample contains galaxies over nearly full range of metallicity probed by the observations ($8.2 < \oh < 8.9$), the uncorrected Equation~\ref{eq:oiiscale} will significantly under-predict the SFR for metal-rich ($\oh > 8.7$) galaxies because of their high dust content. If a reliable dust-correction is available, the dust-corrected equation~\ref{eq:oiiscale} then provides SFRs with an average dispersion of 0.2~dex over the full metallicity range probed.}

For those cases where no dust correction can be performed, i.e., only one Balmer line is detected in the spectrum, we derive for \ha\, (based on 61 individual galaxies with $8.2 < \oh < 8.9$)
\begin{equation}
\log({SFR_{\rm{H}\alpha}}) = 0.92 + 1.04\cdot\log\left(L({\rm{H}\alpha_{\rm{,uncor,42}}})\right)
\end{equation}
with a dispersion of 0.25 dex. For \hb\, the equivalent relation is
\begin{equation}
\log({SFR_{\rm{H}\alpha}}) = 1.47 + 0.98\cdot\log\left(L({\rm{H}\beta_{\rm{,uncor,42}}})\right)
\end{equation}
with a scatter of 0.36 dex from 57 galaxies. For the Balmer lines, dust-corrected relations are by construction equivalent to the one of \citet{1998ARA&A..36..189K}.

For the \oiii($\lambda$5007)\,line, the scaling relations are
\begin{equation}
\begin{split}
\log({SFR_{\rm{H}\alpha}}) = 0.98 + 0.81\cdot\log\left(L{\oiii_{\rm{,uncor,42}}}\right)\\
\log({SFR_{\rm{H}\alpha}}) = 0.71 + 0.92\cdot\log\left(L{\oiii_{\rm{,dered,42}}}\right)
\end{split}
\end{equation}
with a dispersion of 0.42~dex for the uncorrected, and 0.28~dex in the dereddened case (from 57 individual galaxies in total).  {The relatively large dispersion, even after removing the effect of dust reddening, illustrates that \oiii\, is the least reliable SF-tracer considered in this study.} In the following, we use the emission-line that yields the smallest uncertainty in SFR when propagating 1$\sigma$ dispersions accordingly when necessary (typically first \ha, then \hb, \oii, and \oiii\, only if no other emission line is detected).

It is immediately clear that the scaling relations detailed above depend on the distribution of galaxy properties in the reference sample. Or, in other words, we consider these scaling relations most useful and accurate for low-mass, high-redshift, star-forming galaxies which have the average color excess of our sample ($A_V\sim0.8$\,mag or $E_{B-V}\sim0.25$\,mag, Sect.~\ref{sec:av}).

\section{Physical properties of the GRB host population}
\label{sec:prop}

\subsection{GRB hosts in the BPT diagram}

\begin{figure}
\includegraphics[angle=0, width=0.99\columnwidth]{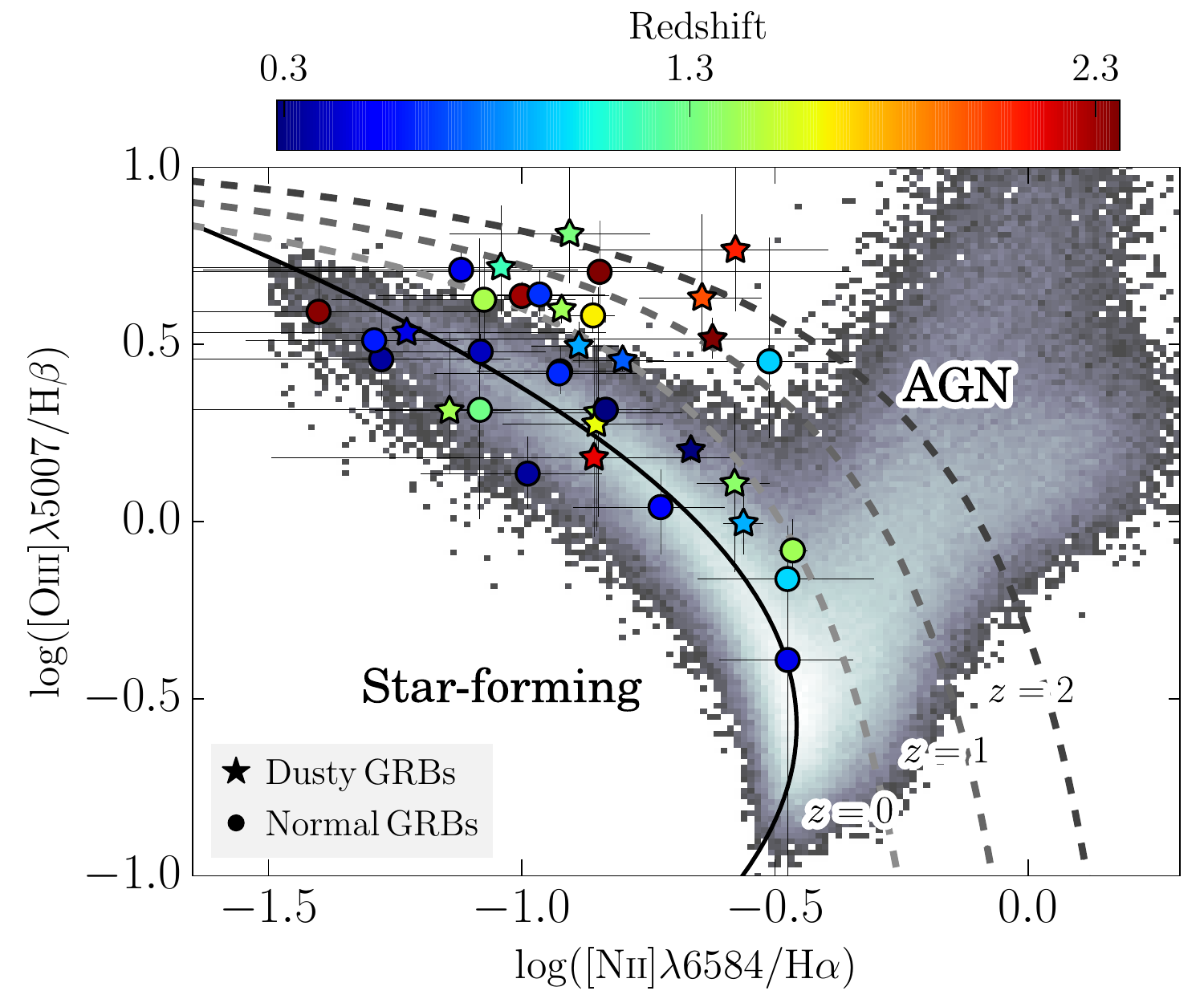}
\caption{Baldwin-Phillips-Terlevich (BPT) classification diagram for GRB hosts. The background sample is the logarithmic space density of local ($z\sim0$) SDSS galaxies with line-fluxes from the MPA/JHU catalog\footnotemark. Colored circles are 34 individual GRB-selected galaxies with redshifts indicated by the color bar. The increasingly darker dashed gray lines are the redshift-dependent differentiation lines between star-forming galaxies and AGN of \citet{2013ApJ...774L..10K} at the indicated redshifts. The black solid line is the ridge line, i.e., the line with the highest density of SDSS galaxies. Dusty/normal GRBs are indicated by stars/circles.}
\label{fig:bpt}
\end{figure}

\begin{figure}
\includegraphics[angle=0, width=0.99\columnwidth]{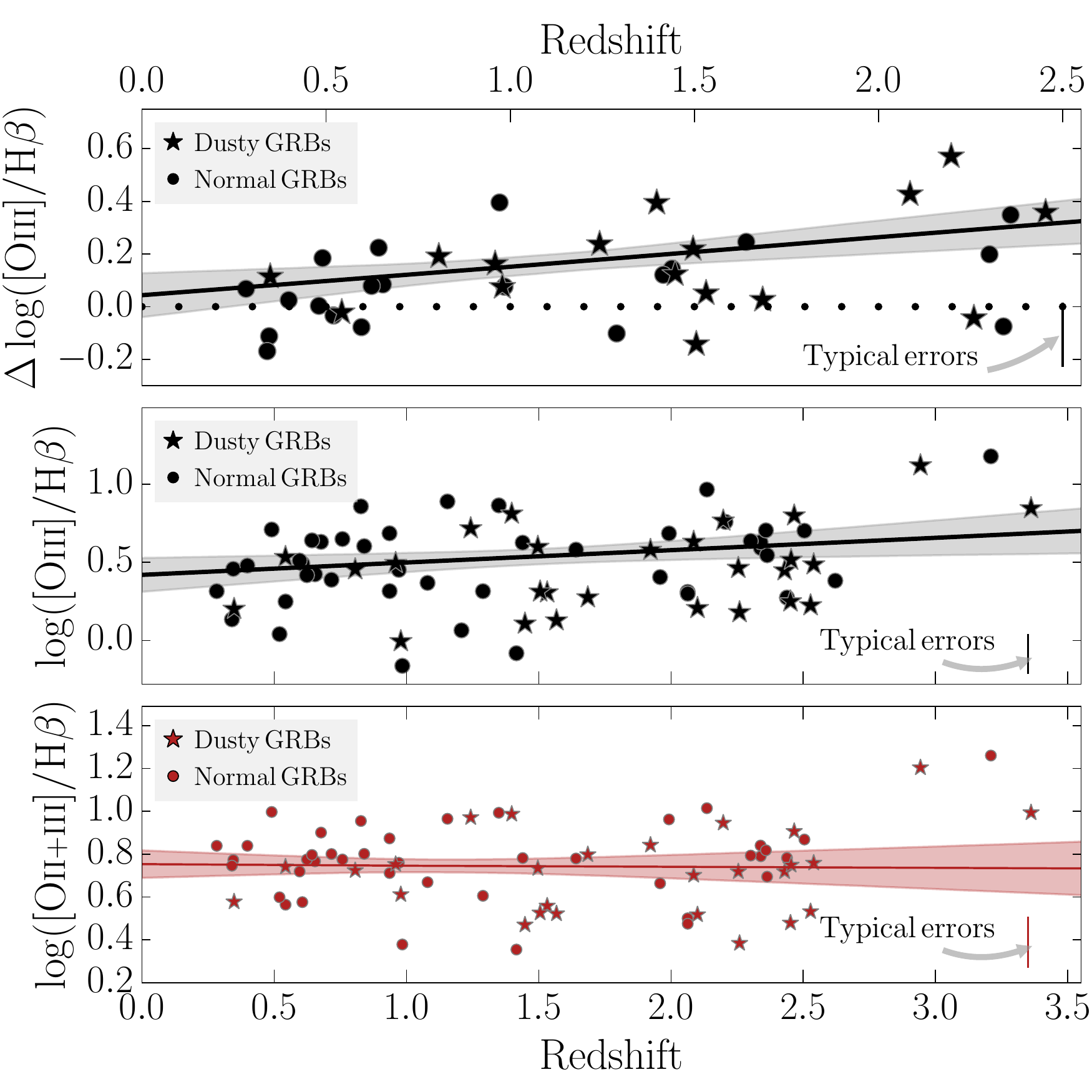}
\caption{Minimum difference of the GRB hosts from the ridge-line (illustrated by the dots) of $z\sim0$ SDSS star-forming galaxies (top panel). We show \oiii/\hb\, in black and (\oiii+\oii)/\hb\, in red as a function of $z$ in the middle and bottom panel, respectively. In all panels, solid lines denote the most-likely linear regression, with regions of 68\,\% probability in shaded areas. Dusty/normal GRBs are indicated by stars/circles. Error bars are omitted to enhance clarity and replaced by typical errors in the bottom-right corner.}
\label{fig:bptdist}
\end{figure}

Figure~\ref{fig:bpt} shows the GRB hosts in the Baldwin-Phillips-Terlevich (BPT) diagram \citep{1981PASP...93....5B} that allows emission-line objects to be separated into star-forming galaxies and AGNs based on the \nii/\ha\, and \oiii$\lambda$5007/\hb\,ratios \citep[e.g.,][]{2001ApJ...556..121K}. In our sample, 34 GRB hosts have adequate constraints on the four emission-line fluxes to establish their position in the BPT diagram. As we are interested in line-flux ratios, we use here (and in the following whenever line-ratios are important) only the statistical errors from Tables~\ref{tab:balmerlines}~and~\ref{tab:forbiddenlines}.

The redshift range of the GRB hosts extends from $z=0.3$ to $z=2.5$. As background sample, we plot the logarithmic space density of 330000 local ($z\sim0$) SDSS \citep{2009ApJS..182..543A} galaxies from the MPA-JHU catalogue \citep[e.g.,][]{2003MNRAS.346.1055K}, as well as the redshift-dependent classification lines between AGN and star-forming galaxies as gray dashed lines \citep{2013ApJ...774L..10K}. The ridge line of SDSS galaxies \citep{2008MNRAS.385..769B} is shown in black.

Our GRB hosts occupy a different phase-space than SDSS galaxies in the BPT diagram because they are mostly above the ridge line which denotes the highest density of local star-forming galaxies. A similar offset is well-observed in many high redshift galaxy samples \citep[e.g.,][]{2005ApJ...635.1006S, 2007ApJ...669..776K, 2014ApJ...795..165S} and also for galaxies hosting H-poor super-luminous supernovae \citep{2014arXiv1409.8331L}. This offset is often attributed to harder ionization fields, higher ionization parameters or changes in the ISM properties \citep[e.g.,][]{2008MNRAS.385..769B, 2013ApJ...774..100K, 2014ApJ...795..165S}. 

 {There is some evidence of an evolution of the position of the targets in the BPT diagram with redshift. The top and middle panel of Fig.~\ref{fig:bptdist} show the minimum difference to the SDSS-based ridge line ($\Delta$(\oiii/\hb)) as well as \oiii/\hb. On average $\Delta$(\oiii/\hb) and \oiii/\hb\, increase by $0.11\pm0.02$ and $0.08\pm0.03$ dex per unit redshift. These values are derived as the median of all stochastic samples (Sect.~\ref{sec:statana}) with errors representing the region of 68\,\% likelihood of all trials. The fractions of samples producing no evolution in \oiii\,/\hb, and $\Delta$(\oiii/\hb) are $p=0.015$ and $p=0.0018$, respectively. A rising trend is not seen (bottom panel of Figure~\ref{fig:bptdist}) in the ratio of the summed oxygen line flux (\oiii\,+\oii)\,to \hb\, ($0.01\pm0.04$ dex per unit redshift).} 

 {Changes in the metallicity of the nebular gas are unlikely to play the dominant role in the displacement between high-redshift and local galaxies as this would move galaxies primarily along the ridge line. Photoionization models from \citet{2014ApJ...795..165S}, for example, predict very little contribution from metallicity changes to the observed offset in the BPT diagram. In addition, the metallicity distributions of the sample are comparable at $z<1$ and $1<z<2$ (Sect.~\ref{sec:metprop}), indicating that the larger difference of galaxies at higher redshift is not driven by a decreasing gas-phase oxygen abundance.}

 {Our observations are thus consistent with the hypothesis} that the difference in the location in the BPT diagram between GRB hosts and $z\sim0$ star-forming galaxies is caused by an increase in the ionization fraction, i.e., for a given metallicity a larger percentage of the total oxygen abundance is present in higher ionization states at higher redshifts. This could be caused by a harder ionization field originating for example from hot O-type stars \citep{2014ApJ...795..165S} that emit a large number of photons capable of ionizing oxygen into \oiii.

\footnotetext{http://www.mpa-garching.mpg.de/SDSS}

\subsection{Ionization evolution}

\begin{figure}
\includegraphics[angle=0, width=0.99\columnwidth]{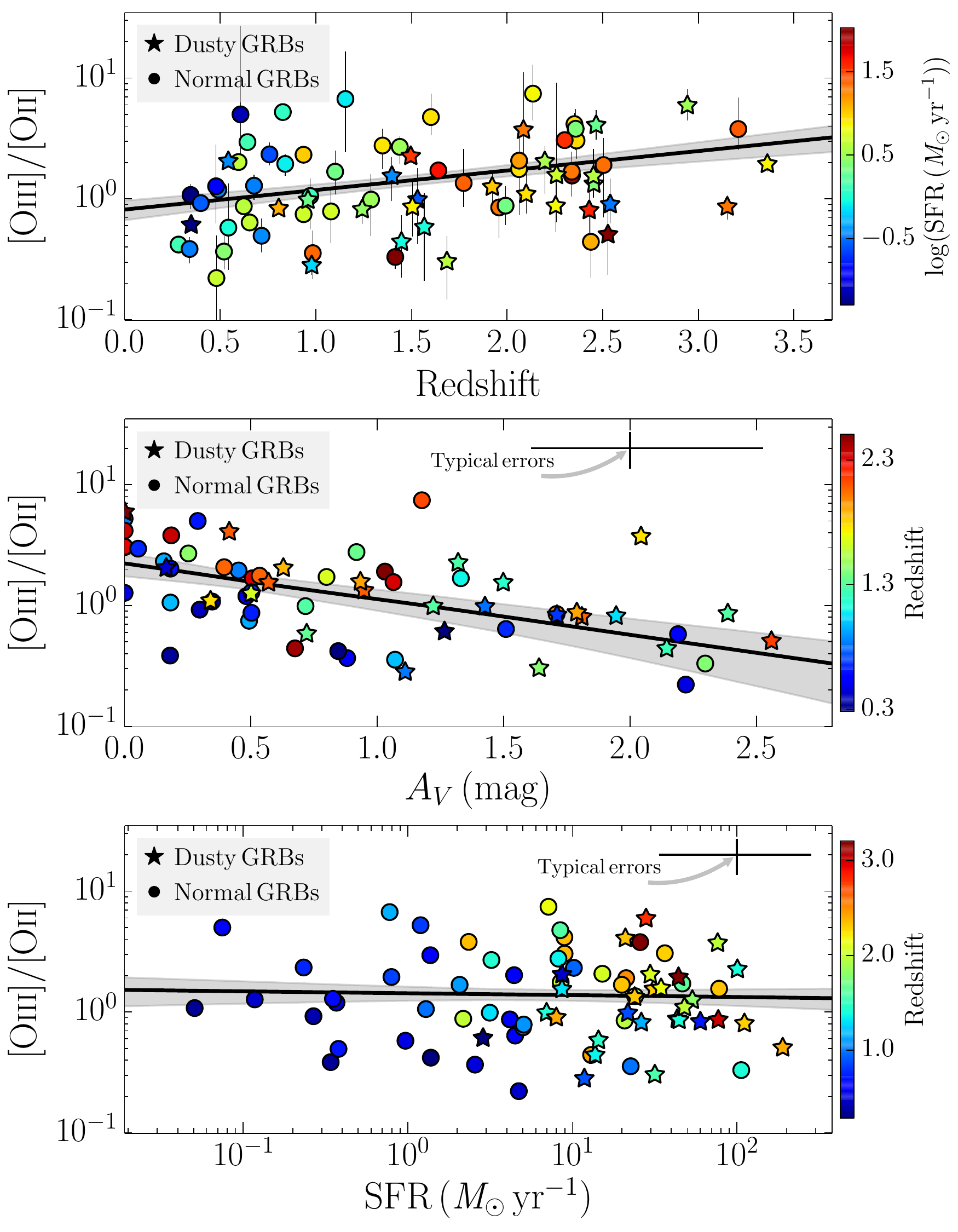}
\caption{\oiii/\oii\, as a function of (from top to bottom): redshift, galaxy $A_V$, and SFR. The color of individual data points are according to the color bars. In all panels, solid lines denote the most-likely linear regression, with regions of 68\,\% probability in shaded areas. Dusty/normal GRBs are indicated by stars/circles.}
\label{fig:ionization}
\end{figure}

The ratio of \oiii/\oii, which we plot in Fig.~\ref{fig:ionization} against redshift, $A_V$, and SFR, is often used as a proxy for the ionization parameter. A rising \oiii/\oii\, ratio with redshift is evident from  {79 events with adequate wavelength coverage to constrain both \oiii\,and \oii}. \oiii/\oii\, increases from 0.9 at $z\sim0.3$ by a factor of 3 to $\sim3$ at $z\sim3.3$.  {The fraction of iterations returning no correlation between the \oiii/\oii\, ratio and redshift is $p=0.0018$ for the full sample and $p=0.0039$ when restricting the redshift range to $z < 2.7$}. 

The predicted \oiii/\oii\,ratio at $z=3.3$ is remarkably similar to the ratio of the composite LBG spectrum at $z\sim3.5$ \citep{2008A&A...488..463M, 2014A&A...563A..58T}. The evolution in \oiii/\oii\, is again unlikely to be caused by metallicity effects (lower metallicity galaxies have higher \oiii/\oii\, ratios) because there is no statistically significant difference seen in metallicity between the $z<1$ and $1<z<2$ sample (Sect.~\ref{sec:met}).

Even though both the ionization parameter and SFR are correlated with redshift, the bottom panel in Fig.~\ref{fig:ionization} shows no obvious relation between \oiii/\oii\, and SFR (Sect.~\ref{sec:sfr}). It does show (middle panel in Fig.~\ref{fig:ionization}) an anti-correlation with dust content {($A_V = 0.35 - 0.90\cdot\log\left(\oiii/\oii\right)$, with $p=0.037$)}. The dust content\footnote{Here, and in the following, $E_{B-V}$ or $A_V$ always refers to the rest-frame color excess/visual attenuation at the redshift of the target galaxy.} as parametrized through a color-excess $E_{B-V}$ or visual attenuation $A_V$ of a galaxy (Sect.~\ref{sec:av}) correlates with stellar mass $M_{\star}$, where $A_V=1$\,mag corresponds roughly to $M_{\star}\sim10^{10}$\,\Msun\,\citep[e.g.,][]{2014ApJ...792...75Z}. 

The \oiii/\oii\, ratio thus increases with redshift, whereas it decreases with $A_V$ (or stellar mass), plausibly an effect of the efficient absorption of ionizing photons on dust. The net effect makes \oiii/\oii\, independent of SFR in our sample.

\subsection{Dust properties - distribution and redshift evolution of galaxy $A_V$}
\label{sec:av}

\begin{figure}
\includegraphics[angle=0, width=0.99\columnwidth]{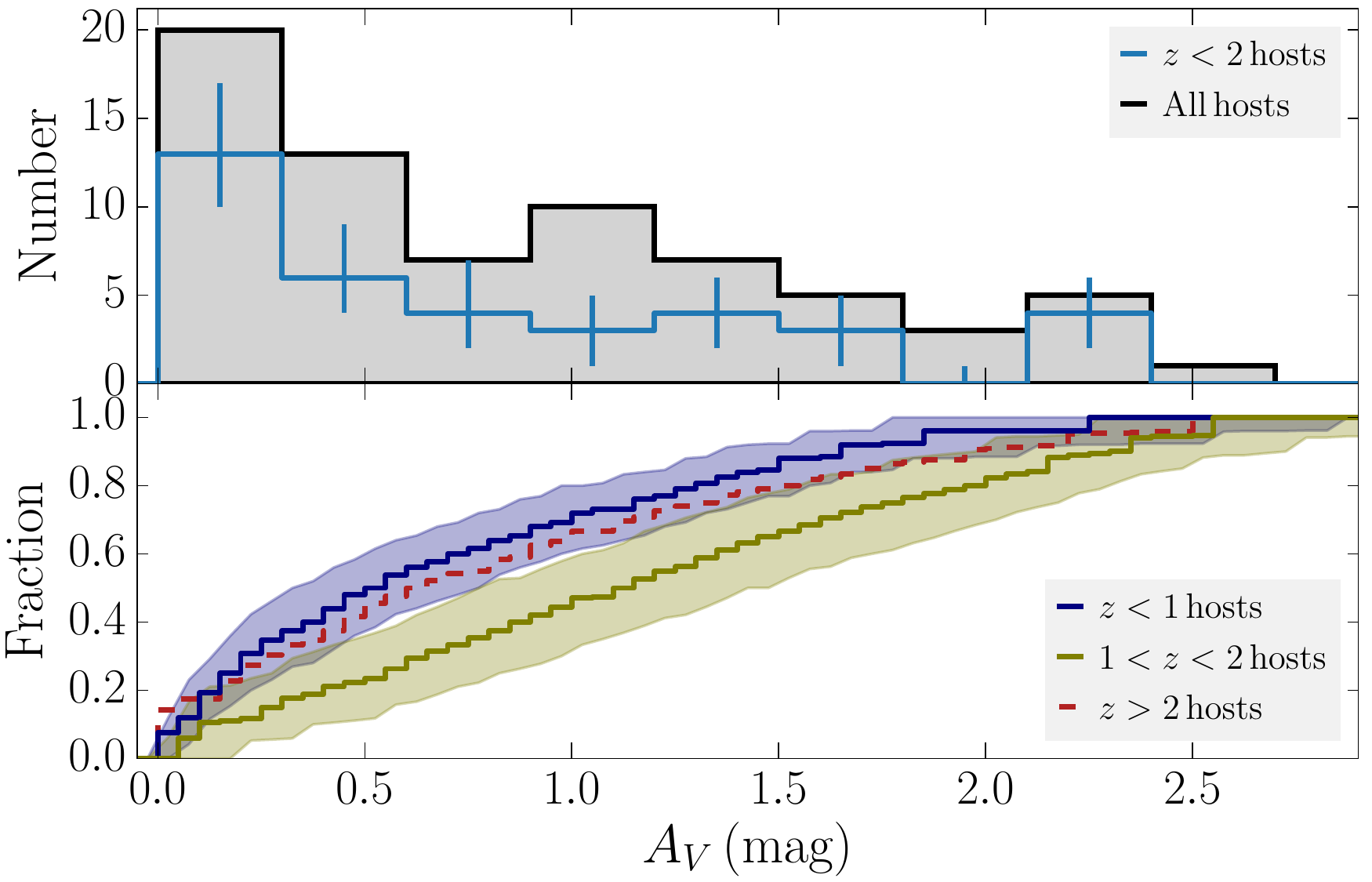}
\caption{The blue solid histogram and error-bars in the top panel shows the $A_V$ distribution for $z < 2$ GRB from the statistical analysis (Sect.~\ref{sec:dusty} and Sect.~\ref{sec:statana}). The black histogram represents all raw measurements without corrections. The lower panel shows the cumulative distribution of $A_V$ and its uncertainty from $z<1$ hosts (dark blue) and $1<z<2$ hosts (olive). The red dashed line shows $z>2$ hosts (error region not shown).}
\label{fig:ebvhist}
\end{figure}

We measure the rest-frame attenuation of light from dust through comparing the observed Balmer decrement $r_{\rm{obs}}$ with the theoretical expectation $r_{\rm{theo}}$ for a given electron temperature. To exploit the broad wavelength coverage and detection of more than two Balmer lines, we simultaneously minimize the various expected ratios against $A_V$ for the four strongest Balmer lines. At an electron temperature of $10^4$\,K and density of $10^{2}$\,cm$^{-3}$ to $10^{4}$\,cm$^{-3}$ these ratios are \citep{1989agna.book.....O} $\rm{H}\alpha/\rm{H}\beta = 2.87$, $\rm{H}\alpha/\rm{H}\gamma = 6.17$, and $\rm{H}\alpha/\rm{H}\delta = 11.24$. In a Milky-Way-type extinction law\footnote{We choose a MW-type extinction law to compare directly with other studies, and note that the difference to extinction laws from the SMC or LMC is small in the wavelength range of H$\delta$ to H$\alpha$.} \citep[][]{1992ApJ...395..130P} with $R_V=3.08$, the wavelengths of the four lines correspond to $A_{\rm{H}\alpha} = 0.79$\,mag, $A_{\rm{H}\beta}$ = 1.16\,mag, $A_{\rm{H}\gamma}$ = 1.31\,mag, $A_{\rm{H}\delta} = 1.381$\,mag for $A_V$ = 1\,mag. For the Balmer decrement of \ha\, over \hb\, follows

\begin{equation}
A_V = \frac{-2.5\cdot\log({r_{\rm{obs}}}/{r_{\rm{theo}}})} {A_{{\rm{H}\alpha}}-A_{{\rm{H}\beta}}},
\end{equation}
and similarly for the other ratios. In the calculation of $A_V$, we use a prior of $A_V \geq 0$\,mag, and provide the median of the a-posteriori probability and 1$\sigma$ equivalent parameter range as error to represent its distribution. In total, we constrain $A_V$ for 71 events in the sample (Table~\ref{tab:physprop}), the distribution of which is shown in Fig.~\ref{fig:ebvhist}. 

 {After applying the statistical analysis and cuts of Sects.~\ref{sec:dusty}~and~\ref{sec:statana}, the median $A_V$ of our sample is $A_V^{\rm med}=0.8^{+0.4}_{-0.3}$\,mag} and seems to be somewhat evolving with redshift (Fig.~\ref{fig:ebvhist}). At redshifts lower than $z<1$, there is a lower number of $A_V>1.5$\,mag hosts when compared to the $z>1$ sample, and we measure $A_V^{\rm{med}}=0.55_{-0.18}^{+0.24}$\,mag. This value is fully consistent with the value reported in \citet{2009ApJ...691..182S}. $A_V^{\rm{med}}$ increases to $A_V^{\rm{med}}=1.2\pm0.3$\,mag at $1<z<2$. It is $A_V^{\rm{med}}=0.7_{-0.2}^{+0.3}$\,mag at $z > 2$, but we note that the galaxies are significantly brighter than comparison samples of GRB hosts in this redshift range.

The significance of this evolution in $A_V$ with redshift however is relatively modest: comparing the two distributions centered around $z\sim0.6$ ($z<1$) and $z\sim1.5$ ($1<z<2$) with each other (Fig.~\ref{fig:ebvhist}), the null-hypothesis that the two distributions are drawn from the same parent sample can be rejected with a significance of 2$\sigma$. At even higher redshifts further selection effects play an important role as well: high $A_V$ hosts are not appearing because of the decreasing sensitivity SFR at high $A_V$ (Fig.~\ref{fig:sens}). 

{The relative sparsity of high $A_V$ events at low redshift is however not easily explained by selection effects. At the average mass of the SF-weighted galaxy population of $M_{\star}\sim10^{10}$\,\Msun\, at $z\sim0.6$ \citep{2013ApJ...778..128P, 2014A&A...565A.112H}, the average reddening for galaxies is $E_{B-V}\sim0.35$\,{mag} \citep[][and references therein]{2014ApJ...792...75Z}. For GRBs to trace star formation representatively, half of the sample would be expected above this value and half below, while we observe ($82\pm8$)\% of all hosts below $E_{B-V}=0.35$\,{mag} at $z<1$.}

Although there are indications of an increasing dust-content in GRB hosts with redshift, the low significance prevents us from reaching stronger conclusions. Interestingly, however, a very similar trend - highest visual extinctions at $z\sim1.5-2$ - exists in the complete sample of $A_V$ values derived from afterglow data \citep{2013MNRAS.432.1231C} and in the evolution of the far UV attenuation in field galaxies \citep{2012A&A...539A..31C}. It is also consistent with the somewhat higher fraction of dusty GRBs at higher redshift (Sect.~\ref{sec:dusty}).

\subsection{Star formation rates}
\label{sec:sfr}

\begin{figure}
\includegraphics[angle=0, width=0.99\columnwidth]{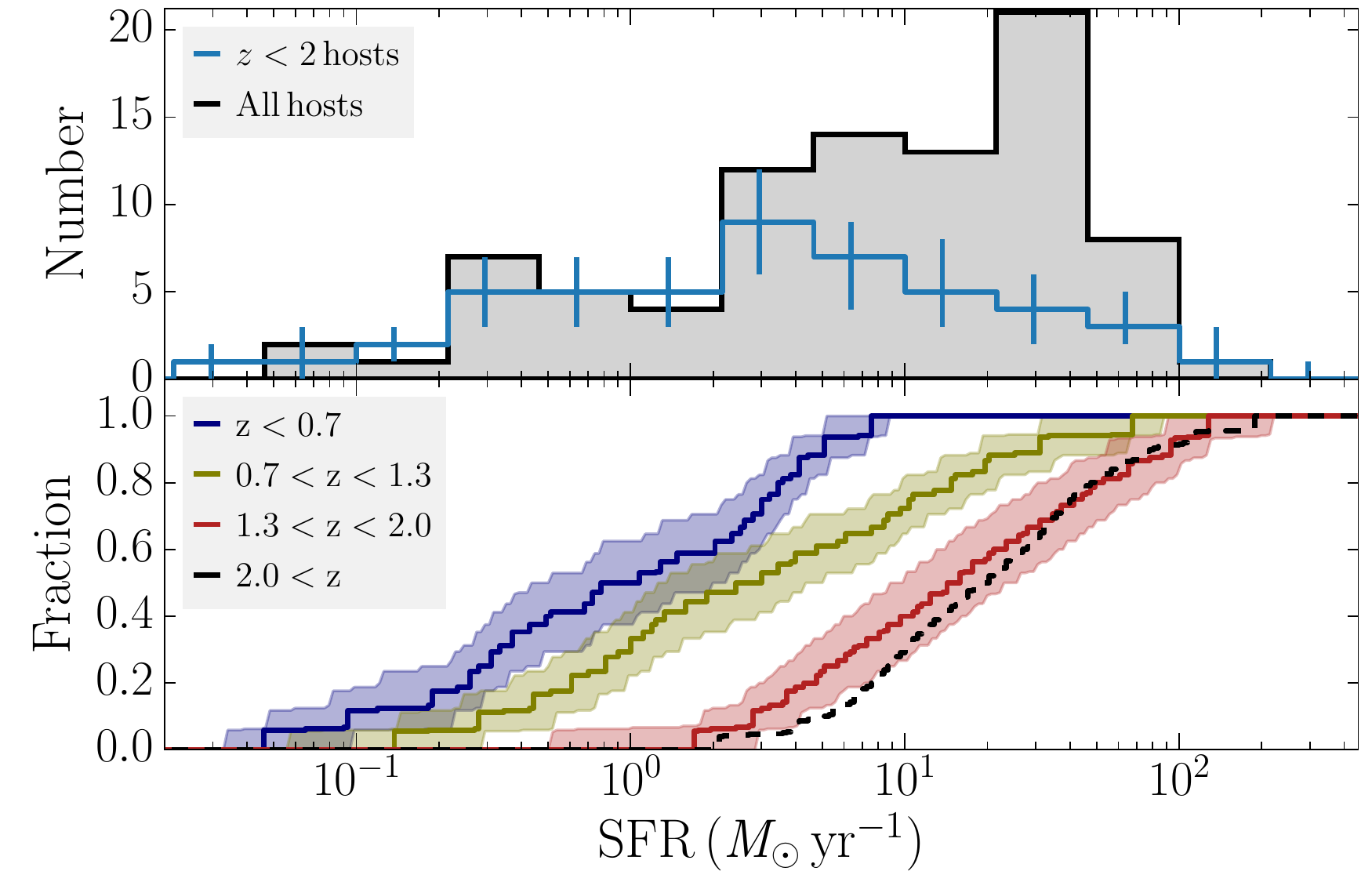}
\caption{In the top panel, the blue solid histogram with error bars shows the SFR distribution and its uncertainty of $z < 2$ GRBs, where we have applied the correction for the over-proportionality of dusty GRBs in the sample (Sects.~\ref{sec:dusty}~and~\ref{sec:statana}). Black data show all raw measurements without corrections. The lower panel shows cumulative distributions of SFR and its uncertainty of $z<0.7$ hosts (dark blue), $0.7<z<1.3$ hosts (olive) and, $1.3<z<2.0$ hosts (red), in the same way corrected as the blue histogram. The black dashed line denotes hosts at $z > 2$.}
\label{fig:sfrhist}
\end{figure}

\begin{figure}
\includegraphics[angle=0, width=0.99\columnwidth]{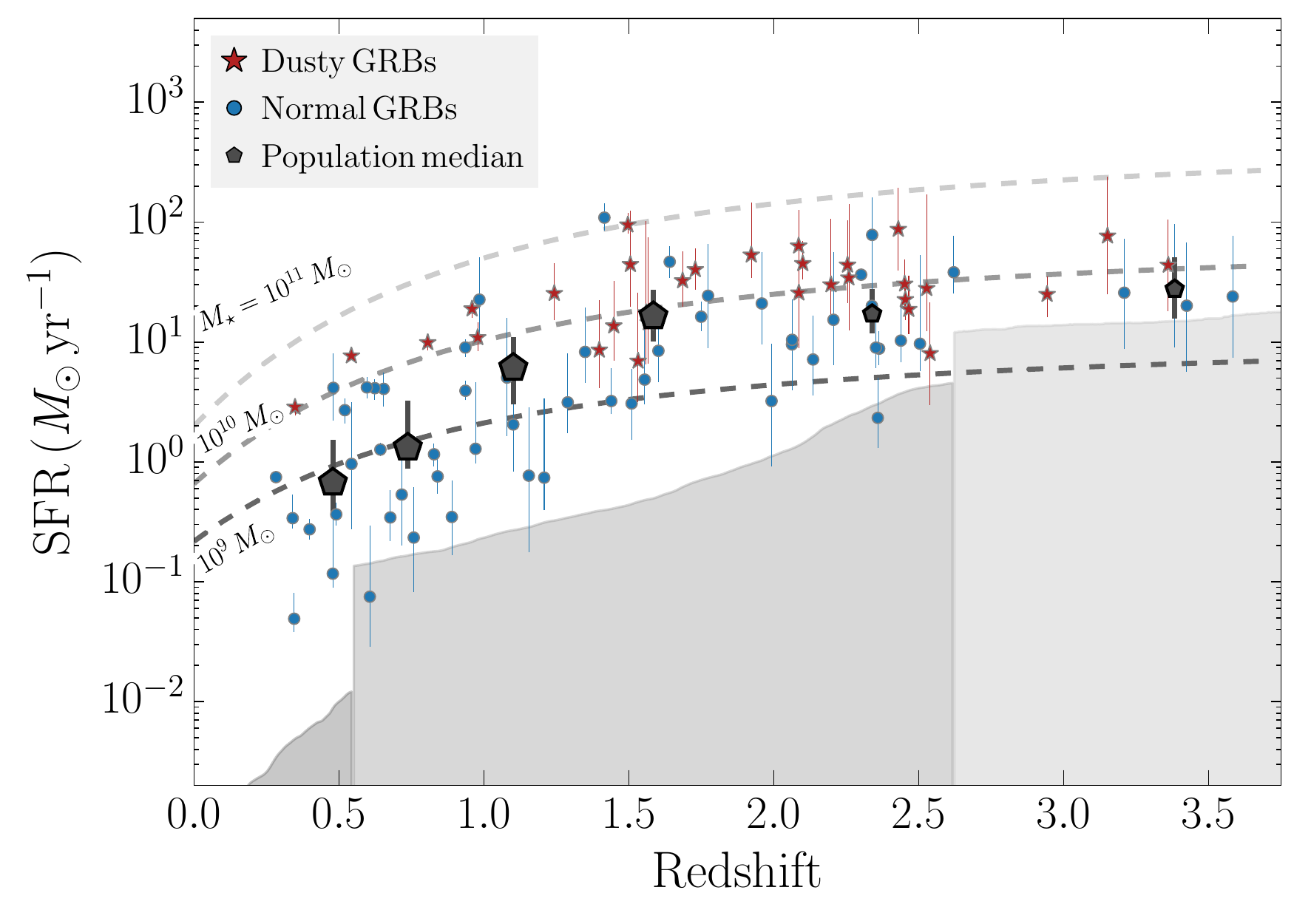}
\caption{SFR as a function of redshift. Blue circles and red stars represent individual galaxies. Large or small pentagons are the median values binned in redshift with bootstrapped errors at $z<2$ and $z>2$, respectively.  {Here, we have applied the correction for the over-proportionality of dusty GRBs in the sample (Sects.~\ref{sec:dusty}~and~\ref{sec:statana})}. The gray shaded areas represent the average sensitivity as detailed in Sect.~\ref{sec:sens} using $A_V=1$\,mag (Sect.~\ref{sec:av}). The three dashed lines show typical SFRs for differently massive galaxies from the main sequence \citep{2014ApJS..214...15S} for a qualitative estimate of the evolution of field galaxies.}
\label{fig:sfrz}
\end{figure}

As introduced in Sect.~\ref{sec:newsc}, we can use the flux of nebular lines to derive constraints on the current star formation rate (SFR). \ha, in particular, is a common tracer of SF on short timescales of a few million years, and its dependence on the uncertainty in the dust-obscuration is only modest. 

All SFRs are given in Table~\ref{tab:physprop} and are corrected for the measured $A_V$ if available (Sect.~\ref{sec:av}), or derived from scaling relations that intrinsically include the effect of dust (Sect.~\ref{sec:newsc}). Here we also add the systematic uncertainty of the flux calibration (Tables~\ref{tab:balmerlines}~and~\ref{tab:forbiddenlines}). In the majority of all cases, however, the error budget is dominated by the uncertainty in the dust correction.

Generally, all of the hosts in the sample are forming stars at a level of at least 0.02\,\Msunyr. Even though we are sensitive to galaxies with a very low SFR at $z<0.5$, we detect none (\citealp[but see][for a quiescent galaxy at the position of GRB~050219A]{2014A&A...572A..47R}). Figure~\ref{fig:sfrhist} shows the distribution of SFRs for 89 hosts, out of which 68 are based on \ha, ten on \hb, eight on \oii, and three on \oiii. The distribution is very broad, spanning nearly four orders of magnitude between 0.05\,\Msunyr\, up to around 200\,\Msunyr\, and peaks at 20\,\Msunyr.  {A large number of galaxies with high SFRs either host dusty GRBs or are at $z>2$ (where selection effects are important). The peak of the corrected SFR distribution for $z<2$ GRBs (blue lines in Fig.~\ref{fig:sfrhist}) is therefore an order of magnitude lower (2\,\Msunyr).} 

\subsubsection{Redshift evolution of star formation rates}

As shown in Figs.~\ref{fig:sfrhist} and \ref{fig:sfrz}, there is clear evolution of the typical SFR of GRB hosts with redshift. While at $z\sim0.5$, the median SFR of a GRB host is $\sim0.6$\,\Msunyr\, \citep[see also][]{2009ApJ...691..182S}, it rapidly increases by a factor of 25 to $\sim$15\,\Msunyr\, at $z\sim2$. Above $z>2$, no further strong rise is evident in the data, and the median SFR levels out at around $\sim$20\,\Msunyr. 

 {This behavior changes only slightly if we take into account that our selection might miss the faintest $\sim$20\% of hosts (total of $\sim5$) at $1<z<2$ (Sect.~\ref{sec:seleffects} and Fig.~\ref{fig:selection}), or that we have initially excluded four spectra of galaxies that could potentially be in this redshift range (Sect.~\ref{sec:tarob})}. 

 {The 30-40\% quantile of SFRs at $z > 1.3$ is still an order of magnitude larger than the median SFR at low redshifts (Fig.~\ref{fig:sfrhist}, lower panel). To be consistent at the 90\% confidence level with no evolution in SFR, we would need to have missed by chance more than 40 GRBs (i.e., twice the sample size in the $1.3<z<2.0$ range) that are hosted by galaxies with $SFR\lesssim$2\,\Msunyr.}

At least at $z>2.5$ there is an obvious effect of sensitivity (shaded area in Fig.~\ref{fig:sfrz}): only the most star-forming hosts ($>$20\,\Msunyr) have emission lines bright enough to be picked up by X-Shooter. We hence sample only the brightest end of the SFR distribution, and the comparison to the $z<2$ data becomes less meaningful. It is however clear that there is no further order-of-magnitude increase in the median SFR of GRB-selected galaxies above $z > 2$. 

The strong evolution in the SFR is not particularly surprising: indeed it is similar to the redshift dependence of the UV luminosity of normal galaxies parametrized with e.g., the normalization of the main sequence \citep[e.g.,][]{2011ApJ...742...96W, 2012ApJ...754L..29W}. To illustrate the evolution of field galaxies, we also plot redshift tracks for galaxies {on the main sequence} with stellar masses between $M_{\star}=10^{9}$\,\Msun\, and $10^{11}$\,\Msun\, in dashed lines in Fig.~\ref{fig:sfrz}. They indicate qualitative (but not quantitative) agreement in the relative rise in SFR with redshift between the different samples. There are significant differences, in particular because of a relative lack of galaxies in the high SFR regime at low redshift.

\subsubsection{Sparsity of high SFR GRB hosts at $z<1$}

\begin{figure}
\includegraphics[angle=0, width=0.99\columnwidth]{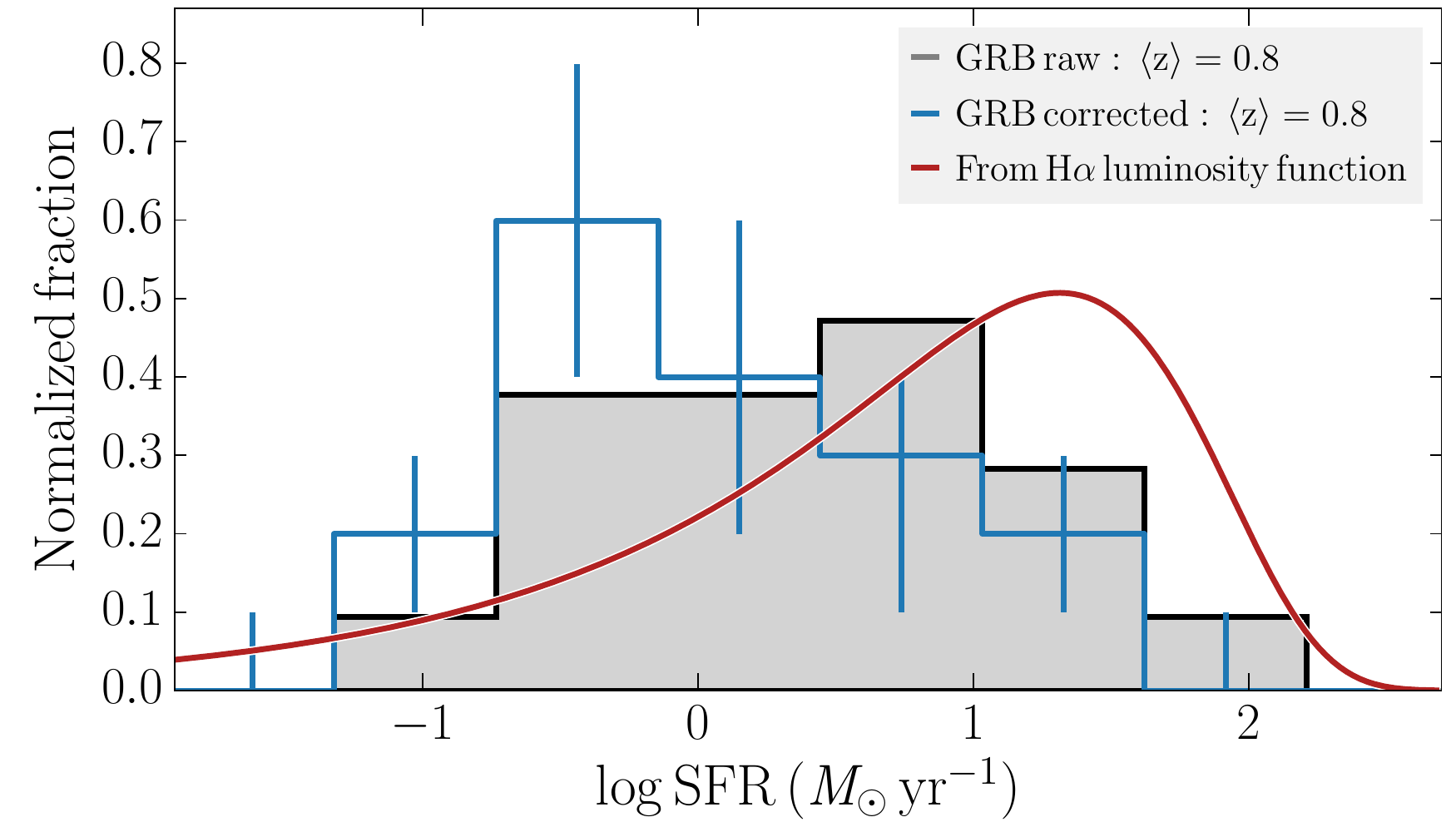}
\caption{ {SFR-weighted \ha-luminosity function at $z=0.8$ \citep{2011ApJ...726..109L} in red compared to the histogram of SFRs from GRB hosts. The blue-solid histogram with error bars shows the SFR distribution and its uncertainty of our $0.55 < z < 1.05$ GRBs, where we have applied the correction for the over-proportionality of dusty GRBs in the sample (Sects.~\ref{sec:dusty}~and~\ref{sec:statana}). Black data are raw measurements in the respective redshift range without corrections. Shown curves are normalized to an area of 1.}}
\label{fig:sfrhalpha}
\end{figure}

 {We illustrate the sparsity of high SFR galaxies hosting GRBs at low redshifts in Fig.~\ref{fig:sfrhalpha} using the GRB host sample between $z=0.55$ and $z=1.05$ ($\langle z \rangle = 0.81$) and the dust-corrected luminosity function (LF) of \ha\, from \citet{2011ApJ...726..109L} at a similar redshift (Schechter parameters $\alpha=1.6$ and $\log(L_*/(\rm{erg\,s^{-1}}))=43.0$). We multiply the LF by the SFR (or, in this case, \ha\,luminosity) since GRBs are expected to follow total SFR and not galaxy numbers.}

 {The SFR-weighted \ha\,-LF then represents the distribution that would be obtained from GRB hosts if they directly traced all star formation. Figure~\ref{fig:sfrhalpha} shows, however, a strong discrepancy between both distributions \citep[see also][]{2013A&A...557A..34B} with the GRB hosts peaking at around an order of magnitude lower SFRs ($\sim1$\,\Msunyr) than the SFR-weighted \ha\,-LF ($\sim10$\,\Msunyr). There is no observational bias in the sample that could explain this discrepancy.}

\subsection{Line broadening}
\label{sec:velo}

\begin{figure}
\includegraphics[angle=0, width=0.99\columnwidth]{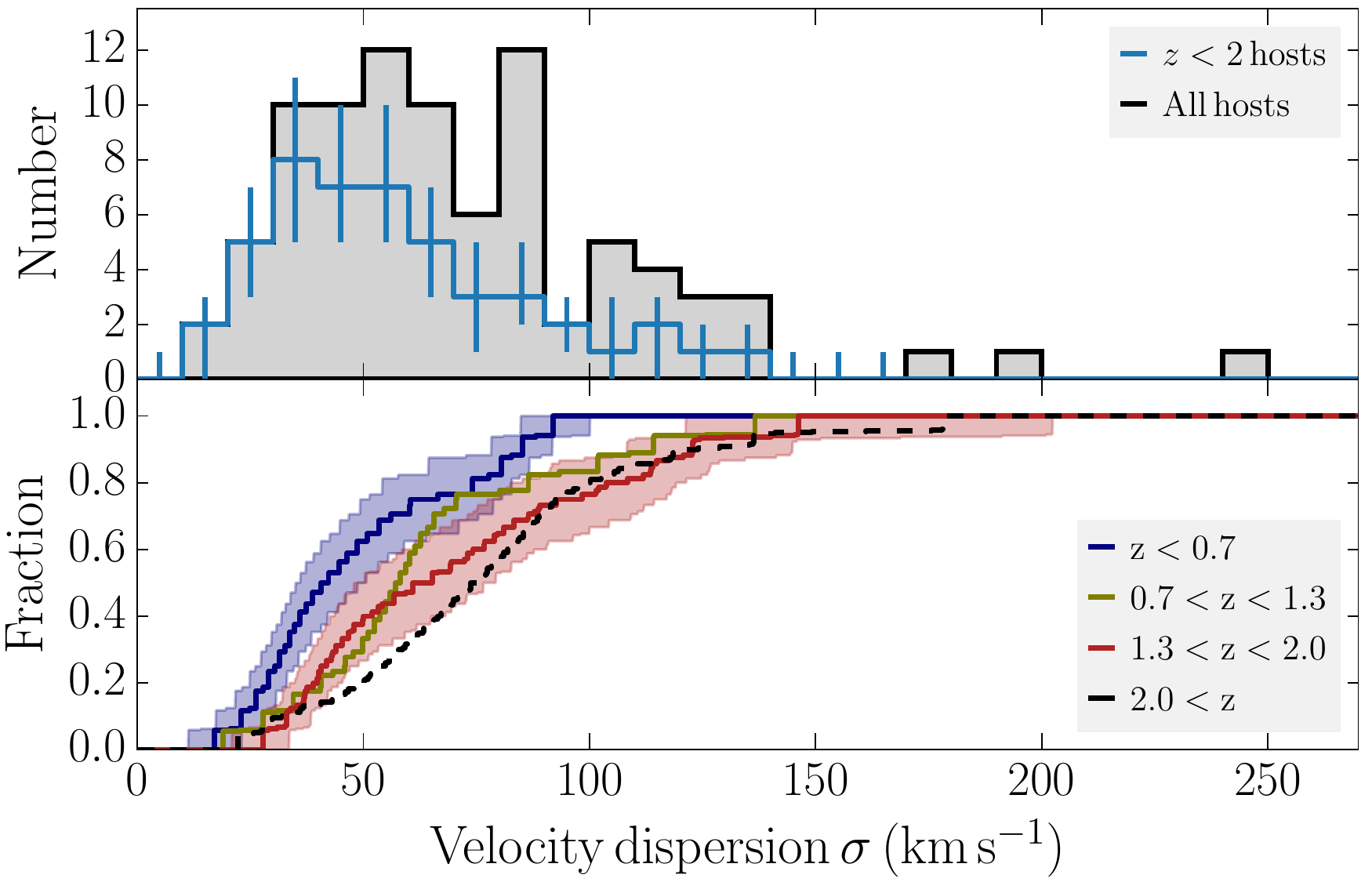}
\caption{In the top panel, the blue histogram with error bars shows the distribution of line broadenings ($\sigma$) and its uncertainty of $z < 2$ GRBs (Sects.~\ref{sec:dusty}~and~\ref{sec:statana}). The black histogram shows raw measurements without corrections. In the bottom panel we plot cumulative distributions of $\sigma$, where black/blue/red colors denote redshift intervals centered around of $z\sim0.5$, $z\sim1.1$, and $z\sim1.6$, respectively. To enhance clarity, the error regions around the $z\sim1.1$ and $z>2$ histogram are not shown.}
\label{fig:velohist}
\end{figure}

\begin{figure}
\includegraphics[angle=0, width=0.99\columnwidth]{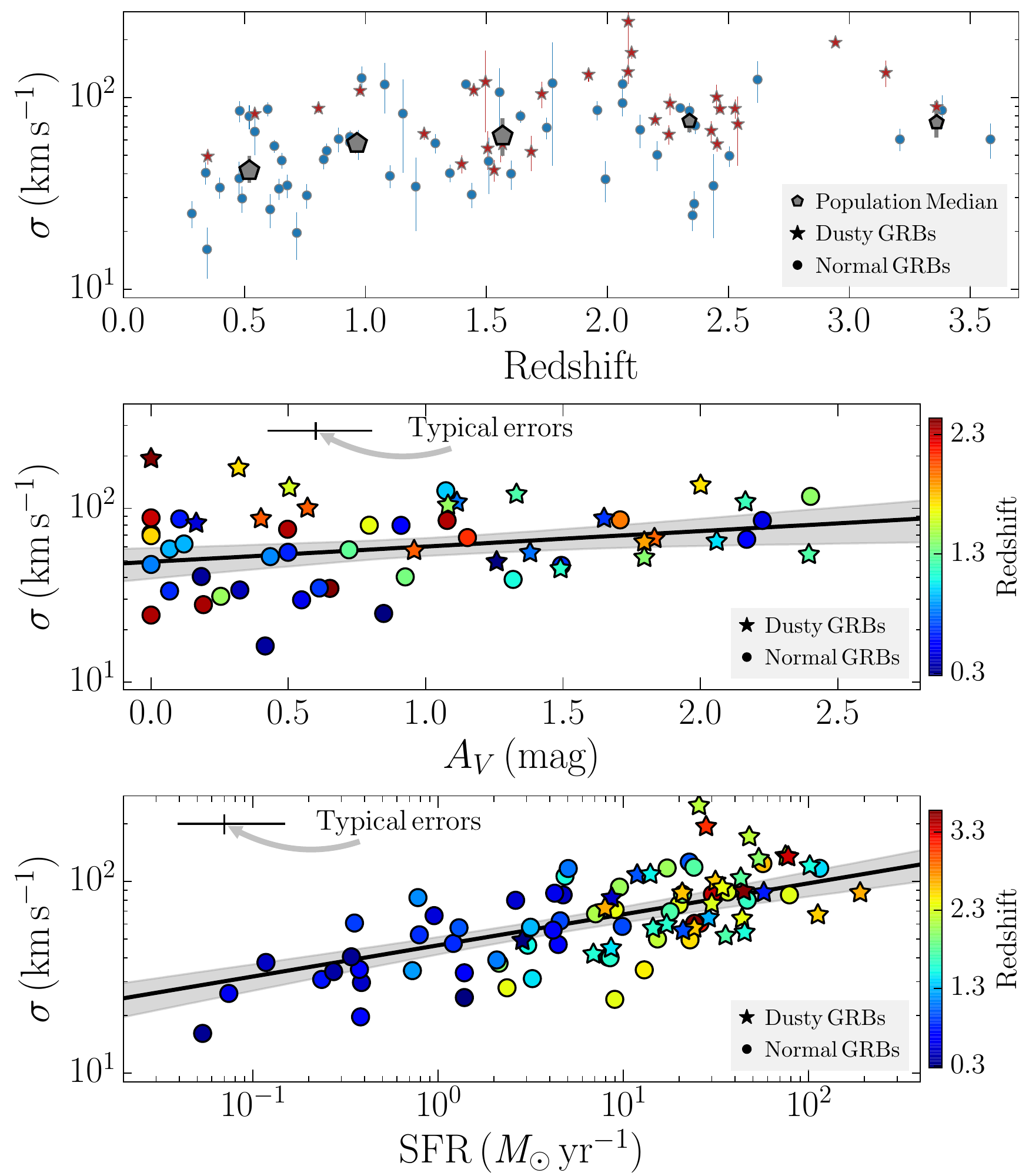}
\caption{Line broadening $\sigma$ as a function of (from top to bottom): redshift, $A_V$, and $\sigma$. In the top panel, large/small pentagons show the median values binned in redshift with bootstrapped errors at $z<2$ and $z>2$, respectively. The color coding of events corresponds to redshift, with the scale given by the color bar. In the lowest two panels, solid lines denote the most-likely linear regression, with error regions (68\,\% probability) in shaded areas.}
\label{fig:veloz}
\end{figure}

The broadening of the nebular lines traces the turbulent and rotational motion of the hot gas within the gravitational potential of the host galaxy. The line-widths are well resolved in all but a few low-redshift cases because of the medium resolving power ($4000\lesssim R \lesssim 10000$) of X-Shooter. The line-broadening from long-slit spectroscopy correlates reasonably well with velocity dispersions measured in spatially resolved data \citep{2011MNRAS.413..643R}, so we directly use the line-width $\sigma$ as a proxy for the gas velocity dispersion $\sigma_{\rm{d}}$. For the bulk of our targets, no evidence for rotation is seen in the spectra, but we caution that the rotational component contributes to the measured line-broadening, in particular at higher stellar masses \citep[e.g.,][]{2009ApJ...706.1364F}.

The virial or dynamical mass $M_{\rm{d}}$ of a galaxy is then a function of galaxy geometry, velocity dispersion $\sigma_{\rm{d}}$, and typical physical extent, or half-light radius $r_{e}$ of the galaxy via the virial theorem \citep[see, e.g.,][]{2001ApJ...554..981P} as $M_{\rm{d}} = C\sigma_{\rm{d}}^2(r_{\rm{e}}/G)$. $C$ is related to the galaxy's geometry, density, and velocity anisotropy, with $C=5$ in the case of an isotropic sphere \citep{2001ApJ...554..981P}, or $C\sim3.4$ for a disk-like geometry \citep{2006ApJ...646..107E}.

No measurements of the effective radius for the bulk of the sample is available to us because they are spatially unresolved from the ground.  {In addition, a systematic uncertainty is present at high redshift because the emission line shape is dominated by luminous \hii\, regions, probing only a part of the gravitational potential. Despite all these uncertainties, the simple velocity broadening is an important tracer of galaxy properties as demonstrated in Figs.~\ref{fig:veloz} and \ref{fig:zpropz}.} 

The peak of the line-width distribution (Fig.~\ref{fig:velohist}) at 40\,\kms\, suggests a dynamical mass on the order of $M_{\rm{d}} \sim 10^{9.4}$\,\Msun, the lowest measured values (20\,\kms) imply $M_{\rm{d}} \sim 10^{8.7}$\,\Msun. At the high $\sigma$ end, there are only three hosts above $\sigma=150$\,\kms, namely GRB~070521 ($z\sim2$ and with a large error on $\sigma$), GRB~090201 ($z\sim2$), and GRB~120118B ($z\sim3$) which indicate high dynamical masses ($M_{\rm{d}} \sim 10^{11}$\,\Msun). These three galaxies are selected through dust-extinguished GRBs and their $\sigma$ is comparable to those of the most luminous galaxies at the respective redshift interval \citep[e.g.,][]{2006ApJ...646..107E}.

With respect to redshift (Fig.~\ref{fig:veloz}), there are indications for an evolution such that lower redshift GRB hosts have -- on average -- a smaller line-broadening. In particular, there is a lack of high $\sigma$ galaxies at $z\lesssim1$ (Fig.~\ref{fig:velohist}, lower panel), while there is no obvious bias that would prevent us from selecting or observing these galaxies in the first place. The average $\sigma$ increases -- qualitatively similar to the evolution of $A_V$ in Sect.~\ref{sec:av} -- from $\sigma_{\rm{med}}=40\,$\kms\, at $z=0.5$ to $60\,\rm{km\,s^{-1}}\lesssim \sigma \lesssim 70\,$\kms\, at $z>1.0$. A mild correlation ($p=0.02$) exists also between $A_V$ and $\sigma$, two quantities which relate to the total mass of a galaxy.
 
More obvious in Fig.~\ref{fig:veloz} is the strong correlation between SFR and $\sigma$ ($<10^{-6}$ of all samples have no correlation). A similar relation between luminosity and velocity dispersion ($L\propto\sigma_{\rm{d}}^{4}$) is well-established in the local Universe. It was first observed in ellipticals by \citet{1976ApJ...204..668F}, and later also in $z\sim2$ star-forming galaxies \citep{2006ApJ...646..107E}. \citet{2005ApJ...618..569M} describe this correlation as a consequence of moderation of star formation activity through momentum-driven winds from supernovae together with radiation pressure. Star formation would drive-out gas above a critical luminosity $L_{M} \propto 4f_{g}c/G\cdot\sigma_{\rm{d}}^{4}$, where $f_g$ is the gas fraction. At the bolometric luminosity $L_M$ a star-forming galaxy then self-regulates its SFR and reaches its maximum luminosity.

The correlation between SFR and $\sigma$ in Fig.~\ref{fig:veloz} is consistent with $SFR \propto \sigma^4$, but there is considerable dispersion in the data. To test for a possible redshift evolution, we fit galaxies below and above $z=1$ with a fixed proportionality of $\sigma^4$. This returns within errors similar normalizations with minor evidence ($1.5\,\sigma$) for the higher-redshift galaxies shifted towards higher SFRs for a given $\sigma$. This would indicate higher gas-fractions in the model of \citet{2005ApJ...618..569M}.

\subsection{Metallicity}
\label{sec:met}

\subsubsection{Metallicity measurements from strong-line diagnostics}

One of the most fundamental -- but also most difficult to measure -- characteristic of galaxies is their metal abundance. While for GRB afterglows, metallicity can be probed directly and accurately in absorption at $z>1.8$ through their DLAs \citep[e.g.,][]{2004A&A...419..927V, 2006NJPh....8..195S, 2007ApJ...666..267P, 2009A&A...506..661L}, for faint (GRB-selected) galaxies it is unavoidable to resort to diagnostic ratios of strong emission lines \citep[e.g.,][]{2006ApJ...644..813E, 2008A&A...488..463M}. These strong line diagnostics use metallicity-dependent ratios of e.g., \oii\, and \oiii\, to \hb\, \citep[e.g.,][]{1979MNRAS.189...95P, 1991ApJ...380..140M, 1994ApJ...420...87Z}, \nii\, to \ha\, \citep{1979A&A....78..200A, 2004MNRAS.348L..59P}, or \nii\, to \oii\, \citep[e.g.,][]{2002ApJS..142...35K} to derive an oxygen abundance $\oh$. For an extensive discussion on the strong line ratios, we refer to \citet{2002ApJS..142...35K}.
 
In particular at high redshift, a reliable determination of the metal content in the hot gas phase becomes challenging through these line ratios \citep[e.g.,][]{2008A&A...488..463M}. As we have seen in the previous paragraphs, the physical conditions, e.g., ionization parameter, SFR, possibly dust content, and gas-fractions are different from those of low-redshift galaxies on which most of the various strong-line diagnostics are calibrated. Currently, only very few directly measured abundances from high-redshift galaxies are available \citep[e.g.,][]{2012MNRAS.427.1973C} and none of the strong-line ratios are thus validated at $z\sim2$. Any metallicity that is derived from these strong-line ratios must thus be taken cautiously \citep[e.g.,][]{2014ApJ...795..165S}.

In addition, we are faced here with the problem that our sample spans a broad range of redshifts, broad range of SFRs and host masses, and a different set of lines is detected for every host. Because of the different calibrations inherent in different strong-line diagnostics, any comparison of them needs thorough consideration \citep[e.g.,][]{2008ApJ...681.1183K}.

Given the aforementioned constraints, we decide to base our metallicity measurements on the single set of calibrators from \citet{2006A&A...459...85N} and \citet{2008A&A...488..463M}, which provide a homogeneous set of line diagnostics for different line ratios. Specifically, we use the five diagnostics based on $R_{23}$, \oii\,and \neiii, \oiii\,and \nii, \ha\,and \nii, \nii\,and \oii\, and simultaneously minimize $\oh$ against the available data. Via the line-flux measurements and errors (i.e., likelihood distributions of fluxes), we construct probability density functions (PDFs) for each diagnostic ratio, as well as one PDF for the combination of them. Section~\ref{sec:metmeas} and Fig.~\ref{fig:zexamples} in the Appendix provide more details on the procedure and individual events. 

We also provide all line-flux measurements and errors in Tables~\ref{tab:balmerlines} and \ref{tab:forbiddenlines}, so oxygen abundances can be reproduced and recalculated on any given metallicity scale with the preferred strong-line diagnostic.

\subsubsection{Sample properties}
\label{sec:metprop}

\begin{figure}
\includegraphics[angle=0, width=0.99\columnwidth]{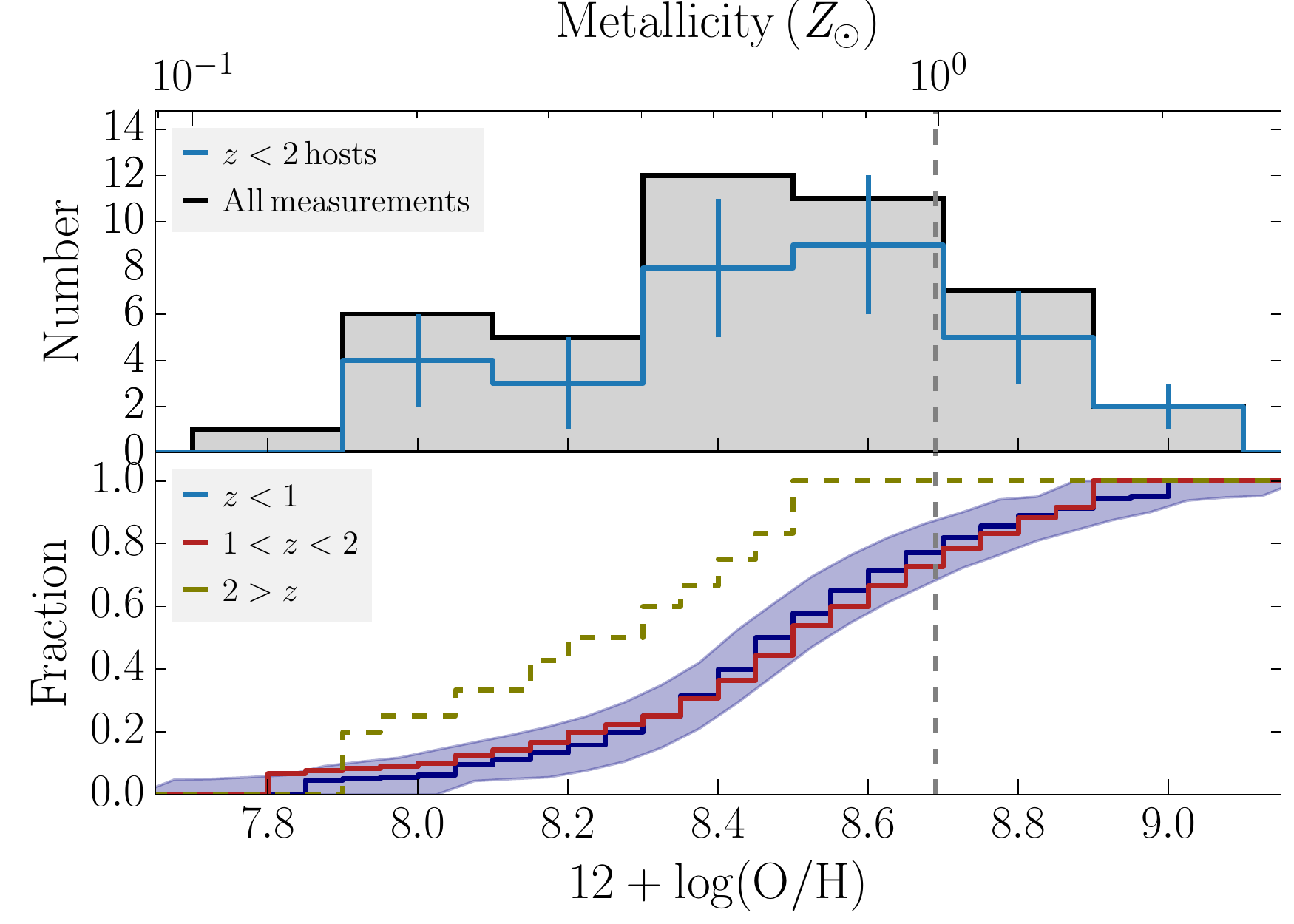}
\caption{The top panel shows the histogram of measured oxygen abundances ($\oh$) derived through strong-line diagnostic ratios in blue following Sects.~\ref{sec:dusty}~and~\ref{sec:statana} and in black for all data. In the bottom panel we plot cumulative distributions of the GRB host metallicities in different redshift intervals, where blue/red/olive colors denote redshift intervals below $z=1$ ($z_{\rm{med}}=0.6$), $1<z<2$ ($z_{\rm{med}}=1.4$), and $2<z$ ($z_{\rm{med}}=2.4$), respectively. $Z_{\odot}$ is indicated by the vertical, gray dashed line. Error-regions of the $1<z<2$ sample have a similar size to the one for $z<1$, but are not shown to enhance clarity.}
\label{fig:zhist}
\end{figure}

In total, we measure well-constrained gas-phase metallicities for 44 events (Table~\ref{tab:physprop}) in a redshift range between $z=0.3$ and $z=3.4$ with a median of $z_{\rm {med}}=1.0$. The oxygen abundances are distributed between $\oh=7.9$ (GRB~120118B) and $\oh=9.0$ (GRB~051117B) with a peak of the distribution at $\oh \sim 8.5$ (Figure~\ref{fig:zhist}). 

Intriguingly, we detect a non-negligible fraction of GRB hosts at high metallicities ($Z\gtrsim Z_{\odot}$, Fig.~\ref{fig:zhist}, \citealp[see also][]{2010ApJ...712L..26L, 2013A&A...556A..23E}). It is thus obvious that there is no strict cut-off in host metallicities at least up to the solar value. We thus directly confirm the predictions from previous studies which indicated that GRBs form in various types of star-forming galaxies, even metal rich ones \citep{2007ApJ...660..504B, 2011A&A...534A.108K, 2013ApJ...778..128P, 2015arXiv150402479P}.

{The number of hosts with a metallicity above solar, however, is relatively small. From the sub-sample of galaxies with metallicity measurements, we derive fractions of $24\pm 10$\%, $26\pm 13$\%, $7_{-4}^{+10}$\% for $z<1$, $1<z<2$, and $2<z$. Under the assumption that the emission-line width $\sigma$ traces oxygen abundance as shown in Sect.~\ref{sec:corrzz}, we can estimate metallicities also for those galaxies for which the strong-line diagnostics can not be applied. For instance, the host of GRB~050525A at $z=0.6$ is faint ($R=25.7$~mag, \citealp{2012ApJ...756..187H}), has a narrow line width and low SFR, and is thus likely metal-poor (the correlations from Sect.~\ref{sec:corrzz} indicate $\oh\sim8.2$). When including all galaxies of this work, the fractions of super-solar metallicity hosts become $16\pm7$\%, $25\pm9$\%, $4_{-2}^{+5}$\% at $z<1$, $1<z<2$, and $2<z$, respectively.}

At the highest redshifts $z\gtrsim3$, emission-line metallicities are traditionally very expensive and challenging to obtain. We derive through observations of \neiii\,for GRBs~110818A ($z=3.36$), 111123A ($z=3.15$), and 120118B ($z=2.94$) metallicities of $\oh=8.25_{-0.25}^{+0.17}$, $\oh=8.01\pm0.28$, and $\oh=7.89_{-0.17}^{+0.23}$, respectively. Comparing this with results on galaxy metallicities at $z\sim3.5$ and above \citep{2008A&A...488..463M, 2011ApJ...739....1L}, we see that those three events are probing the highest measured metallicities of LBGs at the given redshift. This is consistent with the significant line broadening and high SFRs (Sects.~\ref{sec:velo}~and~\ref{sec:sfr}).

\subsubsection{Evolution with redshift}

To test for an evolution of the metal content in the gas-phase with redshift, we divide the galaxies into low ($z<1$, $z_{\rm{med}}=0.6$), medium ($1<z<2$, $z_{\rm{med}}=1.4$), and high redshift ($z>2$, $z_{\rm{med}} = 2.4$) samples and plot histograms of $\oh$ in the bottom panel of Fig.~\ref{fig:zhist}. Remarkably, the cumulative distributions appear similar in the two low-redshift intervals. In contrast, the $z>2$ sample is somewhat shifted towards lower metallicities.

Cosmic-chemical evolution arguably proceeds most rapidly at higher redshifts. Between $z\sim0$ and $z\sim0.7$ the decrease in $\oh$ for a galaxy of a given mass seems relatively modest at around 0.1-0.2~dex \citep{2004ApJ...613..898T, 2005ApJ...635..260S}, even though this difference in redshift corresponds to over 5 Gyr in the evolution of the Universe. The displacement of the mass-metallicity relation at high redshift is likely much larger than this, reaching $\sim$0.5~dex at the high mass ($10^{11}$\,\Msun) end, and 1 dex at $10^{9}$\,\Msun\,at $z\sim3.5$ \citep{2008A&A...488..463M, 2014A&A...563A..58T}. This leads to relatively low gas-phase metallicities in the sample although we are only sensitive to the more luminous GRB hosts at $z>2$.

\subsubsection{Metallicity dependence of GRB hosts}

\begin{figure}
\includegraphics[angle=0, width=0.99\columnwidth]{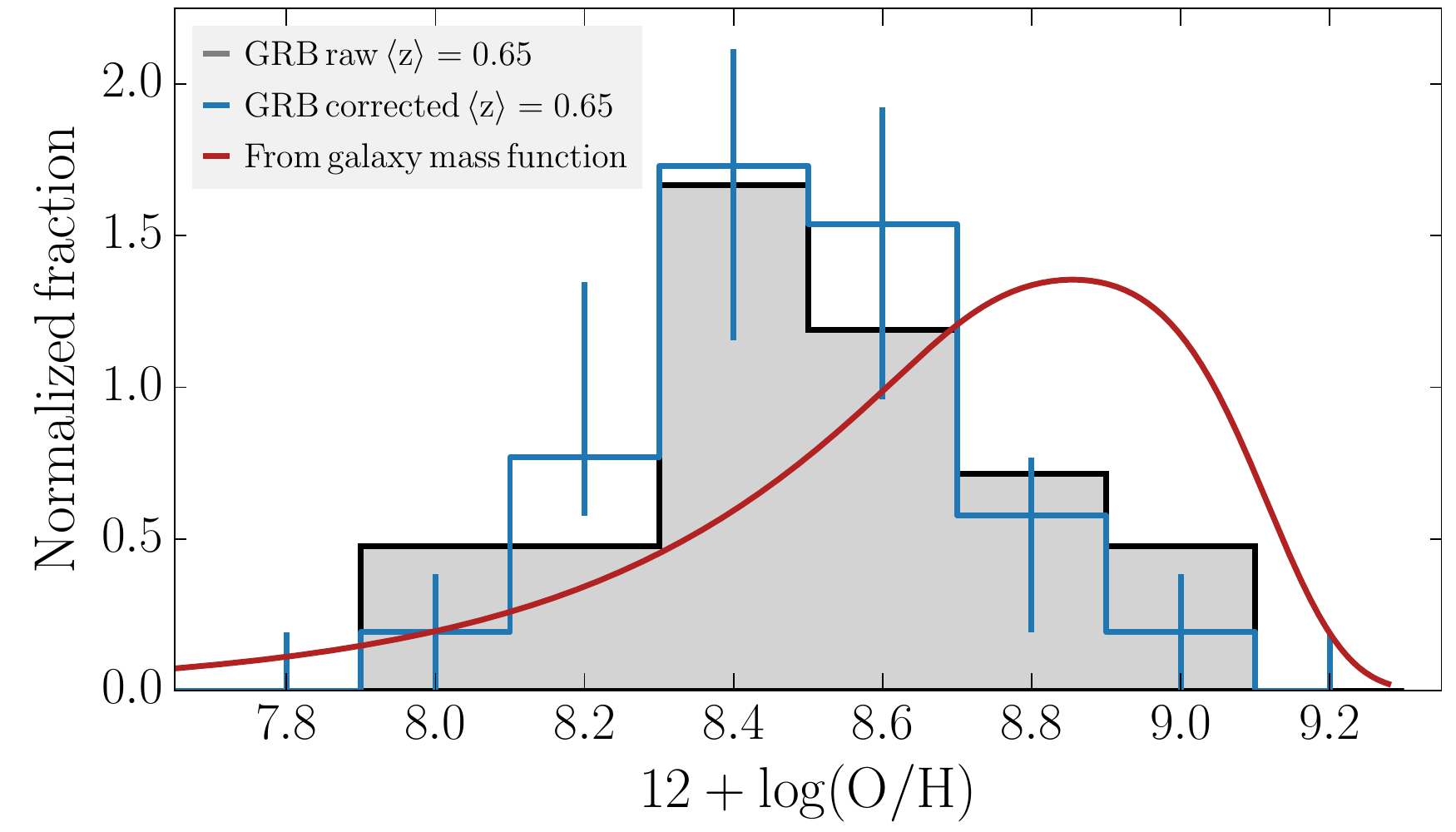}
\caption{ {The blue-solid histogram with error bars shows the metallicity distribution and its uncertainty of 29 $0.3<z<1$ GRB hosts ($\langle z \rangle = 0.65$). We applied a correction for the over-proportionality of dusty GRBs in the sample (Sects.~\ref{sec:dusty}~and~\ref{sec:statana}) and for completeness from the line broadening. Black data show raw measurements. The red line is the expectation from the stellar-mass function at a similar redshift \citep{2014ApJ...783...85T} if GRBs traced star formation without metallicity dependence. Shown curves are normalized to an area of 1.}}
\label{fig:metvsmass}
\end{figure}

To understand the effects of metallicity on GRB progenitors and host galaxies and the implications for their ability to trace star formation, we summarize the primary results from the previous sections: First, GRB hosts with oxygen abundances of $\oh\sim8.5$ are common in our sample. Second, the $z<1$ and $1<z<2$ GRB host metallicity distributions appear to be very similar. Third, there is a modest fraction ($\sim 20\%$ at $z<1$) of GRBs in $Z > Z_{\odot}$ galaxies. And last, the maximum metallicity of the most luminous GRB hosts at $z\sim3$ is $\oh\sim8.4$.

{In particular the relatively low fraction of GRB hosts with very high oxygen abundances points strongly towards a metallicity dependence in low-$z$ GRBs.}  {In Fig.~\ref{fig:metvsmass}, we compare the metallicity distribution derived from 29 $0.3<z<1$ GRB hosts ($\langle z \rangle = 0.6$) with the expectation from the contribution of field galaxies to the global SFR density at a similar redshift. }

As no sufficiently complete sample of field-galaxy metallicities is available in the literature at $z\sim0.6$, this comparison is based on the stellar-mass function from deep photometric surveys. To derive the red lines in Fig.~\ref{fig:metvsmass}, we start with the double Schechter parameterization of the galaxy-mass function {for star-forming galaxies at $0.5<z<0.75$} from \citet{2014ApJ...783...85T}\footnote{{Only subtle differences are introduced in the result when using the mass-function of star-forming galaxies from \citet{2013A&A...556A..55I} in a similar redshift interval.}}. {Passively evolving galaxies are hence implicitly excluded from this comparison.} Using the mean-value of the SFR-stellar mass relation and its dispersion\footnote{{Here, we assume an intrinsic dispersion of 0.3~dex \citep{2014ApJS..214...15S} and a slope of 0.8 in the stellar-mass range below $10^{10}$\,\Msun\, \citep{2009MNRAS.394....3D, 2009A&A...504..751S, 2014MNRAS.437.3516S} and 0.55 above \citep{2012ApJ...754L..29W}.}}, we then calculate the 50\% quartile of SFR for a given stellar mass.
 
 {Finally, we use the parametrization of \citet{2010MNRAS.408.2115M} to convert the SFR, stellar mass pairs into a metallicity in the same scale as our GRB measurements and weight the resulting galaxy-metallicity function with the SFR\footnote{The figure and results remain conceptually unchanged if we use the host's $\oh$ and SFR to derive a stellar mass via \citet{2010MNRAS.408.2115M} to compare it with the SFR-weighted stellar-mass function.}.}

 {The fraction of star formation in galaxies above $Z_{\odot}$ can then be obtained via the integral under the red curve of Fig.~\ref{fig:metvsmass}, and is around 50\,\%\footnote{Similar values are inferred through the analysis of \citet{2013ApJ...778..128P}, \citet{2014A&A...565A.112H} and \citet{2014MNRAS.437.3516S}, which use deep photometric surveys \citep{2009ApJ...702.1393K, 2013A&A...556A..55I} to calculate a galaxy stellar mass of $10^{9.7-10.2}$\,\Msun\, at $z\sim0.6$ above or below which half of all star formation occurs. This implies a metallicity between $Z=0.9\,Z_{\odot}$ and $Z=1.4\,Z_{\odot}$ in our oxygen-abundance scale \citep{2010MNRAS.408.2115M}.}. Similarly, \citet{2011MNRAS.417.1013C} use N-body simulations with semi-analytical galaxy-formation models to predict around $\sim 50\%$ of GRBs in galaxies of super-solar metallicity in cases where GRBs traced all star formation equally (their Fig. 4).}

Our results therefore directly confirm studies that at low redshift, GRB host properties are affected by a tendency of GRBs {to occur in environments with a metallicity below the solar value} \citep[e.g.,][]{2006ApJ...642..636L, 2009ApJ...702..377K, 2010AJ....140.1557L, 2013ApJ...774..119G, 2013ApJ...778..128P, 2013A&A...557A..34B, 2014arXiv1409.7064V, 2015arXiv150304246S, 2015arXiv150402479P}. The scarcity of high $A_V$ and high SFR galaxies as discussed in previous sections further strengthens this interpretation.

{The physical driver behind the environmental bias in the GRB production efficiency could in principle relate to either (or both), metallicity or a high stellar mass. Because of the well-studied impact that metallicity has on stellar evolution, and the lack of an obvious link between GRB formation and a galaxy's stellar mass, we consider a metallicity effect as the most logical explanation for our observations. This interpretation is further supported by the increasing stellar mass of GRB hosts with increasing redshift \citep{2015arXiv150402479P}.}

The observed metallicity dependence is unlikely to play a dominant role at $z\gtrsim3$. Already $z > 2$, the metallicity distribution is somewhat shifted towards lower metallicities although the selection biases against low-luminosity hosts become more severe. At $z\sim3$, galaxies of solar metallicity are scarce even among the most luminous galaxies \citep{2014A&A...563A..58T}. Because the metallicity of star-forming galaxies in general is decreasing, GRBs are expected to form also more frequently in the most luminous galaxies. Therefore, we expect GRB hosts at high redshift to represent star-forming galaxies much better than at $z<1$ \citep[see also][]{2015arXiv150305323G, 2015arXiv150402479P}.

There is already evidence that the implications of the metallicity dependence are weaker at $z\sim2$ \citep{2013ApJ...778..128P, 2015arXiv150304246S}, than in $z<1$ GRB-selected galaxies.

\subsubsection{Correlations of metallicity with other host properties}
\label{sec:corrzz}

\begin{figure}
\includegraphics[angle=0, width=0.99\columnwidth]{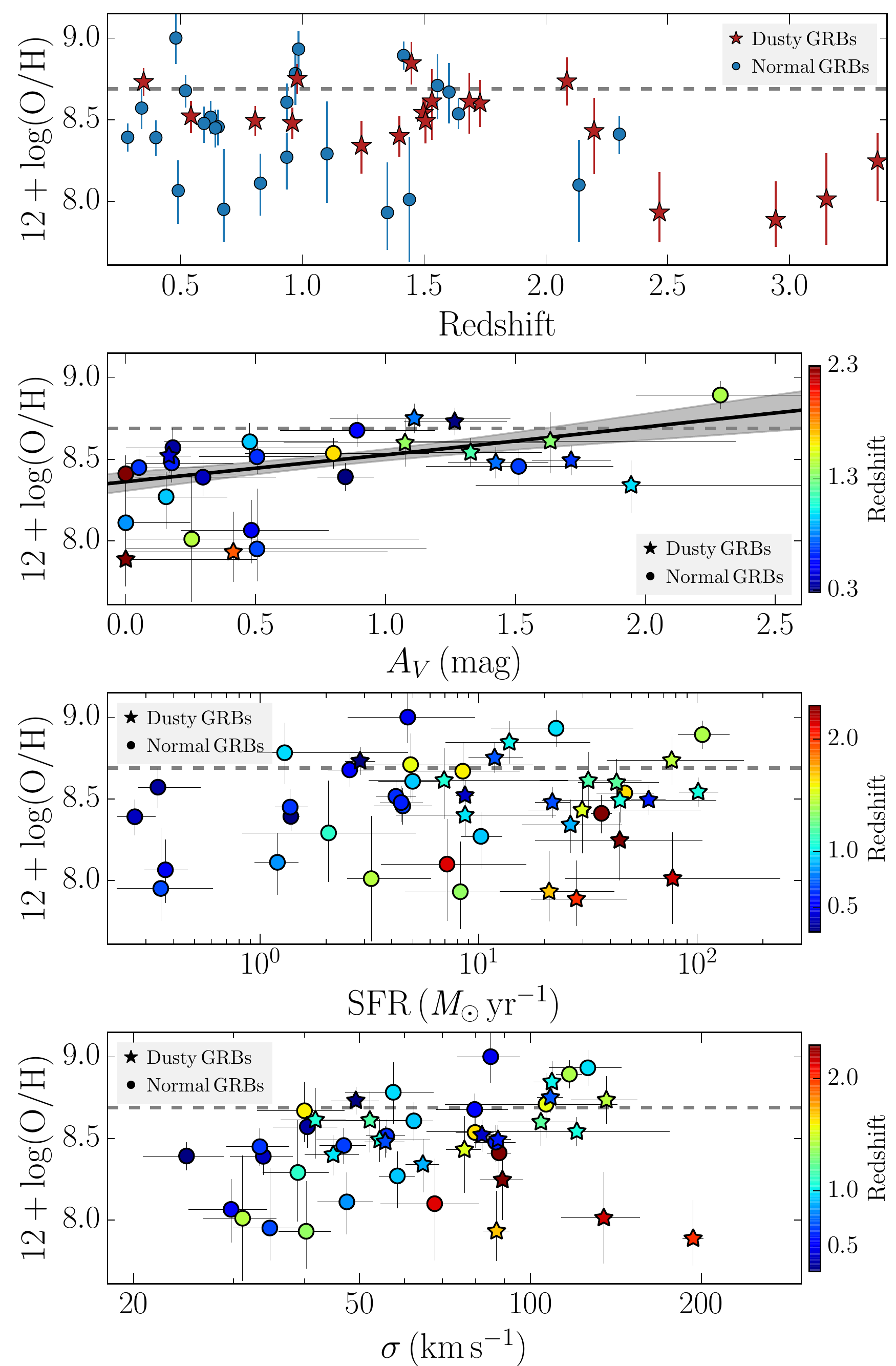}
\caption{Gas-phase metallicity as a function of (from top to bottom): redshift, $A_V$, SFR, and $\sigma$. Black data and error bars represent individual galaxies. The color coding corresponds to redshift, with the scale given by the color bar. $Z_{\odot}$ is indicated by the horizontal, gray dashed line.}
\label{fig:zprop}
\end{figure}

\begin{figure}
\includegraphics[angle=0, width=0.99\columnwidth]{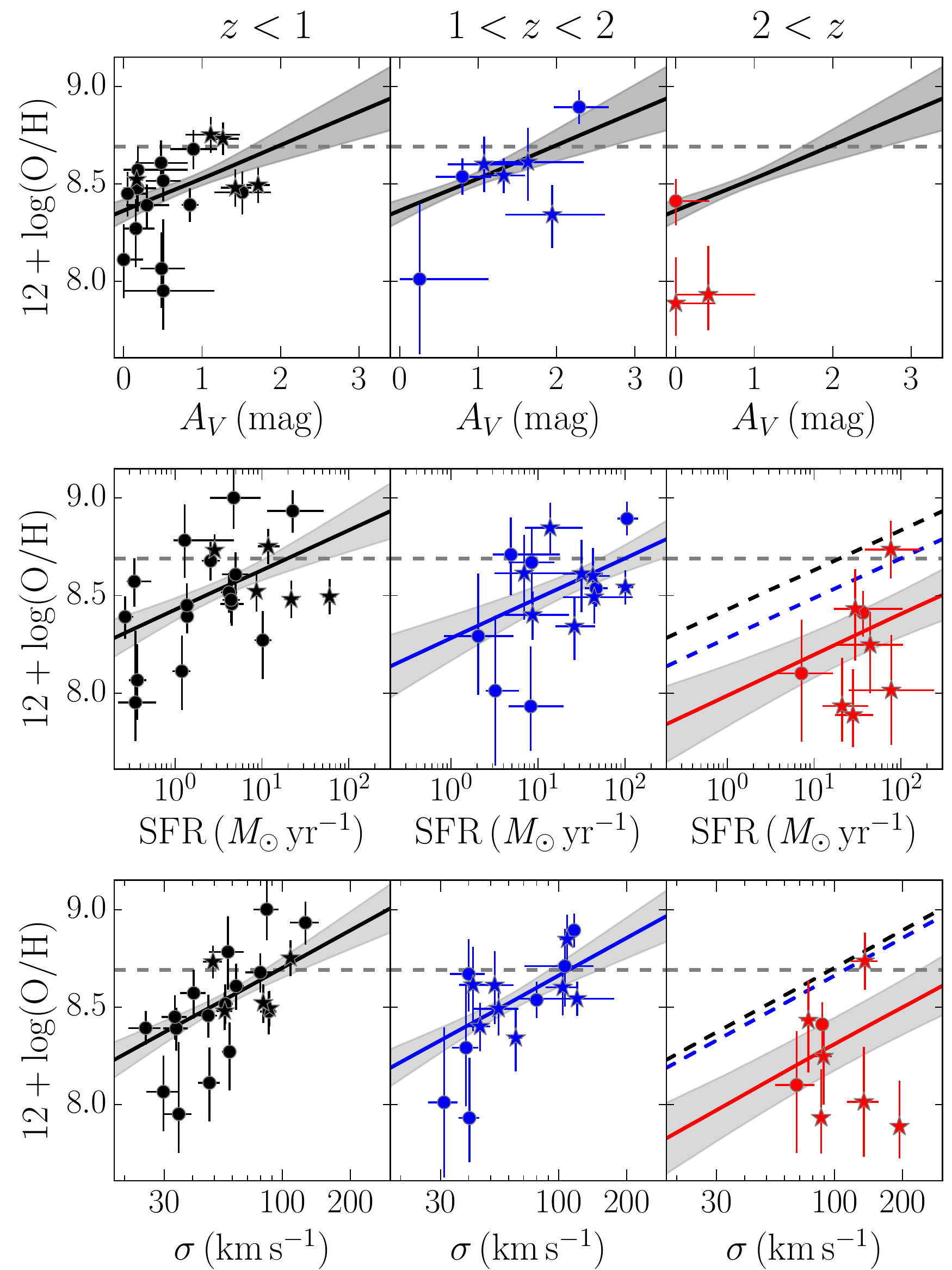}
\caption{Gas-phase metallicity as a function of $z$ and (from top to bottom): $A_V$, SFR, and $\sigma$ as well as $z < 1$, $1<z<2$, and $2<z$ from left to right. Solid lines denote the best linear regression, with regions of 68\,\% probability in shaded areas. The last row shows also the relations from the lower-redshift panels in dashed lines. $Z_{\odot}$ is indicated by the horizontal, gray dashed line.}
\label{fig:zpropz}
\end{figure}

Figure~\ref{fig:zprop} shows the dependence of host metallicity with redshift, dust reddening, SFR, and emission-line width. A correlation exist ($p=0.0007$ of all iterations return no dependence) between $A_V$ and oxygen abundance similar to $\oh = 8.36 + 0.17 \cdot A_V/(\rm{mag})$. This relation is driven by $z<2$ events, as only a few $z>2$ hosts have sufficient data to constrain both $A_V$ and $\oh$\,at the same time. At somewhat higher redshift, a correlation between metallicity and dust content is also seen directly along GRB sight-lines as traced by the GRB afterglow and its DLA \citep[][]{2013A&A...560A..26Z}. 

There are no strong correlations between SFR or $\sigma$ and $\oh$ in the full host sample, but this is the result of the large redshift range present in Fig.~\ref{fig:zprop}. In Fig.~\ref{fig:zpropz} we plot oxygen abundance versus $A_V$, SFR, and $\sigma$ divided into three redshift bins, $z<1$ with $z_{\rm{med}}\sim0.6$, $1<z<2$ with $z_{\rm{med}}\sim1.4$, and $2<z$ with $z_{\rm{med}}\sim2.4$. A mild correlation exists between SFR and $\oh$ similar to $\oh\propto {SFR}^{0.20\pm0.06}$ ($p=0.004$). Using a common slope, but a different normalization in the three redshift ranges, we fit all data simultaneously and derive

\begin{equation}
\label{eq:ohsfr}
\begin{split}
\oh_{z<1} = 8.63 + 0.20\cdot \log({SFR/10\,M_\odot\,\rm{yr}^{-1}}) \\
\oh_{1<z<2} = 8.49 + 0.20\cdot \log({SFR/10\,M_\odot\,\rm{yr}^{-1}})  \\
\oh_{2<z} = 8.20 + 0.20\cdot \log({SFR/10\,M_\odot\,\rm{yr}^{-1}})  \\
\end{split}
\end{equation}

 {with uncertainties of 0.05 in the intercept in the earlier two, and 0.11 in the $z>2$ case. The root mean square (RMS) dispersion in the $z<1$ and $1<z<2$ redshift range is 0.25~dex each.}

The decrease in $\log(\rm{O/H})$ for a given SFR between $z\sim0.6$ and $z\sim1.4$ is $0.14\pm0.08$~dex, but $0.4\pm0.1$~dex for the $z > 2$ sample (Fig.~\ref{fig:zpropz}). This is consistent with the relatively slow evolution of the mass-metallicity relation at lower redshift, as well as a more rapid change at $z > 1.5$. 

The tightest and most significant correlation ($p<0.0001$) is present between $\oh$ and $\sigma$ as $\oh\propto\sigma^{0.64\pm0.16}$ at $z<2$. We again fit all data together with a common slope and free normalization in the three redshift intervals and obtain for $\sigma$

\begin{equation}
\label{eq:ohsigma}
\begin{split}
\oh_{z<1} = 8.70 + 0.64\cdot \log(\sigma/\rm{100\,km\,s^{-1}}) \\
\oh_{1<z<2} = 8.66 + 0.64\cdot \log(\sigma/\rm{100\,km\,s^{-1}}) \\
\oh_{2<z} = 8.31 + 0.64\cdot \log(\sigma/\rm{100\,km\,s^{-1}})
\end{split}
\end{equation}

with uncertainties of 0.05/0.06 in the intercept in the earlier two, and 0.12 in the $z>2$ case. The behavior is qualitatively very similar to relation between SFR and oxygen abundance but with a lower scatter, i.e., a tighter correlation. 

Low- and medium-redshift samples return comparable best-fit relations. In both cases $\sigma=100$\,\kms\, roughly corresponds to a metallicity of $Z\approx Z_{\odot}$. At $z > 2$, this value shifts downward by $0.4\pm0.1$~dex to an oxygen abundance of $\oh \approx 8.3$, but is affected by large measurement uncertainties.  {The RMS dispersion in the $z<1$ and $1<z<2$ redshift range is only 0.19 dex and 0.20~dex, respectively. Given that the typical error in the metallicity measurement via the strong-line diagnostic is already 0.11/0.16~dex, this implies that the very simple measurement of $\sigma$ provides already a good estimate of the gas-phase metallicity.}

Taken at face value, there is no statistically significant correlation in the highest-redshift bin ($z>2$) between SFR and $\oh$\, as well as $\sigma$ and $\oh$. Because of the large measurement uncertainties however, the data are consistent within errors with the lower-redshift slope, but, as argued above, a lower normalization.

\section{Summary} 
\label{sec:conc}

We present data and initial results from X-Shooter emission-line spectroscopy of star-forming galaxies selected through long GRBs in a redshift range $0.1<z<3.6$. In total, we discuss observations of 96 individual targets which allow us to trace the physical properties of GRB hosts through cosmic time. The galaxies have a median redshift of $z_{\rm med}=1.6$ and most of them are in the range $0.5 < z < 1.5$ (35\%) and $1.5 < z < 2.5$ (41\%). We carefully test for selection effects in the sample, and statistically correct for a somewhat higher fraction of dusty GRBs ($35\,\pm5\%$) when compared to the one of representative GRB samples ($20-30\,\%$). 

Our observation yield 16 new GRB redshifts, which increase the redshift completeness of previous samples to 87\% (TOUGH) and 97\% (BAT6), with median redshifts of $z_{\rm{med}} = 2.06\pm0.18$ and $z_{\rm{med}} = 1.67\pm0.15$, respectively.

After excluding a very nearby event (GRB~100316D), a possibly short GRB (GRB~100816A), and five galaxies without clear emission lines we arrive at a total number of 89 emission-line spectra from galaxies selected by GRBs. Based on these data, we provide new scaling relations to derive SFRs from emission lines of \oii, \hb, and \oiii\, even in the absence of a reliable dust correction (but with a large scatter). We measure galaxy-integrated star formation rates (SFRs), visual dust-attenuations ($A_V$), line broadening ($\sigma$), and oxygen abundances ($\oh$) for our targets (Table~\ref{tab:physprop}). This yields the largest and most comprehensive sample of emission-line spectroscopy of GRB hosts available to date and allows us to study the distribution of the physical parameters and their evolution with redshift. 

Our main results can be summarized as follows:
\begin{enumerate}
\item The distribution of $A_V$ for long GRB hosts is broad with the bulk of the targets at $A_V<1.5\,\rm{mag}$ and median $A_V^{\rm med} \sim 0.75$\,mag at $z<2$. We observe a mild trend with redshift such that GRB hosts at $z = 1.5$ tend to have larger dust content ($A_V^{\rm med} \sim 1.2$\,mag) and line broadening ($\sigma_{\rm{med}}\sim70$\,kms) on average than low-redshift hosts ($A_V^{\rm med} \sim 0.6$\,mag, and $\sigma_{\rm{med}}\sim40$\,kms). The significance of each of these results is only around $2\sigma$, but the evolution in $A_V$ is remarkably consistent with a similar behavior observed for GRB afterglows \citep{2013MNRAS.432.1231C}.
\item There is a strong evolution of the average GRB host properties with redshift. The median SFR evolves from $SFR_{\rm{med}} = 0.6$\,\Msunyr\,at $z\sim0.6$ up to $SFR_{\rm{med}} = 15$\,\Msunyr\,at $z\sim2$ above which it does not increase significantly any further. 
\item The ratio of $\oiii/\oii$ of the galaxies increases with redshift as well, which moves the star-forming galaxies away from the location of local galaxies in the BPT diagram. The value changes from  $\oiii/\oii\sim0.9$ at $z\sim0.3$ by a factor of three to $\oiii/\oii\sim3$ at $z\sim3$. While roughly similar at $z\sim0.6$ and $z\sim1.4$, the metallicity distribution of GRB hosts starts to shift to lower metallicities above $z=2$.
\item We find a strong correlation between $\sigma$ and SFR broadly consistent with ${SFR}\propto\sigma^4$ as known from the local Faber-Jackson relation. There is no statistically significant evidence in the data that this proportionality evolves with redshift. 
\item Further correlations exist between oxygen abundance and visual attenuation ($\oh \propto 0.17\cdot A_V$), SFR  ($\oh \propto {SFR}^{0.2}$), and $\sigma$ ($\oh \propto\sigma^{0.6}$). In the last two cases, we observe a relatively similar behavior at $z < 1$ and $1<z<2$, but an evolution towards lower metallicities at $z > 2$ by $\sim 0.4$~dex for a given SFR or $\sigma$.

\item We detect several hosts with high oxygen abundances which rule out a strict metallicity cutoff for GRB hosts below the solar values (in the adopted scale). At the same time, however, the fraction of GRBs hosted in $Z>Z_{\odot}$ galaxies at $z<1$ is smaller ($18\pm7\,\%$) than what would be expected from the contribution of similarly metal-rich galaxies to the total cosmic SFR ($\gtrsim 50\%$). A mechanisms is thus at place that quenches GRB formation at the highest metallicities. 
\end{enumerate}

At $z=3$, most of the star formation takes place in galaxies of $Z\lesssim0.5\cdot Z_\odot$. At the same time GRB hosts with similar metallicities are relatively common in our sample at lower redshift. Under the conditions that the probability of producing GRBs below a certain metallicity is constant (and not a strong function of metallicity itself) then by $z\sim3$ GRBs host will probe a large fraction of the total star formation. In absence of further secondary environmental factors, GRB hosts would then provide an extensive picture of high-redshift, star-forming galaxies. This conclusion is similar to the one reached through studying metallicities of GRB-DLAs \citep{2008ApJ...683..321F, 2014arXiv1410.3510A}, the UV-luminosity function at a similar redshift \citep{2005MNRAS.362..245J, 2015arXiv150304246S, 2015arXiv150305323G}, cosmological simulations \citep{2010MNRAS.408..647C} or modeling the mass distribution of GRB hosts \citep{2009ApJ...702..377K}.

Our results can be understood in the cosmological context and a picture drawn by recent studies \citep{2008AJ....135.1136M, 2009ApJ...702..377K, 2013A&A...557A..34B, 2013ApJ...778..128P, 2014arXiv1409.7064V, 2014ApJS..213...15W, 2014arXiv1406.1503T, 2015arXiv150402479P,  2015arXiv150304246S} in which the GRB host properties at lower redshift ($z < 1-2$) are driven by the GRB's tendency to occur in lower-metallicity galaxies without fully avoiding metal-rich ones. The scarcity of high metallicity, high $A_V$, and high SFR galaxies at low redshift strongly supports this interpretation.

\begin{acknowledgements}
We are grateful to the referee and the editor, Rubina Kotak, for comments and for providing an extensive and constructive report, which has helped to increase the quality and strength of the manuscript significantly. We thank the \textit{Swift} team for building and operating such an excellent facility. We also acknowledge the astronomical and technical support at Paranal observatory without which none of the presented spectra would have been taken. The Dark Cosmology Centre is funded by the Danish National Research Foundation. The research leading to these results has received funding from the European Research Council under the European Union's Seventh Framework Program (FP7/2007-2013)/ERC Grant agreement no. EGGS-278202. Part of the funding for GROND (both hardware as well as personnel) was generously granted from the Leibniz-Prize to Prof. G. Hasinger (DFG grant HA 1850/28-1). The data presented here were obtained in part with ALFOSC, which is provided by the Instituto de Astrofisica de Andalucia (IAA) under a joint agreement with the University of Copenhagen and NOTSA. TK thanks A. M\"{u}ller, L. Guzman, and L. Watson for providing the vibrant scientific atmosphere in Vitacura where some of this work was carried out. DM acknowledges the Instrument center for Danish astrophysics (IDA) for support. SSch acknowledges support from CONICYT-Chile FONDECYT 3140534, Basal- CATA PFB-06/2007, and Project IC120009 ''Millennium Institute of Astrophysics (MAS)'' of Iniciativa Cientifica Milenio del Ministerio de Economia, Fomento y Turismo. AR acknowledges support from PRIN-INAF 2012/13. Support for DAP was provided by NASA through Hubble Fellowship grant HST-HF-51296.01-A awarded by the Space Telescope Science Institute (STScI), which is operated by the Association of Universities  for Research in Astronomy (AURA), Inc., for NASA, under contract NAS 5-26555, and by an award issued by JPL/Caltech. EP acknowledges support from grants ASI-INAF I/088/06/0 and PRIN INAF 2011. PS acknowledges support through the Sofja Kovalevskaja Award from the Alexander von Humboldt Foundation of Germany. JFG acknowledges support through the Sofja Kovalevskaja Award to P. Schady from the Alexander von Humboldt Foundation of Germany. SK and ANG acknowledge support by DFG grant Kl 766/16-1. SSav acknowledges support from the Bundesministerium f\"ur Wirtschaft and Technologie through DLR (Deutsches Zentrum f\"ur Luft- und Raumfahrt e.V.) FKZ 50 OR 1211. Based in part on observations collected with the 2.2~m MPG telescope on La Silla  as part of the program CN2014B-102.
\end{acknowledgements}

\bibliography{./bibtex/refs}

\begin{thebibliography}{233}
\expandafter\ifx\csname natexlab\endcsname\relax\def\natexlab#1{#1}\fi

\bibitem[{{Abazajian} {et~al.}(2009){Abazajian}, {Adelman-McCarthy},
  {Ag{\"u}eros}, {Allam}, {Allende Prieto}, {An}, {Anderson}, {Anderson},
  {Annis}, {Bahcall}, \& et~al.}]{2009ApJS..182..543A}
{Abazajian}, K.~N., {Adelman-McCarthy}, J.~K., {Ag{\"u}eros}, M.~A., {et~al.}
  2009, \apjs, 182, 543

\bibitem[{{Ackermann} {et~al.}(2014){Ackermann}, {Ajello}, {Asano}, {Atwood},
  {Axelsson}, {Baldini}, {Ballet}, {Barbiellini}, {Baring}, {Bastieri},
  {Bechtol}, {Bellazzini}, {Bissaldi}, {Bonamente}, {Bregeon}, {Brigida},
  {Bruel}, {Buehler}, {Burgess}, {Buson}, {Caliandro}, {Cameron}, {Caraveo},
  {Cecchi}, {Chaplin}, {Charles}, {Chekhtman}, {Cheung}, {Chiang}, {Chiaro},
  {Ciprini}, {Claus}, {Cleveland}, {Cohen-Tanugi}, {Collazzi}, {Cominsky},
  {Connaughton}, {Conrad}, {Cutini}, {D'Ammando}, {de Angelis}, {DeKlotz}, {de
  Palma}, {Dermer}, {Desiante}, {Diekmann}, {Di Venere}, {Drell},
  {Drlica-Wagner}, {Favuzzi}, {Fegan}, {Ferrara}, {Finke}, {Fitzpatrick},
  {Focke}, {Franckowiak}, {Fukazawa}, {Funk}, {Fusco}, {Gargano}, {Gehrels},
  {Germani}, {Gibby}, {Giglietto}, {Giles}, {Giordano}, {Giroletti}, {Godfrey},
  {Granot}, {Grenier}, {Grove}, {Gruber}, {Guiriec}, {Hadasch}, {Hanabata},
  {Harding}, {Hayashida}, {Hays}, {Horan}, {Hughes}, {Inoue}, {Jogler},
  {J{\'o}hannesson}, {Johnson}, {Kawano}, {Kn{\"o}dlseder}, {Kocevski}, {Kuss},
  {Lande}, {Larsson}, {Latronico}, {Longo}, {Loparco}, {Lovellette}, {Lubrano},
  {Mayer}, {Mazziotta}, {McEnery}, {Michelson}, {Mizuno}, {Moiseev}, {Monzani},
  {Moretti}, {Morselli}, {Moskalenko}, {Murgia}, {Nemmen}, {Nuss}, {Ohno},
  {Ohsugi}, {Okumura}, {Omodei}, {Orienti}, {Paneque}, {Pelassa}, {Perkins},
  {Pesce-Rollins}, {Petrosian}, {Piron}, {Pivato}, {Porter}, {Racusin},
  {Rain{\`o}}, {Rando}, {Razzano}, {Razzaque}, {Reimer}, {Reimer}, {Ritz},
  {Roth}, {Ryde}, {Sartori}, {Parkinson}, {Scargle}, {Schulz}, {Sgr{\`o}},
  {Siskind}, {Sonbas}, {Spandre}, {Spinelli}, {Tajima}, {Takahashi}, {Thayer},
  {Thayer}, {Thompson}, {Tibaldo}, {Tinivella}, {Torres}, {Tosti}, {Troja},
  {Usher}, {Vandenbroucke}, {Vasileiou}, {Vianello}, {Vitale}, {Winer}, {Wood},
  {Yamazaki}, {Younes}, {Yu}, {Zhu}, {Bhat}, {Briggs}, {Byrne}, {Foley},
  {Goldstein}, {Jenke}, {Kippen}, {Kouveliotou}, {McBreen}, {Meegan},
  {Paciesas}, {Preece}, {Rau}, {Tierney}, {van der Horst}, {von Kienlin},
  {Wilson-Hodge}, {Xiong}, {Cusumano}, {La Parola}, \&
  {Cummings}}]{2014Sci...343...42A}
{Ackermann}, M., {Ajello}, M., {Asano}, K., {et~al.} 2014, Science, 343, 42

\bibitem[{{Aihara} {et~al.}(2011){Aihara}, {Allende Prieto}, {An}, {Anderson},
  {Aubourg}, {Balbinot}, {Beers}, {Berlind}, {Bickerton}, {Bizyaev}, {Blanton},
  {Bochanski}, {Bolton}, {Bovy}, {Brandt}, {Brinkmann}, {Brown}, {Brownstein},
  {Busca}, {Campbell}, {Carr}, {Chen}, {Chiappini}, {Comparat}, {Connolly},
  {Cortes}, {Croft}, {Cuesta}, {da Costa}, {Davenport}, {Dawson}, {Dhital},
  {Ealet}, {Ebelke}, {Edmondson}, {Eisenstein}, {Escoffier}, {Esposito},
  {Evans}, {Fan}, {Femen{\'{\i}}a Castell{\'a}}, {Font-Ribera}, {Frinchaboy},
  {Ge}, {Gillespie}, {Gilmore}, {Gonz{\'a}lez Hern{\'a}ndez}, {Gott}, {Gould},
  {Grebel}, {Gunn}, {Hamilton}, {Harding}, {Harris}, {Hawley}, {Hearty}, {Ho},
  {Hogg}, {Holtzman}, {Honscheid}, {Inada}, {Ivans}, {Jiang}, {Johnson},
  {Jordan}, {Jordan}, {Kazin}, {Kirkby}, {Klaene}, {Knapp}, {Kneib},
  {Kochanek}, {Koesterke}, {Kollmeier}, {Kron}, {Lampeitl}, {Lang}, {Le Goff},
  {Lee}, {Lin}, {Long}, {Loomis}, {Lucatello}, {Lundgren}, {Lupton}, {Ma},
  {MacDonald}, {Mahadevan}, {Maia}, {Makler}, {Malanushenko}, {Malanushenko},
  {Mandelbaum}, {Maraston}, {Margala}, {Masters}, {McBride}, {McGehee},
  {McGreer}, {M{\'e}nard}, {Miralda-Escud{\'e}}, {Morrison}, {Mullally},
  {Muna}, {Munn}, {Murayama}, {Myers}, {Naugle}, {Neto}, {Nguyen}, {Nichol},
  {O'Connell}, {Ogando}, {Olmstead}, {Oravetz}, {Padmanabhan},
  {Palanque-Delabrouille}, {Pan}, {Pandey}, {P{\^a}ris}, {Percival},
  {Petitjean}, {Pfaffenberger}, {Pforr}, {Phleps}, {Pichon}, {Pieri}, {Prada},
  {Price-Whelan}, {Raddick}, {Ramos}, {Reyl{\'e}}, {Rich}, {Richards}, {Rix},
  {Robin}, {Rocha-Pinto}, {Rockosi}, {Roe}, {Rollinde}, {Ross}, {Ross},
  {Rossetto}, {S{\'a}nchez}, {Sayres}, {Schlegel}, {Schlesinger}, {Schmidt},
  {Schneider}, {Sheldon}, {Shu}, {Simmerer}, {Simmons}, {Sivarani}, {Snedden},
  {Sobeck}, {Steinmetz}, {Strauss}, {Szalay}, {Tanaka}, {Thakar}, {Thomas},
  {Tinker}, {Tofflemire}, {Tojeiro}, {Tremonti}, {Vandenberg}, {Vargas
  Maga{\~n}a}, {Verde}, {Vogt}, {Wake}, {Wang}, {Weaver}, {Weinberg}, {White},
  {White}, {Yanny}, {Yasuda}, {Yeche}, \& {Zehavi}}]{2011ApJS..193...29A}
{Aihara}, H., {Allende Prieto}, C., {An}, D., {et~al.} 2011, \apjs, 193, 29

\bibitem[{{Alloin} {et~al.}(1979){Alloin}, {Collin-Souffrin}, {Joly}, \&
  {Vigroux}}]{1979A&A....78..200A}
{Alloin}, D., {Collin-Souffrin}, S., {Joly}, M., \& {Vigroux}, L. 1979, \aap,
  78, 200

\bibitem[{{Appenzeller} {et~al.}(1998){Appenzeller}, {Fricke}, {F{\"u}rtig},
  {G{\"a}ssler}, {H{\"a}fner}, {Harke}, {Hess}, {Hummel}, {J{\"u}rgens},
  {Kudritzki}, {Mantel}, {Meisl}, {Muschielok}, {Nicklas}, {Rupprecht},
  {Seifert}, {Stahl}, {Szeifert}, \& {Tarantik}}]{1998Msngr..94....1A}
{Appenzeller}, I., {Fricke}, K., {F{\"u}rtig}, W., {et~al.} 1998, The
  Messenger, 94, 1

\bibitem[{{Arabsalmani} {et~al.}(2015){Arabsalmani}, {M{\o}ller}, {Fynbo},
  {Christensen}, {Freudling}, {Savaglio}, \& {Zafar}}]{2014arXiv1410.3510A}
{Arabsalmani}, M., {M{\o}ller}, P., {Fynbo}, J.~P.~U., {et~al.} 2015, \mnras,
  446, 990

\bibitem[{{Arnouts} {et~al.}(1999){Arnouts}, {Cristiani}, {Moscardini},
  {Matarrese}, {Lucchin}, {Fontana}, \& {Giallongo}}]{1999MNRAS.310..540A}
{Arnouts}, S., {Cristiani}, S., {Moscardini}, L., {et~al.} 1999, \mnras, 310,
  540

\bibitem[{{Asplund} {et~al.}(2009){Asplund}, {Grevesse}, {Sauval}, \&
  {Scott}}]{2009ARA&A..47..481A}
{Asplund}, M., {Grevesse}, N., {Sauval}, A.~J., \& {Scott}, P. 2009, \araa, 47,
  481

\bibitem[{{Baldwin} {et~al.}(1981){Baldwin}, {Phillips}, \&
  {Terlevich}}]{1981PASP...93....5B}
{Baldwin}, J.~A., {Phillips}, M.~M., \& {Terlevich}, R. 1981, \pasp, 93, 5

\bibitem[{{Berger} {et~al.}(2007){Berger}, {Fox}, {Kulkarni}, {Frail}, \&
  {Djorgovski}}]{2007ApJ...660..504B}
{Berger}, E., {Fox}, D.~B., {Kulkarni}, S.~R., {Frail}, D.~A., \& {Djorgovski},
  S.~G. 2007, \apj, 660, 504

\bibitem[{{Bloom} {et~al.}(2002){Bloom}, {Kulkarni}, \&
  {Djorgovski}}]{2002AJ....123.1111B}
{Bloom}, J.~S., {Kulkarni}, S.~R., \& {Djorgovski}, S.~G. 2002, \aj, 123, 1111

\bibitem[{{Boissier} {et~al.}(2013){Boissier}, {Salvaterra}, {Le Floc'h},
  {Basa}, {Buat}, {Prantzos}, {Vergani}, \& {Savaglio}}]{2013A&A...557A..34B}
{Boissier}, S., {Salvaterra}, R., {Le Floc'h}, E., {et~al.} 2013, \aap, 557,
  A34

\bibitem[{{Brinchmann} {et~al.}(2008){Brinchmann}, {Pettini}, \&
  {Charlot}}]{2008MNRAS.385..769B}
{Brinchmann}, J., {Pettini}, M., \& {Charlot}, S. 2008, \mnras, 385, 769

\bibitem[{{Bufano} {et~al.}(2012){Bufano}, {Pian}, {Sollerman}, {Benetti},
  {Pignata}, {Valenti}, {Covino}, {D'Avanzo}, {Malesani}, {Cappellaro}, {Della
  Valle}, {Fynbo}, {Hjorth}, {Mazzali}, {Reichart}, {Starling}, {Turatto},
  {Vergani}, {Wiersema}, {Amati}, {Bersier}, {Campana}, {Cano},
  {Castro-Tirado}, {Chincarini}, {D'Elia}, {de Ugarte Postigo}, {Deng},
  {Ferrero}, {Filippenko}, {Goldoni}, {Gorosabel}, {Greiner}, {Hammer},
  {Jakobsson}, {Kaper}, {Kawabata}, {Klose}, {Levan}, {Maeda}, {Masetti},
  {Milvang-Jensen}, {Mirabel}, {M{\o}ller}, {Nomoto}, {Palazzi}, {Piranomonte},
  {Salvaterra}, {Stratta}, {Tagliaferri}, {Tanaka}, {Tanvir}, \&
  {Wijers}}]{2012ApJ...753...67B}
{Bufano}, F., {Pian}, E., {Sollerman}, J., {et~al.} 2012, \apj, 753, 67

\bibitem[{{Burrows} {et~al.}(2005){Burrows}, {Hill}, {Nousek}, {Kennea},
  {Wells}, {Osborne}, {Abbey}, {Beardmore}, {Mukerjee}, {Short}, {Chincarini},
  {Campana}, {Citterio}, {Moretti}, {Pagani}, {Tagliaferri}, {Giommi},
  {Capalbi}, {Tamburelli}, {Angelini}, {Cusumano}, {Br{\"a}uninger}, {Burkert},
  \& {Hartner}}]{2005SSRv..120..165B}
{Burrows}, D.~N., {Hill}, J.~E., {Nousek}, J.~A., {et~al.} 2005, \ssr, 120, 165

\bibitem[{{Butler} {et~al.}(2014){Butler}, {Watson}, {Kutyrev}, {Lee},
  {Richer}, {Klein}, {Fox}, {Prochaska}, {Bloom}, {Cucchiara}, {Troja},
  {Littlejohns}, {Ramirez-Ruiz}, {de}, {Georgiev}, {Gonzalez}, {Roman-Zuniga},
  {Gehrels}, \& {Moseley}}]{2014GCN..15732...1B}
{Butler}, N., {Watson}, A.~M., {Kutyrev}, A., {et~al.} 2014, GCN, 15732

\bibitem[{{Campisi} {et~al.}(2011){Campisi}, {Tapparello}, {Salvaterra},
  {Mannucci}, \& {Colpi}}]{2011MNRAS.417.1013C}
{Campisi}, M.~A., {Tapparello}, C., {Salvaterra}, R., {Mannucci}, F., \&
  {Colpi}, M. 2011, \mnras, 417, 1013

\bibitem[{{Cano} {et~al.}(2014){Cano}, {Xu}, {Malesani}, {Jakobsson}, \&
  {Pursimo}}]{2014GCN..15743...1C}
{Cano}, Z., {Xu}, D., {Malesani}, D., {Jakobsson}, P., \& {Pursimo}, T. 2014,
  GCN, 15743

\bibitem[{{Capak} {et~al.}(2004){Capak}, {Cowie}, {Hu}, {Barger}, {Dickinson},
  {Fernandez}, {Giavalisco}, {Komiyama}, {Kretchmer}, {McNally}, {Miyazaki},
  {Okamura}, \& {Stern}}]{2004AJ....127..180C}
{Capak}, P., {Cowie}, L.~L., {Hu}, E.~M., {et~al.} 2004, \aj, 127, 180

\bibitem[{{Cenko} {et~al.}(2009){Cenko}, {Kelemen}, {Harrison}, {Fox},
  {Kulkarni}, {Kasliwal}, {Ofek}, {Rau}, {Gal-Yam}, {Frail}, \&
  {Moon}}]{2009ApJ...693.1484C}
{Cenko}, S.~B., {Kelemen}, J., {Harrison}, F.~A., {et~al.} 2009, \apj, 693,
  1484

\bibitem[{{Chabrier}(2003)}]{2003PASP..115..763C}
{Chabrier}, G. 2003, \pasp, 115, 763

\bibitem[{{Chary} {et~al.}(2007){Chary}, {Berger}, \&
  {Cowie}}]{2007ApJ...671..272C}
{Chary}, R., {Berger}, E., \& {Cowie}, L. 2007, \apj, 671, 272

\bibitem[{{Chen} {et~al.}(2009){Chen}, {Perley}, {Pollack}, {Prochaska},
  {Bloom}, {Dessauges-Zavadsky}, {Pettini}, {Lopez}, {Dall'aglio}, \&
  {Becker}}]{2009ApJ...691..152C}
{Chen}, H., {Perley}, D.~A., {Pollack}, L.~K., {et~al.} 2009, \apj, 691, 152

\bibitem[{{Chen}(2012)}]{2012MNRAS.419.3039C}
{Chen}, H.-W. 2012, \mnras, 419, 3039

\bibitem[{{Chisari} {et~al.}(2010){Chisari}, {Tissera}, \&
  {Pellizza}}]{2010MNRAS.408..647C}
{Chisari}, N.~E., {Tissera}, P.~B., \& {Pellizza}, L.~J. 2010, \mnras, 408, 647

\bibitem[{{Christensen} {et~al.}(2004){Christensen}, {Hjorth}, \&
  {Gorosabel}}]{2004A&A...425..913C}
{Christensen}, L., {Hjorth}, J., \& {Gorosabel}, J. 2004, \aap, 425, 913

\bibitem[{{Christensen} {et~al.}(2012){Christensen}, {Laursen}, {Richard},
  {Hjorth}, {Milvang-Jensen}, {Dessauges-Zavadsky}, {Limousin}, {Grillo}, \&
  {Ebeling}}]{2012MNRAS.427.1973C}
{Christensen}, L., {Laursen}, P., {Richard}, J., {et~al.} 2012, \mnras, 427,
  1973

\bibitem[{{Covino} {et~al.}(2013){Covino}, {Melandri}, {Salvaterra}, {Campana},
  {Vergani}, {Bernardini}, {D'Avanzo}, {D'Elia}, {Fugazza}, {Ghirlanda},
  {Ghisellini}, {Gomboc}, {Jin}, {Kr{\"u}hler}, {Malesani}, {Nava},
  {Sbarufatti}, \& {Tagliaferri}}]{2013MNRAS.432.1231C}
{Covino}, S., {Melandri}, A., {Salvaterra}, R., {et~al.} 2013, \mnras, 432,
  1231

\bibitem[{{Cucchiara} {et~al.}(2015){Cucchiara}, {Fumagalli}, {Rafelski},
  {Kocevski}, {Prochaska}, {Cooke}, \& {Becker}}]{2015ApJ...804...51C}
{Cucchiara}, A., {Fumagalli}, M., {Rafelski}, M., {et~al.} 2015, \apj, 804, 51

\bibitem[{{Cucciati} {et~al.}(2012){Cucciati}, {Tresse}, {Ilbert}, {Le
  F{\`e}vre}, {Garilli}, {Le Brun}, {Cassata}, {Franzetti}, {Maccagni},
  {Scodeggio}, {Zucca}, {Zamorani}, {Bardelli}, {Bolzonella}, {Bielby},
  {McCracken}, {Zanichelli}, \& {Vergani}}]{2012A&A...539A..31C}
{Cucciati}, O., {Tresse}, L., {Ilbert}, O., {et~al.} 2012, \aap, 539, A31

\bibitem[{{de Ugarte Postigo} {et~al.}(2013){de Ugarte Postigo}, {Campana},
  {Th{\"o}ne}, {D'Avanzo}, {S{\'a}nchez-Ram{\'{\i}}rez}, {Melandri},
  {Gorosabel}, {Ghirlanda}, {Veres}, {Mart{\'{\i}}n}, {Petitpas}, {Covino},
  {Fynbo}, \& {Levan}}]{2013A&A...557L..18D}
{de Ugarte Postigo}, A., {Campana}, S., {Th{\"o}ne}, C.~C., {et~al.} 2013,
  \aap, 557, L18

\bibitem[{{de Ugarte Postigo} {et~al.}(2011){de Ugarte Postigo}, {Th{\"o}ne},
  {Goldoni}, {Fynbo}, \& {X-shooter GRB Collaboration}}]{2011AN....332..297D}
{de Ugarte Postigo}, A., {Th{\"o}ne}, C.~C., {Goldoni}, P., {Fynbo}, J.~P.~U.,
  \& {X-shooter GRB Collaboration}. 2011, Astronomische Nachrichten, 332, 297

\bibitem[{{de Ugarte Postigo} {et~al.}(2014){de Ugarte Postigo}, {Th{\"o}ne},
  {Rowlinson}, {Garc{\'{\i}}a-Benito}, {Levan}, {Gorosabel}, {Goldoni},
  {Schulze}, {Zafar}, {Wiersema}, {S{\'a}nchez-Ram{\'{\i}}rez}, {Melandri},
  {D'Avanzo}, {Oates}, {D'Elia}, {De Pasquale}, {Kr{\"u}hler}, {van der Horst},
  {Xu}, {Watson}, {Piranomonte}, {Vergani}, {Milvang-Jensen}, {Kaper},
  {Malesani}, {Fynbo}, {Cano}, {Covino}, {Flores}, {Greiss}, {Hammer},
  {Hartoog}, {Hellmich}, {Heuser}, {Hjorth}, {Jakobsson}, {Mottola}, {Sparre},
  {Sollerman}, {Tagliaferri}, {Tanvir}, {Vestergaard}, \&
  {Wijers}}]{2014A&A...563A..62D}
{de Ugarte Postigo}, A., {Th{\"o}ne}, C.~C., {Rowlinson}, A., {et~al.} 2014,
  \aap, 563, A62

\bibitem[{{Dunne} {et~al.}(2009){Dunne}, {Ivison}, {Maddox}, {Cirasuolo},
  {Mortier}, {Foucaud}, {Ibar}, {Almaini}, {Simpson}, \&
  {McLure}}]{2009MNRAS.394....3D}
{Dunne}, L., {Ivison}, R.~J., {Maddox}, S., {et~al.} 2009, \mnras, 394, 3

\bibitem[{{El{\'{\i}}asd{\'o}ttir} {et~al.}(2009){El{\'{\i}}asd{\'o}ttir},
  {Fynbo}, {Hjorth}, {Ledoux}, {Watson}, {Andersen}, {Malesani}, {Vreeswijk},
  {Prochaska}, {Sollerman}, \& {Jaunsen}}]{2009ApJ...697.1725E}
{El{\'{\i}}asd{\'o}ttir}, {\'A}., {Fynbo}, J.~P.~U., {Hjorth}, J., {et~al.}
  2009, \apj, 697, 1725

\bibitem[{{Elliott} {et~al.}(2012){Elliott}, {Greiner}, {Khochfar}, {Schady},
  {Johnson}, \& {Rau}}]{2012A&A...539A.113E}
{Elliott}, J., {Greiner}, J., {Khochfar}, S., {et~al.} 2012, \aap, 539, A113

\bibitem[{{Elliott} {et~al.}(2013){Elliott}, {Kr{\"u}hler}, {Greiner},
  {Savaglio}, {Olivares}, {Rau}, {de Ugarte Postigo},
  {S{\'a}nchez-Ram{\'{\i}}rez}, {Wiersema}, {Schady}, {Kann}, {Filgas},
  {Nardini}, {Berger}, {Fox}, {Gorosabel}, {Klose}, {Levan}, {Nicuesa
  Guelbenzu}, {Rossi}, {Schmidl}, {Sudilovsky}, {Tanvir}, \&
  {Th{\"o}ne}}]{2013A&A...556A..23E}
{Elliott}, J., {Kr{\"u}hler}, T., {Greiner}, J., {et~al.} 2013, \aap, 556, A23

\bibitem[{{Erb} {et~al.}(2006{\natexlab{a}}){Erb}, {Shapley}, {Pettini},
  {Steidel}, {Reddy}, \& {Adelberger}}]{2006ApJ...644..813E}
{Erb}, D.~K., {Shapley}, A.~E., {Pettini}, M., {et~al.} 2006{\natexlab{a}},
  \apj, 644, 813

\bibitem[{{Erb} {et~al.}(2006{\natexlab{b}}){Erb}, {Steidel}, {Shapley},
  {Pettini}, {Reddy}, \& {Adelberger}}]{2006ApJ...646..107E}
{Erb}, D.~K., {Steidel}, C.~C., {Shapley}, A.~E., {et~al.} 2006{\natexlab{b}},
  \apj, 646, 107

\bibitem[{{Faber} \& {Jackson}(1976)}]{1976ApJ...204..668F}
{Faber}, S.~M. \& {Jackson}, R.~E. 1976, \apj, 204, 668

\bibitem[{{Filgas} {et~al.}(2011){Filgas}, {Kr{\"u}hler}, {Greiner}, {Rau},
  {Palazzi}, {Klose}, {Schady}, {Rossi}, {Afonso}, {Antonelli}, {Clemens},
  {Covino}, {D'Avanzo}, {K{\"u}pc{\"u} Yolda{\c s}}, {Nardini}, {Nicuesa
  Guelbenzu}, {Olivares}, {Updike}, \& {Yolda{\c s}}}]{2011A&A...526A.113F}
{Filgas}, R., {Kr{\"u}hler}, T., {Greiner}, J., {et~al.} 2011, \aap, 526, A113

\bibitem[{{F{\"o}rster Schreiber} {et~al.}(2009){F{\"o}rster Schreiber},
  {Genzel}, {Bouch{\'e}}, {Cresci}, {Davies}, {Buschkamp}, {Shapiro},
  {Tacconi}, {Hicks}, {Genel}, {Shapley}, {Erb}, {Steidel}, {Lutz},
  {Eisenhauer}, {Gillessen}, {Sternberg}, {Renzini}, {Cimatti}, {Daddi},
  {Kurk}, {Lilly}, {Kong}, {Lehnert}, {Nesvadba}, {Verma}, {McCracken},
  {Arimoto}, {Mignoli}, \& {Onodera}}]{2009ApJ...706.1364F}
{F{\"o}rster Schreiber}, N.~M., {Genzel}, R., {Bouch{\'e}}, N., {et~al.} 2009,
  \apj, 706, 1364

\bibitem[{{Friis} {et~al.}(2015){Friis}, {De Cia}, {Kr{\"u}hler}, {Fynbo},
  {Ledoux}, {Vreeswijk}, {Watson}, {Malesani}, {Gorosabel}, {Starling},
  {Jakobsson}, {Varela}, {Wiersema}, {Drachmann}, {Trotter}, {Th{\"o}ne}, {de
  Ugarte Postigo}, {D'Elia}, {Elliott}, {Maturi}, {Goldoni}, {Greiner},
  {Haislip}, {Kaper}, {Knust}, {LaCluyze}, {Milvang-Jensen}, {Reichart},
  {Schulze}, {Sudilovsky}, {Tanvir}, \& {Vergani}}]{2014arXiv1409.6315F}
{Friis}, M., {De Cia}, A., {Kr{\"u}hler}, T., {et~al.} 2015, \mnras, 451, 167

\bibitem[{{Fruchter} {et~al.}(2006){Fruchter}, {Levan}, {Strolger},
  {Vreeswijk}, {Thorsett}, {Bersier}, {Burud}, {Castro Cer{\'o}n},
  {Castro-Tirado}, {Conselice}, {Dahlen}, {Ferguson}, {Fynbo}, {Garnavich},
  {Gibbons}, {Gorosabel}, {Gull}, {Hjorth}, {Holland}, {Kouveliotou}, {Levay},
  {Livio}, {Metzger}, {Nugent}, {Petro}, {Pian}, {Rhoads}, {Riess}, {Sahu},
  {Smette}, {Tanvir}, {Wijers}, \& {Woosley}}]{2006Natur.441..463F}
{Fruchter}, A.~S., {Levan}, A.~J., {Strolger}, L., {et~al.} 2006, \nat, 441,
  463

\bibitem[{{Fynbo} {et~al.}(2003){Fynbo}, {Jakobsson}, {M{\"o}ller}, {Hjorth},
  {Thomsen}, {Andersen}, {Fruchter}, {Gorosabel}, {Holland}, {Ledoux},
  {Pedersen}, {Rhoads}, {Weidinger}, \& {Wijers}}]{2003A&A...406L..63F}
{Fynbo}, J.~P.~U., {Jakobsson}, P., {M{\"o}ller}, P., {et~al.} 2003, \aap, 406,
  L63

\bibitem[{{Fynbo} {et~al.}(2014){Fynbo}, {Kr{\"u}hler}, {Leighly}, {Ledoux},
  {Vreeswijk}, {Schulze}, {Noterdaeme}, {Watson}, {Wijers}, {Bolmer}, {Cano},
  {Christensen}, {Covino}, {D'Elia}, {Flores}, {Friis}, {Goldoni}, {Greiner},
  {Hammer}, {Hjorth}, {Jakobsson}, {Japelj}, {Kaper}, {Klose}, {Knust},
  {Leloudas}, {Levan}, {Malesani}, {Milvang-Jensen}, {M{\o}ller}, {Nicuesa
  Guelbenzu}, {Oates}, {Pian}, {Schady}, {Sparre}, {Tagliaferri}, {Tanvir},
  {Th{\"o}ne}, {de Ugarte Postigo}, {Vergani}, {Wiersema}, {Xu}, \&
  {Zafar}}]{2014arXiv1409.4975F}
{Fynbo}, J.~P.~U., {Kr{\"u}hler}, T., {Leighly}, K., {et~al.} 2014, \aap, 572,
  A12

\bibitem[{{Fynbo} {et~al.}(2008){Fynbo}, {Prochaska}, {Sommer-Larsen},
  {Dessauges-Zavadsky}, \& {M{\o}ller}}]{2008ApJ...683..321F}
{Fynbo}, J.~P.~U., {Prochaska}, J.~X., {Sommer-Larsen}, J.,
  {Dessauges-Zavadsky}, M., \& {M{\o}ller}, P. 2008, \apj, 683, 321

\bibitem[{{Fynbo} {et~al.}(2006){Fynbo}, {Starling}, {Ledoux}, {Wiersema},
  {Th{\"o}ne}, {Sollerman}, {Jakobsson}, {Hjorth}, {Watson}, {Vreeswijk},
  {M{\o}ller}, {Rol}, {Gorosabel}, {N{\"a}r{\"a}nen}, {Wijers},
  {Bj{\"o}rnsson}, {Castro Cer{\'o}n}, {Curran}, {Hartmann}, {Holland},
  {Jensen}, {Levan}, {Limousin}, {Kouveliotou}, {Nelemans}, {Pedersen},
  {Priddey}, \& {Tanvir}}]{2006A&A...451L..47F}
{Fynbo}, J.~P.~U., {Starling}, R.~L.~C., {Ledoux}, C., {et~al.} 2006, \aap,
  451, L47

\bibitem[{{Galama} {et~al.}(1998){Galama}, {Vreeswijk}, {van Paradijs},
  {Kouveliotou}, {Augusteijn}, {B{\"o}hnhardt}, {Brewer}, {Doublier},
  {Gonzalez}, {Leibundgut}, {Lidman}, {Hainaut}, {Patat}, {Heise}, {in't Zand},
  {Hurley}, {Groot}, {Strom}, {Mazzali}, {Iwamoto}, {Nomoto}, {Umeda},
  {Nakamura}, {Young}, {Suzuki}, {Shigeyama}, {Koshut}, {Kippen}, {Robinson},
  {de Wildt}, {Wijers}, {Tanvir}, {Greiner}, {Pian}, {Palazzi}, {Frontera},
  {Masetti}, {Nicastro}, {Feroci}, {Costa}, {Piro}, {Peterson}, {Tinney},
  {Boyle}, {Cannon}, {Stathakis}, {Sadler}, {Begam}, \&
  {Ianna}}]{1998Natur.395..670G}
{Galama}, T.~J., {Vreeswijk}, P.~M., {van Paradijs}, J., {et~al.} 1998, \nat,
  395, 670

\bibitem[{{Gawiser} {et~al.}(2006){Gawiser}, {van Dokkum}, {Herrera}, {Maza},
  {Castander}, {Infante}, {Lira}, {Quadri}, {Toner}, {Treister}, {Urry},
  {Altmann}, {Assef}, {Christlein}, {Coppi}, {Dur{\'a}n}, {Franx}, {Galaz},
  {Huerta}, {Liu}, {L{\'o}pez}, {M{\'e}ndez}, {Moore}, {Rubio}, {Ruiz}, {Toft},
  \& {Yi}}]{2006ApJS..162....1G}
{Gawiser}, E., {van Dokkum}, P.~G., {Herrera}, D., {et~al.} 2006, \apjs, 162, 1

\bibitem[{{Gehrels} {et~al.}(2004){Gehrels}, {Chincarini}, {Giommi}, {Mason},
  {Nousek}, {Wells}, {White}, {Barthelmy}, {Burrows}, {Cominsky}, {Hurley},
  {Marshall}, {M{\'e}sz{\'a}ros}, {Roming}, {Angelini}, {Barbier}, {Belloni},
  {Campana}, {Caraveo}, {Chester}, {Citterio}, {Cline}, {Cropper}, {Cummings},
  {Dean}, {Feigelson}, {Fenimore}, {Frail}, {Fruchter}, {Garmire}, {Gendreau},
  {Ghisellini}, {Greiner}, {Hill}, {Hunsberger}, {Krimm}, {Kulkarni}, {Kumar},
  {Lebrun}, {Lloyd-Ronning}, {Markwardt}, {Mattson}, {Mushotzky}, {Norris},
  {Osborne}, {Paczynski}, {Palmer}, {Park}, {Parsons}, {Paul}, {Rees},
  {Reynolds}, {Rhoads}, {Sasseen}, {Schaefer}, {Short}, {Smale}, {Smith},
  {Stella}, {Tagliaferri}, {Takahashi}, {Tashiro}, {Townsley}, {Tueller},
  {Turner}, {Vietri}, {Voges}, {Ward}, {Willingale}, {Zerbi}, \&
  {Zhang}}]{2004ApJ...611.1005G}
{Gehrels}, N., {Chincarini}, G., {Giommi}, P., {et~al.} 2004, \apj, 611, 1005

\bibitem[{{Gehrels} {et~al.}(2009){Gehrels}, {Ramirez-Ruiz}, \&
  {Fox}}]{2009ARA&A..47..567G}
{Gehrels}, N., {Ramirez-Ruiz}, E., \& {Fox}, D.~B. 2009, \araa, 47, 567

\bibitem[{{Gendre} {et~al.}(2013){Gendre}, {Stratta}, {Atteia}, {Basa},
  {Bo{\"e}r}, {Coward}, {Cutini}, {D'Elia}, {Howell}, {Klotz}, \&
  {Piro}}]{2013ApJ...766...30G}
{Gendre}, B., {Stratta}, G., {Atteia}, J.~L., {et~al.} 2013, \apj, 766, 30

\bibitem[{{Gil de Paz} {et~al.}(2003){Gil de Paz}, {Madore}, \&
  {Pevunova}}]{2003ApJS..147...29G}
{Gil de Paz}, A., {Madore}, B.~F., \& {Pevunova}, O. 2003, \apjs, 147, 29

\bibitem[{{Gilbank} {et~al.}(2010){Gilbank}, {Baldry}, {Balogh}, {Glazebrook},
  \& {Bower}}]{2010MNRAS.405.2594G}
{Gilbank}, D.~G., {Baldry}, I.~K., {Balogh}, M.~L., {Glazebrook}, K., \&
  {Bower}, R.~G. 2010, \mnras, 405, 2594

\bibitem[{{Goad} {et~al.}(2012){Goad}, {Osborne}, {Beardmore}, \&
  {Evans}}]{2012GCN..12922...1G}
{Goad}, M.~R., {Osborne}, J.~P., {Beardmore}, A.~P., \& {Evans}, P.~A. 2012,
  GCN, 12922

\bibitem[{{Goldoni} {et~al.}(2006){Goldoni}, {Royer}, {Fran{\c c}ois},
  {Horrobin}, {Blanc}, {Vernet}, {Modigliani}, \&
  {Larsen}}]{2006SPIE.6269E..80G}
{Goldoni}, P., {Royer}, F., {Fran{\c c}ois}, P., {et~al.} 2006, in SPIE Conf.
  Ser., Vol. 6269

\bibitem[{{Gorosabel} {et~al.}(2012){Gorosabel}, {Huelamo}, {Sanchez-Ramirez},
  {de Ugarte Postigo}, {Kr\"uhler}, {Castro-Tirado}, {Tello}, \&
  {Jelinek}}]{2012GCN..13591...1G}
{Gorosabel}, J., {Huelamo}, N., {Sanchez-Ramirez}, R., {et~al.} 2012, GCN,
  13591

\bibitem[{{Gorosabel} {et~al.}(2005){Gorosabel}, {P{\'e}rez-Ram{\'{\i}}rez},
  {Sollerman}, {de Ugarte Postigo}, {Fynbo}, {Castro-Tirado}, {Jakobsson},
  {Christensen}, {Hjorth}, {J{\'o}hannesson}, {Guziy}, {Castro Cer{\'o}n},
  {Bj{\"o}rnsson}, {Sokolov}, {Fatkhullin}, \& {Nilsson}}]{2005A&A...444..711G}
{Gorosabel}, J., {P{\'e}rez-Ram{\'{\i}}rez}, D., {Sollerman}, J., {et~al.}
  2005, \aap, 444, 711

\bibitem[{{Graham} \& {Fruchter}(2013)}]{2013ApJ...774..119G}
{Graham}, J.~F. \& {Fruchter}, A.~S. 2013, \apj, 774, 119

\bibitem[{{Greiner} {et~al.}(2008){Greiner}, {Bornemann}, {Clemens}, {Deuter},
  {Hasinger}, {Honsberg}, {Huber}, {Huber}, {Krauss}, {Kr{\"u}hler},
  {K{\"u}pc{\"u} Yolda{\c s}}, {Mayer-Hasselwander}, {Mican}, {Primak},
  {Schrey}, {Steiner}, {Szokoly}, {Th{\"o}ne}, {Yolda{\c s}}, {Klose}, {Laux},
  \& {Winkler}}]{2008PASP..120..405G}
{Greiner}, J., {Bornemann}, W., {Clemens}, C., {et~al.} 2008, \pasp, 120, 405

\bibitem[{{Greiner} {et~al.}(2015{\natexlab{a}}){Greiner}, {Fox}, {Schady},
  {Kr{\"u}hler}, {Trenti}, {Cikota}, {Bolmer}, {Elliott}, {Delvaux}, {Perna},
  {Afonso}, {Kann}, {Klose}, {Savaglio}, {Schmidl}, {Schweyer}, {Tanga}, \&
  {Varela}}]{2015arXiv150305323G}
{Greiner}, J., {Fox}, D.~B., {Schady}, P., {et~al.} 2015{\natexlab{a}}, \apj,
  809, 76

\bibitem[{{Greiner} {et~al.}(2011){Greiner}, {Kr{\"u}hler}, {Klose}, {Afonso},
  {Clemens}, {Filgas}, {Hartmann}, {K{\"u}pc{\"u} Yolda{\c s}}, {Nardini},
  {Olivares E.}, {Rau}, {Rossi}, {Schady}, \& {Updike}}]{2011A&A...526A..30G}
{Greiner}, J., {Kr{\"u}hler}, T., {Klose}, S., {et~al.} 2011, \aap, 526, A30

\bibitem[{{Greiner} {et~al.}(2013){Greiner}, {Kr{\"u}hler}, {Nardini},
  {Filgas}, {Moin}, {de Breuck}, {Montenegro-Montes}, {Lundgren}, {Klose},
  {fonso}, {Bertoldi}, {Elliott}, {Kann}, {Knust}, {Menten}, {Nicuesa
  Guelbenzu}, {Olivares E.}, {Rau}, {Rossi}, {Schady}, {Schmidl}, {Siringo},
  {Spezzi}, {Sudilovsky}, {Tingay}, {Updike}, {Wang}, {Weiss}, {Wieringa}, \&
  {Wyrowski}}]{2013A&A...560A..70G}
{Greiner}, J., {Kr{\"u}hler}, T., {Nardini}, M., {et~al.} 2013, \aap, 560, A70

\bibitem[{{Greiner} {et~al.}(2015{\natexlab{b}}){Greiner}, {Mazzali}, {Kann},
  {Kr{\"u}hler}, {Pian}, {Prentice}, {Olivares E.}, {Rossi}, {Klose},
  {Taubenberger}, {Knust}, {Afonso}, {Ashall}, {Bolmer}, {Delvaux}, {Diehl},
  {Elliott}, {Filgas}, {Fynbo}, {Graham}, {Guelbenzu}, {Kobayashi}, {Leloudas},
  {Savaglio}, {Schady}, {Schmidl}, {Schweyer}, {Sudilovsky}, {Tanga}, {Updike},
  {van Eerten}, \& {Varela}}]{2015Greinersubm}
{Greiner}, J., {Mazzali}, P.~A., {Kann}, D.~A., {et~al.} 2015{\natexlab{b}},
  \nat, 523, 189

\bibitem[{{Greiner} {et~al.}(2014){Greiner}, {Yu}, {Kr{\"u}hler}, {Frederiks},
  {Beloborodov}, {Bhat}, {Bolmer}, {van Eerten}, {Aptekar}, {Elliott},
  {Golenetskii}, {Graham}, {Hurley}, {Kann}, {Klose}, {Nicuesa Guelbenzu},
  {Rau}, {Schady}, {Schmidl}, {Sudilovsky}, {Svinkin}, {Tanga}, {Ulanov},
  {Varela}, {von Kienlin}, \& {Zhang}}]{2014A&A...568A..75G}
{Greiner}, J., {Yu}, H.-F., {Kr{\"u}hler}, T., {et~al.} 2014, \aap, 568, A75

\bibitem[{{Groot} {et~al.}(1998){Groot}, {Galama}, {van Paradijs},
  {Kouveliotou}, {Wijers}, {Bloom}, {Tanvir}, {Vanderspek}, {Greiner},
  {Castro-Tirado}, {Gorosabel}, {von Hippel}, {Lehnert}, {Kuijken}, {Hoekstra},
  {Metcalfe}, {Howk}, {Conselice}, {Telting}, {Rutten}, {Rhoads}, {Cole},
  {Pisano}, {Naber}, \& {Schwarz}}]{1998ApJ...493L..27G}
{Groot}, P.~J., {Galama}, T.~J., {van Paradijs}, J., {et~al.} 1998, \apjl, 493,
  L27

\bibitem[{{Guetta} \& {Della Valle}(2007)}]{2007ApJ...657L..73G}
{Guetta}, D. \& {Della Valle}, M. 2007, \apjl, 657, L73

\bibitem[{{Guidorzi} \& {Mundell}(2012)}]{2012GCN..13651...1G}
{Guidorzi}, C. \& {Mundell}, C.~G. 2012, GCN, 13651

\bibitem[{{Hirschi} {et~al.}(2005){Hirschi}, {Meynet}, \&
  {Maeder}}]{2005A&A...443..581H}
{Hirschi}, R., {Meynet}, G., \& {Maeder}, A. 2005, \aap, 443, 581

\bibitem[{{Hjorth} {et~al.}(2012){Hjorth}, {Malesani}, {Jakobsson}, {Jaunsen},
  {Fynbo}, {Gorosabel}, {Kr{\"u}hler}, {Levan}, {Micha{\l}owski},
  {Milvang-Jensen}, {M{\o}ller}, {Schulze}, {Tanvir}, \&
  {Watson}}]{2012ApJ...756..187H}
{Hjorth}, J., {Malesani}, D., {Jakobsson}, P., {et~al.} 2012, \apj, 756, 187

\bibitem[{{Hjorth} {et~al.}(2003){Hjorth}, {Sollerman}, {M{\o}ller}, {Fynbo},
  {Woosley}, {Kouveliotou}, {Tanvir}, {Greiner}, {Andersen}, {Castro-Tirado},
  {Castro Cer{\'o}n}, {Fruchter}, {Gorosabel}, {Jakobsson}, {Kaper}, {Klose},
  {Masetti}, {Pedersen}, {Pedersen}, {Pian}, {Palazzi}, {Rhoads}, {Rol}, {van
  den Heuvel}, {Vreeswijk}, {Watson}, \& {Wijers}}]{2003Natur.423..847H}
{Hjorth}, J., {Sollerman}, J., {M{\o}ller}, P., {et~al.} 2003, \nat, 423, 847

\bibitem[{{Hunt} {et~al.}(2011){Hunt}, {Palazzi}, {Rossi}, {Savaglio},
  {Cresci}, {Klose}, {Micha{\l}owski}, \& {Pian}}]{2011ApJ...736L..36H}
{Hunt}, L., {Palazzi}, E., {Rossi}, A., {et~al.} 2011, \apjl, 736, L36

\bibitem[{{Hunt} {et~al.}(2014){Hunt}, {Palazzi}, {Micha{\l}owski}, {Rossi},
  {Savaglio}, {Basa}, {Berta}, {Bianchi}, {Covino}, {D'Elia}, {Ferrero},
  {G{\"o}tz}, {Greiner}, {Klose}, {Le Borgne}, {Le Floc'h}, {Pian},
  {Piranomonte}, {Schady}, \& {Vergani}}]{2014A&A...565A.112H}
{Hunt}, L.~K., {Palazzi}, E., {Micha{\l}owski}, M.~J., {et~al.} 2014, \aap,
  565, A112

\bibitem[{{Ilbert} {et~al.}(2006){Ilbert}, {Arnouts}, {McCracken},
  {Bolzonella}, {Bertin}, {Le F{\`e}vre}, {Mellier}, {Zamorani}, {Pell{\`o}},
  {Iovino}, {Tresse}, {Le Brun}, {Bottini}, {Garilli}, {Maccagni}, {Picat},
  {Scaramella}, {Scodeggio}, {Vettolani}, {Zanichelli}, {Adami}, {Bardelli},
  {Cappi}, {Charlot}, {Ciliegi}, {Contini}, {Cucciati}, {Foucaud}, {Franzetti},
  {Gavignaud}, {Guzzo}, {Marano}, {Marinoni}, {Mazure}, {Meneux}, {Merighi},
  {Paltani}, {Pollo}, {Pozzetti}, {Radovich}, {Zucca}, {Bondi}, {Bongiorno},
  {Busarello}, {de La Torre}, {Gregorini}, {Lamareille}, {Mathez}, {Merluzzi},
  {Ripepi}, {Rizzo}, \& {Vergani}}]{2006A&A...457..841I}
{Ilbert}, O., {Arnouts}, S., {McCracken}, H.~J., {et~al.} 2006, \aap, 457, 841

\bibitem[{{Ilbert} {et~al.}(2013){Ilbert}, {McCracken}, {Le F{\`e}vre},
  {Capak}, {Dunlop}, {Karim}, {Renzini}, {Caputi}, {Boissier}, {Arnouts},
  {Aussel}, {Comparat}, {Guo}, {Hudelot}, {Kartaltepe}, {Kneib}, {Krogager},
  {Le Floc'h}, {Lilly}, {Mellier}, {Milvang-Jensen}, {Moutard}, {Onodera},
  {Richard}, {Salvato}, {Sanders}, {Scoville}, {Silverman}, {Taniguchi},
  {Tasca}, {Thomas}, {Toft}, {Tresse}, {Vergani}, {Wolk}, \&
  {Zirm}}]{2013A&A...556A..55I}
{Ilbert}, O., {McCracken}, H.~J., {Le F{\`e}vre}, O., {et~al.} 2013, \aap, 556,
  A55

\bibitem[{{Jakobsson} {et~al.}(2005){Jakobsson}, {Bj{\"o}rnsson}, {Fynbo},
  {J{\'o}hannesson}, {Hjorth}, {Thomsen}, {M{\o}ller}, {Watson}, {Jensen},
  {{\"O}stlin}, {Gorosabel}, \& {Gudmundsson}}]{2005MNRAS.362..245J}
{Jakobsson}, P., {Bj{\"o}rnsson}, G., {Fynbo}, J.~P.~U., {et~al.} 2005, \mnras,
  362, 245

\bibitem[{{Jakobsson} {et~al.}(2004){Jakobsson}, {Hjorth}, {Fynbo}, {Watson},
  {Pedersen}, {Bj{\"o}rnsson}, \& {Gorosabel}}]{2004ApJ...617L..21J}
{Jakobsson}, P., {Hjorth}, J., {Fynbo}, J.~P.~U., {et~al.} 2004, \apjl, 617,
  L21

\bibitem[{{Jakobsson} {et~al.}(2012){Jakobsson}, {Hjorth}, {Malesani},
  {Chapman}, {Fynbo}, {Tanvir}, {Milvang-Jensen}, {Vreeswijk}, {Letawe}, \&
  {Starling}}]{2012ApJ...752...62J}
{Jakobsson}, P., {Hjorth}, J., {Malesani}, D., {et~al.} 2012, \apj, 752, 62

\bibitem[{{Jaunsen} {et~al.}(2008){Jaunsen}, {Rol}, {Watson}, {Malesani},
  {Fynbo}, {Milvang-Jensen}, {Hjorth}, {Vreeswijk}, {Ovaldsen}, {Wiersema},
  {Tanvir}, {Gorosabel}, {Levan}, {Schirmer}, \&
  {Castro-Tirado}}]{2008ApJ...681..453J}
{Jaunsen}, A.~O., {Rol}, E., {Watson}, D.~J., {et~al.} 2008, \apj, 681, 453

\bibitem[{{Kajisawa} {et~al.}(2009){Kajisawa}, {Ichikawa}, {Tanaka}, {Konishi},
  {Yamada}, {Akiyama}, {Suzuki}, {Tokoku}, {Uchimoto}, {Yoshikawa}, {Ouchi},
  {Iwata}, {Hamana}, \& {Onodera}}]{2009ApJ...702.1393K}
{Kajisawa}, M., {Ichikawa}, T., {Tanaka}, I., {et~al.} 2009, \apj, 702, 1393

\bibitem[{{Kann} {et~al.}(2015){Kann}, {Schady}, {Olivares E.}, {Klose},
  {Rossi}, {Perley}, {Kr\'uhler}, {Greiner}, {Nicuesa~Guelbenzu}, {Elliot}, \&
  {Knust}}]{2015Kannsubm}
{Kann}, D.~A., {Schady}, P., {Olivares E.}, F., {et~al.} 2015, \aap, submitted

\bibitem[{{Kauffmann} {et~al.}(2003){Kauffmann}, {Heckman}, {Tremonti},
  {Brinchmann}, {Charlot}, {White}, {Ridgway}, {Brinkmann}, {Fukugita}, {Hall},
  {Ivezi{\'c}}, {Richards}, \& {Schneider}}]{2003MNRAS.346.1055K}
{Kauffmann}, G., {Heckman}, T.~M., {Tremonti}, C., {et~al.} 2003, \mnras, 346,
  1055

\bibitem[{{Kelly} {et~al.}(2014){Kelly}, {Filippenko}, {Modjaz}, \&
  {Kocevski}}]{2014ApJ...789...23K}
{Kelly}, P.~L., {Filippenko}, A.~V., {Modjaz}, M., \& {Kocevski}, D. 2014,
  \apj, 789, 23

\bibitem[{{Kennicutt}(1998)}]{1998ARA&A..36..189K}
{Kennicutt}, Jr., R.~C. 1998, \araa, 36, 189

\bibitem[{{Kewley} \& {Dopita}(2002)}]{2002ApJS..142...35K}
{Kewley}, L.~J. \& {Dopita}, M.~A. 2002, \apjs, 142, 35

\bibitem[{{Kewley} {et~al.}(2013{\natexlab{a}}){Kewley}, {Dopita}, {Leitherer},
  {Dav{\'e}}, {Yuan}, {Allen}, {Groves}, \& {Sutherland}}]{2013ApJ...774..100K}
{Kewley}, L.~J., {Dopita}, M.~A., {Leitherer}, C., {et~al.} 2013{\natexlab{a}},
  \apj, 774, 100

\bibitem[{{Kewley} {et~al.}(2001){Kewley}, {Dopita}, {Sutherland}, {Heisler},
  \& {Trevena}}]{2001ApJ...556..121K}
{Kewley}, L.~J., {Dopita}, M.~A., {Sutherland}, R.~S., {Heisler}, C.~A., \&
  {Trevena}, J. 2001, \apj, 556, 121

\bibitem[{{Kewley} \& {Ellison}(2008)}]{2008ApJ...681.1183K}
{Kewley}, L.~J. \& {Ellison}, S.~L. 2008, \apj, 681, 1183

\bibitem[{{Kewley} {et~al.}(2004){Kewley}, {Geller}, \&
  {Jansen}}]{2004AJ....127.2002K}
{Kewley}, L.~J., {Geller}, M.~J., \& {Jansen}, R.~A. 2004, \aj, 127, 2002

\bibitem[{{Kewley} {et~al.}(2013{\natexlab{b}}){Kewley}, {Maier}, {Yabe},
  {Ohta}, {Akiyama}, {Dopita}, \& {Yuan}}]{2013ApJ...774L..10K}
{Kewley}, L.~J., {Maier}, C., {Yabe}, K., {et~al.} 2013{\natexlab{b}}, \apjl,
  774, L10

\bibitem[{{Kissler-Patig} {et~al.}(2008){Kissler-Patig}, {Pirard}, {Casali},
  {Moorwood}, {Ageorges}, {Alves de Oliveira}, {Baksai}, {Bedin}, {Bendek},
  {Biereichel}, {Delabre}, {Dorn}, {Esteves}, {Finger}, {Gojak}, {Huster},
  {Jung}, {Kiekebush}, {Klein}, {Koch}, {Lizon}, {Mehrgan}, {Petr-Gotzens},
  {Pritchard}, {Selman}, \& {Stegmeier}}]{2008A&A...491..941K}
{Kissler-Patig}, M., {Pirard}, J.-F., {Casali}, M., {et~al.} 2008, \aap, 491,
  941

\bibitem[{{Kistler} {et~al.}(2009){Kistler}, {Y{\"u}ksel}, {Beacom}, {Hopkins},
  \& {Wyithe}}]{2009ApJ...705L.104K}
{Kistler}, M.~D., {Y{\"u}ksel}, H., {Beacom}, J.~F., {Hopkins}, A.~M., \&
  {Wyithe}, J.~S.~B. 2009, \apjl, 705, L104

\bibitem[{{Kocevski} \& {West}(2011)}]{2011ApJ...735L...8K}
{Kocevski}, D. \& {West}, A.~A. 2011, \apjl, 735, L8

\bibitem[{{Kocevski} {et~al.}(2009){Kocevski}, {West}, \&
  {Modjaz}}]{2009ApJ...702..377K}
{Kocevski}, D., {West}, A.~A., \& {Modjaz}, M. 2009, \apj, 702, 377

\bibitem[{{Kriek} {et~al.}(2007){Kriek}, {van Dokkum}, {Franx}, {Illingworth},
  {Coppi}, {F{\"o}rster Schreiber}, {Gawiser}, {Labb{\'e}}, {Lira},
  {Marchesini}, {Quadri}, {Rudnick}, {Taylor}, {Urry}, \& {van der
  Werf}}]{2007ApJ...669..776K}
{Kriek}, M., {van Dokkum}, P.~G., {Franx}, M., {et~al.} 2007, \apj, 669, 776

\bibitem[{{Kr{\"u}hler} {et~al.}(2012{\natexlab{a}}){Kr{\"u}hler}, {Fynbo},
  {Geier}, {Hjorth}, {Malesani}, {Milvang-Jensen}, {Levan}, {Sparre}, {Watson},
  \& {Zafar}}]{2012A&A...546A...8K}
{Kr{\"u}hler}, T., {Fynbo}, J.~P.~U., {Geier}, S., {et~al.} 2012{\natexlab{a}},
  \aap, 546, A8

\bibitem[{{Kr{\"u}hler} {et~al.}(2011){Kr{\"u}hler}, {Greiner}, {Schady},
  {Savaglio}, {Afonso}, {Clemens}, {Elliott}, {Filgas}, {Gruber}, {Kann},
  {Klose}, {K{\"u}pc{\"u}-Yolda{\c s}}, {McBreen}, {Olivares}, {Pierini},
  {Rau}, {Rossi}, {Nardini}, {Nicuesa Guelbenzu}, {Sudilovsky}, \&
  {Updike}}]{2011A&A...534A.108K}
{Kr{\"u}hler}, T., {Greiner}, J., {Schady}, P., {et~al.} 2011, \aap, 534, A108

\bibitem[{{Kr{\"u}hler} {et~al.}(2008){Kr{\"u}hler}, {K{\"u}pc{\"u} Yolda{\c
  s}}, {Greiner}, {Clemens}, {McBreen}, {Primak}, {Savaglio}, {Yolda{\c s}},
  {Szokoly}, \& {Klose}}]{2008ApJ...685..376K}
{Kr{\"u}hler}, T., {K{\"u}pc{\"u} Yolda{\c s}}, A., {Greiner}, J., {et~al.}
  2008, \apj, 685, 376

\bibitem[{{Kr{\"u}hler} {et~al.}(2013){Kr{\"u}hler}, {Ledoux}, {Fynbo},
  {Vreeswijk}, {Schmidl}, {Malesani}, {Christensen}, {De Cia}, {Hjorth},
  {Jakobsson}, {Kann}, {Kaper}, {Vergani}, {Afonso}, {Covino}, {de Ugarte
  Postigo}, {D'Elia}, {Filgas}, {Goldoni}, {Greiner}, {Hartoog},
  {Milvang-Jensen}, {Nardini}, {Piranomonte}, {Rossi},
  {S{\'a}nchez-Ram{\'{\i}}rez}, {Schady}, {Schulze}, {Sudilovsky}, {Tanvir},
  {Tagliaferri}, {Watson}, {Wiersema}, {Wijers}, \& {Xu}}]{2013A&A...557A..18K}
{Kr{\"u}hler}, T., {Ledoux}, C., {Fynbo}, J.~P.~U., {et~al.} 2013, \aap, 557,
  A18

\bibitem[{{Kr{\"u}hler} {et~al.}(2012{\natexlab{b}}){Kr{\"u}hler}, {Malesani},
  {Milvang-Jensen}, {Fynbo}, {Hjorth}, {Jakobsson}, {Levan}, {Sparre},
  {Tanvir}, \& {Watson}}]{2012ApJ...758...46K}
{Kr{\"u}hler}, T., {Malesani}, D., {Milvang-Jensen}, B., {et~al.}
  2012{\natexlab{b}}, \apj, 758, 46

\bibitem[{{Kr\"uhler} {et~al.}(2012){Kr\"uhler}, {Nicuesa Guelbenzu}, \&
  {Greiner}}]{2012GCN..14049...1K}
{Kr\"uhler}, T., {Nicuesa Guelbenzu}, A., \& {Greiner}, J. 2012, GCN, 14049

\bibitem[{{Kumar} \& {Zhang}(2015)}]{2014arXiv1410.0679K}
{Kumar}, P. \& {Zhang}, B. 2015, \physrep, 561, 1

\bibitem[{{Lamb} \& {Reichart}(2000)}]{2000ApJ...536....1L}
{Lamb}, D.~Q. \& {Reichart}, D.~E. 2000, \apj, 536, 1

\bibitem[{{Laskar} {et~al.}(2011){Laskar}, {Berger}, \&
  {Chary}}]{2011ApJ...739....1L}
{Laskar}, T., {Berger}, E., \& {Chary}, R.-R. 2011, \apj, 739, 1

\bibitem[{{Le Floc'h} {et~al.}(2006){Le Floc'h}, {Charmandaris}, {Forrest},
  {Mirabel}, {Armus}, \& {Devost}}]{2006ApJ...642..636L}
{Le Floc'h}, E., {Charmandaris}, V., {Forrest}, W.~J., {et~al.} 2006, \apj,
  642, 636

\bibitem[{{Le Floc'h} {et~al.}(2003){Le Floc'h}, {Duc}, {Mirabel}, {Sanders},
  {Bosch}, {Diaz}, {Donzelli}, {Rodrigues}, {Courvoisier}, {Greiner},
  {Mereghetti}, {Melnick}, {Maza}, \& {Minniti}}]{2003A&A...400..499L}
{Le Floc'h}, E., {Duc}, P.-A., {Mirabel}, I.~F., {et~al.} 2003, \aap, 400, 499

\bibitem[{{Ledoux} {et~al.}(2009){Ledoux}, {Vreeswijk}, {Smette}, {Fox},
  {Petitjean}, {Ellison}, {Fynbo}, \& {Savaglio}}]{2009A&A...506..661L}
{Ledoux}, C., {Vreeswijk}, P.~M., {Smette}, A., {et~al.} 2009, \aap, 506, 661

\bibitem[{{Leloudas} {et~al.}(2015){Leloudas}, {Schulze}, {Kr{\"u}hler},
  {Gorosabel}, {Christensen}, {Mehner}, {de Ugarte Postigo}, {Amor{\'{\i}}n},
  {Th{\"o}ne}, {Anderson}, {Bauer}, {Gallazzi}, {He{\l}miniak}, {Hjorth},
  {Ibar}, {Malesani}, {Morell}, {Vinko}, \& {Wheeler}}]{2014arXiv1409.8331L}
{Leloudas}, G., {Schulze}, S., {Kr{\"u}hler}, T., {et~al.} 2015, \mnras, 449,
  917

\bibitem[{{Levan} {et~al.}(2006){Levan}, {Fruchter}, {Rhoads}, {Mobasher},
  {Tanvir}, {Gorosabel}, {Rol}, {Kouveliotou}, {Dell'Antonio}, {Merrill},
  {Bergeron}, {Castro Cer{\'o}n}, {Masetti}, {Vreeswijk}, {Antonelli},
  {Bersier}, {Castro-Tirado}, {Fynbo}, {Garnavich}, {Holland}, {Hjorth},
  {Nugent}, {Pian}, {Smette}, {Thomsen}, {Thorsett}, \&
  {Wijers}}]{2006ApJ...647..471L}
{Levan}, A., {Fruchter}, A., {Rhoads}, J., {et~al.} 2006, \apj, 647, 471

\bibitem[{{Levan} {et~al.}(2014{\natexlab{a}}){Levan}, {Tanvir}, {Fruchter},
  {Hjorth}, {Pian}, {Mazzali}, {Hounsell}, {Perley}, {Cano}, {Graham}, {Cenko},
  {Fynbo}, {Kouveliotou}, {Pe'er}, {Misra}, \&
  {Wiersema}}]{2014ApJ...792..115L}
{Levan}, A.~J., {Tanvir}, N.~R., {Fruchter}, A.~S., {et~al.}
  2014{\natexlab{a}}, \apj, 792, 115

\bibitem[{{Levan} {et~al.}(2014{\natexlab{b}}){Levan}, {Tanvir}, {Starling},
  {Wiersema}, {Page}, {Perley}, {Schulze}, {Wynn}, {Chornock}, {Hjorth},
  {Cenko}, {Fruchter}, {O'Brien}, {Brown}, {Tunnicliffe}, {Malesani},
  {Jakobsson}, {Watson}, {Berger}, {Bersier}, {Cobb}, {Covino}, {Cucchiara},
  {de Ugarte Postigo}, {Fox}, {Gal-Yam}, {Goldoni}, {Gorosabel}, {Kaper},
  {Kr{\"u}hler}, {Karjalainen}, {Osborne}, {Pian},
  {S{\'a}nchez-Ram{\'{\i}}rez}, {Schmidt}, {Skillen}, {Tagliaferri},
  {Th{\"o}ne}, {Vaduvescu}, {Wijers}, \& {Zauderer}}]{2014ApJ...781...13L}
{Levan}, A.~J., {Tanvir}, N.~R., {Starling}, R.~L.~C., {et~al.}
  2014{\natexlab{b}}, \apj, 781, 13

\bibitem[{{Levesque}(2014)}]{2014PASP..126....1L}
{Levesque}, E.~M. 2014, \pasp, 126, 1

\bibitem[{{Levesque} {et~al.}(2010{\natexlab{a}}){Levesque}, {Berger},
  {Kewley}, \& {Bagley}}]{2010AJ....139..694L}
{Levesque}, E.~M., {Berger}, E., {Kewley}, L.~J., \& {Bagley}, M.~M.
  2010{\natexlab{a}}, \aj, 139, 694

\bibitem[{{Levesque} {et~al.}(2011){Levesque}, {Berger}, {Soderberg}, \&
  {Chornock}}]{2011ApJ...739...23L}
{Levesque}, E.~M., {Berger}, E., {Soderberg}, A.~M., \& {Chornock}, R. 2011,
  \apj, 739, 23

\bibitem[{{Levesque} {et~al.}(2010{\natexlab{b}}){Levesque}, {Kewley},
  {Berger}, \& {Zahid}}]{2010AJ....140.1557L}
{Levesque}, E.~M., {Kewley}, L.~J., {Berger}, E., \& {Zahid}, H.~J.
  2010{\natexlab{b}}, \aj, 140, 1557

\bibitem[{{Levesque} {et~al.}(2010{\natexlab{c}}){Levesque}, {Kewley},
  {Graham}, \& {Fruchter}}]{2010ApJ...712L..26L}
{Levesque}, E.~M., {Kewley}, L.~J., {Graham}, J.~F., \& {Fruchter}, A.~S.
  2010{\natexlab{c}}, \apjl, 712, L26

\bibitem[{{Littlejohns} {et~al.}(2015){Littlejohns}, {Butler}, {Cucchiara},
  {Watson}, {Fox}, {Lee}, {Kutyrev}, {Richer}, {Klein}, {Prochaska}, {Bloom},
  {Troja}, {Ramirez-Ruiz}, {de Diego}, {Georgiev}, {Gonz{\'a}lez},
  {Rom{\'a}n-Z{\'u}{\~n}iga}, {Gehrels}, \& {Moseley}}]{2014arXiv1412.6530L}
{Littlejohns}, O.~M., {Butler}, N.~R., {Cucchiara}, A., {et~al.} 2015, \mnras,
  449, 2919

\bibitem[{{Ly} {et~al.}(2011){Ly}, {Lee}, {Dale}, {Momcheva}, {Salim},
  {Staudaher}, {Moore}, \& {Finn}}]{2011ApJ...726..109L}
{Ly}, C., {Lee}, J.~C., {Dale}, D.~A., {et~al.} 2011, \apj, 726, 109

\bibitem[{{MacFadyen} \& {Woosley}(1999)}]{1999ApJ...524..262M}
{MacFadyen}, A.~I. \& {Woosley}, S.~E. 1999, \apj, 524, 262

\bibitem[{{Maiolino} {et~al.}(2008){Maiolino}, {Nagao}, {Grazian}, {Cocchia},
  {Marconi}, {Mannucci}, {Cimatti}, {Pipino}, {Ballero}, {Calura}, {Chiappini},
  {Fontana}, {Granato}, {Matteucci}, {Pastorini}, {Pentericci}, {Risaliti},
  {Salvati}, \& {Silva}}]{2008A&A...488..463M}
{Maiolino}, R., {Nagao}, T., {Grazian}, A., {et~al.} 2008, \aap, 488, 463

\bibitem[{{Malesani} {et~al.}(2012){Malesani}, {Kr\"uhler}, {de Ugarte
  Postigo}, {Xu}, {Tanvir}, {Gorosabel}, {Jakobsson}, {Geier}, \&
  {Pursimo}}]{2012GCN..13639...1M}
{Malesani}, D., {Kr\"uhler}, T., {de Ugarte Postigo}, A., {et~al.} 2012, GCN,
  13639

\bibitem[{{Mannucci} {et~al.}(2010){Mannucci}, {Cresci}, {Maiolino}, {Marconi},
  \& {Gnerucci}}]{2010MNRAS.408.2115M}
{Mannucci}, F., {Cresci}, G., {Maiolino}, R., {Marconi}, A., \& {Gnerucci}, A.
  2010, \mnras, 408, 2115

\bibitem[{{Mannucci} {et~al.}(2011){Mannucci}, {Salvaterra}, \&
  {Campisi}}]{2011MNRAS.414.1263M}
{Mannucci}, F., {Salvaterra}, R., \& {Campisi}, M.~A. 2011, \mnras, 414, 1263

\bibitem[{{Maselli} {et~al.}(2012){Maselli}, {Barthelmy}, {Baumgartner},
  {Burrows}, {Kennea}, {Kuin}, {Markwardt}, {Marshall}, {Palmer}, {Sbarufatti},
  {Siegel}, \& {Troja}}]{2012GCN..14045...1M}
{Maselli}, A., {Barthelmy}, S.~D., {Baumgartner}, W.~H., {et~al.} 2012, GCN,
  14045

\bibitem[{{McBreen} {et~al.}(2010){McBreen}, {Kr{\"u}hler}, {Rau}, {Greiner},
  {Kann}, {Savaglio}, {Afonso}, {Clemens}, {Filgas}, {Klose}, {K{\"u}pc{\"u}
  Yolda{\c s}}, {Olivares E.}, {Rossi}, {Szokoly}, {Updike}, \& {Yolda{\c
  s}}}]{2010A&A...516A..71M}
{McBreen}, S., {Kr{\"u}hler}, T., {Rau}, A., {et~al.} 2010, \aap, 516, A71

\bibitem[{{McGaugh}(1991)}]{1991ApJ...380..140M}
{McGaugh}, S.~S. 1991, \apj, 380, 140

\bibitem[{{Melandri} {et~al.}(2014){Melandri}, {Pian}, {D'Elia}, {D'Avanzo},
  {Della Valle}, {Mazzali}, {Tagliaferri}, {Cano}, {Levan}, {M{$\Delta$}oller},
  {Amati}, {Bernardini}, {Bersier}, {Bufano}, {Campana}, {Castro-Tirado},
  {Covino}, {Ghirlanda}, {Hurley}, {Malesani}, {Masetti}, {Palazzi},
  {Piranomonte}, {Rossi}, {Salvaterra}, {Starling}, {Tanaka}, {Tanvir}, \&
  {Vergani}}]{2014A&A...567A..29M}
{Melandri}, A., {Pian}, E., {D'Elia}, V., {et~al.} 2014, \aap, 567, A29

\bibitem[{{Melandri} {et~al.}(2012){Melandri}, {Sbarufatti}, {D'Avanzo},
  {Salvaterra}, {Campana}, {Covino}, {Vergani}, {Nava}, {Ghisellini},
  {Ghirlanda}, {Fugazza}, {Mangano}, {Capalbi}, \&
  {Tagliaferri}}]{2012MNRAS.421.1265M}
{Melandri}, A., {Sbarufatti}, B., {D'Avanzo}, P., {et~al.} 2012, \mnras, 421,
  1265

\bibitem[{{Micha{\l}owski} {et~al.}(2012){Micha{\l}owski}, {Kamble}, {Hjorth},
  {Malesani}, {Reinfrank}, {Bonavera}, {Castro Cer{\'o}n}, {Ibar}, {Dunlop},
  {Fynbo}, {Garrett}, {Jakobsson}, {Kaplan}, {Kr{\"u}hler}, {Levan},
  {Massardi}, {Pal}, {Sollerman}, {Tanvir}, {van der Horst}, {Watson}, \&
  {Wiersema}}]{2012ApJ...755...85M}
{Micha{\l}owski}, M.~J., {Kamble}, A., {Hjorth}, J., {et~al.} 2012, \apj, 755,
  85

\bibitem[{{Milvang-Jensen} {et~al.}(2012){Milvang-Jensen}, {Fynbo}, {Malesani},
  {Hjorth}, {Jakobsson}, \& {M{\o}ller}}]{2012ApJ...756...25M}
{Milvang-Jensen}, B., {Fynbo}, J.~P.~U., {Malesani}, D., {et~al.} 2012, \apj,
  756, 25

\bibitem[{{Modigliani} {et~al.}(2010){Modigliani}, {Goldoni}, {Royer},
  {Haigron}, {Guglielmi}, {Fran{\c c}ois}, {Horrobin}, {Bristow}, {Vernet},
  {Moehler}, {Kerber}, {Ballester}, {Mason}, \&
  {Christensen}}]{2010SPIE.7737E..56M}
{Modigliani}, A., {Goldoni}, P., {Royer}, F., {et~al.} 2010, in SPIE Conf.
  Series, Vol. 7737, SPIE Conf. Series

\bibitem[{{Modjaz} {et~al.}(2008){Modjaz}, {Kewley}, {Kirshner}, {Stanek},
  {Challis}, {Garnavich}, {Greene}, {Kelly}, \& {Prieto}}]{2008AJ....135.1136M}
{Modjaz}, M., {Kewley}, L., {Kirshner}, R.~P., {et~al.} 2008, \aj, 135, 1136

\bibitem[{{Morgan} {et~al.}(2014){Morgan}, {Perley}, {Cenko}, {Bloom},
  {Cucchiara}, {Richards}, {Filippenko}, {Haislip}, {LaCluyze}, {Corsi},
  {Melandri}, {Cobb}, {Gomboc}, {Horesh}, {James}, {Li}, {Mundell}, {Reichart},
  \& {Steele}}]{2014MNRAS.440.1810M}
{Morgan}, A.~N., {Perley}, D.~A., {Cenko}, S.~B., {et~al.} 2014, \mnras, 440,
  1810

\bibitem[{{Murray} {et~al.}(2005){Murray}, {Quataert}, \&
  {Thompson}}]{2005ApJ...618..569M}
{Murray}, N., {Quataert}, E., \& {Thompson}, T.~A. 2005, \apj, 618, 569

\bibitem[{{Nagao} {et~al.}(2006){Nagao}, {Maiolino}, \&
  {Marconi}}]{2006A&A...459...85N}
{Nagao}, T., {Maiolino}, R., \& {Marconi}, A. 2006, \aap, 459, 85

\bibitem[{{Nakauchi} {et~al.}(2013){Nakauchi}, {Kashiyama}, {Suwa}, \&
  {Nakamura}}]{2013ApJ...778...67N}
{Nakauchi}, D., {Kashiyama}, K., {Suwa}, Y., \& {Nakamura}, T. 2013, \apj, 778,
  67

\bibitem[{{Nicuesa} {et~al.}(2010){Nicuesa}, {Kruehler}, \&
  {Greiner}}]{2010GCN..10835...1N}
{Nicuesa}, A., {Kruehler}, T., \& {Greiner}, J. 2010, GCN, 10835

\bibitem[{{Niino} {et~al.}(2012){Niino}, {Hashimoto}, {Aoki}, {Hattori},
  {Yabe}, \& {Nomoto}}]{2012PASJ...64..115N}
{Niino}, Y., {Hashimoto}, T., {Aoki}, K., {et~al.} 2012, \pasj, 64, 115

\bibitem[{{Noll} {et~al.}(2012){Noll}, {Kausch}, {Barden}, {Jones}, {Szyszka},
  {Kimeswenger}, \& {Vinther}}]{2012A&A...543A..92N}
{Noll}, S., {Kausch}, W., {Barden}, M., {et~al.} 2012, \aap, 543, A92

\bibitem[{{Oates} {et~al.}(2010){Oates}, {Markwardt}, {Norris}, {Evans}, \&
  {Littlejohns}}]{2010GCNR..300....1O}
{Oates}, S.~R., {Markwardt}, C.~B., {Norris}, J., {Evans}, P.~A., \&
  {Littlejohns}, O. 2010, GCN Report, 300

\bibitem[{{Oke} {et~al.}(1995){Oke}, {Cohen}, {Carr}, {Cromer}, {Dingizian},
  {Harris}, {Labrecque}, {Lucinio}, {Schaal}, {Epps}, \&
  {Miller}}]{1995PASP..107..375O}
{Oke}, J.~B., {Cohen}, J.~G., {Carr}, M., {et~al.} 1995, \pasp, 107, 375

\bibitem[{{Osterbrock}(1989)}]{1989agna.book.....O}
{Osterbrock}, D.~E. 1989, {Astrophysics of gaseous nebulae and active galactic
  nuclei}

\bibitem[{{Ovaldsen} {et~al.}(2007){Ovaldsen}, {Jaunsen}, {Fynbo}, {Hjorth},
  {Th{\"o}ne}, {F{\'e}ron}, {Xu}, {Selj}, \& {Teuber}}]{2007ApJ...662..294O}
{Ovaldsen}, J.-E., {Jaunsen}, A.~O., {Fynbo}, J.~P.~U., {et~al.} 2007, \apj,
  662, 294

\bibitem[{{Pagel} {et~al.}(1979){Pagel}, {Edmunds}, {Blackwell}, {Chun}, \&
  {Smith}}]{1979MNRAS.189...95P}
{Pagel}, B.~E.~J., {Edmunds}, M.~G., {Blackwell}, D.~E., {Chun}, M.~S., \&
  {Smith}, G. 1979, \mnras, 189, 95

\bibitem[{{Pei}(1992)}]{1992ApJ...395..130P}
{Pei}, Y.~C. 1992, \apj, 395, 130

\bibitem[{{Perley} {et~al.}(2013{\natexlab{a}}){Perley}, {Bloom}, \&
  {Prochaska}}]{2013EAS....61..391P}
{Perley}, D.~A., {Bloom}, J.~S., \& {Prochaska}, J.~X. 2013{\natexlab{a}}, in
  EAS Publ. Ser., Vol.~61, EAS Publ. Ser., ed. A.~J. {Castro-Tirado},
  J.~{Gorosabel}, \& I.~H. {Park}, 391--395

\bibitem[{{Perley} {et~al.}(2009){Perley}, {Cenko}, {Bloom}, {Chen}, {Butler},
  {Kocevski}, {Prochaska}, {Brodwin}, {Glazebrook}, {Kasliwal}, {Kulkarni},
  {Lopez}, {Ofek}, {Pettini}, {Soderberg}, \& {Starr}}]{2009AJ....138.1690P}
{Perley}, D.~A., {Cenko}, S.~B., {Bloom}, J.~S., {et~al.} 2009, \aj, 138, 1690

\bibitem[{{Perley} {et~al.}(2014){Perley}, {Cenko}, {Corsi}, {Tanvir}, {Levan},
  {Kann}, {Sonbas}, {Wiersema}, {Zheng}, {Zhao}, {Bai}, {Bremer},
  {Castro-Tirado}, {Chang}, {Clubb}, {Frail}, {Fruchter}, {G{\"o}{\u g}{\"u}{\c
  s}}, {Greiner}, {G{\"u}ver}, {Horesh}, {Filippenko}, {Klose}, {Mao},
  {Morgan}, {Pozanenko}, {Schmidl}, {Stecklum}, {Tanga}, {Volnova}, {Volvach},
  {Wang}, {Winters}, \& {Xin}}]{2014ApJ...781...37P}
{Perley}, D.~A., {Cenko}, S.~B., {Corsi}, A., {et~al.} 2014, \apj, 781, 37

\bibitem[{{Perley} {et~al.}(2012){Perley}, {Cenko}, {Morgan}, \&
  {Kr\"uhler}}]{2012GCN..14056...1P}
{Perley}, D.~A., {Cenko}, S.~B., {Morgan}, A.~N., \& {Kr\"uhler}, T. 2012, GCN,
  14056

\bibitem[{{Perley} {et~al.}(2015{\natexlab{a}}){Perley}, {Kr{\"u}hler},
  {Schulze}, {de Ugarte Postigo}, {Hjorth}, {Berger}, {Cenko}, {Chary},
  {Cucchiara}, {Ellis}, {Fong}, {Fynbo}, {Gorosabel}, {Greiner}, {Jakobsson},
  {Kim}, {Laskar}, {Levan}, {Micha{\l}owski}, {Milvang-Jensen}, {Tanvir},
  {Th{\"o}ne}, \& {Wiersema}}]{2015arXiv150402482P}
{Perley}, D.~A., {Kr{\"u}hler}, T., {Schulze}, S., {et~al.} 2015{\natexlab{a}},
  ApJ, submitted [\eprint[arXiv]{1504.02482}]

\bibitem[{{Perley} {et~al.}(2013{\natexlab{b}}){Perley}, {Levan}, {Tanvir},
  {Cenko}, {Bloom}, {Hjorth}, {Kr{\"u}hler}, {Filippenko}, {Fruchter}, {Fynbo},
  {Jakobsson}, {Kalirai}, {Milvang-Jensen}, {Morgan}, {Prochaska}, \&
  {Silverman}}]{2013ApJ...778..128P}
{Perley}, D.~A., {Levan}, A.~J., {Tanvir}, N.~R., {et~al.} 2013{\natexlab{b}},
  \apj, 778, 128

\bibitem[{{Perley} {et~al.}(2015{\natexlab{b}}){Perley}, {Perley}, {Hjorth},
  {Micha{\l}owski}, {Cenko}, {Jakobsson}, {Kr{\"u}hler}, {Levan}, {Malesani},
  \& {Tanvir}}]{2015ApJ...801..102P}
{Perley}, D.~A., {Perley}, R.~A., {Hjorth}, J., {et~al.} 2015{\natexlab{b}},
  \apj, 801, 102

\bibitem[{{Perley} {et~al.}(2015{\natexlab{c}}){Perley}, {Tanvir}, {Hjorth},
  {Laskar}, {Berger}, {Chary}, {de Ugarte Postigo}, {Fynbo}, {Kr{\"u}hler},
  {Levan}, {Micha{\l}owski}, \& {Schulze}}]{2015arXiv150402479P}
{Perley}, D.~A., {Tanvir}, N.~R., {Hjorth}, J., {et~al.} 2015{\natexlab{c}},
  ApJ, submitted [\eprint[arXiv]{1504.02479}]

\bibitem[{{Pettini} \& {Pagel}(2004)}]{2004MNRAS.348L..59P}
{Pettini}, M. \& {Pagel}, B.~E.~J. 2004, \mnras, 348, L59

\bibitem[{{Pettini} {et~al.}(2001){Pettini}, {Shapley}, {Steidel}, {Cuby},
  {Dickinson}, {Moorwood}, {Adelberger}, \& {Giavalisco}}]{2001ApJ...554..981P}
{Pettini}, M., {Shapley}, A.~E., {Steidel}, C.~C., {et~al.} 2001, \apj, 554,
  981

\bibitem[{{Pian} {et~al.}(2006){Pian}, {Mazzali}, {Masetti}, {Ferrero},
  {Klose}, {Palazzi}, {Ramirez-Ruiz}, {Woosley}, {Kouveliotou}, {Deng},
  {Filippenko}, {Foley}, {Fynbo}, {Kann}, {Li}, {Hjorth}, {Nomoto}, {Patat},
  {Sauer}, {Sollerman}, {Vreeswijk}, {Guenther}, {Levan}, {O'Brien}, {Tanvir},
  {Wijers}, {Dumas}, {Hainaut}, {Wong}, {Baade}, {Wang}, {Amati}, {Cappellaro},
  {Castro-Tirado}, {Ellison}, {Frontera}, {Fruchter}, {Greiner}, {Kawabata},
  {Ledoux}, {Maeda}, {M{\o}ller}, {Nicastro}, {Rol}, \&
  {Starling}}]{2006Natur.442.1011P}
{Pian}, E., {Mazzali}, P.~A., {Masetti}, N., {et~al.} 2006, \nat, 442, 1011

\bibitem[{{Piranomonte} {et~al.}(2015){Piranomonte}, {Japelj}, {Vergani},
  {Savaglio}, {Palazzi}, {Covino}, {Flores}, {Goldoni}, {Cupani},
  {Kr{\"u}hler}, {Mannucci}, {Onori}, {Rossi}, {D'Elia}, {Pian}, {D'Avanzo},
  {Gomboc}, {Hammer}, {Randich}, {Fiore}, {Stella}, \&
  {Tagliaferri}}]{2015Silviasubm}
{Piranomonte}, S., {Japelj}, J., {Vergani}, S.~D., {et~al.} 2015, \mnras, 452,
  3293

\bibitem[{{Piranomonte} {et~al.}(2011){Piranomonte}, {Vergani}, {Malesani},
  {Fynbo}, {Wiersema}, \& {Kaper}}]{2011GCN..12164...1P}
{Piranomonte}, S., {Vergani}, S.~D., {Malesani}, D., {et~al.} 2011, GCN, 12164

\bibitem[{{Planck Collaboration}(2014)}]{2014A&A...571A..16P}
{Planck Collaboration}. 2014, \aap, 571, A16

\bibitem[{{Podsiadlowski} {et~al.}(2004){Podsiadlowski}, {Mazzali}, {Nomoto},
  {Lazzati}, \& {Cappellaro}}]{2004ApJ...607L..17P}
{Podsiadlowski}, P., {Mazzali}, P.~A., {Nomoto}, K., {Lazzati}, D., \&
  {Cappellaro}, E. 2004, \apjl, 607, L17

\bibitem[{{Prochaska} {et~al.}(2004){Prochaska}, {Bloom}, {Chen}, {Hurley},
  {Melbourne}, {Dressler}, {Graham}, {Osip}, \& {Vacca}}]{2004ApJ...611..200P}
{Prochaska}, J.~X., {Bloom}, J.~S., {Chen}, H.-W., {et~al.} 2004, \apj, 611,
  200

\bibitem[{{Prochaska} {et~al.}(2007){Prochaska}, {Chen}, {Dessauges-Zavadsky},
  \& {Bloom}}]{2007ApJ...666..267P}
{Prochaska}, J.~X., {Chen}, H.-W., {Dessauges-Zavadsky}, M., \& {Bloom}, J.~S.
  2007, \apj, 666, 267

\bibitem[{{Prochaska} {et~al.}(2009){Prochaska}, {Sheffer}, {Perley}, {Bloom},
  {Lopez}, {Dessauges-Zavadsky}, {Chen}, {Filippenko}, {Ganeshalingam}, {Li},
  {Miller}, \& {Starr}}]{2009ApJ...691L..27P}
{Prochaska}, J.~X., {Sheffer}, Y., {Perley}, D.~A., {et~al.} 2009, \apjl, 691,
  L27

\bibitem[{{Racusin} {et~al.}(2011){Racusin}, {Barthelmy}, {Beardmore},
  {Burrows}, {Campana}, {Evans}, {Holland}, {Kennea}, {Kuin}, {Littlejohns},
  {Markwardt}, {Marshall}, {Page}, {Palmer}, {Saxton}, {Sbarufatti}, {Siegel},
  {Stroh}, {Swenson}, {Troja}, \& {Zhang}}]{2011GCN..12600...1R}
{Racusin}, J.~L., {Barthelmy}, S.~D., {Beardmore}, A.~P., {et~al.} 2011, GCN,
  12600

\bibitem[{{Reddy} \& {Steidel}(2009)}]{2009ApJ...692..778R}
{Reddy}, N.~A. \& {Steidel}, C.~C. 2009, \apj, 692, 778

\bibitem[{{Richard} {et~al.}(2011){Richard}, {Jones}, {Ellis}, {Stark},
  {Livermore}, \& {Swinbank}}]{2011MNRAS.413..643R}
{Richard}, J., {Jones}, T., {Ellis}, R., {et~al.} 2011, \mnras, 413, 643

\bibitem[{{Robertson} \& {Ellis}(2012)}]{2012ApJ...744...95R}
{Robertson}, B.~E. \& {Ellis}, R.~S. 2012, \apj, 744, 95

\bibitem[{{Rossi} {et~al.}(2012){Rossi}, {Klose}, {Ferrero}, {Greiner},
  {Arnold}, {Gonsalves}, {Hartmann}, {Updike}, {Kann}, {Kr{\"u}hler},
  {Palazzi}, {Savaglio}, {Schulze}, {Afonso}, {Amati}, {Castro-Tirado},
  {Clemens}, {Filgas}, {Gorosabel}, {Hunt}, {K{\"u}pc{\"u} Yolda{\c s}},
  {Masetti}, {Nardini}, {Nicuesa Guelbenzu}, {Olivares}, {Pian}, {Rau},
  {Schady}, {Schmidl}, {Yolda{\c s}}, \& {de Ugarte
  Postigo}}]{2012A&A...545A..77R}
{Rossi}, A., {Klose}, S., {Ferrero}, P., {et~al.} 2012, \aap, 545, A77

\bibitem[{{Rossi} {et~al.}(2011){Rossi}, {Nicuesa Guelbenzu}, \&
  {Greiner}}]{2011GCN..12605...1R}
{Rossi}, A., {Nicuesa Guelbenzu}, A., \& {Greiner}, J. 2011, GCN, 12605

\bibitem[{{Rossi} {et~al.}(2014){Rossi}, {Piranomonte}, {Savaglio}, {Palazzi},
  {Micha{\l}owski}, {Klose}, {Hunt}, {Amati}, {Elliott}, {Greiner}, {Guidorzi},
  {Japelj}, {Kann}, {Lo Faro}, {Nicuesa Guelbenzu}, {Schulze}, {Vergani},
  {Arnold}, {Covino}, {D'Elia}, {Ferrero}, {Filgas}, {Goldoni}, {K{\"u}pc{\"u}
  Yolda{\c s}}, {Le Borgne}, {Pian}, {Schady}, \&
  {Stratta}}]{2014A&A...572A..47R}
{Rossi}, A., {Piranomonte}, S., {Savaglio}, S., {et~al.} 2014, \aap, 572, A47

\bibitem[{{Salvaterra} {et~al.}(2012){Salvaterra}, {Campana}, {Vergani},
  {Covino}, {D'Avanzo}, {Fugazza}, {Ghirlanda}, {Ghisellini}, {Melandri},
  {Nava}, {Sbarufatti}, {Flores}, {Piranomonte}, \&
  {Tagliaferri}}]{2012ApJ...749...68S}
{Salvaterra}, R., {Campana}, S., {Vergani}, S.~D., {et~al.} 2012, \apj, 749, 68

\bibitem[{{Santini} {et~al.}(2009){Santini}, {Fontana}, {Grazian}, {Salimbeni},
  {Fiore}, {Fontanot}, {Boutsia}, {Castellano}, {Cristiani}, {de Santis},
  {Gallozzi}, {Giallongo}, {Menci}, {Nonino}, {Paris}, {Pentericci}, \&
  {Vanzella}}]{2009A&A...504..751S}
{Santini}, P., {Fontana}, A., {Grazian}, A., {et~al.} 2009, \aap, 504, 751

\bibitem[{{Savaglio}(2006)}]{2006NJPh....8..195S}
{Savaglio}, S. 2006, New Journal of Physics, 8, 195

\bibitem[{{Savaglio} {et~al.}(2009){Savaglio}, {Glazebrook}, \& {Le
  Borgne}}]{2009ApJ...691..182S}
{Savaglio}, S., {Glazebrook}, K., \& {Le Borgne}, D. 2009, \apj, 691, 182

\bibitem[{{Savaglio} {et~al.}(2005){Savaglio}, {Glazebrook}, {Le Borgne},
  {Juneau}, {Abraham}, {Chen}, {Crampton}, {McCarthy}, {Carlberg}, {Marzke},
  {Roth}, {J{\o}rgensen}, \& {Murowinski}}]{2005ApJ...635..260S}
{Savaglio}, S., {Glazebrook}, K., {Le Borgne}, D., {et~al.} 2005, \apj, 635,
  260

\bibitem[{{Saxton} {et~al.}(2012){Saxton}, {Cummings}, {Kuin}, {Lien},
  {Littlejohns}, {Markwardt}, {Marshall}, {Mountford}, {Oates}, {Page},
  {Palmer}, {Sbarufatti}, {Siegel}, {Swenson}, \&
  {Ukwatta}}]{2012GCN..12980...1S}
{Saxton}, C.~J., {Cummings}, J.~R., {Kuin}, N.~P.~M., {et~al.} 2012, GCN, 12980

\bibitem[{{Schady} {et~al.}(2015){Schady}, {Kr{\"u}hler}, {Greiner}, {Graham},
  {Kann}, {Bolmer}, {Delvaux}, {Elliott}, {Klose}, {Knust}, {Nicuesa
  Guelbenzu}, {Rau}, {Rossi}, {Savaglio}, {Schmidl}, {Schweyer}, {Sudilovsky},
  {Tanga}, {Tanvir}, {Varela}, \& {Wiseman}}]{2015Patsubm}
{Schady}, P., {Kr{\"u}hler}, T., {Greiner}, J., {et~al.} 2015, \aap, 579, A126

\bibitem[{{Schady} {et~al.}(2014){Schady}, {Savaglio}, {M{\"u}ller},
  {Kr{\"u}hler}, {Dwelly}, {Palazzi}, {Hunt}, {Greiner}, {Linz},
  {Micha{\l}owski}, {Pierini}, {Piranomonte}, {Vergani}, \&
  {Gear}}]{2014arXiv1408.5076S}
{Schady}, P., {Savaglio}, S., {M{\"u}ller}, T., {et~al.} 2014, \aap, 570, A52

\bibitem[{{Schlafly} \& {Finkbeiner}(2011)}]{2011ApJ...737..103S}
{Schlafly}, E.~F. \& {Finkbeiner}, D.~P. 2011, \apj, 737, 103

\bibitem[{{Schulze} {et~al.}(2015){Schulze}, {Chapman}, {Hjorth}, {Levan},
  {Jakobsson}, {Bj{\"o}rnsson}, {Perley}, {Kr{\"u}hler}, {Gorosabel}, {Tanvir},
  {de Ugarte Postigo}, {Fynbo}, {Milvang-Jensen}, {M{\o}ller}, \&
  {Watson}}]{2015arXiv150304246S}
{Schulze}, S., {Chapman}, R., {Hjorth}, J., {et~al.} 2015, \apj, 808, 73

\bibitem[{{Schulze} {et~al.}(2014){Schulze}, {Malesani}, {Cucchiara}, {Tanvir},
  {Kr{\"u}hler}, {de Ugarte Postigo}, {Leloudas}, {Lyman}, {Bersier},
  {Wiersema}, {Perley}, {Schady}, {Gorosabel}, {Anderson}, {Castro-Tirado},
  {Cenko}, {De Cia}, {Ellerbroek}, {Fynbo}, {Greiner}, {Hjorth}, {Kann},
  {Kaper}, {Klose}, {Levan}, {Mart{\'{\i}}n}, {O'Brien}, {Page}, {Pignata},
  {Rapaport}, {S{\'a}nchez-Ram{\'{\i}}rez}, {Sollerman}, {Smith}, {Sparre},
  {Th{\"o}ne}, {Watson}, {Xu}, {Bauer}, {Bayliss}, {Bj{\"o}rnsson}, {Bremer},
  {Cano}, {Covino}, {D'Elia}, {Frail}, {Geier}, {Goldoni}, {Hartoog},
  {Jakobsson}, {Korhonen}, {Lee}, {Milvang-Jensen}, {Nardini}, {Nicuesa
  Guelbenzu}, {Oguri}, {Pandey}, {Petitpas}, {Rossi}, {Sandberg}, {Schmidl},
  {Tagliaferri}, {Tilanus}, {Winters}, {Wright}, \&
  {Wuyts}}]{2014A&A...566A.102S}
{Schulze}, S., {Malesani}, D., {Cucchiara}, A., {et~al.} 2014, \aap, 566, A102

\bibitem[{{Shapley} {et~al.}(2005){Shapley}, {Coil}, {Ma}, \&
  {Bundy}}]{2005ApJ...635.1006S}
{Shapley}, A.~E., {Coil}, A.~L., {Ma}, C.-P., \& {Bundy}, K. 2005, \apj, 635,
  1006

\bibitem[{{Sobral} {et~al.}(2014){Sobral}, {Best}, {Smail}, {Mobasher},
  {Stott}, \& {Nisbet}}]{2014MNRAS.437.3516S}
{Sobral}, D., {Best}, P.~N., {Smail}, I., {et~al.} 2014, \mnras, 437, 3516

\bibitem[{{Soderberg} {et~al.}(2007){Soderberg}, {Nakar}, {Cenko}, {Cameron},
  {Frail}, {Kulkarni}, {Fox}, {Berger}, {Gal-Yam}, {Moon}, {Price}, {Anderson},
  {Schmidt}, {Salvo}, {Rich}, {Rau}, {Ofek}, {Chevalier}, {Hamuy}, {Harrison},
  {Kumar}, {MacFadyen}, {McCarthy}, {Park}, {Peterson}, {Phillips}, {Rauch},
  {Roth}, \& {Shectman}}]{2007ApJ...661..982S}
{Soderberg}, A.~M., {Nakar}, E., {Cenko}, S.~B., {et~al.} 2007, \apj, 661, 982

\bibitem[{{Sollerman} {et~al.}(2005){Sollerman}, {{\"O}stlin}, {Fynbo},
  {Hjorth}, {Fruchter}, \& {Pedersen}}]{2005NewA...11..103S}
{Sollerman}, J., {{\"O}stlin}, G., {Fynbo}, J.~P.~U., {et~al.} 2005, \na, 11,
  103

\bibitem[{{Sonbas} {et~al.}(2012){Sonbas}, {Baumgartner}, {Burrows}, {Gehrels},
  {Hoversten}, {Mangano}, {Pagani}, {Page}, \& {Palmer}}]{2012GCN..12920...1S}
{Sonbas}, E., {Baumgartner}, W.~H., {Burrows}, D.~N., {et~al.} 2012, GCN, 12920

\bibitem[{{Sparre} {et~al.}(2011){Sparre}, {Sollerman}, {Fynbo}, {Malesani},
  {Goldoni}, {de Ugarte Postigo}, {Covino}, {D'Elia}, {Flores}, {Hammer},
  {Hjorth}, {Jakobsson}, {Kaper}, {Leloudas}, {Levan}, {Milvang-Jensen},
  {Schulze}, {Tagliaferri}, {Tanvir}, {Watson}, {Wiersema}, \&
  {Wijers}}]{2011ApJ...735L..24S}
{Sparre}, M., {Sollerman}, J., {Fynbo}, J.~P.~U., {et~al.} 2011, \apjl, 735,
  L24

\bibitem[{{Speagle} {et~al.}(2014){Speagle}, {Steinhardt}, {Capak}, \&
  {Silverman}}]{2014ApJS..214...15S}
{Speagle}, J.~S., {Steinhardt}, C.~L., {Capak}, P.~L., \& {Silverman}, J.~D.
  2014, \apjs, 214, 15

\bibitem[{{Stanek} {et~al.}(2006){Stanek}, {Gnedin}, {Beacom}, {Gould},
  {Johnson}, {Kollmeier}, {Modjaz}, {Pinsonneault}, {Pogge}, \&
  {Weinberg}}]{2006AcA....56..333S}
{Stanek}, K.~Z., {Gnedin}, O.~Y., {Beacom}, J.~F., {et~al.} 2006, \actaa, 56,
  333

\bibitem[{{Starling} {et~al.}(2011){Starling}, {Wiersema}, {Levan}, {Sakamoto},
  {Bersier}, {Goldoni}, {Oates}, {Rowlinson}, {Campana}, {Sollerman}, {Tanvir},
  {Malesani}, {Fynbo}, {Covino}, {D'Avanzo}, {O'Brien}, {Page}, {Osborne},
  {Vergani}, {Barthelmy}, {Burrows}, {Cano}, {Curran}, {de Pasquale}, {D'Elia},
  {Evans}, {Flores}, {Fruchter}, {Garnavich}, {Gehrels}, {Gorosabel}, {Hjorth},
  {Holland}, {van der Horst}, {Hurkett}, {Jakobsson}, {Kamble}, {Kouveliotou},
  {Kuin}, {Kaper}, {Mazzali}, {Nugent}, {Pian}, {Stamatikos}, {Th{\"o}ne}, \&
  {Woosley}}]{2011MNRAS.411.2792S}
{Starling}, R.~L.~C., {Wiersema}, K., {Levan}, A.~J., {et~al.} 2011, \mnras,
  411, 2792

\bibitem[{{Steidel} {et~al.}(1996){Steidel}, {Giavalisco}, {Pettini},
  {Dickinson}, \& {Adelberger}}]{1996ApJ...462L..17S}
{Steidel}, C.~C., {Giavalisco}, M., {Pettini}, M., {Dickinson}, M., \&
  {Adelberger}, K.~L. 1996, \apjl, 462, L17

\bibitem[{{Steidel} {et~al.}(2014){Steidel}, {Rudie}, {Strom}, {Pettini},
  {Reddy}, {Shapley}, {Trainor}, {Erb}, {Turner}, {Konidaris}, {Kulas}, {Mace},
  {Matthews}, \& {McLean}}]{2014ApJ...795..165S}
{Steidel}, C.~C., {Rudie}, G.~C., {Strom}, A.~L., {et~al.} 2014, \apj, 795, 165

\bibitem[{{Storey} \& {Zeippen}(2000)}]{2000MNRAS.312..813S}
{Storey}, P.~J. \& {Zeippen}, C.~J. 2000, \mnras, 312, 813

\bibitem[{{Svensson} {et~al.}(2010){Svensson}, {Levan}, {Tanvir}, {Fruchter},
  \& {Strolger}}]{2010MNRAS.405...57S}
{Svensson}, K.~M., {Levan}, A.~J., {Tanvir}, N.~R., {Fruchter}, A.~S., \&
  {Strolger}, L.-G. 2010, \mnras, 405, 57

\bibitem[{{Svensson} {et~al.}(2012){Svensson}, {Levan}, {Tanvir}, {Perley},
  {Michalowski}, {Page}, {Bloom}, {Cenko}, {Hjorth}, {Jakobsson}, {Watson}, \&
  {Wheatley}}]{2012MNRAS.421...25S}
{Svensson}, K.~M., {Levan}, A.~J., {Tanvir}, N.~R., {et~al.} 2012, \mnras, 421,
  25

\bibitem[{{Tanvir} {et~al.}(2004){Tanvir}, {Barnard}, {Blain}, {Fruchter},
  {Kouveliotou}, {Natarajan}, {Ramirez-Ruiz}, {Rol}, {Smith}, {Tilanus}, \&
  {Wijers}}]{2004MNRAS.352.1073T}
{Tanvir}, N.~R., {Barnard}, V.~E., {Blain}, A.~W., {et~al.} 2004, \mnras, 352,
  1073

\bibitem[{{Tanvir} {et~al.}(2012){Tanvir}, {Levan}, {Fruchter}, {Fynbo},
  {Hjorth}, {Wiersema}, {Bremer}, {Rhoads}, {Jakobsson}, {O'Brien}, {Stanway},
  {Bersier}, {Natarajan}, {Greiner}, {Watson}, {Castro-Tirado}, {Wijers},
  {Starling}, {Misra}, {Graham}, \& {Kouveliotou}}]{2012ApJ...754...46T}
{Tanvir}, N.~R., {Levan}, A.~J., {Fruchter}, A.~S., {et~al.} 2012, \apj, 754,
  46

\bibitem[{{Tanvir} {et~al.}(2008){Tanvir}, {Levan}, {Rol}, {Starling},
  {Gorosabel}, {Priddey}, {Malesani}, {Jakobsson}, {O'Brien}, {Jaunsen},
  {Hjorth}, {Fynbo}, {Melandri}, {Gomboc}, {Milvang-Jensen}, {Fruchter},
  {Jarvis}, {Fernandes}, \& {Wold}}]{2008MNRAS.388.1743T}
{Tanvir}, N.~R., {Levan}, A.~J., {Rol}, E., {et~al.} 2008, \mnras, 388, 1743

\bibitem[{{Th{\"o}ne} {et~al.}(2013){Th{\"o}ne}, {Fynbo}, {Goldoni}, {de
  Ugarte}, {Campana}, {Vergani}, {Covino}, {Kr{\"u}hler}, {Kaper}, {Tanvir},
  {Zafar}, {D'Elia}, {Gorosabel}, {Greiner}, {Groot}, {Hammer}, {Jakobsson},
  {Klose}, {Levan}, {Milvang-Jensen}, {Nicuesa}, {Palazzi}, {Piranomonte},
  {Tagliaferri}, {Watson}, {Wiersema}, \& {Wijers}}]{2013MNRAS.428.3590T}
{Th{\"o}ne}, C.~C., {Fynbo}, J.~P.~U., {Goldoni}, P., {et~al.} 2013, \mnras,
  428, 3590

\bibitem[{{Tody}(1993)}]{1993ASPC...52..173T}
{Tody}, D. 1993, in ASPC Ser., Vol.~52, Astronomical Data Analysis Software and
  Systems II, 173

\bibitem[{{Tomczak} {et~al.}(2014){Tomczak}, {Quadri}, {Tran}, {Labb{\'e}},
  {Straatman}, {Papovich}, {Glazebrook}, {Allen}, {Brammer}, {Kacprzak},
  {Kawinwanichakij}, {Kelson}, {McCarthy}, {Mehrtens}, {Monson}, {Persson},
  {Spitler}, {Tilvi}, \& {van Dokkum}}]{2014ApJ...783...85T}
{Tomczak}, A.~R., {Quadri}, R.~F., {Tran}, K.-V.~H., {et~al.} 2014, \apj, 783,
  85

\bibitem[{{Tremonti} {et~al.}(2004){Tremonti}, {Heckman}, {Kauffmann},
  {Brinchmann}, {Charlot}, {White}, {Seibert}, {Peng}, {Schlegel}, {Uomoto},
  {Fukugita}, \& {Brinkmann}}]{2004ApJ...613..898T}
{Tremonti}, C.~A., {Heckman}, T.~M., {Kauffmann}, G., {et~al.} 2004, \apj, 613,
  898

\bibitem[{{Trenti} {et~al.}(2015){Trenti}, {Perna}, \&
  {Jimenez}}]{2014arXiv1406.1503T}
{Trenti}, M., {Perna}, R., \& {Jimenez}, R. 2015, \apj, 802, 103

\bibitem[{{Trenti} {et~al.}(2012){Trenti}, {Perna}, {Levesque}, {Shull}, \&
  {Stocke}}]{2012ApJ...749L..38T}
{Trenti}, M., {Perna}, R., {Levesque}, E.~M., {Shull}, J.~M., \& {Stocke},
  J.~T. 2012, \apjl, 749, L38

\bibitem[{{Troja} {et~al.}(2012){Troja}, {Barthelmy}, {Beardmore}, {Burrows},
  {Gehrels}, {Gronwall}, {Holland}, {Kennea}, {Littlejohns}, {Page}, \&
  {Palmer}}]{2012GCN..13588...1T}
{Troja}, E., {Barthelmy}, S.~D., {Beardmore}, A.~P., {et~al.} 2012, GCN, 13588

\bibitem[{{Troja} {et~al.}(2014){Troja}, {Beardmore}, {Bernardini}, {Burrows},
  {D'Elia}, {Evans}, {Malesani}, {Marshall}, {Maselli}, {O'Brien}, {Page},
  {Palmer}, \& {Siegel}}]{2014GCN..15728...1T}
{Troja}, E., {Beardmore}, A.~P., {Bernardini}, M.~G., {et~al.} 2014, GCN, 15728

\bibitem[{{Troncoso} {et~al.}(2014){Troncoso}, {Maiolino}, {Sommariva},
  {Cresci}, {Mannucci}, {Marconi}, {Meneghetti}, {Grazian}, {Cimatti},
  {Fontana}, {Nagao}, \& {Pentericci}}]{2014A&A...563A..58T}
{Troncoso}, P., {Maiolino}, R., {Sommariva}, V., {et~al.} 2014, \aap, 563, A58

\bibitem[{{van der Horst} {et~al.}(2009){van der Horst}, {Kouveliotou},
  {Gehrels}, {Rol}, {Wijers}, {Cannizzo}, {Racusin}, \&
  {Burrows}}]{2009ApJ...699.1087V}
{van der Horst}, A.~J., {Kouveliotou}, C., {Gehrels}, N., {et~al.} 2009, \apj,
  699, 1087

\bibitem[{{van der Horst} {et~al.}(2014){van der Horst}, {Paragi}, {de Bruyn},
  {Granot}, {Kouveliotou}, {Wiersema}, {Starling}, {Curran}, {Wijers},
  {Rowlinson}, {Anderson}, {Fender}, {Yang}, \& {Strom}}]{2014MNRAS.444.3151V}
{van der Horst}, A.~J., {Paragi}, Z., {de Bruyn}, A.~G., {et~al.} 2014, \mnras,
  444, 3151

\bibitem[{{Vergani} {et~al.}(2011){Vergani}, {Flores}, {Covino}, {Fugazza},
  {Gorosabel}, {Levan}, {Puech}, {Salvaterra}, {Tello}, {de Ugarte Postigo},
  {D'Avanzo}, {D'Elia}, {Fern{\'a}ndez}, {Fynbo}, {Ghirlanda},
  {Jel{\'{\i}}nek}, {Lundgren}, {Malesani}, {Palazzi}, {Piranomonte},
  {Rodrigues}, {S{\'a}nchez-Ram{\'{\i}}rez}, {Terr{\'o}n}, {Th{\"o}ne},
  {Antonelli}, {Campana}, {Castro-Tirado}, {Goldoni}, {Hammer}, {Hjorth},
  {Jakobsson}, {Kaper}, {Melandri}, {Milvang-Jensen}, {Sollerman},
  {Tagliaferri}, {Tanvir}, {Wiersema}, \& {Wijers}}]{2011A&A...535A.127V}
{Vergani}, S.~D., {Flores}, H., {Covino}, S., {et~al.} 2011, \aap, 535, A127

\bibitem[{{Vergani} {et~al.}(2015){Vergani}, {Salvaterra}, {Japelj}, {Le
  Floc'h}, {D'Avanzo}, {Fernandez-Soto}, {Kr{\"u}hler}, {Melandri}, {Boissier},
  {Covino}, {Puech}, {Greiner}, {Hunt}, {Perley}, {Petitjean}, {Hammer},
  {Levan}, {Mannucci}, {Campana}, {Flores}, {Gomboc}, \&
  {Tagliaferri}}]{2014arXiv1409.7064V}
{Vergani}, S.~D., {Salvaterra}, R., {Japelj}, J., {et~al.} 2015, \aap, in press
  [\eprint[arXiv]{1409.7064}]

\bibitem[{{Vernet} {et~al.}(2011){Vernet}, {Dekker}, {D'Odorico}, {Kaper},
  {Kjaergaard}, {Hammer}, {Randich}, {Zerbi}, {Groot}, {Hjorth}, {Guinouard},
  {Navarro}, {Adolfse}, {Albers}, {Amans}, {Andersen}, {Andersen}, {Binetruy},
  {Bristow}, {Castillo}, {Chemla}, {Christensen}, {Conconi}, {Conzelmann},
  {Dam}, {de Caprio}, {de Ugarte Postigo}, {Delabre}, {di Marcantonio},
  {Downing}, {Elswijk}, {Finger}, {Fischer}, {Flores}, {Fran{\c c}ois},
  {Goldoni}, {Guglielmi}, {Haigron}, {Hanenburg}, {Hendriks}, {Horrobin},
  {Horville}, {Jessen}, {Kerber}, {Kern}, {Kiekebusch}, {Kleszcz}, {Klougart},
  {Kragt}, {Larsen}, {Lizon}, {Lucuix}, {Mainieri}, {Manuputy}, {Martayan},
  {Mason}, {Mazzoleni}, {Michaelsen}, {Modigliani}, {Moehler}, {M{\o}ller},
  {Norup S{\o}rensen}, {N{\o}rregaard}, {P{\'e}roux}, {Patat}, {Pena}, {Pragt},
  {Reinero}, {Rigal}, {Riva}, {Roelfsema}, {Royer}, {Sacco}, {Santin},
  {Schoenmaker}, {Spano}, {Sweers}, {Ter Horst}, {Tintori}, {Tromp}, {van
  Dael}, {van der Vliet}, {Venema}, {Vidali}, {Vinther}, {Vola}, {Winters},
  {Wistisen}, {Wulterkens}, \& {Zacchei}}]{2011A&A...536A.105V}
{Vernet}, J., {Dekker}, H., {D'Odorico}, S., {et~al.} 2011, \aap, 536, A105

\bibitem[{{Virgili} {et~al.}(2013){Virgili}, {Mundell}, {Pal'shin}, {Guidorzi},
  {Margutti}, {Melandri}, {Harrison}, {Kobayashi}, {Chornock}, {Henden},
  {Updike}, {Cenko}, {Tanvir}, {Steele}, {Cucchiara}, {Gomboc}, {Levan},
  {Cano}, {Mottram}, {Clay}, {Bersier}, {Kopa{\v c}}, {Japelj}, {Filippenko},
  {Li}, {Svinkin}, {Golenetskii}, {Hartmann}, {Milne}, {Williams}, {O'Brien},
  {Fox}, \& {Berger}}]{2013ApJ...778...54V}
{Virgili}, F.~J., {Mundell}, C.~G., {Pal'shin}, V., {et~al.} 2013, \apj, 778,
  54

\bibitem[{{Vreeswijk} {et~al.}(2004){Vreeswijk}, {Ellison}, {Ledoux}, {Wijers},
  {Fynbo}, {M{\o}ller}, {Henden}, {Hjorth}, {Masi}, {Rol}, {Jensen}, {Tanvir},
  {Levan}, {Castro Cer{\'o}n}, {Gorosabel}, {Castro-Tirado}, {Fruchter},
  {Kouveliotou}, {Burud}, {Rhoads}, {Masetti}, {Palazzi}, {Pian}, {Pedersen},
  {Kaper}, {Gilmore}, {Kilmartin}, {Buckle}, {Seigar}, {Hartmann}, {Lindsay},
  \& {van den Heuvel}}]{2004A&A...419..927V}
{Vreeswijk}, P.~M., {Ellison}, S.~L., {Ledoux}, C., {et~al.} 2004, \aap, 419,
  927

\bibitem[{{Vreeswijk} {et~al.}(2001){Vreeswijk}, {Fruchter}, {Kaper}, {Rol},
  {Galama}, {van Paradijs}, {Kouveliotou}, {Wijers}, {Pian}, {Palazzi},
  {Masetti}, {Frontera}, {Savaglio}, {Reinsch}, {Hessman}, {Beuermann},
  {Nicklas}, \& {van den Heuvel}}]{2001ApJ...546..672V}
{Vreeswijk}, P.~M., {Fruchter}, A., {Kaper}, L., {et~al.} 2001, \apj, 546, 672

\bibitem[{{Vreeswijk} {et~al.}(2007){Vreeswijk}, {Ledoux}, {Smette}, {Ellison},
  {Jaunsen}, {Andersen}, {Fruchter}, {Fynbo}, {Hjorth}, {Kaufer}, {M{\o}ller},
  {Petitjean}, {Savaglio}, \& {Wijers}}]{2007A&A...468...83V}
{Vreeswijk}, P.~M., {Ledoux}, C., {Smette}, A., {et~al.} 2007, \aap, 468, 83

\bibitem[{{Wang} \& {Dai}(2014)}]{2014ApJS..213...15W}
{Wang}, F.~Y. \& {Dai}, Z.~G. 2014, \apjs, 213, 15

\bibitem[{{Whitaker} {et~al.}(2012){Whitaker}, {van Dokkum}, {Brammer}, \&
  {Franx}}]{2012ApJ...754L..29W}
{Whitaker}, K.~E., {van Dokkum}, P.~G., {Brammer}, G., \& {Franx}, M. 2012,
  \apjl, 754, L29

\bibitem[{{Wiersema}(2011)}]{2011MNRAS.414.2793W}
{Wiersema}, K. 2011, \mnras, 414, 2793

\bibitem[{{Wiersema} {et~al.}(2012{\natexlab{a}}){Wiersema}, {Curran},
  {Kr{\"u}hler}, {Melandri}, {Rol}, {Starling}, {Tanvir}, {van der Horst},
  {Covino}, {Fynbo}, {Goldoni}, {Gorosabel}, {Hjorth}, {Klose}, {Mundell},
  {O'Brien}, {Palazzi}, {Wijers}, {D'Elia}, {Evans}, {Filgas}, {Gomboc},
  {Greiner}, {Guidorzi}, {Kaper}, {Kobayashi}, {Kouveliotou}, {Levan}, {Rossi},
  {Rowlinson}, {Steele}, {de Ugarte Postigo}, \&
  {Vergani}}]{2012MNRAS.426....2W}
{Wiersema}, K., {Curran}, P.~A., {Kr{\"u}hler}, T., {et~al.}
  2012{\natexlab{a}}, \mnras, 426, 2

\bibitem[{{Wiersema} {et~al.}(2012{\natexlab{b}}){Wiersema}, {Goldoni},
  {Fynbo}, {Tanvir}, {Malesani}, {de Ugarte Postigo}, {Vergani}, {Covino}, \&
  {Flores}}]{2012GCN..12991...1W}
{Wiersema}, K., {Goldoni}, P., {Fynbo}, J.~P.~U., {et~al.} 2012{\natexlab{b}},
  GCN, 12991

\bibitem[{{Wijers} {et~al.}(1998){Wijers}, {Bloom}, {Bagla}, \&
  {Natarajan}}]{1998MNRAS.294L..13W}
{Wijers}, R.~A.~M.~J., {Bloom}, J.~S., {Bagla}, J.~S., \& {Natarajan}, P. 1998,
  \mnras, 294, L13

\bibitem[{{Wolfe} {et~al.}(2005){Wolfe}, {Gawiser}, \&
  {Prochaska}}]{2005ARA&A..43..861W}
{Wolfe}, A.~M., {Gawiser}, E., \& {Prochaska}, J.~X. 2005, \araa, 43, 861

\bibitem[{{Woosley}(1993)}]{1993ApJ...405..273W}
{Woosley}, S.~E. 1993, \apj, 405, 273

\bibitem[{{Wuyts} {et~al.}(2011){Wuyts}, {F{\"o}rster Schreiber}, {van der
  Wel}, {Magnelli}, {Guo}, {Genzel}, {Lutz}, {Aussel}, {Barro}, {Berta},
  {Cava}, {Graci{\'a}-Carpio}, {Hathi}, {Huang}, {Kocevski}, {Koekemoer},
  {Lee}, {Le Floc'h}, {McGrath}, {Nordon}, {Popesso}, {Pozzi}, {Riguccini},
  {Rodighiero}, {Saintonge}, \& {Tacconi}}]{2011ApJ...742...96W}
{Wuyts}, S., {F{\"o}rster Schreiber}, N.~M., {van der Wel}, A., {et~al.} 2011,
  \apj, 742, 96

\bibitem[{{Xu} {et~al.}(2013){Xu}, {de Ugarte Postigo}, {Leloudas},
  {Kr{\"u}hler}, {Cano}, {Hjorth}, {Malesani}, {Fynbo}, {Th{\"o}ne},
  {S{\'a}nchez-Ram{\'{\i}}rez}, {Schulze}, {Jakobsson}, {Kaper}, {Sollerman},
  {Watson}, {Cabrera-Lavers}, {Cao}, {Covino}, {Flores}, {Geier}, {Gorosabel},
  {Hu}, {Milvang-Jensen}, {Sparre}, {Xin}, {Zhang}, {Zheng}, \&
  {Zou}}]{2013ApJ...776...98X}
{Xu}, D., {de Ugarte Postigo}, A., {Leloudas}, G., {et~al.} 2013, \apj, 776, 98

\bibitem[{{Yoon} {et~al.}(2006){Yoon}, {Langer}, \&
  {Norman}}]{2006A&A...460..199Y}
{Yoon}, S.-C., {Langer}, N., \& {Norman}, C. 2006, \aap, 460, 199

\bibitem[{{Zafar} \& {Watson}(2013)}]{2013A&A...560A..26Z}
{Zafar}, T. \& {Watson}, D. 2013, \aap, 560, A26

\bibitem[{{Zafar} {et~al.}(2011){Zafar}, {Watson}, {Fynbo}, {Malesani},
  {Jakobsson}, \& {de Ugarte Postigo}}]{2011A&A...532A.143Z}
{Zafar}, T., {Watson}, D., {Fynbo}, J.~P.~U., {et~al.} 2011, \aap, 532, A143

\bibitem[{{Zahid} {et~al.}(2014){Zahid}, {Kashino}, {Silverman}, {Kewley},
  {Daddi}, {Renzini}, {Rodighiero}, {Nagao}, {Arimoto}, {Sanders},
  {Kartaltepe}, {Lilly}, {Maier}, {Geller}, {Capak}, {Carollo}, {Chu},
  {Hasinger}, {Ilbert}, {Kajisawa}, {Koekemoer}, {Kovac{\#728}}, {Le
  F{\`e}vre}, {Masters}, {McCracken}, {Onodera}, {Scoville}, {Strazzullo},
  {Sugiyama}, {Taniguchi}, \& {The COSMOS Team}}]{2014ApJ...792...75Z}
{Zahid}, H.~J., {Kashino}, D., {Silverman}, J.~D., {et~al.} 2014, \apj, 792, 75

\bibitem[{{Zahid} {et~al.}(2011){Zahid}, {Kewley}, \&
  {Bresolin}}]{2011ApJ...730..137Z}
{Zahid}, H.~J., {Kewley}, L.~J., \& {Bresolin}, F. 2011, \apj, 730, 137

\bibitem[{{Zaritsky} {et~al.}(1994){Zaritsky}, {Kennicutt}, \&
  {Huchra}}]{1994ApJ...420...87Z}
{Zaritsky}, D., {Kennicutt}, Jr., R.~C., \& {Huchra}, J.~P. 1994, \apj, 420, 87

\end{thebibliography}

\begin{appendix}

\section{Notes on individual targets}
\label{app:newz}

\subsection{GRB 111129A}

The X-Shooter spectrum of the GRB~111129A afterglow \citep{2011GCN..12600...1R} was taken on 2011-Nov-30, approximately 8.3\,h after the initial \textit{Swift} trigger. Above a red continuum \citep{2011GCN..12605...1R} we detect a single emission line (significance of $5.4\sigma$) at a wavelength of 7756\,\AA. Interpreting this as \oii, the redshift of GRB~111129A is $z=1.0796$. If it were any of the other strong lines, we would expect to detect at least one other emission line in the spectrum, which we do not. At $z=1.0796$, however, there are indications for \oiii\, and \hb\,at the $2\sigma$ level in the NIR arm (\ha\, is in the $JH$-bandgap and thus not seen).

\subsection{GRB 120211A}

In deep Keck imaging taken on 2013-Feb-10, we detect an $I_{\rm{AB}}=25.0\pm0.2$\,mag host galaxy candidate within the XRT error circle \citep{2012GCN..12922...1G} of GRB~120211A \citep{2012GCN..12920...1S}. The spectral continuum of the putative host galaxy is detected also at high significance in the X-Shooter spectrum from 2013-Mar-20 in a wavelength range between 9000\,\AA\,and 4100\,\AA. Below 4100\,\AA\, the flux sharply drops to zero. This apparent break is consistent with the onset of Ly$\alpha$ absorption at $z\sim2.4$. Consistent with this redshift, there are minor indications at the $2\sigma$-significance level for \oiii$\lambda$5007 emission at $z=2.346$ located in a region of high sky emission.

\subsection{GRB 120224A}

In the X-Shooter spectrum  \citep{2012GCN..12991...1W} of an afterglow candidate of GRB~120224A \citep{2012GCN..12980...1S} taken on 2012-Feb-25, we detect a red continuum in the VIS and NIR arms, without obvious emission lines. Later GROND imaging reveals this to be a very red ($r-K_{\rm{AB}}=3.7$\,mag), and constant source, a plausible GRB host candidate. The spectral continuum drops below the noise floor blueward of 7000\,\AA. We rule out that the break is due to the Ly$\alpha$ at $z=4.8$ because the photometric colors, in particular the $g$-band detection ($g'=25.2\pm0.2$) disfavors such a high redshift. A photometric redshift analysis gives 90\% confidence interval of $0.9 < z_{\rm{phot}} < 1.3$. Indeed, there is a small redshift window at $z\sim1.1$ in which all strong emission lines (\oii, \oiii\, and \ha) would be hidden in telluric bands. This would provide an explanation for the non-detection of emission lines in our spectrum. The drop to zero flux in the spectrum would in this case be caused by intrinsic redness of the galaxy and the Balmer break. Significantly higher or lower redshifts are unlikely: The stellar mass of the galaxy derived from the photometry exceeds $10^{11}$\Msun\,already at $z > 1.5$, making the non-detection of emission lines hard to explain. Similarly, at $z \lesssim 0.6$, we would have expected to detect at least \ha\,in the spectrum even for a dusty galaxy.

\subsection{GRB 120805A}

The galaxy coincident with the afterglow \citep{2012GCN..13651...1G, 2012GCN..13591...1G} of GRB~120805A \citep{2012GCN..13588...1T} is identified through NOT \citep{2012GCN..13639...1M} and late GROND and VLT/HAWK-I observations at a brightness of $r=24.1\pm0.1$\,mag. It was observed with X-Shooter spectroscopically on 2012-Aug-15. The spectral continuum is detected at high significance in the UVB/VIS arm between 5000\,\AA\, and 9500\,\AA. No signal is recorded blueward of 5000\,\AA\, or in the NIR arms. The shape of the spectral continuum and the GROND photometry are consistent with a redshift of $z\sim3.1$. The Ly$\alpha$ break explains the drop in flux at around 5000\,\AA, and the absence of strong emission lines in the spectrum is caused by telluric absorption bands. At this redshift, the host is luminous: $M_{\rm1700\AA\,}\sim-21.9$\,mag corresponds to $\sim2.5\cdot M^{*}$ at $z\sim3$ \citep[e.g.,][]{2009ApJ...692..778R}.

\subsection{GRB 121209A}

The $R_{\rm{AB}}=24.1\pm0.1$\,mag host galaxy of GRB~121209A \citep{2012GCN..14045...1M} was identified already very early \citep{2012GCN..14049...1K, 2012GCN..14056...1P}, and observed with X-Shooter on 2012-Dec-13. The galaxy-continuum is detected significantly with X-Shooter between 4900\,\AA\, and 9500\,\AA\, but not redward and with Keck LRIS between the atmospheric cutoff and 10300\,\AA\,\citep{2012GCN..14056...1P}. Using all available late photometry from VLT/FORS2 and VLT/HAWK-I, the best-fit photometric redshift for this galaxy is $z_{\rm{phot}} = 2.1\pm0.3$, in agreement with earlier limits \citep{2012GCN..14056...1P}. Despite the bright continuum, no emission lines are seen. This is also consistent with $z_{\rm{phot}}\sim1.9$ because all strong emission lines (\oiii, \oii, and \ha) would be located in telluric absorption bands at this redshift. 

\subsection{GRB 140114A}

At the position of the optical afterglow \citep{2014GCN..15732...1B, 2014GCN..15743...1C} of GRB~140114A \citep{2014GCN..15728...1T}, we detect an $R_{\rm{AB}}=24.4\pm0.2\,$mag galaxy with NOT/ALFOSC that we interpret as the GRB's host galaxy. Our X-Shooter spectroscopy from 2014-Mar-28 reveals no emission lines, but the galaxy continuum between 4850\,\AA\, and 9800\,\AA. A sharp drop in flux blueward of this wavelength range is best described by the Ly$\alpha$ break at $z=3.0$. The relatively high redshift is also consistent with the absence of emission lines because the strong forbidden lines of \oii\, and \oiii\,are in telluric absorption bands and \ha\, outside of the wavelength response of X-Shooter.

\section{Details on metallicity measurements and notable individual events}
\label{sec:metmeas}

To derive oxygen abundances via the strong-line diagnostic ratios from \citet{2006A&A...459...85N} and \citet{2008A&A...488..463M}, we take into account the dust attenuation, its uncertainty (\nii\, versus \ha\, or \neiii\,versus \oii\, have a negligible dependence on $A_V$) and the systematic scatter in the respective relation. We use \nii/\ha\, if available to discriminate between the upper and lower $R_{23}$ branch. Table~\ref{tab:physprop} provides the calculated metallicities and errors (as the range between 16\% and 84\% of cumulative probability distribution). These oxygen abundances need to be taken with a grain of salt, in particular at higher redshifts, as indicated in the main text. 

In some cases, in particular at the highest redshifts ($z>2.5$), dust corrections to the line fluxes are not available because of lacking Balmer lines. In those cases, however, metallicities are derived using line ratios that are only marginally sensitive to dust corrections such as \neiii\,over \oii. The missing constraints on $E_{B-V}$ are thus not a strong concern for the metallicity measurement. The systematic and statistical error of the \neiii\,over \oii\,ratio is large, and it does not provide strong additional constraints once \nii\,-based line diagnostics are available. It is, however, the only accessible method at $z>3$ so we use it with the caveat that it is the least tested and used strong-line diagnostic ratio. It did provide a consistent metallicity in those cases where we could test it with other methods.

\begin{figure*}
\begin{subfigure}{.33\textwidth}
  \includegraphics[width=0.999\linewidth]{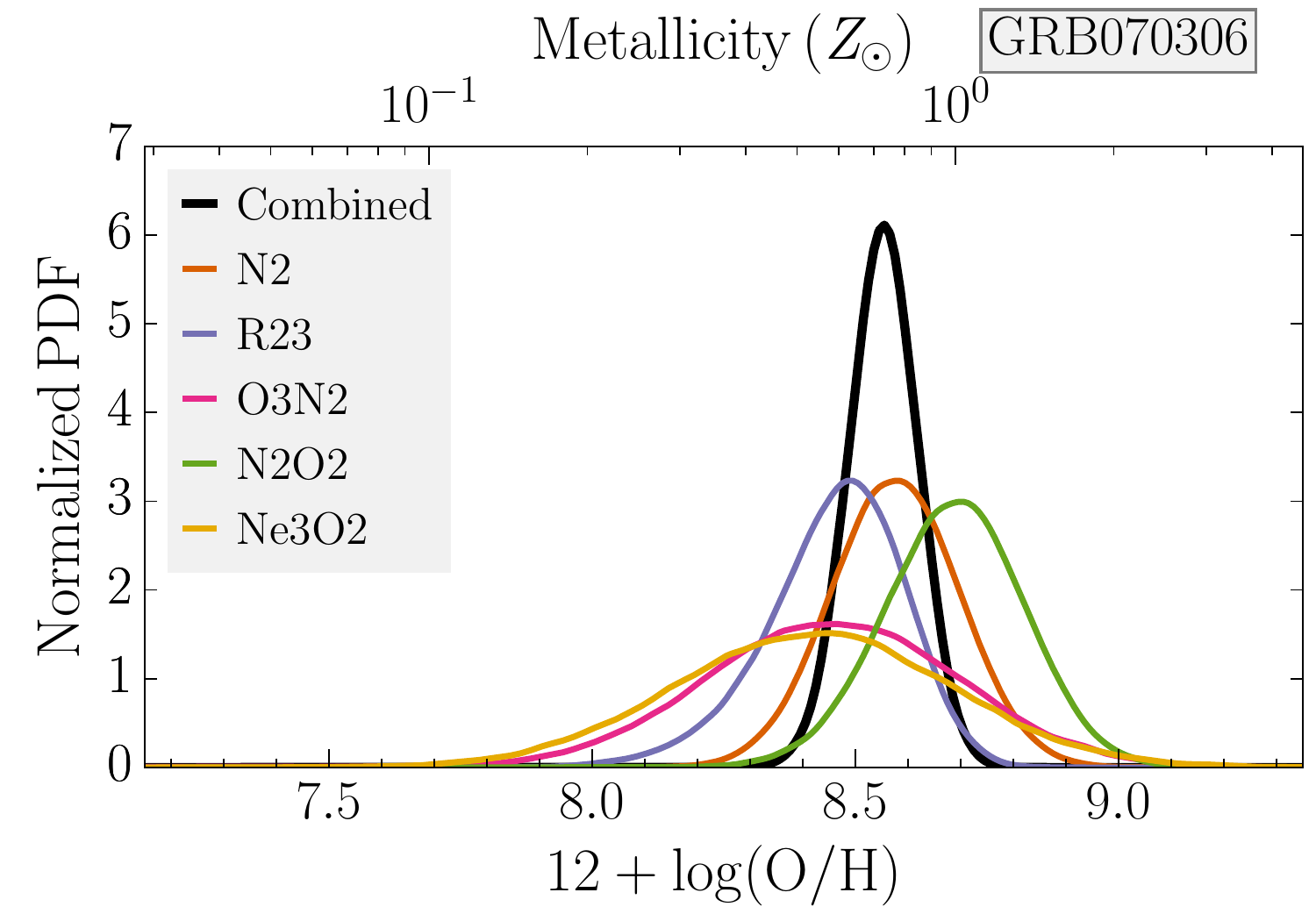}
\end{subfigure}
\begin{subfigure}{.33\textwidth}
  \includegraphics[width=0.999\linewidth]{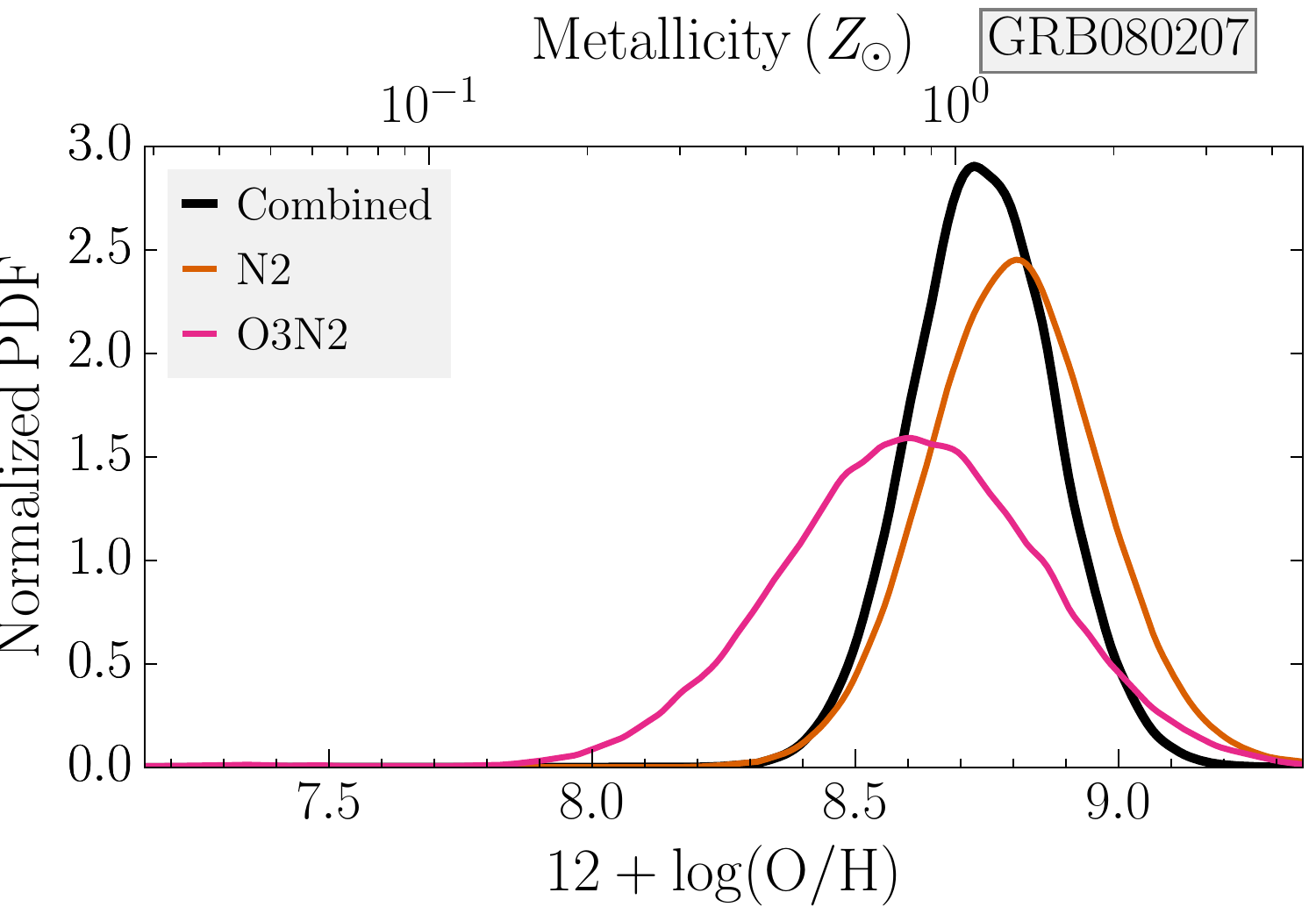}
\end{subfigure}
\begin{subfigure}{.33\textwidth}
  \includegraphics[width=0.999\linewidth]{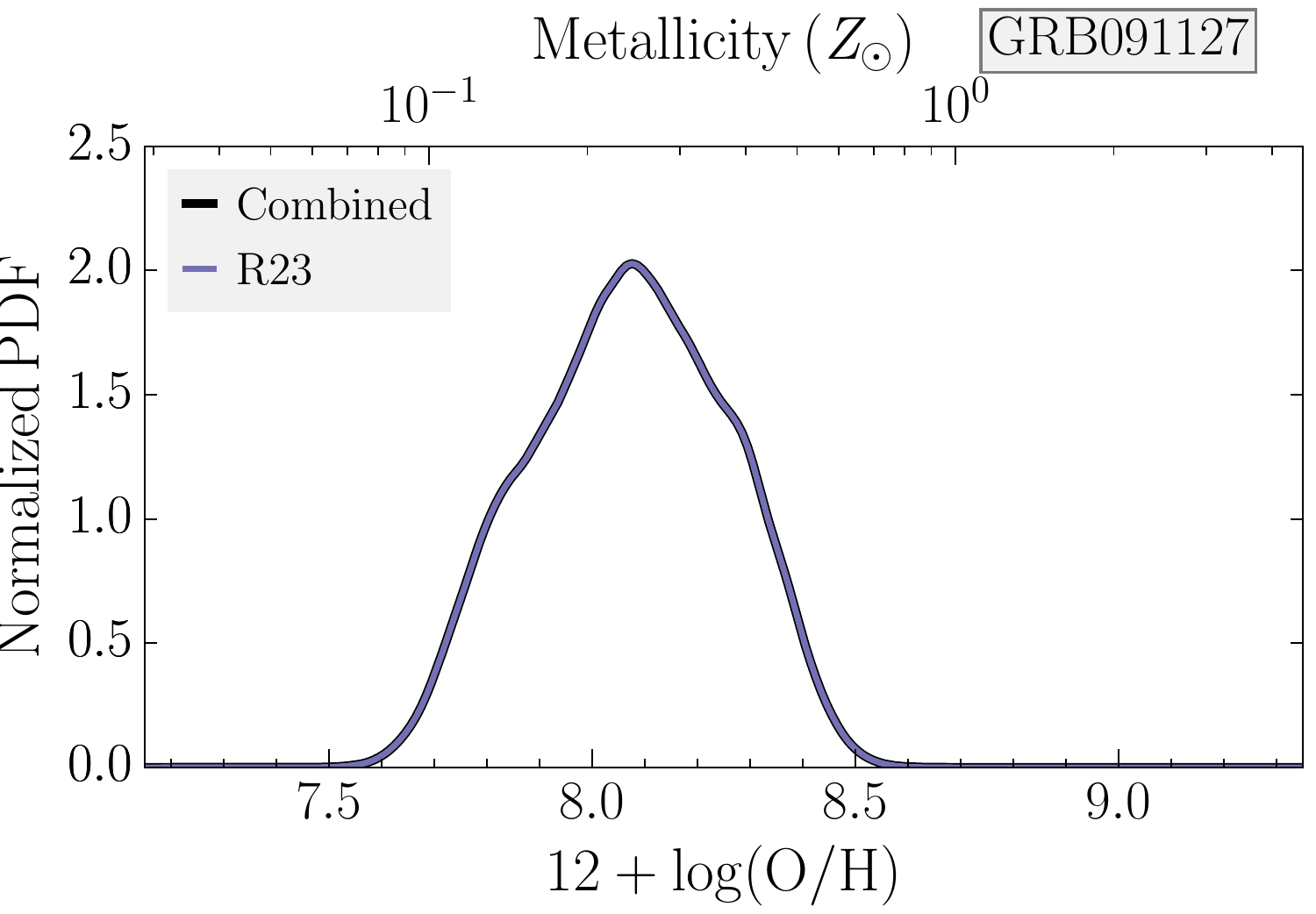}
\end{subfigure}
\begin{subfigure}{.33\textwidth}
  \includegraphics[width=0.999\linewidth]{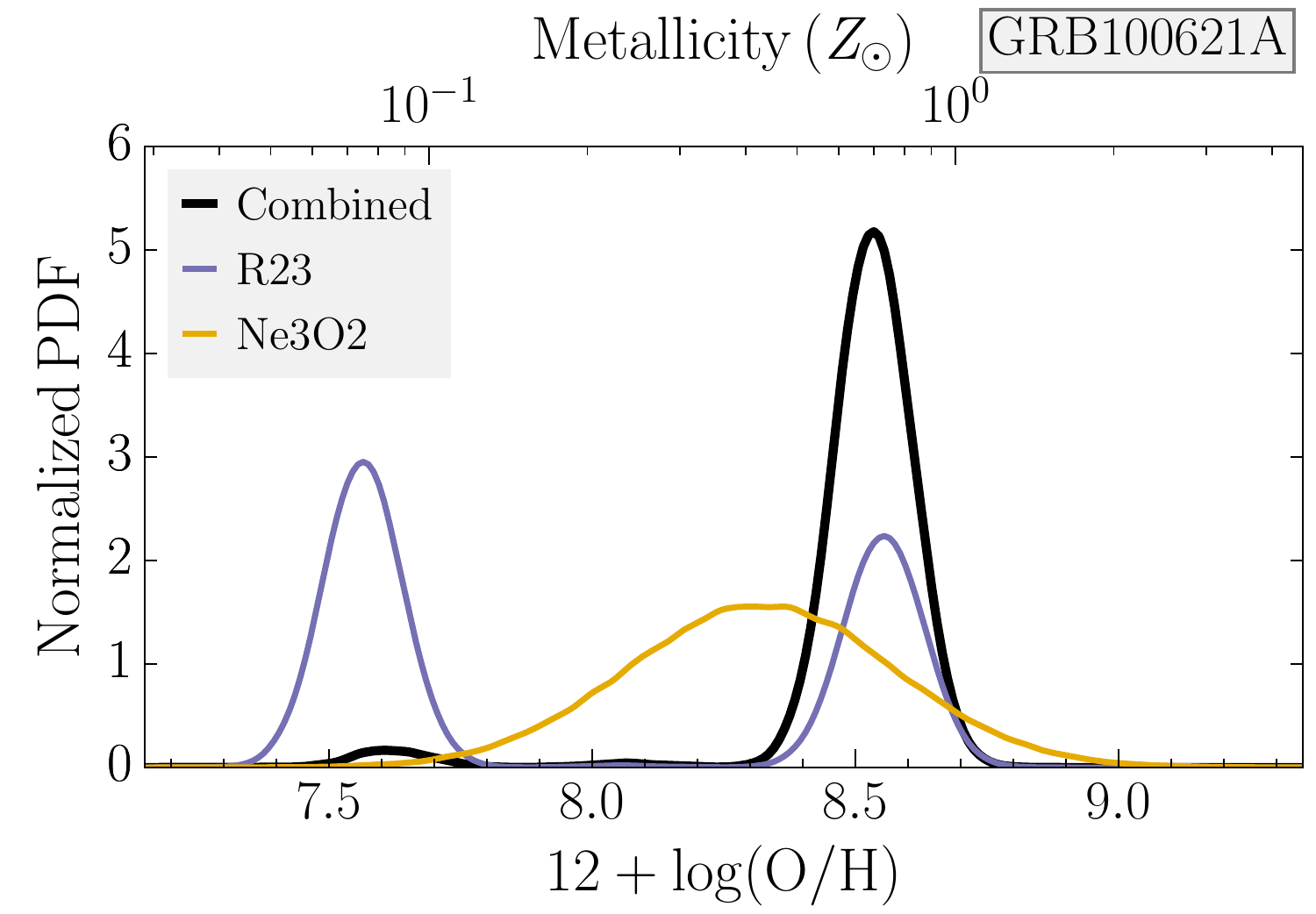}
\end{subfigure}
\begin{subfigure}{.33\textwidth}
  \includegraphics[width=0.999\linewidth]{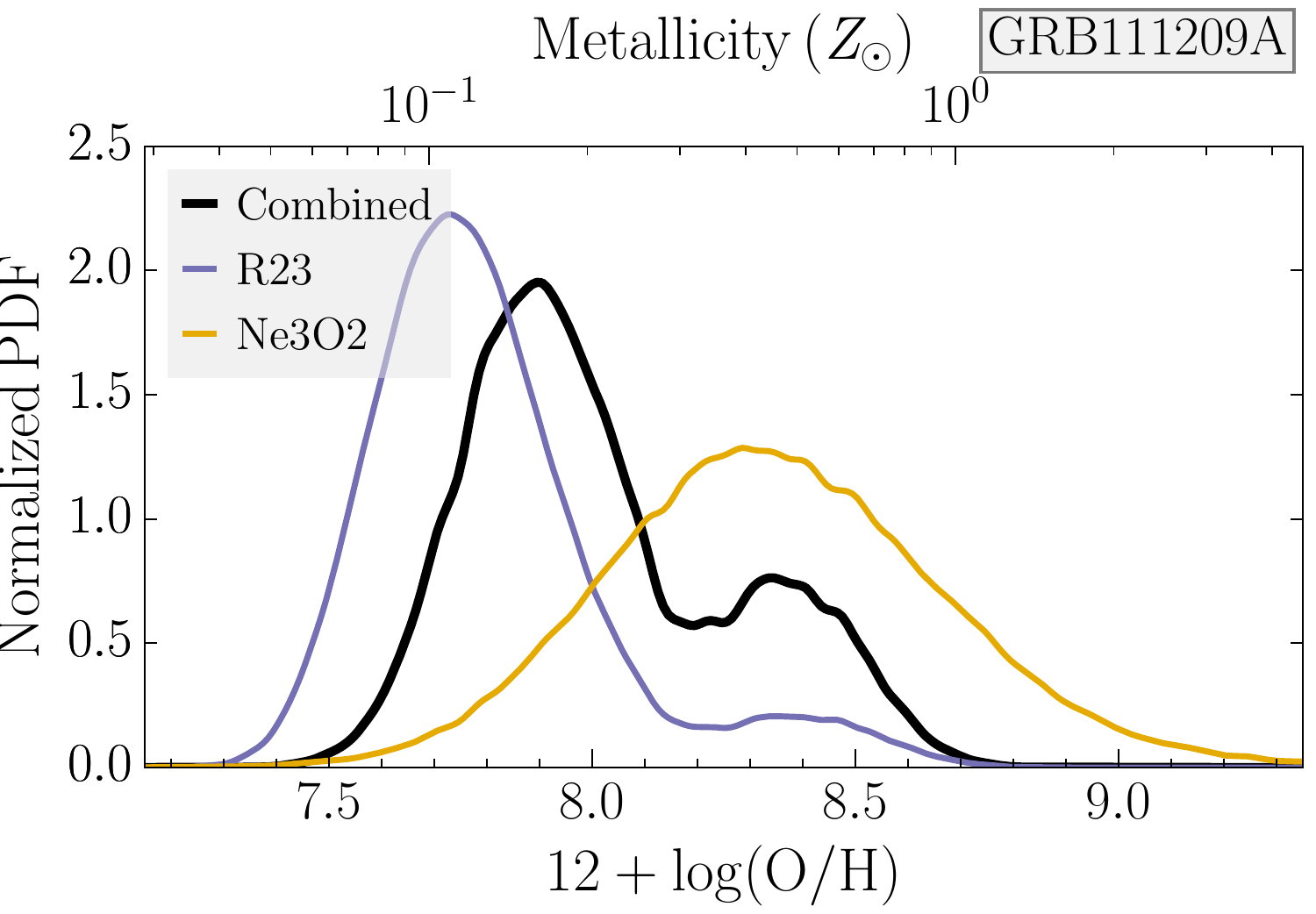}
\end{subfigure}
\begin{subfigure}{.33\textwidth}
  \includegraphics[width=0.999\linewidth]{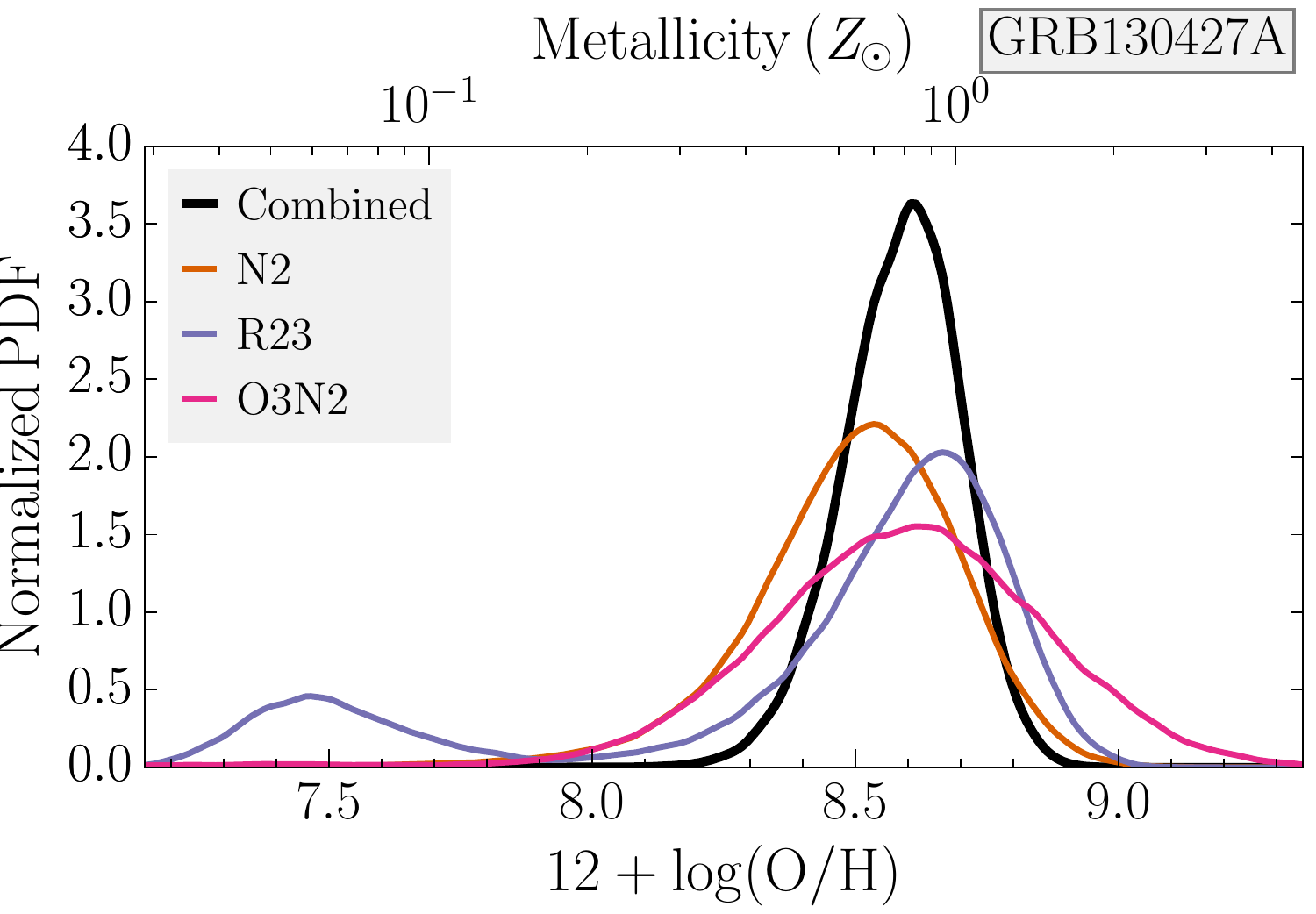}
\end{subfigure}
\caption{Example of metallicity measurements using different strong line diagnostics \citep{2006A&A...459...85N, 2008A&A...488..463M}. Every colored line represents the normalized probability distribution of one strong-line diagnostic. The black line is the probability distribution of $\oh$ simultaneously minimized against all available measured line ratios.}
\label{fig:zexamples}
\end{figure*}

Figure~\ref{fig:zexamples} shows several representative examples of metallicity determinations for six galaxies at different redshifts. We specifically pick GRB hosts or afterglows that were extensively discussed in the recent literature: GRB~070306 \citep{2008ApJ...681..453J, 2011A&A...534A.108K}, GRB~080207 \citep{2011ApJ...736L..36H, 2012MNRAS.421...25S}, GRB~091127 \citep{2011A&A...535A.127V}, GRB~100621A \citep{2013A&A...560A..70G}, GRB~111209A \citep{2013ApJ...766...30G, 2014ApJ...781...13L, 2015Greinersubm, 2015Kannsubm} and GRB~130427A \citep[e.g.,][]{2013ApJ...776...98X, 2014ApJ...781...37P}. We will discuss them briefly in the following section as they are illustrative for the metallicity measurement procedure and interesting individually.

\subsection{GRB~070306} GRB~070306 ($z=1.497$) is a well-studied dusty GRB \citep{2008ApJ...681..453J}, with a host galaxy that is detected throughout the electromagnetic spectrum \citep{2014A&A...565A.112H, 2014arXiv1408.5076S, 2015ApJ...801..102P}. The X-Shooter spectrum shows numerous emission lines, which allow the Balmer decrement to be reasonably well measured ($E_{B-V} = 0.43_{-0.07}^{+0.08}$\,mag). The \nii/\ha\,line-flux ratio uniquely puts the $R_{23}$ solution onto the upper branch. The individual emission line diagnostics have overlapping probability distributions, and a simultaneous fit to all five line ratios results in a metallicity of $\oh=8.54\pm0.09$, which is $0.8\pm0.2$ times the solar value (Fig.~\ref{fig:zexamples}).

\subsection{GRB 080207} GRB~080207 ($z=2.086$) is also a well-discussed dark GRB \citep{2011ApJ...736L..36H, 2012MNRAS.421...25S}. Its host is detected by various facilities from the optical to radio \citep{2012A&A...545A..77R, 2013ApJ...778..128P}, and has a very high stellar mass ($M_{\star}\sim10^{11}$\,\Msun), extremely red optical/NIR colors and is vigorously forming stars ($SFR_{\rm{H\alpha}}\sim 90\,$\Msunyr). Because of the large $A_V\sim2$, \oii\, is not detected and the applied strong-line diagnostics are thus based exclusively on \nii. We measure an oxygen abundance around solar, $\oh = 8.74\pm0.15$. This is the GRB host with the highest metallicity at $z>2$ in the sample. Its metallicity, however, is averaged over the multiple knots that are detected at the spatial resolution of HST \citep{2012MNRAS.421...25S}, which in theory could host a range of metallicities. 

\subsection{GRB~091127} GRB~091127 ($z=0.490$) is a typical low-redshift GRB with a bright afterglow and faint host \citep{2011A&A...535A.127V,2011A&A...526A.113F}. It is an example of a galaxy having $R_{23}$ in the overlap region between the two metallicity solutions. Even though the forbidden oxygen and Balmer lines are detected at good S/N, the lack of detection of either \neiii\,or \nii\, limits the number of useful line-diagnostics to $R_{23}$. Given that the line ratios put the host of GRB~091127 at around the turn-over point of the two branches, the probability distribution is not double valued, but rather broad because $\oh$ is not very sensitive to variations in $R_{23}$ in this region. We derive $\oh=8.07_{-0.20}^{+0.18}$ for the host of GRB~091127.

\subsection{GRB~100621A} GRB~100621A ($z=0.543$) is a low-$z$ GRB with a faint and dust-reddened afterglow \citep{2013A&A...560A..70G}. The host spectrum has, in addition to well-detected \oii, \oiii, \ha, and \hb\, emission lines, a significant detection of \neiii\,(but not \nii). The degeneracy of the double-peaked $R_{23}$ probability distribution in metallicity is clearly evident in Fig.~\ref{fig:zexamples}. It is, however, broken through \neiii. Even though the individual \neiii-based metallicity is not well constrained, in combination with $R_{23}$ we can derive $\oh$ accurately to $\oh=8.52\pm0.10$ with only a very small remaining probability for the lower branch solution of the $R_{23}$ diagnostic.

\subsection{GRB~111209A} GRB~111209A ($z=0.677$) is one of the longest GRBs ever detected. The extreme temporal properties were used to postulate a new class of GRBs \citep{2013ApJ...766...30G, 2014ApJ...781...13L}, a result questioned by \citet{2013ApJ...778...54V}. The ultra-long duration is possibly indicative of a different progenitor channel \citep{2013ApJ...778...67N}, and the associated luminous supernova is well-explained in a magnetar scenario \citep{2015Kannsubm, 2015Greinersubm}. Based on the detection of \oii, \oiii\, and \neiii, in addition to the Balmer lines, we measure $\oh=7.95_{-0.17}^{+0.30}$ or between 10\% and 40\% of the solar value in the adopted metallicity scale. This puts the galaxy at the low end of metallicities compared to many GRBs of more normal duration at a similar redshift (Fig.~\ref{fig:zhist}). Together with the low reddening, narrow line width, compact morphology and low luminosity \citep{2014ApJ...781...13L} it shares many similarities with Blue Compact Dwarfs \citep[e.g.,][and references therein]{2003ApJS..147...29G} or hosts of super-luminous supernovae \citep[e.g.,][]{2014arXiv1409.8331L}. The galaxy hosting GRB~111209A is, however, not extremely metal-poor, and its spectroscopic properties do not provide a stark contrast to long GRBs with a more common duration (see, e.g., GRB~091127 above), a conclusion similar to the one of \citet{2014ApJ...781...13L}.

\subsection{GRB~130427A} GRB~130427A ($z=0.340$) is one of the brightest GRBs of all time due to its low redshift and is extremely well studied from the radio to GeV energies \citep[e.g.,][]{2014ApJ...781...37P, 2014Sci...343...42A, 2014MNRAS.444.3151V}. It is a very rare example of a low-redshift GRB with an energy budget comparable to common high-redshift GRBs and therefore elucidates cosmological GRBs and their connection to SNe through a local analogue \citep{2013ApJ...776...98X, 2014ApJ...792..115L, 2014A&A...567A..29M}. The host characteristics are relatively typical for the given redshift: With a $SFR_{\rm{H\alpha}}= 0.3^{+0.2}_{-0.1}\,$\Msunyr and metallicity of $\oh=8.57_{-0.13}^{+0.12}$ it is well within the distribution of host properties for low-redshift GRBs.

\end{appendix}
\end{document}